\documentclass[a4paper, traditabstract, longauth]{aa} 

\usepackage{amsmath}
\usepackage{graphicx}
\usepackage{color}
\usepackage[breaklinks, colorlinks, citecolor=blue]{hyperref}
\usepackage{natbib}
\usepackage{longtable,lscape}
\usepackage{txfonts}
\usepackage{amsfonts}
\usepackage{amssymb}
\usepackage{newlfont}
\usepackage{textcomp}
\usepackage{times}
\usepackage{units}
\usepackage{mathbbol}
\usepackage{amsbsy}
\usepackage{multirow}
\usepackage{xcolor}

\def\setsymbol#1#2{\expandafter\def\csname #1\endcsname{#2}}
\def\getsymbol#1{\csname #1\endcsname}

\def\Planck{\textit{Planck}}

\def\all2013resultspapers{\nocite{planck2013-p01, planck2013-p02, planck2013-p02a, planck2013-p02d, planck2013-p02b, planck2013-p03, planck2013-p03c, planck2013-p03f, planck2013-p03d, planck2013-p03e, planck2013-p01a, planck2013-p06, planck2013-p03a, planck2013-pip88, planck2013-p08, planck2013-p11, planck2013-p12, planck2013-p13, planck2013-p14, planck2013-p15, planck2013-p05b, planck2013-p17, planck2013-p09, planck2013-p09a, planck2013-p20, planck2013-p19, planck2013-pipaberration, planck2013-p05, planck2013-p05a, planck2013-pip56, planck2013-p06b}}

\newbox\tablebox    \newdimen\tablewidth
\def\leaderfil{\leaders\hbox to 5pt{\hss.\hss}\hfil}
\def\endPlancktable{\tablewidth=\columnwidth 
    $$\hss\copy\tablebox\hss$$
    \vskip-\lastskip\vskip -2pt}

\def\tablenote#1 #2\par{\begingroup \parindent=0.8em
    \abovedisplayshortskip=0pt\belowdisplayshortskip=0pt
    \noindent
    $$\hss\vbox{\hsize\tablewidth \hangindent=\parindent \hangafter=1 \noindent
    \hbox to \parindent{$^#1$\hss}\strut#2\strut\par}\hss$$
    \endgroup}
\def\doubleline{\vskip 3pt\hrule \vskip 1.5pt \hrule \vskip 5pt}

\def\L2{\ifmmode L_2\else $L_2$\fi}

\def\DeltaT{\ifmmode \Delta T\else $\Delta T$\fi}
\def\deltat{\ifmmode \Delta t\else $\Delta t$\fi}
\def\fknee{\ifmmode f_{\rm knee}\else $f_{\rm knee}$\fi}
\def\Fmax{\ifmmode F_{\rm max}\else $F_{\rm max}$\fi}
\def\solar{\ifmmode{\rm M}_{\mathord\odot}\else${\rm M}_{\mathord\odot}$\fi}
\def\Msolar{\ifmmode{\rm M}_{\mathord\odot}\else${\rm M}_{\mathord\odot}$\fi}
\def\Lsolar{\ifmmode{\rm L}_{\mathord\odot}\else${\rm L}_{\mathord\odot}$\fi}

\def\inv{\ifmmode^{-1}\else$^{-1}$\fi}
\def\mo{\ifmmode^{-1}\else$^{-1}$\fi}
\def\sup#1{\ifmmode ^{\rm #1}\else $^{\rm #1}$\fi}
\def\expo#1{\ifmmode \times 10^{#1}\else $\times 10^{#1}$\fi}
\def\,{\thinspace}
\def\lsim{\mathrel{\raise .4ex\hbox{\rlap{$<$}\lower 1.2ex\hbox{$\sim$}}}}
\def\gsim{\mathrel{\raise .4ex\hbox{\rlap{$>$}\lower 1.2ex\hbox{$\sim$}}}}

\def\simprop{\mathrel{\raise .4ex\hbox{\rlap{$\propto$}\lower 1.2ex\hbox{$\sim$}}}}
\def\deg{\ifmmode^\circ\else$^\circ$\fi}
\def\pdeg{\ifmmode $\setbox0=\hbox{$^{\circ}$}\rlap{\hskip.11\wd0 .}$^{\circ}
          \else \setbox0=\hbox{$^{\circ}$}\rlap{\hskip.11\wd0 .}$^{\circ}$\fi}
\def\arcs{\ifmmode {^{\scriptstyle\prime\prime}}
          \else $^{\scriptstyle\prime\prime}$\fi}
\def\arcm{\ifmmode {^{\scriptstyle\prime}}
          \else $^{\scriptstyle\prime}$\fi}
\newdimen\sa  \newdimen\sb
\def\parcs{\sa=.07em \sb=.03em
     \ifmmode \hbox{\rlap{.}}^{\scriptstyle\prime\kern -\sb\prime}\hbox{\kern -\sa}
     \else \rlap{.}$^{\scriptstyle\prime\kern -\sb\prime}$\kern -\sa\fi}
\def\parcm{\sa=.08em \sb=.03em
     \ifmmode \hbox{\rlap{.}\kern\sa}^{\scriptstyle\prime}\hbox{\kern-\sb}
     \else \rlap{.}\kern\sa$^{\scriptstyle\prime}$\kern-\sb\fi}
\def\ra[#1 #2 #3.#4]{#1\sup{h}#2\sup{m}#3\sup{s}\llap.#4}
\def\dec[#1 #2 #3.#4]{#1\deg#2\arcm#3\arcs\llap.#4}
\def\deco[#1 #2 #3]{#1\deg#2\arcm#3\arcs}
\def\rra[#1 #2]{#1\sup{h}#2\sup{m}}

\def\dots{\relax\ifmmode \ldots\else $\ldots$\fi}
\def\WHzsr{\ifmmode $W\,Hz\mo\,sr\mo$\else W\,Hz\mo\,sr\mo\fi}
\def\mHz{\ifmmode $\,mHz$\else \,mHz\fi}
\def\GHz{\ifmmode $\,GHz$\else \,GHz\fi}
\def\mKs{\ifmmode $\,mK\,s$^{1/2}\else \,mK\,s$^{1/2}$\fi}
\def\muKs{\ifmmode \,\mu$K\,s$^{1/2}\else \,$\mu$K\,s$^{1/2}$\fi}
\def\muKRJs{\ifmmode \,\mu$K$_{\rm RJ}$\,s$^{1/2}\else \,$\mu$K$_{\rm RJ}$\,s$^{1/2}$\fi}
\def\muKHz{\ifmmode \,\mu$K\,Hz$^{-1/2}\else \,$\mu$K\,Hz$^{-1/2}$\fi}
\def\MJysr{\ifmmode \,$MJy\,sr\mo$\else \,MJy\,sr\mo\fi}
\def\MJysrmK{\ifmmode \,$MJy\,sr\mo$\,mK$_{\rm CMB}\mo\else \,MJy\,sr\mo\,mK$_{\rm CMB}\mo$\fi}
\def\microns{\ifmmode \,\mu$m$\else \,$\mu$m\fi}

\def\muK{\ifmmode \,\mu$K$\else \,$\mu$\hbox{K}\fi}
\def\microK{\ifmmode \,\mu$K$\else \,$\mu$\hbox{K}\fi}
\def\muW{\ifmmode \,\mu$W$\else \,$\mu$\hbox{W}\fi}
\def\kms{\ifmmode $\,km\,s$^{-1}\else \,km\,s$^{-1}$\fi}
\def\kmsMpc{\ifmmode $\,\kms\,Mpc\mo$\else \,\kms\,Mpc\mo\fi}
\setsymbol{LFI:center:frequency:70GHz:units}{70.3\,GHz}
\setsymbol{LFI:center:frequency:44GHz:units}{44.1\,GHz}
\setsymbol{LFI:center:frequency:30GHz:units}{28.5\,GHz}

\setsymbol{LFI:center:frequency:70GHz}{70.3}
\setsymbol{LFI:center:frequency:44GHz}{44.1}
\setsymbol{LFI:center:frequency:30GHz}{28.5}

\setsymbol{LFI:center:frequency:LFI18:Rad:M:units}{71.7\GHz}
\setsymbol{LFI:center:frequency:LFI19:Rad:M:units}{67.5\GHz}
\setsymbol{LFI:center:frequency:LFI20:Rad:M:units}{69.2\GHz}
\setsymbol{LFI:center:frequency:LFI21:Rad:M:units}{70.4\GHz}
\setsymbol{LFI:center:frequency:LFI22:Rad:M:units}{71.5\GHz}
\setsymbol{LFI:center:frequency:LFI23:Rad:M:units}{70.8\GHz}
\setsymbol{LFI:center:frequency:LFI24:Rad:M:units}{44.4\GHz}
\setsymbol{LFI:center:frequency:LFI25:Rad:M:units}{44.0\GHz}
\setsymbol{LFI:center:frequency:LFI26:Rad:M:units}{43.9\GHz}
\setsymbol{LFI:center:frequency:LFI27:Rad:M:units}{28.3\GHz}
\setsymbol{LFI:center:frequency:LFI28:Rad:M:units}{28.8\GHz}
\setsymbol{LFI:center:frequency:LFI18:Rad:S:units}{70.1\GHz}
\setsymbol{LFI:center:frequency:LFI19:Rad:S:units}{69.6\GHz}
\setsymbol{LFI:center:frequency:LFI20:Rad:S:units}{69.5\GHz}
\setsymbol{LFI:center:frequency:LFI21:Rad:S:units}{69.5\GHz}
\setsymbol{LFI:center:frequency:LFI22:Rad:S:units}{72.8\GHz}
\setsymbol{LFI:center:frequency:LFI23:Rad:S:units}{71.3\GHz}
\setsymbol{LFI:center:frequency:LFI24:Rad:S:units}{44.1\GHz}
\setsymbol{LFI:center:frequency:LFI25:Rad:S:units}{44.1\GHz}
\setsymbol{LFI:center:frequency:LFI26:Rad:S:units}{44.1\GHz}
\setsymbol{LFI:center:frequency:LFI27:Rad:S:units}{28.5\GHz}
\setsymbol{LFI:center:frequency:LFI28:Rad:S:units}{28.2\GHz}

\setsymbol{LFI:center:frequency:LFI18:Rad:M}{71.7}
\setsymbol{LFI:center:frequency:LFI19:Rad:M}{67.5}
\setsymbol{LFI:center:frequency:LFI20:Rad:M}{69.2}
\setsymbol{LFI:center:frequency:LFI21:Rad:M}{70.4}
\setsymbol{LFI:center:frequency:LFI22:Rad:M}{71.5}
\setsymbol{LFI:center:frequency:LFI23:Rad:M}{70.8}
\setsymbol{LFI:center:frequency:LFI24:Rad:M}{44.4}
\setsymbol{LFI:center:frequency:LFI25:Rad:M}{44.0}
\setsymbol{LFI:center:frequency:LFI26:Rad:M}{43.9}
\setsymbol{LFI:center:frequency:LFI27:Rad:M}{28.3}
\setsymbol{LFI:center:frequency:LFI28:Rad:M}{28.8}
\setsymbol{LFI:center:frequency:LFI18:Rad:S}{70.1}
\setsymbol{LFI:center:frequency:LFI19:Rad:S}{69.6}
\setsymbol{LFI:center:frequency:LFI20:Rad:S}{69.5}
\setsymbol{LFI:center:frequency:LFI21:Rad:S}{69.5}
\setsymbol{LFI:center:frequency:LFI22:Rad:S}{72.8}
\setsymbol{LFI:center:frequency:LFI23:Rad:S}{71.3}
\setsymbol{LFI:center:frequency:LFI24:Rad:S}{44.1}
\setsymbol{LFI:center:frequency:LFI25:Rad:S}{44.1}
\setsymbol{LFI:center:frequency:LFI26:Rad:S}{44.1}
\setsymbol{LFI:center:frequency:LFI27:Rad:S}{28.5}
\setsymbol{LFI:center:frequency:LFI28:Rad:S}{28.2}

\setsymbol{LFI:white:noise:sensitivity:70GHz:units}{134.7\muKs}
\setsymbol{LFI:white:noise:sensitivity:44GHz:units}{164.7\muKs}
\setsymbol{LFI:white:noise:sensitivity:30GHz:units}{143.4\muKs}

\setsymbol{LFI:white:noise:sensitivity:70GHz}{134.7}
\setsymbol{LFI:white:noise:sensitivity:44GHz}{164.7}
\setsymbol{LFI:white:noise:sensitivity:30GHz}{143.4}

\setsymbol{LFI:white:noise:sensitivity:LFI18:Rad:M:units}{512.0\muKs}
\setsymbol{LFI:white:noise:sensitivity:LFI19:Rad:M:units}{581.4\muKs}
\setsymbol{LFI:white:noise:sensitivity:LFI20:Rad:M:units}{590.8\muKs}
\setsymbol{LFI:white:noise:sensitivity:LFI21:Rad:M:units}{455.2\muKs}
\setsymbol{LFI:white:noise:sensitivity:LFI22:Rad:M:units}{492.0\muKs}
\setsymbol{LFI:white:noise:sensitivity:LFI23:Rad:M:units}{507.7\muKs}
\setsymbol{LFI:white:noise:sensitivity:LFI24:Rad:M:units}{462.2\muKs}
\setsymbol{LFI:white:noise:sensitivity:LFI25:Rad:M:units}{413.6\muKs}
\setsymbol{LFI:white:noise:sensitivity:LFI26:Rad:M:units}{478.6\muKs}
\setsymbol{LFI:white:noise:sensitivity:LFI27:Rad:M:units}{277.7\muKs}
\setsymbol{LFI:white:noise:sensitivity:LFI28:Rad:M:units}{312.3\muKs}
\setsymbol{LFI:white:noise:sensitivity:LFI18:Rad:S:units}{465.7\muKs}
\setsymbol{LFI:white:noise:sensitivity:LFI19:Rad:S:units}{555.6\muKs}
\setsymbol{LFI:white:noise:sensitivity:LFI20:Rad:S:units}{623.2\muKs}
\setsymbol{LFI:white:noise:sensitivity:LFI21:Rad:S:units}{564.1\muKs}
\setsymbol{LFI:white:noise:sensitivity:LFI22:Rad:S:units}{534.4\muKs}
\setsymbol{LFI:white:noise:sensitivity:LFI23:Rad:S:units}{542.4\muKs}
\setsymbol{LFI:white:noise:sensitivity:LFI24:Rad:S:units}{399.2\muKs}
\setsymbol{LFI:white:noise:sensitivity:LFI25:Rad:S:units}{392.6\muKs}
\setsymbol{LFI:white:noise:sensitivity:LFI26:Rad:S:units}{418.6\muKs}
\setsymbol{LFI:white:noise:sensitivity:LFI27:Rad:S:units}{302.9\muKs}
\setsymbol{LFI:white:noise:sensitivity:LFI28:Rad:S:units}{285.3\muKs}

\setsymbol{LFI:white:noise:sensitivity:LFI18:Rad:M}{512.0}
\setsymbol{LFI:white:noise:sensitivity:LFI19:Rad:M}{581.4}
\setsymbol{LFI:white:noise:sensitivity:LFI20:Rad:M}{590.8}
\setsymbol{LFI:white:noise:sensitivity:LFI21:Rad:M}{455.2}
\setsymbol{LFI:white:noise:sensitivity:LFI22:Rad:M}{492.0}
\setsymbol{LFI:white:noise:sensitivity:LFI23:Rad:M}{507.7}
\setsymbol{LFI:white:noise:sensitivity:LFI24:Rad:M}{462.2}
\setsymbol{LFI:white:noise:sensitivity:LFI25:Rad:M}{413.6}
\setsymbol{LFI:white:noise:sensitivity:LFI26:Rad:M}{478.6}
\setsymbol{LFI:white:noise:sensitivity:LFI27:Rad:M}{277.7}
\setsymbol{LFI:white:noise:sensitivity:LFI28:Rad:M}{312.3}
\setsymbol{LFI:white:noise:sensitivity:LFI18:Rad:S}{465.7}
\setsymbol{LFI:white:noise:sensitivity:LFI19:Rad:S}{555.6}
\setsymbol{LFI:white:noise:sensitivity:LFI20:Rad:S}{623.2}
\setsymbol{LFI:white:noise:sensitivity:LFI21:Rad:S}{564.1}
\setsymbol{LFI:white:noise:sensitivity:LFI22:Rad:S}{534.4}
\setsymbol{LFI:white:noise:sensitivity:LFI23:Rad:S}{542.4}
\setsymbol{LFI:white:noise:sensitivity:LFI24:Rad:S}{399.2}
\setsymbol{LFI:white:noise:sensitivity:LFI25:Rad:S}{392.6}
\setsymbol{LFI:white:noise:sensitivity:LFI26:Rad:S}{418.6}
\setsymbol{LFI:white:noise:sensitivity:LFI27:Rad:S}{302.9}
\setsymbol{LFI:white:noise:sensitivity:LFI28:Rad:S}{285.3}

\setsymbol{LFI:knee:frequency:70GHz:units}{29.5\mHz}
\setsymbol{LFI:knee:frequency:44GHz:units}{56.2\mHz}
\setsymbol{LFI:knee:frequency:30GHz:units}{113.7\mHz}

\setsymbol{LFI:knee:frequency:70GHz}{29.5}
\setsymbol{LFI:knee:frequency:44GHz}{56.2}
\setsymbol{LFI:knee:frequency:30GHz}{113.7}

\setsymbol{LFI:knee:frequency:LFI18:Rad:M:units}{16.3\mHz}
\setsymbol{LFI:knee:frequency:LFI19:Rad:M:units}{15.1\mHz}
\setsymbol{LFI:knee:frequency:LFI20:Rad:M:units}{18.7\mHz}
\setsymbol{LFI:knee:frequency:LFI21:Rad:M:units}{37.2\mHz}
\setsymbol{LFI:knee:frequency:LFI22:Rad:M:units}{12.7\mHz}
\setsymbol{LFI:knee:frequency:LFI23:Rad:M:units}{34.6\mHz}
\setsymbol{LFI:knee:frequency:LFI24:Rad:M:units}{46.2\mHz}
\setsymbol{LFI:knee:frequency:LFI25:Rad:M:units}{24.9\mHz}
\setsymbol{LFI:knee:frequency:LFI26:Rad:M:units}{67.6\mHz}
\setsymbol{LFI:knee:frequency:LFI27:Rad:M:units}{187.4\mHz}
\setsymbol{LFI:knee:frequency:LFI28:Rad:M:units}{122.2\mHz}
\setsymbol{LFI:knee:frequency:LFI18:Rad:S:units}{17.7\mHz}
\setsymbol{LFI:knee:frequency:LFI19:Rad:S:units}{22.0\mHz}
\setsymbol{LFI:knee:frequency:LFI20:Rad:S:units}{8.7\mHz}
\setsymbol{LFI:knee:frequency:LFI21:Rad:S:units}{25.9\mHz}
\setsymbol{LFI:knee:frequency:LFI22:Rad:S:units}{15.8\mHz}
\setsymbol{LFI:knee:frequency:LFI23:Rad:S:units}{129.8\mHz}
\setsymbol{LFI:knee:frequency:LFI24:Rad:S:units}{100.9\mHz}
\setsymbol{LFI:knee:frequency:LFI25:Rad:S:units}{38.9\mHz}
\setsymbol{LFI:knee:frequency:LFI26:Rad:S:units}{58.9\mHz}
\setsymbol{LFI:knee:frequency:LFI27:Rad:S:units}{104.4\mHz}
\setsymbol{LFI:knee:frequency:LFI28:Rad:S:units}{40.7\mHz}

\setsymbol{LFI:knee:frequency:LFI18:Rad:M}{16.3}
\setsymbol{LFI:knee:frequency:LFI19:Rad:M}{15.1}
\setsymbol{LFI:knee:frequency:LFI20:Rad:M}{18.7}
\setsymbol{LFI:knee:frequency:LFI21:Rad:M}{37.2}
\setsymbol{LFI:knee:frequency:LFI22:Rad:M}{12.7}
\setsymbol{LFI:knee:frequency:LFI23:Rad:M}{34.6}
\setsymbol{LFI:knee:frequency:LFI24:Rad:M}{46.2}
\setsymbol{LFI:knee:frequency:LFI25:Rad:M}{24.9}
\setsymbol{LFI:knee:frequency:LFI26:Rad:M}{67.6}
\setsymbol{LFI:knee:frequency:LFI27:Rad:M}{187.4}
\setsymbol{LFI:knee:frequency:LFI28:Rad:M}{122.2}
\setsymbol{LFI:knee:frequency:LFI18:Rad:S}{17.7}
\setsymbol{LFI:knee:frequency:LFI19:Rad:S}{22.0}
\setsymbol{LFI:knee:frequency:LFI20:Rad:S}{8.7}
\setsymbol{LFI:knee:frequency:LFI21:Rad:S}{25.9}
\setsymbol{LFI:knee:frequency:LFI22:Rad:S}{15.8}
\setsymbol{LFI:knee:frequency:LFI23:Rad:S}{129.8}
\setsymbol{LFI:knee:frequency:LFI24:Rad:S}{100.9}
\setsymbol{LFI:knee:frequency:LFI25:Rad:S}{38.9}
\setsymbol{LFI:knee:frequency:LFI26:Rad:S}{58.9}
\setsymbol{LFI:knee:frequency:LFI27:Rad:S}{104.4}
\setsymbol{LFI:knee:frequency:LFI28:Rad:S}{40.7}

\setsymbol{LFI:slope:70GHz:units}{$-1.03$\mHz}
\setsymbol{LFI:slope:44GHz:units}{$-0.89$\mHz}
\setsymbol{LFI:slope:30GHz:units}{$-0.87$\mHz}

\setsymbol{LFI:slope:70GHz}{$-1.03$}
\setsymbol{LFI:slope:44GHz}{$-0.89$}
\setsymbol{LFI:slope:30GHz}{$-0.87$}

\setsymbol{LFI:slope:LFI18:Rad:M:units}{$-1.04$\mHz}
\setsymbol{LFI:slope:LFI19:Rad:M:units}{$-1.09$\mHz}
\setsymbol{LFI:slope:LFI20:Rad:M:units}{$-0.69$\mHz}
\setsymbol{LFI:slope:LFI21:Rad:M:units}{$-1.56$\mHz}
\setsymbol{LFI:slope:LFI22:Rad:M:units}{$-1.01$\mHz}
\setsymbol{LFI:slope:LFI23:Rad:M:units}{$-0.96$\mHz}
\setsymbol{LFI:slope:LFI24:Rad:M:units}{$-0.83$\mHz}
\setsymbol{LFI:slope:LFI25:Rad:M:units}{$-0.91$\mHz}
\setsymbol{LFI:slope:LFI26:Rad:M:units}{$-0.95$\mHz}
\setsymbol{LFI:slope:LFI27:Rad:M:units}{$-0.87$\mHz}
\setsymbol{LFI:slope:LFI28:Rad:M:units}{$-0.88$\mHz}
\setsymbol{LFI:slope:LFI18:Rad:S:units}{$-1.15$\mHz}
\setsymbol{LFI:slope:LFI19:Rad:S:units}{$-1.00$\mHz}
\setsymbol{LFI:slope:LFI20:Rad:S:units}{$-0.95$\mHz}
\setsymbol{LFI:slope:LFI21:Rad:S:units}{$-0.92$\mHz}
\setsymbol{LFI:slope:LFI22:Rad:S:units}{$-1.01$\mHz}
\setsymbol{LFI:slope:LFI23:Rad:S:units}{$-0.95$\mHz}
\setsymbol{LFI:slope:LFI24:Rad:S:units}{$-0.73$\mHz}
\setsymbol{LFI:slope:LFI25:Rad:S:units}{$-1.16$\mHz}
\setsymbol{LFI:slope:LFI26:Rad:S:units}{$-0.79$\mHz}
\setsymbol{LFI:slope:LFI27:Rad:S:units}{$-0.82$\mHz}
\setsymbol{LFI:slope:LFI28:Rad:S:units}{$-0.91$\mHz}

\setsymbol{LFI:slope:LFI18:Rad:M}{$-1.04$}
\setsymbol{LFI:slope:LFI19:Rad:M}{$-1.09$}
\setsymbol{LFI:slope:LFI20:Rad:M}{$-0.69$}
\setsymbol{LFI:slope:LFI21:Rad:M}{$-1.56$}
\setsymbol{LFI:slope:LFI22:Rad:M}{$-1.01$}
\setsymbol{LFI:slope:LFI23:Rad:M}{$-0.96$}
\setsymbol{LFI:slope:LFI24:Rad:M}{$-0.83$}
\setsymbol{LFI:slope:LFI25:Rad:M}{$-0.91$}
\setsymbol{LFI:slope:LFI26:Rad:M}{$-0.95$}
\setsymbol{LFI:slope:LFI27:Rad:M}{$-0.87$}
\setsymbol{LFI:slope:LFI28:Rad:M}{$-0.88$}
\setsymbol{LFI:slope:LFI18:Rad:S}{$-1.15$}
\setsymbol{LFI:slope:LFI19:Rad:S}{$-1.00$}
\setsymbol{LFI:slope:LFI20:Rad:S}{$-0.95$}
\setsymbol{LFI:slope:LFI21:Rad:S}{$-0.92$}
\setsymbol{LFI:slope:LFI22:Rad:S}{$-1.01$}
\setsymbol{LFI:slope:LFI23:Rad:S}{$-0.95$}
\setsymbol{LFI:slope:LFI24:Rad:S}{$-0.73$}
\setsymbol{LFI:slope:LFI25:Rad:S}{$-1.16$}
\setsymbol{LFI:slope:LFI26:Rad:S}{$-0.79$}
\setsymbol{LFI:slope:LFI27:Rad:S}{$-0.82$}
\setsymbol{LFI:slope:LFI28:Rad:S}{$-0.91$}

\setsymbol{LFI:FWHM:70GHz:units}{13\parcm01}
\setsymbol{LFI:FWHM:44GHz:units}{27\parcm92}
\setsymbol{LFI:FWHM:30GHz:units}{32\parcm65}

\setsymbol{LFI:FWHM:70GHz}{13.01}
\setsymbol{LFI:FWHM:44GHz}{27.92}
\setsymbol{LFI:FWHM:30GHz}{32.65}

\setsymbol{LFI:FWHM:LFI18:units}{13\parcm39}
\setsymbol{LFI:FWHM:LFI19:units}{13\parcm01}
\setsymbol{LFI:FWHM:LFI20:units}{12\parcm75}
\setsymbol{LFI:FWHM:LFI21:units}{12\parcm74}
\setsymbol{LFI:FWHM:LFI22:units}{12\parcm87}
\setsymbol{LFI:FWHM:LFI23:units}{13\parcm27}
\setsymbol{LFI:FWHM:LFI24:units}{22\parcm98}
\setsymbol{LFI:FWHM:LFI25:units}{30\parcm46}
\setsymbol{LFI:FWHM:LFI26:units}{30\parcm31}
\setsymbol{LFI:FWHM:LFI27:units}{32\parcm65}
\setsymbol{LFI:FWHM:LFI28:units}{32\parcm66}

\setsymbol{LFI:FWHM:LFI18}{13.39}
\setsymbol{LFI:FWHM:LFI19}{13.01}
\setsymbol{LFI:FWHM:LFI20}{12.75}
\setsymbol{LFI:FWHM:LFI21}{12.74}
\setsymbol{LFI:FWHM:LFI22}{12.87}
\setsymbol{LFI:FWHM:LFI23}{13.27}
\setsymbol{LFI:FWHM:LFI24}{22.98}
\setsymbol{LFI:FWHM:LFI25}{30.46}
\setsymbol{LFI:FWHM:LFI26}{30.31}
\setsymbol{LFI:FWHM:LFI27}{32.65}
\setsymbol{LFI:FWHM:LFI28}{32.66}

\setsymbol{LFI:FWHM:uncertainty:LFI18:units}{0.170\arcm}
\setsymbol{LFI:FWHM:uncertainty:LFI19:units}{0.174\arcm}
\setsymbol{LFI:FWHM:uncertainty:LFI20:units}{0.170\arcm}
\setsymbol{LFI:FWHM:uncertainty:LFI21:units}{0.156\arcm}
\setsymbol{LFI:FWHM:uncertainty:LFI22:units}{0.164\arcm}
\setsymbol{LFI:FWHM:uncertainty:LFI23:units}{0.171\arcm}
\setsymbol{LFI:FWHM:uncertainty:LFI24:units}{0.652\arcm}
\setsymbol{LFI:FWHM:uncertainty:LFI25:units}{1.075\arcm}
\setsymbol{LFI:FWHM:uncertainty:LFI26:units}{1.131\arcm}
\setsymbol{LFI:FWHM:uncertainty:LFI27:units}{1.266\arcm}
\setsymbol{LFI:FWHM:uncertainty:LFI28:units}{1.287\arcm}

\setsymbol{LFI:FWHM:uncertainty:LFI18}{0.170}
\setsymbol{LFI:FWHM:uncertainty:LFI19}{0.174}
\setsymbol{LFI:FWHM:uncertainty:LFI20}{0.170}
\setsymbol{LFI:FWHM:uncertainty:LFI21}{0.156}
\setsymbol{LFI:FWHM:uncertainty:LFI22}{0.164}
\setsymbol{LFI:FWHM:uncertainty:LFI23}{0.171}
\setsymbol{LFI:FWHM:uncertainty:LFI24}{0.652}
\setsymbol{LFI:FWHM:uncertainty:LFI25}{1.075}
\setsymbol{LFI:FWHM:uncertainty:LFI26}{1.131}
\setsymbol{LFI:FWHM:uncertainty:LFI27}{1.266}
\setsymbol{LFI:FWHM:uncertainty:LFI28}{1.287}

\setsymbol{HFI:center:frequency:100GHz:units}{100\,GHz}
\setsymbol{HFI:center:frequency:143GHz:units}{143\,GHz}
\setsymbol{HFI:center:frequency:217GHz:units}{217\,GHz}
\setsymbol{HFI:center:frequency:353GHz:units}{353\,GHz}
\setsymbol{HFI:center:frequency:545GHz:units}{545\,GHz}
\setsymbol{HFI:center:frequency:857GHz:units}{857\,GHz}

\setsymbol{HFI:center:frequency:100GHz}{100}
\setsymbol{HFI:center:frequency:143GHz}{143}
\setsymbol{HFI:center:frequency:217GHz}{217}
\setsymbol{HFI:center:frequency:353GHz}{353}
\setsymbol{HFI:center:frequency:545GHz}{545}
\setsymbol{HFI:center:frequency:857GHz}{857}

\setsymbol{HFI:Ndetectors:100GHz}{8}
\setsymbol{HFI:Ndetectors:143GHz}{11}
\setsymbol{HFI:Ndetectors:217GHz}{12}
\setsymbol{HFI:Ndetectors:353GHz}{12}
\setsymbol{HFI:Ndetectors:545GHz}{3}
\setsymbol{HFI:Ndetectors:857GHz}{4}

\setsymbol{HFI:FWHM:Maps:100GHz:units}{9\parcm88}
\setsymbol{HFI:FWHM:Maps:143GHz:units}{7\parcm18}
\setsymbol{HFI:FWHM:Maps:217GHz:units}{4\parcm87}
\setsymbol{HFI:FWHM:Maps:353GHz:units}{4\parcm65}
\setsymbol{HFI:FWHM:Maps:545GHz:units}{4\parcm72}
\setsymbol{HFI:FWHM:Maps:857GHz:units}{4\parcm39}
\setsymbol{HFI:FWHM:Maps:100GHz}{9.88}
\setsymbol{HFI:FWHM:Maps:143GHz}{7.18}
\setsymbol{HFI:FWHM:Maps:217GHz}{4.87}
\setsymbol{HFI:FWHM:Maps:353GHz}{4.65}
\setsymbol{HFI:FWHM:Maps:545GHz}{4.72}
\setsymbol{HFI:FWHM:Maps:857GHz}{4.39}

\setsymbol{HFI:beam:ellipticity:Maps:100GHz}{1.15}
\setsymbol{HFI:beam:ellipticity:Maps:143GHz}{1.01}
\setsymbol{HFI:beam:ellipticity:Maps:217GHz}{1.06}
\setsymbol{HFI:beam:ellipticity:Maps:353GHz}{1.05}
\setsymbol{HFI:beam:ellipticity:Maps:545GHz}{1.14}
\setsymbol{HFI:beam:ellipticity:Maps:857GHz}{1.19}

\setsymbol{HFI:FWHM:Mars:100GHz:units}{9\parcm37}
\setsymbol{HFI:FWHM:Mars:143GHz:units}{7\parcm04}
\setsymbol{HFI:FWHM:Mars:217GHz:units}{4\parcm68}
\setsymbol{HFI:FWHM:Mars:353GHz:units}{4\parcm43}
\setsymbol{HFI:FWHM:Mars:545GHz:units}{3\parcm80}
\setsymbol{HFI:FWHM:Mars:857GHz:units}{3\parcm67}

\setsymbol{HFI:FWHM:Mars:100GHz}{9.37}
\setsymbol{HFI:FWHM:Mars:143GHz}{7.04}
\setsymbol{HFI:FWHM:Mars:217GHz}{4.68}
\setsymbol{HFI:FWHM:Mars:353GHz}{4.43}
\setsymbol{HFI:FWHM:Mars:545GHz}{3.80}
\setsymbol{HFI:FWHM:Mars:857GHz}{3.67}

\setsymbol{HFI:beam:ellipticity:Mars:100GHz}{1.18}
\setsymbol{HFI:beam:ellipticity:Mars:143GHz}{1.03}
\setsymbol{HFI:beam:ellipticity:Mars:217GHz}{1.14}
\setsymbol{HFI:beam:ellipticity:Mars:353GHz}{1.09}
\setsymbol{HFI:beam:ellipticity:Mars:545GHz}{1.25}
\setsymbol{HFI:beam:ellipticity:Mars:857GHz}{1.03}

\setsymbol{HFI:CMB:relative:calibration:100GHz}{$\lsim 1\%$}
\setsymbol{HFI:CMB:relative:calibration:143GHz}{$\lsim 1\%$}
\setsymbol{HFI:CMB:relative:calibration:217GHz}{$\lsim 1\%$}
\setsymbol{HFI:CMB:relative:calibration:353GHz}{$\lsim 1\%$}
\setsymbol{HFI:CMB:relative:calibration:545GHz}{}
\setsymbol{HFI:CMB:relative:calibration:857GHz}{}

\setsymbol{HFI:CMB:absolute:calibration:100GHz}{$\lsim 2\%$}
\setsymbol{HFI:CMB:absolute:calibration:143GHz}{$\lsim 2\%$}
\setsymbol{HFI:CMB:absolute:calibration:217GHz}{$\lsim 2\%$}
\setsymbol{HFI:CMB:absolute:calibration:353GHz}{$\lsim 2\%$}
\setsymbol{HFI:CMB:absolute:calibration:545GHz}{}
\setsymbol{HFI:CMB:absolute:calibration:857GHz}{}

\setsymbol{HFI:FIRAS:gain:calibration:accuracy:statistical:100GHz}{}
\setsymbol{HFI:FIRAS:gain:calibration:accuracy:statistical:143GHz}{}
\setsymbol{HFI:FIRAS:gain:calibration:accuracy:statistical:217GHz}{}
\setsymbol{HFI:FIRAS:gain:calibration:accuracy:statistical:353GHz}{2.5\%}
\setsymbol{HFI:FIRAS:gain:calibration:accuracy:statistical:545GHz}{1\%}
\setsymbol{HFI:FIRAS:gain:calibration:accuracy:statistical:857GHz}{0.5\%}

\setsymbol{HFI:FIRAS:gain:calibration:accuracy:systematic:100GHz}{}
\setsymbol{HFI:FIRAS:gain:calibration:accuracy:systematic:143GHz}{}
\setsymbol{HFI:FIRAS:gain:calibration:accuracy:systematic:217GHz}{}
\setsymbol{HFI:FIRAS:gain:calibration:accuracy:systematic:353GHz}{}
\setsymbol{HFI:FIRAS:gain:calibration:accuracy:systematic:545GHz}{7\%}
\setsymbol{HFI:FIRAS:gain:calibration:accuracy:systematic:857GHz}{7\%}

\setsymbol{HFI:FIRAS:zero:point:accuracy:100GHz:units}{0.8\MJysr}
\setsymbol{HFI:FIRAS:zero:point:accuracy:143GHz:units}{}
\setsymbol{HFI:FIRAS:zero:point:accuracy:217GHz:units}{}
\setsymbol{HFI:FIRAS:zero:point:accuracy:353GHz:units}{1.4\MJysr}
\setsymbol{HFI:FIRAS:zero:point:accuracy:545GHz:units}{2.2\MJysr}
\setsymbol{HFI:FIRAS:zero:point:accuracy:857GHz:units}{1.7\MJysr}

\setsymbol{HFI:FIRAS:zero:point:accuracy:100GHz}{0.8}
\setsymbol{HFI:FIRAS:zero:point:accuracy:143GHz}{}
\setsymbol{HFI:FIRAS:zero:point:accuracy:217GHz}{}
\setsymbol{HFI:FIRAS:zero:point:accuracy:353GHz}{1.4}
\setsymbol{HFI:FIRAS:zero:point:accuracy:545GHz}{2.2}
\setsymbol{HFI:FIRAS:zero:point:accuracy:857GHz}{1.7}

\setsymbol{HFI:unit:conversion:100GHz:units}{0.2415\MJysrmK}
\setsymbol{HFI:unit:conversion:143GHz:units}{0.3694\MJysrmK}
\setsymbol{HFI:unit:conversion:217GHz:units}{0.4811\MJysrmK}
\setsymbol{HFI:unit:conversion:353GHz:units}{0.2883\MJysrmK}
\setsymbol{HFI:unit:conversion:545GHz:units}{0.05826\MJysrmK}
\setsymbol{HFI:unit:conversion:857GHz:units}{0.002238\MJysrmK}

\setsymbol{HFI:unit:conversion:100GHz}{0.2415}
\setsymbol{HFI:unit:conversion:143GHz}{0.3694}
\setsymbol{HFI:unit:conversion:217GHz}{0.4811}
\setsymbol{HFI:unit:conversion:353GHz}{0.2883}
\setsymbol{HFI:unit:conversion:545GHz}{0.05826}
\setsymbol{HFI:unit:conversion:857GHz}{0.002238}

\setsymbol{HFI:colour:correction:alpha=-2:V1.01:100GHz}{0.9893}
\setsymbol{HFI:colour:correction:alpha=-2:V1.01:143GHz}{0.9759}
\setsymbol{HFI:colour:correction:alpha=-2:V1.01:217GHz}{1.0007}
\setsymbol{HFI:colour:correction:alpha=-2:V1.01:353GHz}{1.0028}
\setsymbol{HFI:colour:correction:alpha=-2:V1.01:545GHz}{1.0019}
\setsymbol{HFI:colour:correction:alpha=-2:V1.01:857GHz}{0.9889}

\setsymbol{HFI:colour:correction:alpha=0:V1.01:100GHz}{1.0008}
\setsymbol{HFI:colour:correction:alpha=0:V1.01:143GHz}{1.0148}
\setsymbol{HFI:colour:correction:alpha=0:V1.01:217GHz}{0.9909}
\setsymbol{HFI:colour:correction:alpha=0:V1.01:353GHz}{0.9888}
\setsymbol{HFI:colour:correction:alpha=0:V1.01:545GHz}{0.9878}
\setsymbol{HFI:colour:correction:alpha=0:V1.01:857GHz}{1.0014}

\providecommand{\sorthelp}[1]{}

\begin{document}

\title{\textit{Planck} 2013 results. XXII. Constraints on inflation}
\author{\small
Planck Collaboration:
P.~A.~R.~Ade\inst{90}
\and
N.~Aghanim\inst{62}
\and
C.~Armitage-Caplan\inst{96}
\and
M.~Arnaud\inst{75}
\and
M.~Ashdown\inst{72, 6}
\and
F.~Atrio-Barandela\inst{19}
\and
J.~Aumont\inst{62}
\and
C.~Baccigalupi\inst{89}
\and
A.~J.~Banday\inst{99, 10}
\and
R.~B.~Barreiro\inst{69}
% ...
}
\institute{\small
APC, AstroParticule et Cosmologie, Universit\'{e} Paris Diderot, CNRS/IN2P3, CEA/lrfu, Observatoire de Paris, Sorbonne Paris Cit\'{e}, 10, rue Alice Domon et L\'{e}onie Duquet, 75205 Paris Cedex 13, France\\
\and
Aalto University Mets\"{a}hovi Radio Observatory, Mets\"{a}hovintie 114, FIN-02540 Kylm\"{a}l\"{a}, Finland\\
\and
African Institute for Mathematical Sciences, 6-8 Melrose Road, Muizenberg, Cape Town, South Africa\\
\and
Agenzia Spaziale Italiana Science Data Center, Via del Politecnico snc, 00133, Roma, Italy\\
\and
Agenzia Spaziale Italiana, Viale Liegi 26, Roma, Italy\\
\and
Astrophysics Group, Cavendish Laboratory, University of Cambridge, J J Thomson Avenue, Cambridge CB3 0HE, U.K.\\
\and
Astrophysics \& Cosmology Research Unit, School of Mathematics, Statistics \& Computer Science, University of KwaZulu-Natal, Westville Campus, Private Bag X54001, Durban 4000, South Africa\\
\and
Atacama Large Millimeter/submillimeter Array, ALMA Santiago Central Offices, Alonso de Cordova 3107, Vitacura, Casilla 763 0355, Santiago, Chile\\
\and
CITA, University of Toronto, 60 St. George St., Toronto, ON M5S 3H8, Canada\\
\and
CNRS, IRAP, 9 Av. colonel Roche, BP 44346, F-31028 Toulouse cedex 4, France\\
\and
California Institute of Technology, Pasadena, California, U.S.A.\\
\and
Centre for Theoretical Cosmology, DAMTP, University of Cambridge, Wilberforce Road, Cambridge CB3 0WA, U.K.\\
\and
Centro de Estudios de F\'{i}sica del Cosmos de Arag\'{o}n (CEFCA), Plaza San Juan, 1, planta 2, E-44001, Teruel, Spain\\
\and
Computational Cosmology Center, Lawrence Berkeley National Laboratory, Berkeley, California, U.S.A.\\
\and
Consejo Superior de Investigaciones Cient\'{\i}ficas (CSIC), Madrid, Spain\\
\and
DSM/Irfu/SPP, CEA-Saclay, F-91191 Gif-sur-Yvette Cedex, France\\
\and
DTU Space, National Space Institute, Technical University of Denmark, Elektrovej 327, DK-2800 Kgs. Lyngby, Denmark\\
\and
D\'{e}partement de Physique Th\'{e}orique, Universit\'{e} de Gen\`{e}ve, 24, Quai E. Ansermet,1211 Gen\`{e}ve 4, Switzerland\\
\and
Departamento de F\'{\i}sica Fundamental, Facultad de Ciencias, Universidad de Salamanca, 37008 Salamanca, Spain\\
\and
Departamento de F\'{\i}sica, Universidad de Oviedo, Avda. Calvo Sotelo s/n, Oviedo, Spain\\
\and
Department of Astronomy and Astrophysics, University of Toronto, 50 Saint George Street, Toronto, Ontario, Canada\\
\and
Department of Astrophysics/IMAPP, Radboud University Nijmegen, P.O. Box 9010, 6500 GL Nijmegen, The Netherlands\\
\and
Department of Electrical Engineering and Computer Sciences, University of California, Berkeley, California, U.S.A.\\
\and
Department of Physics \& Astronomy, University of British Columbia, 6224 Agricultural Road, Vancouver, British Columbia, Canada\\
\and
Department of Physics and Astronomy, Dana and David Dornsife College of Letter, Arts and Sciences, University of Southern California, Los Angeles, CA 90089, U.S.A.\\
\and
Department of Physics and Astronomy, University College London, London WC1E 6BT, U.K.\\
\and
Department of Physics and Astronomy, University of Sussex, Brighton BN1 9QH, U.K.\\
\and
Department of Physics, Florida State University, Keen Physics Building, 77 Chieftan Way, Tallahassee, Florida, U.S.A.\\
\and
Department of Physics, Gustaf H\"{a}llstr\"{o}min katu 2a, University of Helsinki, Helsinki, Finland\\
\and
Department of Physics, Princeton University, Princeton, New Jersey, U.S.A.\\
\and
Department of Physics, University of California, Berkeley, California, U.S.A.\\
\and
Department of Physics, University of California, One Shields Avenue, Davis, California, U.S.A.\\
\and
Department of Physics, University of California, Santa Barbara, California, U.S.A.\\
\and
Department of Physics, University of Illinois at Urbana-Champaign, 1110 West Green Street, Urbana, Illinois, U.S.A.\\
\and
Dipartimento di Fisica e Astronomia G. Galilei, Universit\`{a} degli Studi di Padova, via Marzolo 8, 35131 Padova, Italy\\
\and
Dipartimento di Fisica e Scienze della Terra, Universit\`{a} di Ferrara, Via Saragat 1, 44122 Ferrara, Italy\\
\and
Dipartimento di Fisica, Universit\`{a} La Sapienza, P. le A. Moro 2, Roma, Italy\\
\and
Dipartimento di Fisica, Universit\`{a} degli Studi di Milano, Via Celoria, 16, Milano, Italy\\
\and
Dipartimento di Fisica, Universit\`{a} degli Studi di Trieste, via A. Valerio 2, Trieste, Italy\\
\and
Dipartimento di Fisica, Universit\`{a} di Roma Tor Vergata, Via della Ricerca Scientifica, 1, Roma, Italy\\
\and
Discovery Center, Niels Bohr Institute, Blegdamsvej 17, Copenhagen, Denmark\\
\and
Dpto. Astrof\'{i}sica, Universidad de La Laguna (ULL), E-38206 La Laguna, Tenerife, Spain\\
\and
European Southern Observatory, ESO Vitacura, Alonso de Cordova 3107, Vitacura, Casilla 19001, Santiago, Chile\\
\and
European Space Agency, ESAC, Planck Science Office, Camino bajo del Castillo, s/n, Urbanizaci\'{o}n Villafranca del Castillo, Villanueva de la Ca\~{n}ada, Madrid, Spain\\
\and
European Space Agency, ESTEC, Keplerlaan 1, 2201 AZ Noordwijk, The Netherlands\\
\and
Haverford College Astronomy Department, 370 Lancaster Avenue, Haverford, Pennsylvania, U.S.A.\\
\and
Helsinki Institute of Physics, Gustaf H\"{a}llstr\"{o}min katu 2, University of Helsinki, Helsinki, Finland\\
\and
INAF - Osservatorio Astronomico di Padova, Vicolo dell'Osservatorio 5, Padova, Italy\\
\and
INAF - Osservatorio Astronomico di Roma, via di Frascati 33, Monte Porzio Catone, Italy\\
\and
INAF - Osservatorio Astronomico di Trieste, Via G.B. Tiepolo 11, Trieste, Italy\\
\and
INAF Istituto di Radioastronomia, Via P. Gobetti 101, 40129 Bologna, Italy\\
\and
INAF/IASF Bologna, Via Gobetti 101, Bologna, Italy\\
\and
INAF/IASF Milano, Via E. Bassini 15, Milano, Italy\\
\and
INFN, Sezione di Bologna, Via Irnerio 46, I-40126, Bologna, Italy\\
\and
INFN, Sezione di Roma 1, Universit\`{a} di Roma Sapienza, Piazzale Aldo Moro 2, 00185, Roma, Italy\\
\and
IPAG: Institut de Plan\'{e}tologie et d'Astrophysique de Grenoble, Universit\'{e} Joseph Fourier, Grenoble 1 / CNRS-INSU, UMR 5274, Grenoble, F-38041, France\\
\and
IUCAA, Post Bag 4, Ganeshkhind, Pune University Campus, Pune 411 007, India\\
\and
Imperial College London, Astrophysics group, Blackett Laboratory, Prince Consort Road, London, SW7 2AZ, U.K.\\
\and
Infrared Processing and Analysis Center, California Institute of Technology, Pasadena, CA 91125, U.S.A.\\
\and
Institut N\'{e}el, CNRS, Universit\'{e} Joseph Fourier Grenoble I, 25 rue des Martyrs, Grenoble, France\\
\and
Institut Universitaire de France, 103, bd Saint-Michel, 75005, Paris, France\\
\and
Institut d'Astrophysique Spatiale, CNRS (UMR8617) Universit\'{e} Paris-Sud 11, B\^{a}timent 121, Orsay, France\\
\and
Institut d'Astrophysique de Paris, CNRS (UMR7095), 98 bis Boulevard Arago, F-75014, Paris, France\\
\and
Institute for Space Sciences, Bucharest-Magurale, Romania\\
\and
Institute of Astronomy and Astrophysics, Academia Sinica, Taipei, Taiwan\\
\and
Institute of Astronomy, University of Cambridge, Madingley Road, Cambridge CB3 0HA, U.K.\\
\and
Institute of Theoretical Astrophysics, University of Oslo, Blindern, Oslo, Norway\\
\and
Instituto de Astrof\'{\i}sica de Canarias, C/V\'{\i}a L\'{a}ctea s/n, La Laguna, Tenerife, Spain\\
\and
Instituto de F\'{\i}sica de Cantabria (CSIC-Universidad de Cantabria), Avda. de los Castros s/n, Santander, Spain\\
\and
Jet Propulsion Laboratory, California Institute of Technology, 4800 Oak Grove Drive, Pasadena, California, U.S.A.\\
\and
Jodrell Bank Centre for Astrophysics, Alan Turing Building, School of Physics and Astronomy, The University of Manchester, Oxford Road, Manchester, M13 9PL, U.K.\\
\and
Kavli Institute for Cosmology Cambridge, Madingley Road, Cambridge, CB3 0HA, U.K.\\
\and
LAL, Universit\'{e} Paris-Sud, CNRS/IN2P3, Orsay, France\\
\and
LERMA, CNRS, Observatoire de Paris, 61 Avenue de l'Observatoire, Paris, France\\
\and
Laboratoire AIM, IRFU/Service d'Astrophysique - CEA/DSM - CNRS - Universit\'{e} Paris Diderot, B\^{a}t. 709, CEA-Saclay, F-91191 Gif-sur-Yvette Cedex, France\\
\and
Laboratoire Traitement et Communication de l'Information, CNRS (UMR 5141) and T\'{e}l\'{e}com ParisTech, 46 rue Barrault F-75634 Paris Cedex 13, France\\
\and
Laboratoire de Physique Subatomique et de Cosmologie, Universit\'{e} Joseph Fourier Grenoble I, CNRS/IN2P3, Institut National Polytechnique de Grenoble, 53 rue des Martyrs, 38026 Grenoble cedex, France\\
\and
Laboratoire de Physique Th\'{e}orique, Universit\'{e} Paris-Sud 11 \& CNRS, B\^{a}timent 210, 91405 Orsay, France\\
\and
Lawrence Berkeley National Laboratory, Berkeley, California, U.S.A.\\
\and
Leung Center for Cosmology and Particle Astrophysics, National Taiwan University, Taipei 10617, Taiwan\\
\and
Max-Planck-Institut f\"{u}r Astrophysik, Karl-Schwarzschild-Str. 1, 85741 Garching, Germany\\
\and
McGill Physics, Ernest Rutherford Physics Building, McGill University, 3600 rue University, Montr\'{e}al, QC, H3A 2T8, Canada\\
\and
MilliLab, VTT Technical Research Centre of Finland, Tietotie 3, Espoo, Finland\\
\and
National University of Ireland, Department of Experimental Physics, Maynooth, Co. Kildare, Ireland\\
\and
Niels Bohr Institute, Blegdamsvej 17, Copenhagen, Denmark\\
\and
Observational Cosmology, Mail Stop 367-17, California Institute of Technology, Pasadena, CA, 91125, U.S.A.\\
\and
Optical Science Laboratory, University College London, Gower Street, London, U.K.\\
\and
SB-ITP-LPPC, EPFL, CH-1015, Lausanne, Switzerland\\
\and
SISSA, Astrophysics Sector, via Bonomea 265, 34136, Trieste, Italy\\
\and
School of Physics and Astronomy, Cardiff University, Queens Buildings, The Parade, Cardiff, CF24 3AA, U.K.\\
\and
School of Physics and Astronomy, University of Nottingham, Nottingham NG7 2RD, U.K.\\
\and
Space Research Institute (IKI), Russian Academy of Sciences, Profsoyuznaya Str, 84/32, Moscow, 117997, Russia\\
\and
Space Sciences Laboratory, University of California, Berkeley, California, U.S.A.\\
\and
Special Astrophysical Observatory, Russian Academy of Sciences, Nizhnij Arkhyz, Zelenchukskiy region, Karachai-Cherkessian Republic, 369167, Russia\\
\and
Stanford University, Dept of Physics, Varian Physics Bldg, 382 Via Pueblo Mall, Stanford, California, U.S.A.\\
\and
Sub-Department of Astrophysics, University of Oxford, Keble Road, Oxford OX1 3RH, U.K.\\
\and
Theory Division, PH-TH, CERN, CH-1211, Geneva 23, Switzerland\\
\and
UPMC Univ Paris 06, UMR7095, 98 bis Boulevard Arago, F-75014, Paris, France\\
\and
Universit\'{e} de Toulouse, UPS-OMP, IRAP, F-31028 Toulouse cedex 4, France\\
% ...
}

{\def\reff@jnl#1{{\rm#1\/}}
\def\apj{\reff@jnl{ApJ}}       
\def\apjs{\reff@jnl{ApJS}}     
\def\aaps{\reff@jnl{A\&AS}}    
\def\mnras{\reff@jnl{MNRAS}}   
\def\prd{\reff@jnl{Phys.\ Rev.\ D}}    

\newcommand{\Nside}{\ensuremath{N_{\mathrm{side}}}} 
\newcommand{\Npix}{\ensuremath{N_{\mathrm{pix}}}}   
\newcommand{\Ntau}{\ensuremath{N_{\tau}}}   
\newcommand{\vA}{\mathbf{A}}
\newcommand{\va}{\mathbf{a}}
\newcommand{\vB}{\mathbf{B}}
\newcommand{\vM}{\mathbf{M}}
\newcommand{\vN}{\mathbf{N}}
\newcommand{\vP}{\mathbf{P}}
\newcommand{\vS}{\mathbf{S}}
\newcommand{\vX}{\mathbf{X}}
\newcommand{\vY}{\mathbf{Y}}
\newcommand{\vd}{\mathbf{d}}
\newcommand{\vn}{\mathbf{n}}
\newcommand{\vs}{\mathbf{s}}
\newcommand{\vC}{\mathbf{C}}
\newcommand{\vI}{\mathbf{I}}
\newcommand{\vt}{\mathbf{t}}
\newcommand{\vE}{\mathbf{E}}
\newcommand{\vx}{\mathbf{x}}
\newcommand{\vphi}{\mathbf{\phi}}
\newcommand{\veta}{\mathbf{\eta}}
\newcommand{\vshat}{\vec{\hat{s}}}
\newcommand{\vxhat}{\vec{\hat{x}}}
\newcommand{\vnu}{\vn_{u}}
\newcommand{\vNu}{\vN_{u}}
\newcommand{\tr}{^{\mathrm{T}}} 
\newcommand{\beq}{\begin{equation}}
\newcommand{\eeq}{\end{equation}}
\newcommand{\rsht}{{MASTER}}
\newcommand{\tC}{\widetilde{C}}
\newcommand{\tN}{\widetilde{N}}
\renewcommand{\r}{{\bf{r}}}
\newcommand{\n}{{\bf{n}}}
\renewcommand{\k}{{\bf{k}}}
\newcommand{\fsky}{{f_{\rm sky}}}
\newcommand{\fzero}{{F^{(0)}}}
\newcommand{\fzerol}{{F^{(0)}_{\ell}}}
\newcommand{\npix}{{N_{\rm pix}}}
\newcommand{\nbins}{{n_{\rm bins}}}
\newcommand{\lmax}{{\ell_{\rm max}}}
\newcommand{\nmc}{{N_{\rm MC}}}
\newcommand{\nmcs}{{N_{\rm MC}^{\rm (s)}}}
\newcommand{\nmcn}{{N_{\rm MC}^{\rm (n)}}}
\newcommand{\nmcsn}{{N_{\rm MC}^{\rm (s+n)}}}
\newcommand{\ntau}{{N_{\tau}}}
\newcommand{\nfft}{{N_{\rm FFT}}}
\newcommand{\Cltheory}{{C_\ell^{\rm th}}}
\newcommand{\Ctheory}{{C^{\rm th}}}
\newcommand{\VEV}[1]{\langle#1\rangle}
\newcommand{\boom}{{\sc{BOOMERanG}}}
\newcommand{\bldb}{{Boom-LDB}}
\newcommand{\wjjj}[6]
{{
\left(
\begin{array}{lcr} #1 & #2 & #3 \\#4 & #5 & #6 \end{array}
\right)
}}
\newcommand{\niter}{{N_{\rm iter}}}
\newcommand{\col}[2]{\left[\begin{array}{c}{#1}\\{#2}\end{array}\right]}
\newcommand{\mat}[4]{\left[\begin{array}{cc}{#1}&{#2}\\{#3}&{#4}\end{array}
\right]}
\newcommand{\vsh}{\hat{\vs}}
\newcommand{\vxh}{\hat{\vx}}
\newcommand{\Nmap}{\vN_{\mathrm{map}}}
\newcommand{\Noff}{\vN_{\mathrm{off}}}
\newcommand{\Nmapz}{\Nmap^{(0)}}
\newcommand{\thalf}{\tfrac{1}{2}}
\newcommand{\be}{\begin{equation}}
\newcommand{\ee}{\end{equation}}
\newcommand{\bea}{\begin{eqnarray}}
\newcommand{\eea}{\end{eqnarray}}
\def\nn{\nonumber}
\def\L{\mathcal{L}}

\renewcommand{\dbltopfraction}{1.0}
\renewcommand{\textfraction}{0}

\newenvironment{myitem}%
{\begin{enumerate}\setlength{\itemsep}{0mm}}%
{\end{enumerate}}
\newenvironment{myenum}%
{\begin{enumerate}\setlength{\itemsep}{0mm}}%
{\end{enumerate}}

\def\nd#1#2{{d #1 \over d #2}}
\def\pd#1#2{{\upartial #1 \over \upartial #2}}
\def\spd#1#2#3{{\upartial ^2 #1 \over \upartial #2 \upartial #3}}
\def\sspd#1#2{{\upartial ^2 #1 \over \upartial #2^2}}
\def\tfrac#1#2{{\textstyle\frac{#1}{#2}}}
\def\vect#1{{\mathbf{#1}}}

\newcommand{\bc}{\begin{center}}
\newcommand{\ec}{\end{center}}
\newcommand{\bi}{\begin{itemize}}
\newcommand{\ei}{\end{itemize}}
\newcommand{\ben}{\begin{enumerate}}
\newcommand{\een}{\end{enumerate}}
\newcommand{\R}{\Re\textrm{e}}
\newcommand{\I}{\Im\textrm{m}}
\newcommand{\Ab}{\boldsymbol{A}}
\newcommand{\Mb}{\boldsymbol{M}}
\newcommand{\Tb}{\boldsymbol{T}}

 \newcommand{\mtc}[1]{\mathcal{#1}}
 \newcommand{\Pow}{{\mathcal P}}
 \newcommand{\nad}{n_\mathrm{ad}}
 \newcommand{\nadI}{n_\mathrm{ar}}
 \newcommand{\nadII}{n_\mathrm{as}}
 \newcommand{\niso}{n_\mathrm{iso}}
 \newcommand{\ncor}{n_\mathrm{cor}}
 \newcommand{\etzz}{\eta_{\sigma\sigma}}
 \newcommand{\etzs}{\eta_{\sigma s}}
 \newcommand{\etss}{\eta_{ss}}
 \newcommand{\veps}{\varepsilon}
 \newcommand{\jvc}[1]{{{\textcolor{red}{JV: #1 endJV.}}}}
\newcommand{\ff}[1]{{{\textcolor{blu}{#1}}}}

\newfont{\gwpfont}{cmssq8 scaled 1000}
\newcommand{\rexcess}{{\gwpfont REXCESS}}

\def\xmm{{\it XMM-Newton}}
\def\Mv {M_\mathrm{500}}
\def\msol {\mathrm{M}_{\odot}}
\def\YX {Y_\mathrm{X}} 
\def \Rv {R_{500}} 
\def\keV {\mathrm{keV}} 

\newcommand{\chisq}{\Delta \chi^2_\mathrm{eff}}
\newcommand{\kpiv}{k_{\mathrm{pivot}}}
\newcommand{\Mpl}{M_{\mathrm{pl}}}
\newcommand{\aend}{a_{\mathrm{end}}}
\newcommand{\aeq}{a_{\mathrm{eq}}}
\newcommand{\arh}{a_{\mathrm{th}}}
\newcommand{\wprim}{w_{\mathrm{prim}}}
\newcommand{\wint}{w_{\mathrm{int}}}
\newcommand{\rhorh}{\rho_{\mathrm{th}}}
\newcommand{\rhoeq}{\rho_{\mathrm{eq}}}
\newcommand{\rhoend}{\rho_{\mathrm{end}}}
\newcommand{\vend}{V_{\mathrm{end}}}
\newcommand{\ModeCode}{{\tt ModeCode}}
\newcommand{\MultiNest}{{\tt MultiNest}}
\newcommand{\CAMB}{{\tt CAMB}}
\newcommand{\CosmoMC}{{\tt CosmoMC}}
\newcommand{\codename}{{\ModeCode}}

\def\gtorder{\mathrel{\raise.3ex\hbox{$>$}\mkern-14mu
             \lower0.6ex\hbox{$\sim$}}}
\def\ltorder{\mathrel{\raise.3ex\hbox{$<$}\mkern-14mu
             \lower0.6ex\hbox{$\sim$}}}

\abstract
{ 
\def\ltorder{\mathrel{\raise.3ex\hbox{$<$}\mkern-14mu
             \lower0.6ex\hbox{$\sim$}}}

We analyse the implications of the \Planck\ data for cosmic
inflation. The \Planck\ nominal mission temperature anisotropy measurements,
combined with the {\it WMAP} large-angle polarization, constrain 
the scalar spectral index to be $n_\mathrm{s}=0.9603 \pm 0.0073$, 
ruling out exact scale invariance at over 5$\sigma.$ 
\Planck\ establishes an upper bound on 
the tensor-to-scalar ratio of $r<0.11$ (95\% CL).
The \Planck\ data thus shrink the space of allowed standard inflationary models, 
preferring potentials with $V^{\prime \prime }<0$.
Exponential potential models, the simplest hybrid inflationary models, 
and monomial potential models of degree $n \ge 2$ do not provide a good fit to the data.
\Planck\ does not find statistically significant running of the scalar spectral index, obtaining 
$\mathrm{d} n_\mathrm{s} / \mathrm{d} \ln k = -0.0134 \pm 0.0090$.
We verify these conclusions through a numerical analysis, which makes no slow-roll approximation,
and carry out a Bayesian parameter estimation and model-selection analysis for a number of inflationary models
including monomial, natural, and hilltop potentials. 
For each model, we present the 
\Planck\ constraints on the parameters of the potential and explore several possibilities for the 
post-inflationary entropy generation epoch,
thus obtaining nontrivial data-driven constraints. We also present a direct reconstruction of the observable
range of the inflaton potential. Unless a quartic term is allowed in the potential,
we find results consistent with second-order slow-roll predictions.
We also investigate whether the primordial power spectrum 
contains any features.
We find that models with a parameterized oscillatory feature improve the fit by 
$\Delta \chi^2_\mathrm{eff} \approx 10$; however, Bayesian evidence does not prefer these models.
We constrain several single-field inflation models with generalized Lagrangians by combining power spectrum data with \Planck\ bounds 
on $f_\mathrm{NL}$. 
\Planck\ constrains with unprecedented accuracy the amplitude and possible 
correlation (with the adiabatic mode) of non-decaying isocurvature fluctuations. 
The fractional primordial contributions of cold dark matter (CDM)
isocurvature modes of the types expected in the curvaton and axion scenarios have
upper bounds of 0.25\% and 3.9\% (95\%~CL), respectively.
In models with arbitrarily correlated CDM or neutrino isocurvature modes, an
anticorrelated isocurvature component can improve the $\chi^2_\mathrm{eff}$ by approximately $4$ 
as a result of slightly lowering the theoretical prediction for the $\ell \ltorder 40$ multipoles relative to the higher
multipoles.  Nonetheless, the data are consistent with adiabatic initial conditions.

}

}

\keywords{Cosmology: theory -- early Universe -- inflation}

\authorrunning{Planck Collaboration}
\titlerunning{Constraints on inflation}

\maketitle

\section{Introduction}
\label{sec:introduction}

{
This paper, one of a set associated with the 2013 release 
of data from the \Planck\footnote{\Planck\ (\url{http://www.esa.int/Planck}) is a project of the European Space
Agency (ESA) with instruments provided by two scientific consortia funded by ESA member
states (in particular the lead countries France and Italy), with contributions from NASA
(USA) and telescope reflectors provided by a collaboration between ESA and a scientific
consortium led and funded by Denmark.} mission 
\all2013resultspapers
(\cite{planck2013-p01}--\cite{planck2013-p01a}),
describes the implications of the
\Planck\ measurement of cosmic microwave background (CMB) anisotropies for cosmic inflation.
In this first release only the \Planck\ temperature data resulting from the {\it nominal} mission are used, 
which includes $2.6$ full surveys of the sky.
The interpretation of the CMB polarization as seen by \Planck\ will be presented in a later series of publications.
This paper exploits the data presented in 
\cite{planck2013-p02},
\cite{planck2013-p06},
\cite{planck2013-p08}, and 
\cite{planck2013-p12}. 
Other closely related papers discuss the estimates of cosmological parameters in \cite{planck2013-p11}
and investigations of non-Gaussianity in \cite{planck2013-p09a}.

In the early 1980s inflationary cosmology, which postulates an
epoch of nearly exponential expansion, was proposed in order to
resolve a number of puzzles of standard big bang cosmology 
such as the entropy, flatness, horizon, smoothness, and monopole problems 
\citep{Brout:1977ix, Starobinsky:1980te, Kazanas:1980tx, Sato1981,
Guth:1980zm, Linde:1981mu, Albrecht:1982wi, Linde:1983gd}.
During inflation,
cosmological fluctuations resulting from quantum fluctuations
are generated and can be calculated using the semiclassical
theory of quantum fields in curved spacetime 
\citep{Mukhanov:1981xt, Mukhanov:1982nu,Hawking:1982cz,Guth:1982ec,Starobinsky:1982ee,Bardeen:1983qw,Mukhanov:1985rz}.

Cosmological observations prior to {\Planck} are consistent with 
the simplest models of inflation within the slow-roll paradigm.
Recent observations of the CMB anisotropies 
\citep{Story:2012wx,Bennett:2012fp,Hinshaw:2012fq,Hou:2012xq,Das:2013zf} 
and of large-scale structure \citep{Beutler:2011hx,Padmanabhan:2012hf,Anderson:2012sa} 
indicate that our Universe is very close to spatially flat and has 
primordial density fluctuations that are nearly Gaussian and adiabatic and are described 
by a nearly scale-invariant power spectrum. 
Pre-{\Planck} CMB observations also 
established that the amplitude of primordial gravitational waves, with a nearly scale-invariant spectrum 
\citep{Starobinsky:1979ty,Rubakov:1982df,Fabbri:1983us}, is at most small.

Most of the results in this paper are based on the two-point statistics of the CMB as 
measured by \Planck, 
exploiting the data presented in \citet{planck2013-p08}, \citet{planck2013-p11}, and \citet{planck2013-p12}. 
The \Planck\ results testing the Gaussianity 
of the primordial CMB component are described in the companion papers 
\citet{planck2013-p09}, 
\citet{planck2013-p09a}, and \citet{planck2013-p20}. 
\Planck\ finds values for 
the non-Gaussian $f_\mathrm{NL}$ parameter of the CMB bispectrum consistent with the Gaussian hypothesis 
\citep{planck2013-p09a}. 
This result has important implications for inflation. 
The simplest slow-roll inflationary models 
predict a level of 
$f_\mathrm{NL}$ of the same order as the slow-roll parameters and therefore too small to be detected by \Planck. 

The paper is organized as follows. Section \ref{sec:paradigm} reviews 
inflationary theory, emphasizing in particular those aspects used later in the paper.
In Sect. \ref{sec:methodology} the statistical methodology 
and the \Planck\ likelihood as well as the likelihoods
from the other astrophysical data sets used here are described. Section \ref{sec:slow-roll} 
presents constraints on slow-roll inflation 
and studies their robustness 
under generalizations of the minimal 
assumptions of our baseline cosmological model. 
In Sect.~\ref{sec:modelcomp} Bayesian model comparison of several inflationary models is carried out taking into account
the uncertainty from the end of inflation to the beginning of the radiation dominated era.
Section \ref{sec:numerical} reconstructs the inflationary potential over the range corresponding to the scales observable in the CMB. In 
Sect.~\ref{sec:pk} a penalized likelihood reconstruction of the primordial perturbation spectrum is performed. Section 
\ref{sec:oscillations} reports on a parametric search for oscillations and features in the primordial
scalar power spectrum. Section \ref{sec:putfnl} examines constraints on non-canonical single-field models 
of inflation including the $f_\mathrm{NL}$ measurements from \citet{planck2013-p09a}. 
In Sect. \ref{sec:isometho} 
constraints on isocurvature modes are established, thus 
testing the hypothesis that initial conditions were solely
adiabatic.  We summarize our conclusions in Sect. \ref{sec:conclusions}.  
Appendix A is dedicated to the constraints on slow-roll inflation derived by 
sampling the Hubble flow functions (HFF) in the analytic expressions for the scalar 
and tensor power spectra. Definitions of the most relevant symbols used in this paper 
can be found in Tables~\ref{table:CPDefinitions} and \ref{table:InflationDefinitions}.


}

\section{Lightning review of inflation}
\label{sec:paradigm}

{

\def\ba{\begin{eqnarray}}
\def\ea{\end{eqnarray}}

\begin{table*}[tmb]
\begingroup
\newdimen\tblskip \tblskip=5pt
\caption{Cosmological parameter definitions}
\label{table:CPDefinitions}
\nointerlineskip
\vskip -3mm
\footnotesize
\setbox\tablebox=\vbox{
      \newdimen\digitwidth
      \setbox0=\hbox{\rm 0}
      \digitwidth=\wd0
      \catcode`"=\active
      \def"{\kern\digitwidth}
      \newdimen\signwidth
      \setbox0=\hbox{+}
      \signwidth=\wd0
      \catcode`!=\active
      \def!{\kern\signwidth}
\halign{\hbox to 1.0in{$#$\leaderfil}\tabskip=3em&

\vtop{\hsize=11cm\hangafter=1\hangindent=1.5em\strut\noindent 
#\strut\par}\tabskip=4pt\cr
\noalign{\doubleline}
\omit\hfil Parameter\hfil&\omit\hfil Definition\hfil\cr
\noalign{\vskip 3pt\hrule\vskip 5pt}
\Omega_\mathrm{b}        &Baryon fraction today (compared to critical density)\cr
\Omega_\mathrm{c}           &Cold dark matter fraction today (compared to critical density)\cr
h                   & Current expansion rate (as fraction of 100~km s$^{-1}$ Mpc$^{-1}$)\cr
\theta_{\mathrm MC}              &Approximation to the angular size of sound horizon at last scattering\cr
\tau                  &Thomson scattering optical depth of reionized intergalactic medium \cr
N_\mathrm{eff}           &Effective number of massive and massless neutrinos \cr 
\Sigma m_\nu          &Sum of neutrino masses \cr 
Y_\mathrm{P}            &Fraction of baryonic mass in primordial helium\cr
\Omega_K              &Spatial curvature parameter \cr
w_\mathrm{de}                     &Dark energy equation of state parameter (i.e., $p/\rho $) (assumed constant)\cr
{\cal R}              &Curvature perturbation \cr
{\cal I}              &Isocurvature perturbation \cr
\mathcal{P}_X ={k^3 |X_k|^2}/{2 \pi^2}&Power spectrum of $X$ \cr
A_X &$X$ power spectrum amplitude (at $k_* = 0.05 $ Mpc$^{-1}$)\cr
n_\mathrm{s}                   &Scalar spectrum spectral index (at $k_* = 0.05 $ Mpc$^{-1}$, unless otherwise stated)\cr
\mathrm{d}n_\mathrm{s}/\mathrm{d} \ln k            &Running of scalar spectral index 
(at $k_* = 0.05 $ Mpc$^{-1}$, unless otherwise stated) \cr
\mathrm{d}^2 n_\mathrm{s}/\mathrm{d} \ln k^2            &Running of running of scalar spectral index 
(at $k_* = 0.05 $ Mpc$^{-1}$) \cr
r                     &Tensor-to-scalar power ratio (at $k_* = 0.05 $ Mpc$^{-1}$, unless otherwise stated)\cr
n_\mathrm{t}                   & Tensor spectrum spectral index (at $k_* = 0.05 $ Mpc$^{-1}$)\cr
\mathrm{d}n_\mathrm{t}/\mathrm{d} \ln k    & Running of tensor spectral index (at $k_* = 0.05 $ Mpc$^{-1}$) \cr
\noalign{\vskip 5pt\hrule\vskip 3pt}}}
\endPlancktable
\endgroup
\end{table*}

Before describing cosmic inflation, which was developed in the early 1980s, it
is useful to review the state of theory prior to its introduction. \cite{Lifshitz:1945du} (see also
\cite{Lifshitz:1963ps}) first wrote down and solved the equations for the evolution of linearized 
perturbations about a
homogeneous and isotropic Friedmann-Lema\^itre-Robertson-Walker spacetime within the 
framework of general relativity.
The general framework adopted was based on two assumptions:
\begin{enumerate}
\item[(i)]
The cosmological perturbations can be described by a single-component fluid,
at very early times.

\item[(ii)]
The initial cosmological perturbations were statistically
homogeneous and isotropic, and Gaussian.
\end{enumerate}
These are the simplest---but by no means unique---assumptions for defining a stochastic
process for the initial conditions. Assumption (i), where only a single
{\it adiabatic} mode is excited, is just the simplest possibility. 
In Sect.~\ref{sec:isometho} we shall describe isocurvature perturbations, 
where other available modes are excited,
and report on the constraints established by \Planck.
Assumption (ii) is {\it a priori} more questionable given the understanding
at the time. An appeal can be made to the fact that any physics at weak coupling
could explain (ii), but at the time these assumptions were somewhat {\it ad hoc}.

Even with the strong assumptions (i) and (ii), comparisons with observations cannot be made
without further restrictions on the functional form of the primordial power spectrum of large-scale 
spatial curvature inhomogeneities $\mathcal{R}$,
$\mathcal{P_R} (k) \propto k^{n_\mathrm{s} - 1}$, where $n_\mathrm{s}$ is the (scalar) spectral index.
The notion of a {\it scale-invariant} (i.e., $n_\mathrm{s}=1$) primordial power spectrum
was introduced by \citet{Harrison:1969fb}, \citet{Zeldovich:1972zz}, and \citet{1970ApJ...162..815P} to address 
this problem. 
These authors showed that a scale-invariant
power law was consistent with the crude constraints on large- and small-scale perturbations available
at the time. However, other than its mathematical simplicity,
no compelling theoretical explanation for this Ansatz was put forth. An important current 
question, addressed in Sect.~\ref{sec:slow-roll}, is whether $n_\mathrm{s}=1$ (i.e., exact scale 
invariance) is consistent with 
the data, or whether there is convincing evidence for small deviations from exact scale invariance.
Although the inflationary potential can be tuned to obtain $n_\mathrm{s}=1$, inflationary models 
generically predict 
deviations from $n_\mathrm{s}=1$, usually on the red side (i.e, $n_\mathrm{s}<1$).

\subsection{Cosmic inflation}

Inflation was developed in a series of papers by
\cite{Brout:1977ix},
\cite{Starobinsky:1980te}, 
\cite{Kazanas:1980tx},
\cite{Sato1981},
\cite{Guth:1980zm},
\cite{Linde:1981mu, Linde:1983gd},
and 
\cite{Albrecht:1982wi}. 
By generating an equation of state with a significant negative pressure
(i.e., $w=p/\rho \approx -1$) before the radiation epoch, inflation solves a number of
cosmological conundrums (the monopole, horizon, smoothness, and entropy
problems), which had plagued all cosmological models extrapolating
a matter-radiation equation of state all the way back to the singularity.
Such an equation of state ($p \approx -\rho$) and the resulting nearly exponential expansion 
are obtained from a scalar field, the \emph{inflaton}, with a 
canonical kinetic term (i.e., $\frac{1}{2}(\partial \phi )^2$), slowly rolling 
in the framework of Einstein gravity.

The homogeneous evolution of the inflaton field $\phi$ is governed by the equation of motion 
\begin{equation}
\ddot \phi (t) + 3 H(t) \dot \phi (t) + V_{ \phi } = 0,
\label{InfZero}
\end{equation}
and the Friedmann equation
\begin{equation}
H^2
=\frac{1}{3{M_\mathrm{pl}}^2}\left(\frac{1}{2}\dot \phi ^2 +V (\phi) \right).
\label{InfZeroBis}
\end{equation}
Here $H =\dot a/a$ is the Hubble parameter, the subscript ${\phi}$ denotes 
the derivative with respect to $\phi,$  
$M_\mathrm{pl}=(8\pi G)^{-1/2}$ is the reduced Planck mass, and $V$ is the potential.
(We use units where $c=\hbar =1.$)
The evolution during the stage of quasi-exponential expansion, when
the scalar field rolls slowly down the potential, 
can be approximated by neglecting the second time derivative in 
Eq.~\ref{InfZero} and the kinetic energy term in Eq.~\ref{InfZeroBis}, so that
\begin{align}
3 H(t) \dot \phi (t) &\approx  - V_{ \phi } \, ,  \label{slowrollphi} \\
H^2 &\approx  \frac{V(\phi)}{3{M_\mathrm{pl}}^2} \,. \label{slowrollH}
\end{align}
Necessary conditions for the {\em slow-roll} described above are 
$\epsilon_V \ll 1$ and $|\eta_V| \ll 1$, where 
the slow-roll parameters $\epsilon_V$ and $\eta_V$ are defined as
\begin{align}
\label{epsilon_def}
\epsilon_V &=\frac{M_\mathrm{pl}^2 V_{ \phi}^2}{2 V^2} \,, \\
\label{eta_def}
\eta_V &= \frac{M_\mathrm{pl}^2 V_{ \phi \phi}}{V} \,.
\end{align}

The analogous hierarchy of HFF slow-roll parameters measures instead the deviation
from an exact exponential expansion.
This hierarchy is defined as $\epsilon_1 = - \dot H/H^2$,
$\epsilon_{i+1} \equiv \dot \epsilon_i/(H \epsilon_i)$, with $i \ge 1$. By using 
Eqs.~\ref{slowrollphi} and \ref{slowrollH}, 
we have that 
$\epsilon_1 \approx \epsilon_V$, $\epsilon_2 \approx -2 \eta_V + 4 \epsilon_V$.

\begin{table*}[tmb]
\begingroup
\newdimen\tblskip \tblskip=5pt
\caption{Conventions and definitions for inflation physics}
\label{table:InflationDefinitions}
\nointerlineskip
\vskip -3mm
\footnotesize
\setbox\tablebox=\vbox{
      \newdimen\digitwidth
      \setbox0=\hbox{\rm 0}
      \digitwidth=\wd0
      \catcode`"=\active
      \def"{\kern\digitwidth}
      \newdimen\signwidth
      \setbox0=\hbox{+}
      \signwidth=\wd0
      \catcode`!=\active
      \def!{\kern\signwidth}
\halign{\hbox to 1.2in{$#$\leaderfil}\tabskip=3em&

\vtop{\hsize=11cm\hangafter=1\hangindent=1.5em\strut\noindent 
#\strut\par}\tabskip=4pt\cr
\noalign{\doubleline}
\omit\hfil Parameter\hfil&\omit\hfil Definition\hfil\cr
\noalign{\vskip 3pt\hrule\vskip 5pt}
\phi       & Inflaton \cr
V(\phi)         & Inflaton potential \cr
a          & Scale factor  \cr
t	& Cosmic (proper) time \cr
\delta X   & Fluctuation of $X$ \cr
\dot{X} = \mathrm{d} X/\mathrm{d} t  & Derivative with respect to proper time \cr
X' = \mathrm{d} X /\mathrm{d} \eta  & Derivative with respect to conformal time \cr 
X_\phi  = \partial X /\partial \phi & Partial derivative with respect to $\phi$\cr
M_\mathrm{pl}  & Reduced Planck mass ($=2.435 \times 10^{18}$ GeV) \cr
Q               & Scalar perturbation variable\cr
h^{+,\times}              & Gravitational wave amplitude of $(+,\times)$-polarization component \cr 
X_*             & $X$ evaluated at Hubble exit during inflation of mode with wavenumber $k_*$ \cr
X_{\mathrm e}             & $X$ evaluated at end of inflation \cr
\epsilon _V={M_\mathrm{pl}^2 V_\phi^2}/{2 V^2}         & First slow-roll parameter for $V (\phi)$ \cr
\eta _V={M_\mathrm{pl}^2 V_{\phi \phi}}/{V}           & Second slow-roll parameter for $V(\phi)$ \cr
\xi^2_V={M_\mathrm{pl}^4 V_{\phi} V_{\phi \phi \phi}}/{V^2}  & Third slow-roll parameter for $V (\phi)$ \cr
\varpi^3_V  ={M_\mathrm{pl}^6 V_{\phi}^2 V_{\phi \phi \phi \phi}}/{V^3}  & Fourth slow-roll parameter for $V (\phi)$ \cr      
\epsilon_1 = - {\dot H}/{H^2}     & First Hubble hierarchy parameter \cr
\epsilon_{n+1} ={\dot \epsilon_n}/{H \epsilon_n}  & $(n+1)$th Hubble hierarchy parameter (where $n \ge 1$) \cr
N (t) =\int_{t}^{t_{\mathrm e}} \! \mathrm{d}t \; H& Number of $e$-folds to end of inflation \cr
\delta \sigma          & Curvature field perturbation \cr
\delta s               & Isocurvature field perturbation \cr 
\noalign{\vskip 5pt\hrule\vskip 3pt}}}
\endPlancktable
\endgroup
\end{table*}

\subsection{Quantum generation of fluctuations}
\label{FluctGeneration}

Without quantum fluctuations, inflationary theory would fail. 
Classically,
any initial spatial curvature or gradients in the scalar field, as well as any inhomogeneities
in other fields, would rapidly decay away during the quasi-exponential
expansion. The resulting universe would be too homogeneous and 
isotropic compared with observations. Quantum fluctuations must exist in order to satisfy the uncertainty relations that follow
from the canonical commutation relations of quantum field theory. The quantum fluctuations in the inflaton and in  
the transverse and traceless parts of the metric are amplified by the nearly exponential expansion yielding 
the scalar and tensor primordial power spectra, respectively.

Many essentially equivalent approaches to quantizing the linearized cosmological fluctuations can be found in 
the original literature \citep[see, e.g.,][]{Mukhanov:1981xt,Hawking:1982cz,Guth:1982ec,Starobinsky:1982ee,Bardeen:1983qw}.
A simple formalism, which we shall follow here, was introduced by \citet{Mukhanov:1988jd}, \citet{Mukhanov:1990me},
and \citet{Sasaki:1986hm}. In this approach a gauge-invariant inflaton fluctuation $Q$ is constructed and canonically quantized. 
This gauge-invariant variable $Q$ is the inflaton fluctuation $\delta \phi (t,x)$ in the uniform curvature gauge.
The mode function of the inflaton fluctuations $\delta \phi (t,x)$ obeys the evolution equation
\be
(a \delta \phi_k)'' + 
\left( k^2 - \frac{z''}{z}  \right) (a \delta \phi_k )= 0,
\label{Scalar:Evolution}
\ee
where $z=a \dot \phi/H$. 
The gauge-invariant field fluctuation is directly related to the comoving curvature 
perturbation\footnote{Another important quantity is the curvature perturbation on uniform density hypersurfaces
$\zeta$ (in the Newtonian gauge, $\zeta = - \psi - H \delta \rho/\dot \rho$, where $\psi$ is the generalized gravitational potential),
which is related to the perturbed spatial curvature according to $^{(3)}R = - 4 \nabla^2 \zeta/a^2$. On large 
scales $\zeta \approx {\cal R}$.}
\be
{\cal R} = - H \frac{\delta \phi}{\dot \phi} \,.  
\label{curv-def}
\ee

Analogously, gravitational waves are described by the two polarization states ($+,\times$)
of the transverse traceless parts of the metric fluctuations and are amplified by the expansion of the universe as 
well \citep{Grishchuk:1974ny}.  
The evolution equation for their mode function is
\be
(a h^{+,\times}_k)'' + \left( k^2 - \frac{a''}{a}  \right) (a h^{+,\times}_k)= 0.
\label{Tensor:Evolution}
\ee
Early discussions of the generation of gravitational waves during 
inflation include
\cite{Starobinsky:1979ty}, \cite{Rubakov:1982df}, 
\cite{Fabbri:1983us}, \cite{Abbott:1984fp}, and \cite{Starobinsky:1985ww}.

Because the primordial perturbations are small, of order $10^{-5}$, the linearized 
Eqs.~\ref{Scalar:Evolution} and 
\ref{Tensor:Evolution} provide an accurate description for the generation
and subsequent evolution of the cosmological perturbations during inflation.
In this paper we use two approaches for solving for the cosmological
perturbations. Firstly, we use an approximate treatment based on the {\it slow-roll}
approximation described below. Secondly, we use an almost exact approach based
on numerical integration of the ordinary differential equations \ref{Scalar:Evolution} and
\ref{Tensor:Evolution} for each value of the comoving wavenumber $k.$ For fixed $k$
the evolution may be divided into three epochs: (i) sub-Hubble evolution, (ii) Hubble crossing evolution,
and (iii) super-Hubble evolution. During (i) the wavelength is much smaller than the Hubble length,
and the mode oscillates as it would in a non-expanding universe (i.e., Minkowski space). Therefore
we can proceed with quantization as we would in Minkowski space. We
quantize by singling out the positive frequency solution, as in the Bunch-Davies vacuum~\citep{Bunch:1978yq}.
This epoch is the oscillating regime in the WKB approximation. In epoch (iii), by contrast,
there are two solutions, a growing and a decaying mode, and the evolution becomes independent
of $k.$ We care only about the growing mode. 
On scales much larger than the Hubble radius (i.e., $k \ll a H$), both curvature and tensor fluctuations admit 
solutions constant in time.\footnote{On 
large scales, the curvature fluctuation is constant in time when non-adiabatic pressure terms are negligible. This 
condition is typically violated in multi-field inflationary models.}
All the interesting, or nontrivial, evolution 
takes place between epochs (i) and (iii)---that is, during (ii), a few $e$-folds before and after Hubble 
crossing, and this is the interval where the numerical integration is most useful since the asymptotic expansions
are not valid in this transition region. Two numerical codes are used in this paper, 
{\tt ModeCode}~\citep{2001PhRvD..64l3514A,2003ApJS..148..213P, 2009PhRvD..79j3519M, 2012PhRvD..85j3533E}, 
and the inflation module of \citet{Lesgourgues:2007gp}
as implemented in {\tt CLASS} \citep{Lesgourgues:2011re,Blas:2011rf}.\footnote{\tt \tiny http://zuserver2.star.ucl.ac.uk/\~{ }hiranya/ModeCode/{\rm ,} http://class-code.net}

It is convenient to expand the power spectra of curvature and tensor perturbations on super-Hubble scales as
\begin{align}
\mathcal{P}_{\cal R}(k) &= A_\mathrm{s} \left( \frac{k}{k_*}\right)^{n_\mathrm{s}-1 + 
\frac{1}{2} \, \mathrm{d}n_\mathrm{s}/\mathrm{d}\ln k \ln(k/k_*) + \frac{1}{6} \, 
\mathrm{d}^2n_\mathrm{s}/\mathrm{d}\ln k^2  \left( \ln(k/k_*) \right)^2 + ...}, \label{scalarps}\\
\mathcal{P}_\mathrm{t}(k) &= A_\mathrm{t} \left( \frac{k}{k_*}\right)^{n_\mathrm{t} + \frac{1}{2} \, 
\mathrm{d}n_\mathrm{t}/\mathrm{d}\ln k \ln(k/k_*) + ... }, 
\label{tensorps}
\end{align}
where $A_\mathrm{s} \, (A_\mathrm{t})$ is the scalar (tensor) amplitude and $n_\mathrm{s} \, 
(n_\mathrm{t})$, $\mathrm{d}n_\mathrm{s}/\mathrm{d}\ln k \, (\mathrm{d}n_\mathrm{t}/\mathrm{d}\ln k)$ 
and $\mathrm{d}^2n_\mathrm{s}/\mathrm{d}\ln k^2$ are the scalar (tensor) spectral index, 
the running of the scalar (tensor) spectral index, and the running of the running of the scalar spectral index, respectively. 

The parameters of the scalar and tensor power spectra may be calculated approximately in the framework 
of the slow-roll approximation
by evaluating the following equations at the value of the inflation field $\phi _*$ where the mode
$k_* = a_* H_*$ crosses the Hubble radius for the first time. (For 
a nice review of the slow-roll approximation, see for example \cite{Liddle:1993fq}). The number of $e$-folds before the end of 
inflation, $N_*$, at which the
pivot scale $k_*$ exits from the Hubble radius, is
\begin{equation}
N_* =  \int_{t_*}^{t_{\mathrm e}} \! \mathrm{d}t \; H
\approx \frac{1}{M_\mathrm{pl}^2}
\int_{\phi_*}^{\phi_{\mathrm e}} \!  \mathrm{d}\phi \; \frac{V}{V_\phi} ,
\label{eq:nefolds1}
\end{equation}
where the equality holds in the slow-roll approximation, and subscript $\mathrm{e}$ denotes the end of inflation.

The coefficients of Eqs.~\ref{scalarps} and \ref{tensorps} at 
their respective leading orders in the slow-roll parameters are given by 
\begin{align}
A_\mathrm{s} & \approx  \frac{V}{24 \pi^2 M_\mathrm{pl}^4 \epsilon_V}, \label{eq:as_def} \\
A_\mathrm{t} & \approx  \frac{2 V}{3 \pi^2 M_\mathrm{pl}^4}, \label{eq:at_def} \\
n_\mathrm{s} -1 &\approx 2 \eta _V- 6 \epsilon _V, \label{eq:ns_def} \\
n_\mathrm{t} &\approx - 2 \epsilon_V,\\
\mathrm{d}n_\mathrm{s}/\mathrm{d}\ln k & \approx  +16 \epsilon _V\eta _V- 24 \epsilon^2_V - 2 \xi^2_V, \label{eq:alphas_def} \\
\mathrm{d}n_\mathrm{t}/\mathrm{d}\ln k& \approx  + 4 \epsilon_V \eta_V - 8 \epsilon^2_V, \label{eq:alphat_def} \\
\begin{split}
\mathrm{d}^2n_\mathrm{s}/\mathrm{d}\ln k^2  & \approx  -192 \epsilon^3_V + 192 \epsilon^2_V \eta _V- 
32 \epsilon _V\eta^2_V\\  
& \quad - 24 \epsilon _V\xi^2_V + 2 \eta _V\xi^2_V + 2 \varpi^3_V \label{eq:betas_def},
\end{split}
\end{align}
where the {\em slow-roll parameters} $\epsilon _V$ and $\eta _V$ are defined in Eqs.~\ref{epsilon_def} and \ref{eta_def},
and the higher order parameters are defined as 
\begin{equation}
\xi^2_V =\frac{M_\mathrm{pl}^4 V_{\phi} V_{\phi \phi \phi}}{V^2}
\end{equation}
and
\begin{equation}
\varpi^3 _V=\frac{M_\mathrm{pl}^6 V_{\phi}^2 V_{\phi \phi \phi \phi}}{V^3}.
\end{equation}

In single-field inflation with a standard kinetic term, as discussed here, the tensor
spectrum shape is not independent from the other parameters.
The slow-roll paradigm implies a tensor-to-scalar 
ratio at the pivot scale of
\begin{equation}
r = \frac{\mathcal{P}_{\mathrm t}(k_*)}{\mathcal{P}_{\cal R}(k_*)} \approx  
16 \epsilon _V\approx -8 n_\mathrm{t} \,,
\end{equation}
referred to as the consistency relation.
This consistency relation is also useful to help understand how $r$ is connected to the evolution of the inflaton:
\begin{equation}
\frac{\Delta \phi}{M_\mathrm{pl}} \approx \frac{1}{\sqrt{8}} \int^N_0 \mathrm{d} N \sqrt{r} \,.
\end{equation}
The above relation, called the Lyth bound \citep{Lyth:1996im}, 
implies that an inflaton variation of the order of the Planck mass is needed to produce
$r \gtrsim 0.01$. Such a threshold is useful to classify large- and small-field inflationary models with respect to the Lyth bound. 

\subsection{Ending inflation and the epoch of entropy generation}
\label{EntropyGeneration}

The greatest uncertainty in calculating the perturbation spectrum predicted
from a particular inflationary potential arises
in establishing the correspondence between the comoving wavenumber
today and the inflaton energy density when the mode of that
wavenumber crossed the Hubble radius during inflation \citep{2006JCAP...03..011K}.
This correspondence depends both on the inflationary model and
on the cosmological evolution from the end of inflation to the present.

After the slow-roll stage, $\ddot \phi $ becomes
as important as the {\it cosmological damping} term $3H\dot \phi$. 
Inflation ends gradually as the inflaton picks up kinetic energy
so that $w$ is no longer slightly above $-1$, but rather far from that
value. We may arbitrarily deem that inflation ends when $w=-1/3$ 
(the value dividing the cases of 
an expanding and a contracting comoving Hubble radius), or,
equivalently, at $\epsilon_V\approx 1,$ 
after which the epoch of entropy generation starts.
Because of couplings to other fields, the energy initially in the form
of scalar field vacuum energy is transferred to the other fields by 
perturbative decay (reheating), possibly
preceded by a non-perturbative stage 
(preheating). There is considerable uncertainty about the mechanisms of 
entropy generation, or thermalization, which subsequently lead to a standard $w=1/3$ equation of state for radiation.
 
On the other hand, if we want to identify some $k_*$ today with the value of 
the inflaton field at the time this scale left the Hubble radius,  Eq.~\ref{eq:nefolds1}
needs to be matched to an expression that 
quantifies how much $k_*$ has shrunk relative to the size of the Hubble radius between 
the end of inflation and the time when that mode re-enters the Hubble radius. This quantity depends both 
on the inflationary potential and the details of the entropy 
generation process and is given by
\begin{equation}
\begin{aligned}
N_* \approx &  \; 67
- \ln \left(\frac{k_*}{a_0 H_0}\right) + \frac{1}{4}\ln \left( \frac{V_*}{\Mpl^4}\right)  
+  \frac{1}{4}\ln \left( \frac{V_*}{\rhoend}\right)  \\
& + \frac{1-3w_\mathrm{int}}{12(1+w_\mathrm{int})} 
\ln{\left(\frac{\rhorh}{\rhoend}\right)}  - \frac{1}{12} \ln (g_\mathrm{th}) \; ,
\label{eq:nefolds}
\end{aligned}
\end{equation}
where $\rhoend$ is the energy density at the end of inflation, 
$\rhorh$ is an energy scale by which the universe has thermalized, 
$a_0 H_0$ is the present Hubble radius, 
$V_*$ is the potential energy when $k_*$ left the Hubble radius 
during inflation, $w_\mathrm{int}$ characterizes the effective equation of state 
between the end of inflation and the energy scale specified by $\rhorh$, 
and $g_\mathrm{th}$ is the number of effective bosonic degrees of freedom at the energy scale $\rhorh$.
In predicting the primordial power spectra at observable scales for a specific inflaton potential, 
this uncertainty in the reheating history of the universe becomes relevant and can be taken into 
account by allowing $N_*$ to vary over a range of values.  Note that $\wint$ is not intended to 
provide a detailed model for entropy generation, but rather to parameterize the uncertainty 
regarding the expansion rate of the universe during this intermediate era. Nevertheless, 
constraints on $\wint$ provide observational limits on the uncertain physics during this period. 

The first two terms of Eq.~\ref{eq:nefolds} are model independent, with the second
term being roughly $5$ for $k_* = 0.05~\mathrm{Mpc}^{-1}$.  If thermalization occurs 
rapidly, or if the reheating stage is close to radiation-like, the magnitude of the
last term in Eq.~\ref{eq:nefolds} is less than roughly unity. 
The magnitude of the $\ln (g_{th})/12$ term is negligible, giving a shift
of only 0.58 for the extreme value $g_{th}=10^3.$
For most reasonable inflation models, the fourth term is $\mathcal{O}(1)$ and the third 
term is approximately $ - 10$, motivating the commonly assumed range 
$50 < N_* <60$.  Nonetheless, more extreme values at both ends are in principle 
possible~\citep{Liddle:2003as}.
In the figures of Sect.~\ref{sec:slow-roll} we will mark the range 
$50 < N_* <60$ as a general guide.


\subsection{Perturbations from cosmic inflation at higher order}

To calculate the quantum fluctuations generated during cosmic inflation,
a linearized quantum field theory in a time-dependent background
can be used. 
The leading order is the 
two-point correlation function
\be
\left< {\cal R}(\mathbf{k}_1)~{\cal R}(\mathbf{k}_2)
\right>
=(2\pi )^3~\frac{2\pi^2}{k^3} \mathcal{P}_{\cal R}(k)~\delta^3({\mathbf{k_1}}+{\mathbf{k_2}}),
\ee
but the inflaton self-interactions and the nonlinearity of Einstein gravity give small higher-order
corrections, of which the next-to-leading order is the three point function
\be
\left< {\cal R}(\mathbf{k_1})~{\cal R}(\mathbf{k}_2)~{\cal R}(\mathbf{k}_3)
\right> =(2\pi )^3 B_{{\cal R}} (k_1,k_2,k_3) \delta^3({\mathbf{k_1}}+{\mathbf{k_2}}+{\mathbf{k_3}}) \, , 
\ee
which is in general non-zero.

For single-field inflation with a standard kinetic term in a smooth potential 
(with initial fluctuations in the Bunch-Davies vacuum), the non-Gaussian 
contribution to the curvature perturbation during inflation is ${\cal O} (\epsilon_V \,, \eta_V)$ 
\citep{2003NuPhB.667..119A,2003JHEP...05..013M}, i.e.,  
at an undetectable level smaller than  
other general relativistic contributions, such as the cross-correlation between the integrated Sachs-Wolfe effect and 
weak gravitational lensing of the CMB. For a general scalar field 
Lagrangian, the non-Gaussian contribution can be large enough to be accessible to \Planck\
with $f_\mathrm{NL}$ of order $c_\mathrm{s}^{-2}$ \citep{2007JCAP...01..002C}, 
where $c_\mathrm{s}$ is the sound speed of inflaton fluctuations 
(see Sect.~\ref{sec:putfnl}). Other higher order 
kinetic and spatial derivative 
terms contribute to larger non-Gaussianities. 
For a review of non-Gaussianity generated during inflation, 
see, for example, \citet{Bartolo:2004if} and \citet{2010AdAst2010E..72C} as well as the companion paper \citet{planck2013-p09a}.

\subsection{Multi-field models of cosmic inflation}

Inflation as described so far assumes a single scalar field that drives and terminates
the quasi-exponential expansion and also 
generates the large-scale curvature perturbations. 
When there is more than one field with an effective mass smaller than $H$,    
isocurvature perturbations are also generated during inflation by the same mechanism of 
amplification due to the stretching of the spacetime geometry 
\citep{Axenides:1983hj,Linde:1985yf}. 
Cosmological perturbations in models with an
$M$-component inflaton $\phi_i$ can be analysed by considering
perturbations parallel and perpendicular
to the classical trajectory, as treated for example in \cite{Gordon:2000hv}. 
The definition of
curvature perturbation generalizing Eq.~\ref{curv-def} to the multi-field case is 
\be
\mtc{R} = - H \frac{\sum_{i=1}^M \dot{\phi}_i Q_i}{\dot{\sigma}^2} \,, \ee
where $Q_i$ is the gauge-invariant field fluctuation associated with $\phi_i$
and $\dot{\sigma}^2
\equiv \sum_{i=1}^M \dot{\phi}^2_i$.
The above formula for the
curvature perturbation can also be obtained through the
$\delta N$ formalism, i.e.,
$\mtc{R} = \sum_{i=1}^M (\partial N/\partial \phi_i) Q_i$,
where the number of $e$-folds to the end of inflation $N$ is generalized to
the multi-field case \citep{Starobinsky:1986fxa,Sasaki:1995aw}.
The $M-1$ normal directions are connected to $M-1$
isocurvature perturbations $\delta s_{ij}$ 
according to
\be
\delta s_{ij} = \frac{\dot{\phi}_i Q_j - \dot{\phi}_j Q_i}{\dot{\sigma}}. \ee
If the trajectory of the average field is curved in field space,
then during inflation both curvature and isocurvature fluctuations are
generated with non-vanishing correlations \citep{Langlois:1999dw}.

Isocurvature perturbations can be converted into curvature perturbations 
on large scales, but the opposite does not hold \citep{Mollerach:1989hu}.
If such isocurvature perturbations are not totally converted into 
curvature perturbations,
they can have observable effects on CMB 
anisotropies and on structure formation. In Sect.~\ref{sec:isometho}, 
we present the \Planck\ constraints on 
a combination of curvature and isocurvature initial conditions and the 
implications for important two-field scenarios, such as the curvaton \citep{Lyth:2001nq} and axion \citep{Lyth:1989pb}
models. 

Isocurvature perturbations may lead to a higher level of non-Gaussianity 
compared to a single inflaton with a standard kinetic term 
\citep{GrootNibbelink:2000vx}.
There is no reason to expect 
the inflaton to be a single-component field. The scalar sector of the 
Standard Model, as well as its extensions, contains
more than one scalar field.

}

\section{Methodology \label{sec:methodology}}

{

\subsection{Cosmological model and parameters}

The parameters of the models to be estimated in this paper fall
into three categories:
(i) parameters describing the initial perturbations, i.e., characterizing the particular inflationary 
scenario in question; 
(ii) parameters determining cosmological evolution at late times ($z \lesssim 10^4$); 
and (iii) parameters that 
quantify our uncertainty about the instrument and foreground 
contributions to the angular power spectrum. 
These will be described in Sect. \ref{sec:cmbdata}.

Unless specified otherwise, we assume that the late time cosmology is 
the standard flat six-parameter $\Lambda$CDM model whose energy content 
consists of photons, baryons, cold dark matter, neutrinos (assuming 
$N_\mathrm{eff} = 3.046$ effective species, one of which is taken to be massive,
with a mass of $m_\nu = 0.06$~eV), and a cosmological constant.  The primordial helium fraction,
$Y_\mathrm{P}$, is set as a function of $\Omega_\mathrm{b}h^2$ and $N_\mathrm{eff}$ 
according to the big bang nucleosynthesis consistency condition
 \citep{Ichikawa:2006dt,Hamann:2007sb}, and we fix the CMB mean temperature to $T_0 = 2.7255$~K
 \citep{Fixsen:2009ug}.  
Reionization is modelled to occur instantaneously at a redshift 
$z_\mathrm{re}$, and the optical depth $\tau$ is calculated as a function of $z_\mathrm{re}.$ 
This model can be characterized by four free cosmological 
parameters: $\Omega_\mathrm{b}h^2, \Omega_\mathrm{c}h^2, 
\theta_\mathrm{MC},$ and $\tau ,$ defined in Table \ref{table:CPDefinitions}, in addition to the
parameters describing the initial perturbations.

\subsection{Data}

The primary CMB data used for this paper consist of the \Planck\ CMB temperature likelihood 
supplemented by
the {\it Wilkinson} Microwave Anisotropy Probe ({\it WMAP} henceforth) 
large-scale polarization likelihood (henceforth {\it Planck}+WP), as described in
Sect.~\ref{sec:cmbdata}. The large-angle $E$-mode polarization spectrum is important for
constraining reionization because it breaks the degeneracy in the temperature
data between the primordial power spectrum amplitude and the optical depth to reionization. 
In the analysis constraining cosmic inflation, we restrict ourselves to combining the {\it Planck}
temperature data with various combinations of the following additional data sets: the {\it Planck} 
lensing power spectrum, other CMB data extending the {\it Planck} data to higher $\ell,$ and BAO data. 
For the higher-resolution CMB data we use measurements 
from the Atacama Cosmology Telescope (ACT) and the South Pole Telescope (SPT).
These
complementary data sets are among the most useful to break degeneracies in parameters.
The consequences of including other data sets such as Supernovae Type Ia (SN Ia) 
or the local measurement of the Hubble constant $H_0$ on some of the cosmological models
discussed here can be found in the compilation of cosmological parameters for numerous models
included in the on-line Planck Legacy archive.\footnote{Available at:
\texttt{http://www.sciops.esa.int/index.php\\?project=planck\&page=Planck\_Legacy\_Archive}}
Combining {\it Planck}+WP with various SN Ia data compilations \citep{Conley:2011ku,Suzuki:2011hu} or with
a direct measurement of $H_0$ \citep{Riess:2011yx} does not significantly alter the conclusions for
the simplest slow-roll inflationary models presented below.
The approach adopted here is the same as in the parameters paper 
\cite{planck2013-p11}.

\subsubsection{\Planck\ CMB temperature data \label{sec:cmbdata}}

The \Planck\ CMB likelihood is based on a hybrid approach, which combines a Gaussian 
likelihood approximation derived from temperature pseudo cross-spectra 
at high multipoles
\citep{Hamimeche:2008ai}, with a pixel-based temperature and polarization likelihood at low multipoles.
We summarize the likelihood here. For a detailed description the reader is referred to \cite{planck2013-p08}.

The small-scale \Planck\ temperature likelihood is based on pseudo 
cross-spectra between pairs of maps at 100, 143, 
and 217~GHz, masked to retain 49\%, 31\%, and 31\% of the sky, 
respectively. This results in angular auto- and 
cross-correlation power spectra covering 
multipole ranges of $50 \leq \ell \leq 1200$ at 100~GHz, $50 \leq \ell 
\leq 2000$ at 143~GHz, and $500 \leq \ell \leq 2500$ at 217~GHz as well as 
for the $143\times217$~GHz cross-spectrum. In 
addition to instrumental uncertainties, mitigated here by using only 
cross-spectra among different detectors, 
small-scale foreground and CMB 
secondary anisotropies need to be accounted for.  
The foreground model used in the \Planck\ high-$\ell$ likelihood is 
described in detail 
in~\cite{planck2013-p08} and \cite{planck2013-p11}, and
includes contributions to the cross-frequency power spectra from unresolved radio point sources, the cosmic 
infrared background (CIB), and the thermal and kinetic Sunyaev-Zeldovich 
effects. There are
eleven adjustable {\it nuisance} 
parameters: ($A^\mathrm{PS}_{100}, A^\mathrm{PS}_{143}, 
A^\mathrm{PS}_{217}, r^\mathrm{PS}_{143\times217}$, 
$A^\mathrm{CIB}_{143}, A^\mathrm{CIB}_{217}, 
r^\mathrm{CIB}_{143\times217}, \gamma^\mathrm{CIB}$, $A^\mathrm{tSZ}_{143}, 
A^\mathrm{kSZ}, \xi^\mathrm{tSZ-CIB}$). 
In addition, the calibration 
parameters for the 100 and 217~GHz channels, $c_{100}$ and $c_{217}$, relative 
to the 143~GHz channel, and the dominant beam uncertainty eigenmode 
amplitude $B^1_1$ are left free in the analysis, with other beam uncertainties marginalized analytically.
The \Planck\  high-$\ell $ likelihood therefore includes 14 nuisance parameters.\footnote{
After the Planck March 2013 release, a minor error was found in the ordering of the beam
transfer functions applied to the $217\times217$
cross-spectra in the \Planck\ high-$\ell$ likelihood. An extensive analysis of the corresponding 
revised \Planck\ high-$\ell$ likelihood showed that this error has a negligible impact on cosmological parameters and
is absorbed by small shifts in the foreground parameters. See \cite{planck2013-p11} for more details.}

The low-$\ell$ \Planck\ likelihood combines the \Planck\ temperature data 
with the large scale 9-year {\it WMAP} polarization data for this release. 
The procedure introduced in \cite{Page:2006hz} separates the temperature and 
polarization likelihood under the assumption of negligible noise in the temperature map. 
The temperature likelihood uses Gibbs 
sampling \citep{Eriksen:2006xr}, mapping out the distribution 
of the $\ell<50$ CMB temperature multipoles from a foreground-cleaned combination of the $30-353$ GHz 
maps \citep{planck2013-p06}. The polarization likelihood is pixel-based using the {\it WMAP} 9-year 
polarization maps at 33, 41, and 61~GHz  and includes the temperature-polarization 
cross-correlation \citep{Page:2006hz}. Its angular range is $\ell\le 23$ for $TE$, $EE$, and $BB$.  

\subsubsection{\Planck\ lensing data}
\label{lensing:data:subsection}

The primary CMB anisotropies are distorted by the gravitational 
potential induced by intervening matter. Such lensing, which broadens and 
smooths out the acoustic oscillations, is taken into account as a correction 
to the observed temperature power spectrum. The lensing power spectrum can also 
be recovered by measuring higher-order correlation functions.

Some of our analysis includes the \Planck\ lensing likelihood, 
derived in \cite{planck2013-p12}, which measures the non-Gaussian trispectrum 
of the CMB and is proportional to the power spectrum of the 
lensing potential. 
As described in \cite{planck2013-p12}, this potential is reconstructed using quadratic estimators 
\citep{Okamoto:2003zw}, and its power spectrum is used to estimate the lensing 
deflection power spectrum.
The spectrum is estimated from the 143 and 217~GHz maps, using 
multipoles in the range $40<\ell<400$. The theoretical predictions for the 
lensing potential power spectrum are calculated at linear order.

\subsubsection{ACT and SPT temperature data}
\label{act:spt:subesection}

We include data from ACT and SPT to extend the multipole range of our CMB likelihood.
ACT measures the power spectra and cross spectrum 
of the 148 and 218 GHz channels 
\citep{Das:2013zf}, and covers angular scales $500 < \ell < 10\,000$ at 148 GHz 
and $1500 < \ell < 10\, 000$ at 218 GHz. We use these data in the range $\ell>1000$ in combination with \Planck. 
SPT measures the power spectrum for angular scales $2000 < \ell < 10\,000$ at 95, 150, and 220 GHz \citep{Reichardt:2011yv}. 
The spectrum at larger scales is also measured at 150~GHz \citep{Story:2012wx}, but we do not include this data
in our analysis. To model the foregrounds for ACT and SPT we follow a similar approach to the likelihood described in 
\citet{Dunkley:2013vu}, extending the model used for the \Planck\ high-$\ell$ likelihood. Additional nuisance 
parameters are included to model the Poisson source amplitude, the residual Galactic dust contribution, 
and the inter-frequency calibration parameters. More details are provided in \citet{planck2013-p08} and 
\citet{planck2013-p11}.

\subsubsection{BAO data} 
\label{bao:subsection}

The BAO (Baryon Acoustic Oscillation) angular scale serves 
as a {\it standard ruler} and allows us to map out the expansion
history of the Universe after last scattering. 
The BAO scale, extracted from galaxy redshift 
surveys, provides a constraint on the late-time geometry
and breaks degeneracies with other cosmological parameters. 
Galaxy surveys constrain the ratio 
$D_V(\bar{z})/r_\mathrm{s}$, where $D_V(\bar{z})$ is the spherically 
averaged distance scale to the effective survey redshift $\bar{z}$ and 
$r_\mathrm{s}$ is the sound horizon~\citep{Mehta:2012hh}. 

In this analysis we 
consider a combination of the measurements by the 6dFGRS 
(\cite{Beutler:2011hx}, $\bar{z} = 0.106$), 
SDSS-II~(\citealt{Padmanabhan:2012hf}, $\bar{z} = 0.35$), and 
BOSS CMASS~(\citealt{Anderson:2012sa}, $\bar{z} = 0.57$) surveys, assuming no 
correlation between the three data points. This likelihood is described 
further in \citet{planck2013-p11}.

\subsection{Parameter estimation}

Given a model $\mathcal{M}$ with free parameters $\mathbf{x} \equiv 
\{x_1,\cdots,x_k\}$ and a likelihood function of the data 
$\mathcal{L}(\text{data} | \mathbf{x})$, the (posterior) probability 
density $\mathcal{P}$ as a function of the parameters can be expressed as
\begin{equation}
\mathcal{P}({\mathbf x} | \text{data}, \mathcal{M}) \propto 
\mathcal{L}(\text{data} | \mathbf{x}) \cdot P(\mathbf{x} | \mathcal{M}),
\end{equation}
where $P(\mathbf{x} | \mathcal{M})$ represents the data-independent prior 
probability density.  Unless specified otherwise, we choose wide top-hat 
prior distributions for all cosmological parameters.

We construct the posterior parameter probabilities using the 
Markov Chain Monte Carlo (MCMC) sampler as implemented in the
\CosmoMC~\citep{Lewis:2002ah} or {\tt MontePython}~\citep{Audren:2012wb} 
packages. In some cases, when the calculation of the Bayesian evidence 
(see below) is desired or when the likelihood function deviates 
strongly from a multivariate Gaussian, we use the nested sampling 
algorithm provided by the \MultiNest\ add-on module 
\citep{Feroz:2007kg,Feroz:2008xx} instead of
the Metropolis-Hastings algorithm.

Joint two-dimensional and one-dimensional posterior 
distributions are obtained by marginalization. Numerical values and 
constraints on parameters are quoted in terms of the mean and 68\% central 
Bayesian interval of the respective one-dimensional marginalized posterior 
distribution.

\subsection{Model selection}
\label{sec:modelsel}

Two approaches to model selection are commonly used in statistics.
The first approach examines the logarithm of the likelihood ratio, 
or {\it effective} $\chi ^2,$
\begin{equation}
\Delta \chi^2_\text{eff} \equiv 2 \left[\ln \mathcal{L}_{\text{max}}(\mathcal{M}_1) - \ln \mathcal{L}_{\text{max}}(\mathcal{M}_2) \right],
\end{equation}
between models $\mathcal{M}_1$ and $\mathcal{M}_2$, corrected for the fact that models with more parameters provide a 
better fit due to fitting away noise, even when the more complicated
model is not correct. Various information 
criteria have been proposed based on this idea
\citep{1974ITAC...19..716A,Schwarz:1978xx}; see also 
\citet{Liddle:2007fy}.  These quantities have the advantage of 
being independent of prior choice and fairly easy to calculate.
The second approach is 
Bayesian~\citep{Cox:1946, Jeffreys1998BK,Jaynes2003}, and is based 
on evaluating ratios of the {\it model averaged likelihood}, or Bayesian 
evidence, defined by 
\begin{equation} \label{eq:Evidencefull} 
\mathcal{E}_i=\int \text{d}^kx~ P(\mathbf{x} | 
\mathcal{M}_i)\mathcal{L}(\text{data} | \mathbf{x}). 
\end{equation} 
Evidence ratios, also known as Bayes factors, $B_{12} \equiv \mathcal{E}_1/\mathcal{E}_2$,
are naturally interpreted as {\it betting odds} 
between models.\footnote{Note that since 
the average is performed over the entire support of the prior 
probability density, the evidence depends strongly on
the probability range for the adjustable parameters.  
Whereas in parametric inference, 
the exact extent of the prior ranges often becomes irrelevant as long as 
they are ``wide enough" (i.e., containing the bulk of the high-likelihood 
region in parameter space), the value of the evidence will generally 
depend on precisely how wide the prior range was chosen.} 
Nested sampling algorithms allow rapid 
numerical evaluation of $\mathcal{E}.$  In this paper we will consider both the effective $\chi ^2$ 
and the Bayesian 
evidence.\footnote{After the submission of the 
first version of this paper, uncertainties arising from
the minimization
algorithm in the best fit cosmological parameters and the best fit
likelihood were studied. The uncertainties  
found were $\cal O(10^{-1})$ and therefore do not alter our conclusions. The
values for $\Delta \chi^2$ reported 
have not been updated.}

}

\section{Constraints on slow-roll inflationary models}
\label{sec:slow-roll}

{

In this section we describe constraints on slow-roll inflation using \Planck+WP
data in combination with the likelihoods described in Sects.~\ref{lensing:data:subsection}--\ref{bao:subsection}.
First we concentrate on characterizing the primordial power spectrum using  
\Planck\  and other data.  We start by showing that the empirical pre-inflationary Harrison-Zeldovich (HZ)
spectrum with $n_\mathrm{s}=1$ does not fit the {\it Planck} measurements. 
We further examine whether generalizing the cosmological model, for example by allowing the number
of neutrino species to vary, allowing the helium fraction to vary, or admitting a non-standard reionization 
scenario could reconcile the data with $n_\mathrm{s}=1$. We conclude that $n_\mathrm{s}\ne 1$ is robust. 

We then investigate the \Planck\ constraints on slow-roll inflation, allowing a tilt for the spectral index and 
the presence of tensor modes, and discuss the implications for the simplest standard inflationary models. 
In this section the question is studied using the slow-roll approximation, but later sections move beyond 
the slow-roll approximation. We show that compared to previous experiments, 
\Planck\  significantly narrows  the space of allowed inflationary models. Next we consider evidence for a running
of $n_\mathrm{s}$ and constrain it to be small, 
although we find a preference for negative running at modest statistical significance.
Finally, we comment on the implications for inflation of the 
\Planck\ constraints on possible deviations from spatial flatness.

\subsection{Ruling out exact scale invariance}

The simplest Ansatz for characterizing the statistical properties 
of the primordial cosmological perturbations is the so-called HZ model
proposed by \citet{Harrison:1969fb}, \citet{Zeldovich:1972zz}, and
\citet{1970ApJ...162..815P}.
These authors pointed out that a power spectrum with exact scale invariance
for the Newtonian gravitation potential fitted the data available at the time, but without giving 
any theoretical
justification for this form of the spectrum. 
Under exact scale invariance, which would constitute an
unexplained 
new symmetry, the primordial perturbations
in the Newtonian gravitational potential look statistically 
the same whether they are magnified 
or demagnified.  
In this simple model, vector and tensor perturbations are absent 
and the spectrum of curvature perturbations 
is characterized by a single parameter, the amplitude $A_\mathrm{s}$.  
Inflation, on the other hand, generically breaks this rescaling symmetry.
Although under inflation scale invariance still holds approximately, 
inflation must end. Therefore as different scales are imprinted, the physical conditions must evolve. 

Although a detection of a violation of scale invariance would not definitively prove that inflation is responsible for the 
generation of the primordial perturbations, ruling out the HZ model would 
confirm the expectation of small deviations from scale invariance, 
almost always on the red side, which are generic
to all inflationary models without fine tuning.
We examine in detail the viability of the HZ model 
using statistics to compare to the more general model where the spectral index is
allowed to vary, as motivated by slow-roll inflation.

When the cosmological model with $n_\mathrm{s}=1$ is compared with a model in which $n_\mathrm{s}$ is allowed to vary,
we find that allowing $n_\mathrm{s}$ to deviate from one 
decreases the best fit effective $\chi ^2$ by $27.9$ with respect to the HZ model. Thus
the significance of the finding that $n_\mathrm{s}\ne 1$ is in excess of  5$\sigma$. 
The parameters and maximum likelihood of this comparison are reported in Table~\ref{tab:HZ}.

One might wonder whether $n_\mathrm{s}=1$ could be reconciled with the data by relaxing some 
of the assumptions of the underlying cosmological model. Of particular interest is
exploring those parameters almost degenerate with the spectral index 
such as the effective number of neutrino species
$N_\mathrm{eff}$ and the primordial helium fraction $Y_\mathrm{P}$,
which both alter the damping tail of the temperature
spectrum~\citep{Trotta:2003xg,Hou:2011ec}, 
somewhat mimicking a spectral tilt.  
Assuming a Harrison-Zeldovich spectrum and allowing
$N_\mathrm{eff}$ or $Y_\mathrm{P}$ to float, and thus deviate
from their standard values,
gives almost as good a fit to \Planck+WP data as the $\Lambda$CDM 
model with a varying spectral index, with 
$\Delta \chi_\mathrm{eff}^2 = 2.8$ and $2.2$, respectively.
However, as 
shown in Table~\ref{tab:HZ}, the HZ, HZ+$N_\mathrm{eff}$, and
HZ+$Y_\mathrm{P}$ models require significantly higher baryon densities
and reionization optical depths compared to $\Lambda$CDM.  In the HZ+$Y_\mathrm{P}$
model, one obtains a helium fraction of $Y_\mathrm{P} = 0.3194 \pm
0.013$. This value is incompatible both with direct 
measurements of the primordial helium abundance~\citep{Aver:2011bw} and with 
standard big bang nucleosynthesis \citep{Hamann:2007sb}. 
(For comparison, we note that the value $Y_P=0.2477$ was obtained
as best fit for the $\Lambda $CDM model.)
The HZ+$N_\mathrm{eff}$ model, on the other hand, would imply the presence
of $\Delta N_\mathrm{eff} \approx 1$ new effective neutrino species 
beyond the three known species.  
When BAO measurements are included in the likelihood,
$\Delta \chi_\mathrm{eff}^2$ increases to 39.2 (HZ),
4.6 (HZ+$Y_\mathrm{P}$), and 8.0 (HZ+$N_\mathrm{eff}$), respectively,
for the three models. The significance of this detection is also discussed in \citet{planck2013-p11}.

\begin{table*}
\centering
\begin{tabular}{l|cccc}
\hline
\hline
& HZ & HZ + $Y_\mathrm{P}$ & HZ + $N_\mathrm{eff}$ & $\Lambda$CDM \\
\hline
$10^5 \Omega_\mathrm{b}h^2$ & $2296 \pm 24$ & $2296 \pm 23$ & $2285 \pm 23$ & $2205 \pm 28$ \\ 
$10^4 \Omega_\mathrm{c}h^2$ & $1088 \pm 13$ & $1158 \pm 20$ &  $1298 \pm 43$ & $1199 \pm 27$ \\
$100 \, \theta_\mathrm{MC}$ &\phantom{$\int $} $1.04292 \pm 0.00054$ & $1.04439 \pm 0.00063$ & $1.04052 \pm 0.00067$ & $1.04131 \pm 0.00063$ \\
$\tau$ & $0.125^{+0.016}_{-0.014}$ & $0.109^{+0.013}_{-0.014}$ & $0.105^{+0.014}_{-0.013}$ & $0.089^{+0.012}_{-0.014}$ \\
$\ln \left( 10^{10} A_\mathrm{s} \right)$ & $3.133^{+0.032}_{-0.028}$  & $3.137^{+0.027}_{-0.028}$  & $3.143^{+0.027}_{-0.026}$  & $3.089^{+0.024}_{-0.027}$ \\
$n_{\mathrm s}$ & ---  &  --- &  --- & $0.9603 \pm 0.0073$  \\
$N_\mathrm{eff}$ & ---  & ---  & $3.98 \pm 0.19$ & --- \\
$Y_\mathrm{P}$ & ---  & $0.3194 \pm 0.013$  & ---  & --- \\
\hline
$-2 \Delta \ln({\cal L}_\mathrm{max})$ & 27.9 & 2.2 & 2.8 & 0 \\
\hline
\end{tabular}
\vspace{.4cm}
\caption{\label{tab:HZ}
Constraints on cosmological parameters and best fit $-2 \Delta \ln({\cal L})$ with respect to the standard $\Lambda$CDM model,
using \Planck+WP data, testing the significance of the deviation from the HZ model.} 
\end{table*}

\subsection{Constraining inflationary models using the slow-roll approximation}
\label{akaSsectionFourOne}

We now consider all inflationary models that can be described by the
primordial power spectrum 
parameters consisting of the scalar amplitude, $A_\mathrm{s}$, the
spectral index, $n_\mathrm{s}$, and the tensor-to-scalar
ratio $r$, all defined at the pivot scale $k_*$.  
We assume that the spectral index is independent of the wavenumber $k$. 
Negligible running of the spectral index
is expected if the slow-roll condition is satisfied
and higher order corrections in the slow-roll approximations can be neglected.
In the next subsection we relax this assumption.

Sampling the power spectrum parameters $A_\mathrm{s}$, $n_\mathrm{s}$, and $r$  
is not the only method for constraining slow-roll inflation. 
Another possibility is to sample the Hubble flow functions 
in the analytic expressions for the scalar and tensor power spectra \citep{stewart:1993,Gong:2001he,Leach:2002ar}. 
In the Appendix, we compare the slow-roll inflationary predictions by sampling the HFF with {\it Planck} data 
and show that the results obtained in this way agree with those derived by 
sampling the power spectrum parameters.
This confirms similar studies based on previous data \citep{Hamann:2008pb,Finelli:2009bs}.

The spectral index estimated from {\it Planck}+WP data is 
\begin{equation}
n_\mathrm{s}=0.9603 \pm 0.0073.
\end{equation}
This tight bound on $n_\mathrm{s}$ is crucial for constraining inflation. The {\it Planck}
constraint on $r$ depends slightly on the pivot scales; 
we adopt $k_*=0.002$~Mpc$^{-1}$ to quote our results, with $r_{0.002} < 0.12$ at 95\%~CL. 
This bound improves on the most recent results, including the 
{\it WMAP} 9-year constraint of $r < 0.38$ \citep{Hinshaw:2012fq}, the {\it WMAP} 7-year + ACT limit of
$r < 0.28$ \citep{Sievers:2013wk}, and the {\it WMAP} 7-year + SPT limit of $r < 0.18$ 
\citep{Story:2012wx}. 
The new bound from \Planck\ is consistent with the theoretical limit from temperature anisotropies 
alone \citep{Knox:1994qj}.
When a possible tensor component is included, the spectral index from \Planck+WP does 
not significantly change, with 
$n_\mathrm{s}=0.9624 \pm 0.0075$.

The {\it Planck} constraint on $r$ corresponds to an upper bound on the energy scale of inflation
\begin{equation}
V_* = \frac{3 \pi^2 A_{\mathrm{s}}}{2} \, r \, M_{\mathrm {pl}}^4
= (1.94 \times 10^{16}~{\mathrm{GeV}} )^4  \frac{r_*}{0.12}
\end{equation}
at 95\% CL. This is equivalent to an upper bound on the Hubble parameter during inflation of
$H_*/M_{\mathrm {pl}} < 3.7 \times 10^{-5}$.
In terms of slow-roll parameters, \Planck+WP constraints imply $\epsilon_V < 0.008$ at 95\% CL, and $\eta_V = -0.010^{+0.005}_{-0.011}$.

\begin{table*}
\centering
\begin{tabular}{cc|cccc}
\hline
\hline
Model & Parameter & {\it Planck}+WP & {\it Planck}+WP+lensing & {\it Planck} + WP+high-$\ell$ & {\it Planck}+WP+BAO \\
\hline
\multirow{2}{*}{$\Lambda$CDM + tensor}& $n_\mathrm{s}$ & $0.9624 \pm 0.0075$ & $0.9653 \pm 0.0069$ &
$0.9600 \pm 0.0071$ & $0.9643 + 0.0059$ \\
& $r_{0.002}$ & $< 0.12$ & $ < 0.13$ & $< 0.11$ & $< 0.12$ \\
\hline
& $-2 \Delta \ln{\cal L}_\mathrm{max}$ & 0 & 0 & 0 & -0.31 \\
\hline
\end{tabular}
\vspace{.4cm}
\caption{\label{tab:firstorder}
Constraints on the primordial perturbation parameters in the $\Lambda$CDM+tensor model from \Planck\ combined with other data sets.
The constraints are given at the pivot scale $k_* = 0.002$~Mpc$^{-1}$.}
\end{table*}

The {\it Planck} results on $n_\mathrm{s}$ and $r$ are robust to the addition of external data sets (see Table~\ref{tab:firstorder}).
When the high-$\ell$ CMB ACT + SPT data are added, we obtain $n_{\mathrm{s}}=0.9600 \pm 0.0071$ and $r_{0.002}<0.11$ 
at 95\%~CL. Including the {\it Planck} lensing likelihood we obtain $n_{\mathrm{s}}=0.9653 \pm 0.0069$ and $r_{0.002}<0.13$, 
and adding BAO data gives $n_{\mathrm{s}}=0.9643 \pm 0.0059$ and $r_{0.002}<0.12$.

The above bounds are robust to small changes in the
polarization likelihood at low multipoles. To test this robustness,
instead of using the {\it WMAP} polarization likelihood, we impose a Gaussian
prior $\tau= 0.07 \pm 0.013$ to take into account small shifts
due to uncertainties in residual foreground contamination or instrument
systematic effects in the evaluation of $\tau$, as performed in Appendix B of 
\cite{planck2013-p11}.  We find at most a reduction of 8\% for the upper bound on $r$.

It is useful to plot the inflationary potentials in the $n_\mathrm{s}\textrm{-}r$ plane using 
the first two slow-roll parameters evaluated at the pivot 
scale $k_*=0.002$~Mpc$^{-1}$ \citep{Dodelson:1997hr}.
Given our ignorance of the details of the epoch of entropy generation, 
we assume that the number of {\it e}-folds $N_*$ to the end of inflation
lies in the interval $[50,60]$.
This uncertainty is plotted for those potentials 
predicting an exit from inflation without changing the potential. 

Figure~\ref{fig:nsvsr} shows the \Planck\  constraints in the 
$n_\mathrm{s}\textrm{-}r$ plane 
and indicates the predictions of a number of 
representative inflationary potentials (see \cite{Lyth:1998xn} for a review 
of particle physics models of inflation). 
The sensitivity of {\it Planck} data to high multipoles removes the
degeneracy between $n_\mathrm{s}$ and $r$ found 
using the {\it WMAP} data.
{\it Planck} data favour models with a concave potential. 
As shown in Fig.~\ref{fig:nsvsr}, 
most of the joint 95\% allowed region lies below the convex
potential limit, and concave models with a red tilt in the range
[0.945-0.98] are allowed by {\it Planck} at 95\% CL.
In the following we consider the status of several illustrative and commonly
discussed inflationary potentials in light of the \Planck\ observations.

\begin{figure*}
\includegraphics[width=18cm]{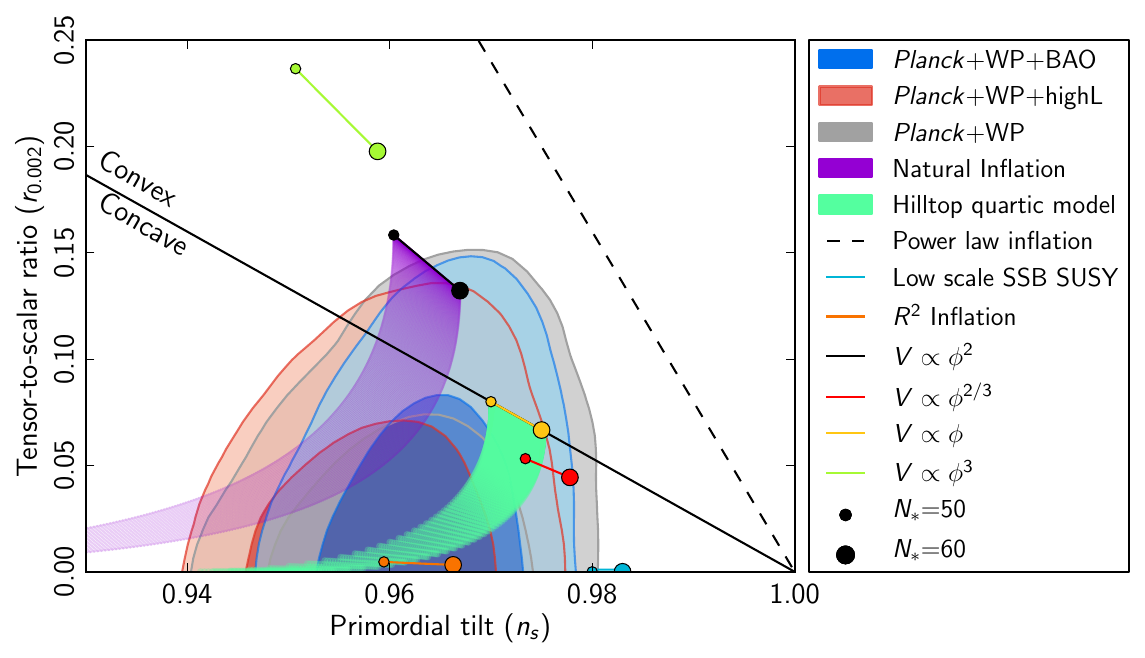}
\caption{Marginalized joint 68\% and 95\%~CL regions for $n_\mathrm{s}$ and $r_{0.002}$
from \Planck\ in combination with other data sets
compared to the theoretical predictions
of selected inflationary models.} 
\label{fig:nsvsr}
\end{figure*}

\subsubsection*{Power law potential and chaotic inflation}

The simplest class of inflationary models is characterized by a single monomial potential of the form
\begin{equation}
V(\phi) = \lambda M_\mathrm{pl}^4 \left( \frac{\phi}{M_\mathrm{pl}} \right)^n \,.
\label{PowerLawPot:Eq}
\end{equation} 
This class of potentials includes the simplest chaotic models, in which inflation starts from 
large values for the inflaton, 
$\phi > M_\mathrm{pl}$. 
Inflation ends when slow-roll is no longer valid, and we assume this to occur at $\epsilon_V=1$. 
According to Eqs.~\ref{epsilon_def}, \ref{eta_def}, and \ref{eq:ns_def}, this class of potentials 
predicts to lowest order in 
slow-roll parameters $n_\mathrm{s} - 1 \approx  -n (n+2) M_\mathrm{pl}^2/\phi_*^2$, $r \approx  8 n^2 M_\mathrm{pl}^2/\phi_*^2$, 
$\phi_*^2 \approx  n M_\mathrm{pl}^2 (4 N_* + n)/2$.
The $\lambda \phi^4$ model lies well outside the joint 99.7\%~CL region in the $n_\mathrm{s}\textrm{-}r$ plane. 
This result confirms previous findings from, for example, \cite{Hinshaw:2012fq}, in which this model lies outside
the 95\% CL for the {\it WMAP} 9-year data and is further excluded by CMB data at smaller scales. 

The model with a quadratic potential, $n=2$ \citep{Linde:1983gd}, often considered the simplest example 
for inflation, now lies outside the joint 95\% CL for the {\it Planck}+WP+high-$\ell$  data for \mbox{$N_* \lesssim 60$} $e$-folds, 
as shown in Fig.~\ref{fig:nsvsr}.

A linear potential with $n=1$ \citep{McAllister:2008hb}, 
motivated by axion monodromy, has $\eta_V=0$ and 
lies  within the 95\%~CL region. 
Inflation with $n=2/3$ \citep{Silverstein:2008sg}, however, also motivated by axion monodromy, 
now lies on the boundary of the joint 95\%~CL region. 
More permissive entropy generation priors allowing $N_* < 50$ could reconcile this model with 
the {\it Planck} data.

\subsubsection*{Exponential potential and power law inflation}

Inflation with an exponential potential 
\begin{equation}
V (\phi) = \Lambda^4 \exp\left(-\lambda \frac{\phi}{M_\mathrm{pl}}\right)
\label{powlaw}
\end{equation}
is called power law inflation \citep{Lucchin:1984yf},
because the exact solution for the scale factor is given by $a(t) \propto t^{2/\lambda^2}$.
This model is incomplete since inflation would not end without an additional mechanism to stop it. Under the assumption
that such a
mechanism exists and leaves predictions for
cosmological perturbations unmodified, this class of models predicts \mbox{$r = -8 (n_{\mathrm{s}}-1)$} and now lies outside the joint 99.7\%~CL contour.

\subsubsection*{Inverse power law potential}

Intermediate inflationary models \citep{Barrow:1990vx,Muslimov:1990be} with inverse power law potentials 
\begin{equation}
V (\phi) = \Lambda^4 \left( \frac{\phi}{M_\mathrm{pl}} \right)^{-\beta}
\label{intermediate}
\end{equation}
lead to inflation with $a(t) \propto \exp(A t^f)$, with $A >0$ and \mbox{$0<f<1$}, where $f=4/(4+\beta)$ and $\beta>0$. 
In intermediate inflation there is no natural end to inflation, but if the exit mechanism leaves the inflationary
predictions for the cosmological perturbations unmodified, this class of models predicts 
$n_{\mathrm{s}}-1 \approx - \beta (\beta-2)/\phi_*^2$ and  
$r \approx - 8 \beta (n_{\mathrm{s}}-1)/(\beta-2)$ at lowest order in the slow-roll approximation
\citep{Barrow:1993zq}.\footnote{See \cite{Starobinsky:2005ab} for the inflationary model producing 
an exactly scale-invariant power spectrum with $r \ne 0$ beyond the slow-roll approximation.} 
Intermediate inflationary models lie outside the joint 95\%~CL contour for any $\beta$.

\subsubsection*{Hilltop models}

In another interesting class of potentials,
the inflaton rolls away from an unstable equilibrium as in the 
first new inflationary models \citep{Albrecht:1982wi,Linde:1981mu}. 
We consider 
\begin{equation}
V(\phi) \approx \Lambda^4 \left( 1 - \frac{\phi^{{p}}}{\mu^{{p}}} + ...\right) \,,
\label{newinf}
\end{equation}
where the ellipsis indicates higher order terms that are negligible during inflation but 
ensure positiveness of the potential later on.
An exponent of $p=2$ is allowed only as a 
large field inflationary model, predicting
\mbox{$n_\mathrm{s} -1 \approx -4 M^2_\mathrm{pl}/\mu^2 + 3 r/8$} 
and $r \approx 32 \phi^2_* M^2_\mathrm{pl}/\mu^4$.
This potential leads to predictions in agreement
with \Planck+WP+BAO joint 95\%~CL contours for super Planckian values of 
$\mu$, i.e., $\mu \gtrsim 9~M_\mathrm{pl}$. 

Models with $p \ge 3$ predict $n_\mathrm{s}-1 \approx -(2/N)(p-1)/(p-2)$ when $r \ll 1$.  
The hilltop potential with $p=3$ lies outside the joint 95\%~CL region for {\it Planck}+WP+BAO data.
The case with $p=4$ is also in tension with {\it Planck}+WP+BAO, but allowed within the joint 95\%~CL 
region for $N_* \gtrsim 50$ when $r \ll 1$. 
For larger values of  $r$ these models provide a better fit to the {\it Planck}+WP+BAO data. 
The $p=4$ hilltop model---without extra terms denoted by the ellipsis in 
Eq.~(37)---is displayed in Fig. 1 in the standard range $50 < N_* < 60$ at different values 
of $\mu$ (this model 
approximates the linear potential for large $\mu/M_\mathrm{pl}$).

\subsubsection*{A simple symmetry breaking potential}

The symmetry breaking potential \citep{Olive:1989nu}
\begin{equation}
V(\phi) = \Lambda^4 \left( 1 - \frac{\phi^2}{\mu^2} \right)^2
\label{LG}
\end{equation}
can be considered as a self-consistent completion of the hilltop model with $p=2$ (although it has a different 
limiting large-field branch for non-zero $r$). This potential leads to predictions in agreement 
with \Planck\ + WP + BAO joint 95\% CL contours for super Planckian values of $\mu$ (i.e. $\mu \gtrsim 13~M_\mathrm{pl})$.

\subsubsection*{Natural inflation}

Another interesting class of potentials is {\it natural} inflation ~\citep{1990PhRvL..65.3233F,Adams:1992bn},
initially motivated by its origin in symmetry breaking in an attempt
to naturally give rise to the extremely flat potentials required
for inflationary cosmology.
In {\it natural inflation} the effective
one-dimensional potential takes the form
\begin{equation}
V(\phi )=\Lambda^4 \left[ 1+\cos \left(\frac{\phi}{f} \right) \right], 
\label{NatInf}
\end{equation}
where $f$ is a scale which determines the slope of the potential (see
also \cite{Binetruy:1986ss} for an earlier motivation of a cosine potential for the inflaton
in the context of superstring theory).
Depending on the value of $f$, the model falls into the large field  
($f \gtrsim 1.5~M_\mathrm{pl}$) or small field ($f \lesssim 1.5~M_\mathrm{pl}$) 
categories. Therefore,  
$n_\mathrm{s} \approx 1-M^2_\mathrm{pl}/f^2$ holds for small $f,$ while 
$n_\mathrm{s} \approx 1-2/N$, $r \approx 8/N$ holds for large $f$, approximating the 
$m^2 \phi^2$ potential in the latter case 
(with $N_* \approx (2 f^2/M^2_\mathrm{pl}) \ln [\sin(\phi_\mathrm{e}/f)/\sin(\phi_*/f)]$). 
This model agrees with \Planck+WP data for $f \gtrsim 5~M_\mathrm{pl}$.

\subsubsection*{Hybrid inflation}

In hybrid inflationary models a second field, $\chi$, coupled to the inflaton, 
undergoes symmetry breaking. The simplest example of this class is
\begin{equation}
V(\phi,\chi)= \Lambda^4 \left(1-\frac{\chi^2}{\mu^2}\right)^2
+ U (\phi) + \frac{g^2}{2} \phi^2 \chi^2 \,.
\label{HybridInf}
\end{equation}
Over most of their parameter space, these models 
behave effectively as single-field models for the inflaton $\phi .$ 
The second field $\chi$ is close to the origin 
during the slow-roll regime for $\phi$, and 
inflation ends either by breakdown of slow roll for the inflaton at $\epsilon_\phi \approx M_\mathrm{pl}^2 
(\mathrm{d} U /\mathrm{d} \phi)^2/(\Lambda^4+ U(\phi ))^2 \approx 1$ or by the waterfall transition of $\chi$.
The simplest models with 
\begin{equation}
U(\phi) = \frac{m^2}{2} \phi^2
\end{equation}
are disfavoured for most of the parameter space \citep{Cortes:2009ej}. 
Models with $m^2 \phi^2/2 \sim \Lambda^4$ are disfavoured due to a high tensor-to-scalar ratio, and models with 
\mbox{$U(\phi) \ll \Lambda^4$} predict a spectral index $n_\mathrm{s}>1$, also disfavoured by the {\it Planck} data. 

We discuss hybrid inflationary models predicting $n_\mathrm{s} < 1$ separately.
As an example, the spontaneously broken SUSY model \citep{Dvali:1994ms} 
\begin{equation}
U(\phi) = \alpha_\mathrm{h} \Lambda^4 \ln \left( \frac{\phi}{\mu} \right) 
\end{equation}
predicts
$n_\mathrm{s} -1 \approx -(1+3 \alpha_\mathrm{h}/2)/N_*$ and $r \approx 8 \alpha_\mathrm{h}/N_*$.
For $\alpha_\mathrm{h} \ll 1$ and $N_* \approx 50$, $n_\mathrm{s} \approx 0.98$ is disfavoured by {\it Planck}+WP+BAO data at more than 95$\%$ CL.
However, more permissive 
entropy generation priors allowing $N_* < 50$ or a non-negligible $\alpha_\mathrm{h}$ 
give models consistent with the {\it Planck} data.

\subsubsection*{$R^2$ inflation}

Inflationary models can also be accommodated within extended theories of gravity. 
These theories can be analysed either in the original (Jordan) frame or in the conformally-related Einstein frame with a Klein-Gordon
scalar field. Due to the invariance of curvature and tensor perturbation power spectra with respect to this conformal transformation,
we can use the same methodology described earlier.

The first inflationary model proposed was of this type and was based on higher order 
gravitational terms in the action \citep{Starobinsky:1980te}
\begin{equation}
S = \int \mathrm{d}^4 x \sqrt{-g} \frac{M^2_\mathrm{pl}}{2} \left( R + \frac{R^2}{6 M^2} \right)\,,
\label{R2}
\end{equation}
with the motivation to include semi-classical quantum effects. 
The predictions for $R^2$ inflation were first studied in \cite{Mukhanov:1981xt} and \cite{Starobinsky:1983zz},
and can be summarized as $n_\mathrm{s} -1 \approx -8 (4 N_*+9)/(4 N_*+3)^2 $ and $r \approx 192/(4 N_*+3)^2$. 
Since $r$ is suppressed by another $1/N_*$ with respect to the scalar tilt, this model 
predicts a tiny amount of gravitational waves. 
This model predicts $n_\mathrm{s} = 0.963$ for $N_* =55$ and is fully consistent with the {\it Planck} constraints.

\subsubsection*{Non-minimally coupled inflaton}

A non-minimal coupling of the inflaton to gravity with the action 
\begin{equation}
S = \int \mathrm{d}^4 x \sqrt{-g} \left[ \frac{M^2_\mathrm{pl} + \xi \phi^2}{2} R - \frac{1}{2} g^{\mu \nu} \partial_\mu \phi 
\partial_\nu \phi 
- \frac{\lambda}{4} \left( \phi^2 -\phi_0^2 \right)^2 \right] 
\label{NMC}
\end{equation}
leads to several interesting consequences, such as a lowering of the tensor-to-scalar ratio. 

The case of a massless self-interacting inflaton ($\phi_0=0$) agrees with the \Planck+WP data for $\xi \ne 0$. 
Within the range $50<N_*<60$, this model is within the \Planck+WP joint 95\%~CL region for $\xi > 0.0019$, improving 
on previous bounds \citep{Tsujikawa:2004my,Okada:2010jf}. 

The amplitude of scalar perturbations is proportional to $\lambda/\xi^2$ for $\xi \gg 1$, and therefore 
the problem of tiny values for the inflaton self-coupling $\lambda$ can be alleviated 
\citep{Spokoiny:1984bd,Lucchin:1985ip,Salopek:1988qh,Fakir:1990eg}. 
The regime $\phi_0 \ll M_\mathrm{pl}$ is allowed and $\phi$ could be the Standard Model Higgs as proposed 
in \citet{Bezrukov:2007ep} at the tree level (see \citet{Barvinsky:2008ia,Bezrukov:2009db} for the inclusion of loop corrections). 
The Higgs case with $\xi \gg 1$ has the same predictions as the $R^2$ model in terms 
of $n_\mathrm{s}$ and $r$ as a function of $N_*$. 
The entropy generation mechanism in the Higgs case can be more efficient than in the $R^2$ case
and therefore predicts a slightly larger $n_\mathrm{s}$ \citep{Bezrukov:2011gp}.
This model is fully consistent with the {\it Planck} constraints. 

The case with $\xi <0$ and $|\xi| \phi_0^2/M^2_\mathrm{pl} \sim 1$ was also recently emphasized in \citet{Linde:2011nh}. 
With the symmetry breaking potential in Eq.~\ref{NMC}, the 
large field case with $\phi > \phi_0$ is disfavoured by {\it Planck} data, whereas the small field case $\phi < \phi_0$ is 
in agreement with the data.

\subsection{Running spectral index}
\label{running:section}

We have shown that the single parameter Harrison-Zeldovich spectrum does not fit
the data and that at least the first two terms $A_\mathrm{s}$ and $n_\mathrm{s}$ in the expansion of the primordial
power spectrum in powers of $\ln (k)$ given in Eq.~\ref{scalarps} are needed. Here we consider
whether the data require the next term known as 
the {\it running of the spectral index}~\citep{Kosowsky:1995aa}, defined as
the derivative of the spectral index with respect to $\ln k$,
$\mathrm{d}n_\mathrm{s \,, t}/\mathrm{d} \ln k$ for scalar or tensor fluctuations.
If the slow-roll approximation holds and the inflaton has reached its attractor
solution, $\mathrm{d}n_\mathrm{s}/\mathrm{d} \ln k$ and $\mathrm{d}n_\mathrm{t}/\mathrm{d} \ln k$ are 
related to the potential slow-roll parameters, as in Eqs.~\ref{eq:alphas_def} and \ref{eq:alphat_def}. 
In slow-roll single-field inflation, the running is second order in the
Hubble slow-roll parameters, for scalar and for tensor perturbations 
\citep{Kosowsky:1995aa,Leach:2002ar}, 
and thus is typically suppressed with respect to $n_\mathrm{s}-1$ and $n_\mathrm{t}$, 
which are first order. 
Given the tight constraints on the first two slow-roll parameters $\epsilon_V$ and $\eta_{V}$ ($\epsilon_1$ and $\epsilon_2$) 
from the present data, typical values of the running to which {\it Planck} is sensitive \citep{Pahud:2007gi}
would generically be dominated by the contribution from the third derivative of the potential, encoded
in $\xi_V^2$ (or $\epsilon_3$). 

While it is easy to see 
that the running is invariant under a change in pivot scale, the same does not
hold for the spectral index and the amplitude of the primordial power
spectrum. It is convenient to choose $k_*$ such that $\mathrm{d} n_{\mathrm s}/\mathrm{d} \ln k$ 
and $n_{\mathrm s}$ are uncorrelated \citep{Cortes:2007ak}.
This approach minimizes the inferred variance of $n_{\mathrm s}$ and
facilitates comparison with constraints on $n_{\mathrm s}$ in the
power law models. Note, however, that the decorrelation pivot scale
$k_*^{\mathrm{dec}}$ depends on both the model and the data set used.

\begin{table*}
\centering
\begin{tabular}{c|c|cccc}
\hline
\hline
Model & Parameter & {\it Planck}+WP & {\it Planck}+WP+lensing & {\it Planck}+WP+high-$\ell$ & {\it Planck}+WP+BAO \\
\hline
\multirow{4}{*}{$\Lambda$CDM + $\mathrm{d} n_\mathrm{s}/\mathrm{d} \ln k$} & 
$n_{\mathrm{s}}$ & $0.9561 \pm 0.0080$ & $0.9615 \pm 0.0072$ & $0.9548 \pm 0.0073$ & $0.9596 \pm 0.0063$ \phantom{$\Big|$}\\
& $\mathrm{d} n_\mathrm{s}/\mathrm{d} \ln k$ & $-0.0134 \pm 0.0090$ & $-0.0094 \pm 0.0085$ & $-0.0149 \pm 0.0085$ & 
$-0.0130 \pm 0.0090$ \\
& $-2 \Delta \ln {\cal L}_\mathrm{max}$ & -1.50 & -0.77 & -2.95 & -1.45 \\
\hline
\multirow{7}{*}{\phantom{$\Bigg|$}+ $\mathrm{d}^2 n_\mathrm{s}/\mathrm{d} \ln k^2$} \phantom{$\Big|$}
&\phantom{$\Big|$}
$n_{\mathrm{s}}$ & $0.9514^{+0.087}_{-0.090}$ & $0.9573^{+0.077}_{-0.079}$ & $0.9476^{+0.086}_{-0.088}$ & $0.9568^{+0.068}_{-0.063}$\phantom{$\Big|$}\\
{$\Lambda$CDM + $\mathrm{d} n_\mathrm{s}/\mathrm{d} \ln k$}& \phantom{$\Big|$}
$\mathrm{d} n_\mathrm{s}/\mathrm{d} \ln k$ & $0.001^{+0.016}_{-0.014}$ & $0.006^{+0.015}_{-0.014}$ & $0.001^{+0.013}_{-0.014}$ & $0.000^{+0.016}_{-0.013}$\phantom{$\Big|$}\\
& $\mathrm{d}^2 n_\mathrm{s}/\mathrm{d} \ln k^2$ \phantom{$\Bigg|$} & $0.020^{+0.016}_{-0.015}$ & $0.019^{+0.018}_{-0.014}$ & $0.022^{+0.016}_{-0.013}$ & $0.017^{+0.016}_{-0.014}$\phantom{$\Big|$} \\
& $-2 \Delta \ln {\cal L}_\mathrm{max}$ & -2.65 & -2.14 & -5.42 & -2.40\\
\hline
\multirow{6}{*}{$\Lambda$CDM + $r$ + $\mathrm{d} n_\mathrm{s}/\mathrm{d} \ln k$}& $n_{\mathrm{s}}$ & 
$0.9583 \pm 0.0081$ & $0.9633 \pm  0.0072$ & $0.9570 \pm 0.0075$ & $0.9607 \pm 0.0063$ \phantom{$\Big|$}\\
& $r$ & $<0.25$ & $<0.26$ & $<0.23$ & $<0.25$ \\
& $\mathrm{d} n_\mathrm{s}/\mathrm{d} \ln k$ & 
$-0.021 \pm 0.012$ & $-0.017 \pm 0.012$ & $-0.022^{+0.011}_{- 0.010}$ & $-0.021^{+0.012}_{-0.010}$ \phantom{$\Big|$}\\
& $-2 \Delta \ln {\cal L}_\mathrm{max}$ & -1.53 & -0.26 & -3.25 & -1.5  \\
\hline
\end{tabular}
\vspace{.4cm}
\caption{\label{tab:alpha}
Constraints on the primordial perturbation parameters for 
$\Lambda$CDM+$\mathrm{d} n_\mathrm{s}/\mathrm{d} \ln k$, $\Lambda$CDM+$\mathrm{d} n_\mathrm{s}/\mathrm{d} \ln k$+$r,$ and 
$\Lambda$CDM+$\mathrm{d} n_\mathrm{s}/\mathrm{d} \ln k$+$\mathrm{d}^2 n_\mathrm{s}/\mathrm{d} \ln k^2$ models from {\it Planck} combined with other data sets.   
Constraints on the spectral index and its dependence on the wavelength are given at the pivot scale of $k_* = 0.05$~Mpc$^{-1}$.} 
\end{table*}

We consider a model parameterizing the power spectrum using
$A_\mathrm{s} (k_*)\,, n_\mathrm{s} (k_*),$ 
and 
$\mathrm{d} n_\mathrm{s}/\mathrm{d} \ln k$, 
where $k_*=0.05~\mathrm{Mpc}^{-1}$. 
The joint constraints on $n_{\mathrm s}$ and $\mathrm{d} n_{\mathrm s}/\mathrm{d} \ln k$ at the decorrelation scale
of $k_*^{\mathrm{dec}} = 0.038$~Mpc$^{-1}$  are shown in
Fig.~\ref{fig:nalpha}. 
The {\it Planck}+WP constraints on the running do not change significantly
when complementary data sets such as {\it Planck} lensing, CMB high-$\ell,$ and BAO data are included. 
We find 
\begin{equation}
\mathrm{d} n_{\mathrm s}/\mathrm{d} \ln k = -0.013 \pm 0.009 \quad\mbox{(68\% CL, \Planck+WP)}\,,
\end{equation}
which is negative at the 1.5$\sigma$ level. This reduces the
uncertainty compared to previous CMB
results. Error bars are reduced by 60\% compared to the {\it WMAP} 9-year 
results \citep{Hinshaw:2012fq}, and by 20--30\%  compared to {\it WMAP} supplemented by 
SPT and ACT data 
\citep{Hou:2012xq,Sievers:2013wk}. {\it Planck} finds a smaller scalar running than 
SPT + {\it WMAP7} \citep{Hou:2012xq}, and larger than ACT + {\it WMAP7} \citep{Sievers:2013wk}. 
The best fit likelihood improves by only  
\mbox{$\Delta \chi^2_\mathrm{eff} \approx 1.5$} ($3$ when high-$\ell$ data are included) 
with respect to the minimal case in which $n_{\mathrm s}$
is scale independent, indicating that the deviation from scale independence is not very significant. 
The constraint for the spectral index in this case is $0.9630 \pm 0.0065$ at 68\% CL 
at the decorrelation pivot scale 
$k_* = 0.038$~Mpc$^{-1}$. This result implies that the third derivative of the potential 
is small, i.e., $|\xi^2_V| \sim 0.007$, but compatible with zero at 95\%~CL, for inflation 
at low energy (i.e., with $\epsilon_V \approx 0$).

\begin{figure}
\includegraphics[height=0.37\textwidth,angle=0]{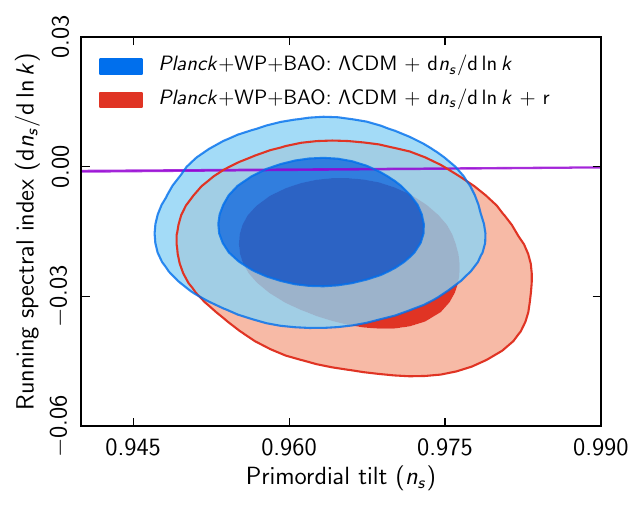}
\caption{Marginalized joint 68\% and 95\% CL for
$(\mathrm{d} n_{\mathrm s}/\mathrm{d} \ln k \,, n_{\mathrm s})$ 
using {\it Planck}+WP+BAO, either marginalizing over $r$ or fixing $r=0$ at 
$k_* = 0.038$~Mpc$^{-1}$. 
The purple strip shows the prediction for single monomial chaotic inflationary 
models with $50 < N_* <60$ for comparison.
}
\label{fig:nalpha} 
\end{figure}

We also test the possibility that the running depends on the wavelength so that 
$\mathrm{d}^2 n_{\mathrm s}/\mathrm{d} \ln k^2$
is nonzero. 
With \Planck+WP data, we find \mbox{$\mathrm{d}^2 n_{\mathrm s}/\mathrm{d} \ln k^2 = 0.020^{+0.016}_{-0.015}$}. 
This result is stable with respect to the addition of complementary data sets, as can be seen from 
Table~\ref{tab:alpha} and Fig.~\ref{fig:nalphabeta}.
When $\mathrm{d}^2 n_{\mathrm s}/\mathrm{d} \ln k^2$ is allowed in the fit, we find a value for the running $\mathrm{d} n_{\mathrm s}/\mathrm{d} \ln k$ 
consistent with zero.

Finally we allow a
non-zero primordial gravitational wave spectrum together with the running. The tensor
spectral index and its running are set by the slow-roll consistency relations to second order, with 
$n_\mathrm{t} = - r (2-r/8 - n_\mathrm{s})/8$ and $\mathrm{d} n_\mathrm{t}/\mathrm{d} \ln k = r (r/8+n_\mathrm{s}-1)/8$. 
{\it Planck} measures the running to be $\mathrm{d} n_{\mathrm s}/\mathrm{d} \ln k = -0.016 \pm 0.010$ when tensors are included (see
Table~\ref{tab:alpha} and Fig.~\ref{fig:nalpha-four}). 
The constraints on the
tensor-to-scalar ratio are relaxed compared to the case with no running, due to an anti-correlation between $r$
and $\mathrm{d} n_\mathrm{s}/\mathrm{d} \ln k$, as shown in Fig.~\ref{fig:nalpha-four} for \Planck+WP+BAO.

Varying both tensors and running, \Planck+WP implications for slow-roll parameters are $\epsilon_V < 0.015$ at 95\% CL, 
$\eta_V = -0.014_{-0.011}^{+0.015}$, and $|\xi^2_V| = 0.009 \pm 0.006$. 

In summary, the {\it Planck} data prefer a negative running for the scalar spectral index 
of order $\mathrm{d} n_\mathrm{s}/\mathrm{d} \ln k \approx -0.015,$ but  
at only the 1.5$\sigma$ significance level. This is for {\it Planck} 
alone and in combination with other astrophysical data sets.
Weak statistical evidence for negative values of
$\mathrm{d} n_\mathrm{s}/\mathrm{d} \ln k$ has been claimed in several 
previous investigations with the {\it WMAP} data and smaller scale CMB data 
\citep[e.g.,][]{spergel2003,2003ApJS..148..213P,Dunkley:2010ge,Hinshaw:2012fq,Hou:2012xq}.

If primordial, negative values for $\mathrm{d} n_\mathrm{s}/\mathrm{d} \ln k$ of order $10^{-2}$ would be interesting for the 
physics of inflation. 
The running of the scalar spectral index is a key prediction for inflationary models. It is strictly zero for power law inflation, 
whose fit to {\it Planck} was shown to be quite poor in the previous section. Chaotic monomial models with $V(\phi) \propto \phi^n$ 
predict 
$\mathrm{d} n_\mathrm{s}/\mathrm{d} \ln k \approx - 8 (n+2)/(4N+n)^2 \approx (n_\mathrm{s}-1)^2$, 
and the same order of magnitude $(10^{-3})$ is quite typical for many slow-roll 
inflationary models, such as natural inflation \citep{Adams:1992bn} or 
hilltop inflation \citep{Boubekeur:2005zm}. It was pointed out that 
a large negative running of $\mathrm{d} n_{\mathrm s}/\mathrm{d} \ln k \lesssim - 10^{-2}$ 
would make it difficult to support the $N_* \approx 50$ $e$-foldings
required from inflation~\citep{Easther:2006tv}, but this holds only without nonzero 
derivatives higher than the third order in the inflationary potential.
Designing inflationary models that predict a negative running of ${\cal O}(10^{-2})$ with an acceptable $n_\mathrm{s}$ and number 
of $e$-folds is not impossible, as the case with modulated oscillations in the inflationary potential demonstrates \citep{Kobayashi:2010pz}.
This occurs, for instance, 
in the axion monodromy model when the instanton contribution is taken into account \citep{McAllister:2008hb}, giving the potential  
\begin{equation}
V(\phi) = \mu^3 \phi + \Lambda^4 \cos \left( \frac{\phi}{f} \right) \,.
\label{amistanton}
\end{equation}

\begin{figure}
\includegraphics[height=0.37\textwidth,angle=0]{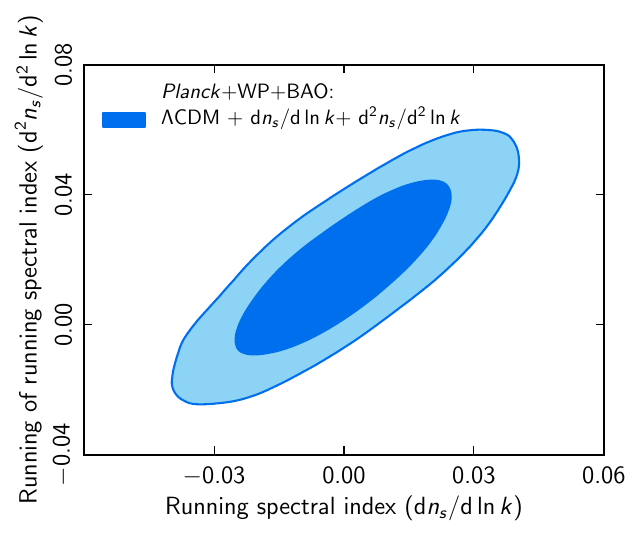}
\caption{Marginalized joint 68\% and 95\%~CL regions for
$(\mathrm{d}^2 n_{\mathrm s}/\mathrm{d} \ln k^2 \,, \mathrm{d} n_{\mathrm s}/\mathrm{d} \ln k)$ using {\it Planck}+WP+BAO.}
\label{fig:nalphabeta}
\end{figure}

\begin{figure}
\includegraphics[height=0.37\textwidth,angle=0]{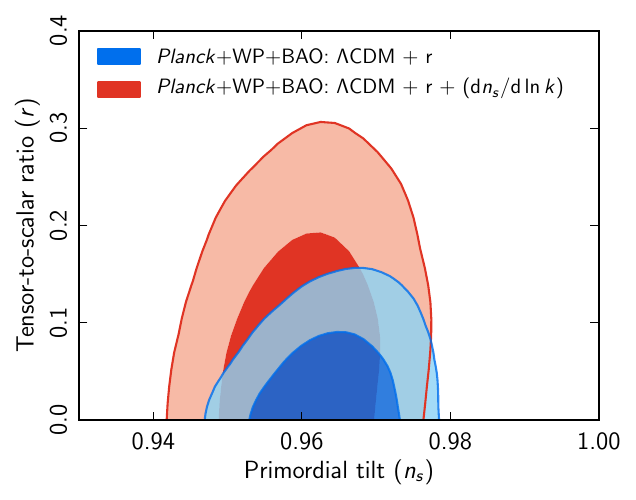}
\caption{Marginalized joint 68\% and 95\%~CL regions for
$(r \,, n_{\mathrm s})$, using {\it Planck}+WP+BAO with and without a running spectral index.}
\label{fig:nalpha-four}
\end{figure}


\subsection{Open inflation}

Most models of inflation predict a nearly flat spatial geometry with small deviations from 
perfect spatial flatness of $|\Omega_K| \sim10^{-5}$. Curvature fluctuations may 
be regarded as local fluctuations in 
the spatial curvature, and even in models of inflation where the perturbations are calculated
about a spatially flat background, the spatial curvature on the largest scales accessible 
to observation now are subject to fluctuations from perfect spatial flatness (i.e., $\Omega_K=0$). 
This prediction for this fluctuation 
is calculated by simply extrapolating the power law spectrum
to the largest scale accessible today, so that $\Omega_K$ as probed by the CMB roughly represents
the local curvature fluctuation averaged over our (causal) horizon volume. Although it has sometimes been claimed that spatial
flatness is a firm prediction of inflation, it was realized early on that spatial flatness
is not an inexorable consequence of inflation and large amounts of spatial curvature
(i.e., large compared to the above prediction) can be introduced in a precise way while 
retaining all the advantages of inflation \citep{Gott:1982zf,Gott:1984ps} through bubble nucleation
by false vacuum decay \citep{Coleman:1980aw}.
This proposal gained credence when it was shown how to calculate the perturbations in this model 
around and beyond the curvature scale 
\citep{Bucher:1994gb,Bucher:1995ga,Yamamoto:1995sw,Tanaka:1994qa}.
See also \cite{Ratra:1994vw,Ratra:1994dm} and \cite{Lyth:1990dh}.  For more refined later calculations see for example 
\citet{Garriga:1997wz,Garriga:1998he}, \citet{Gratton:1999ya},
and references therein. For predictions 
of the tensor perturbations see for example \citet{Bucher:1997xs},
\citet{Sasaki:1997ex}, and 
\citet{Hertog:1999kg}.

An interesting proposal using {\it singular} instantons and not requiring a false vacuum 
may be found in \citet{Hawking:1998bn}, 
and for calculations of the resulting perturbation spectra see 
\cite{Hertog:1999kg} and \cite{Gratton:1999hv}.
Models of this sort have been studied more recently in the context of the string
landscape. (See, for example, \cite{Vilenkin:2006xv} for a nice review.)
Although some proposals for universes with positive curvature within the framework
of inflation have been put forth \citep{Gratton:2001gw}, it is much harder to obtain 
a closed universe with a spatial geometry of positive spatial curvature (i.e., $\Omega_K<0$)
\citep{Linde:2003hc}.

Theoretically, it is of interest to measure $\Omega_K$ to an accuracy of approximately $10^{-4}$
or slightly better to test the prediction of simple flat inflation for this observable.
A statistically significant positive value would suggest that open inflation, perhaps
in the context of the landscape, was at play. A statistically significant negative value could
pose difficulties for the inflationary paradigm. For a recent discussion of 
these questions, see for example \cite{Freivogel:2005vv}, \cite{Kleban:2012ph}, and \cite{Guth:2012ww}.

\begin{figure}
\includegraphics[width=8.8cm]{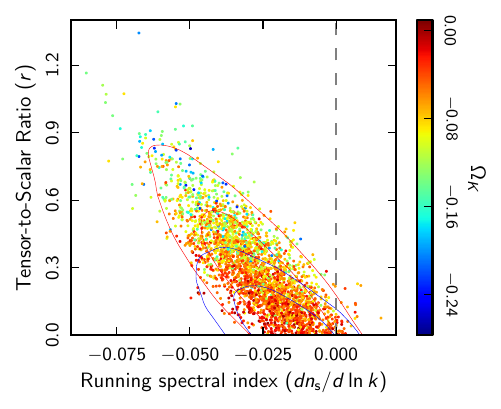}
\caption{Joint posterior for tensors and running of the scalar spectral index marginalizing over other
parameters.  The contours are set at 68\% and 95\%.  The red contours apply for \Planck+WP+high-$\ell$ data. The colour of the scattered 
points indicates the distribution of $\Omega_K$. The blue contours apply when BAO data is also included.  ($\Omega_K$ is then
found to be well constrained close to zero.)  The dashed vertical line shows
the {\it no-running} solution.
\label{fig:openinf}}
\end{figure}

In order to see how much spatial curvature is allowed, we
consider a rather general model including the parameters
$r,$ $n_{\mathrm s},$ and $\mathrm{d}n_{\mathrm s}/\mathrm{d}\ln k$ as well as $\Omega_K$.
We find that 
\mbox{$\Omega _K=-0.058^{+0.046}_{-0.026}$}
with \Planck+WP, and 
$\Omega _K=-0.004\pm0.0036$ with \Planck+WP+BAO.
More details can be found by consulting the parameter tables available
online.\footnote{Available at: \texttt{http://www.sciops.esa.int/index.php\\
?project=planck\&page=Planck\_Legacy\_Archive}}
Figure~\ref{fig:openinf} shows
$r$ and $\mathrm{d}n_{\mathrm s}/\mathrm{d}\ln k$ for this family of models.
We conclude that any possible spatial curvature is small in magnitude
even within this general model and that the spatial curvature scale is
constrained to lie far beyond the horizon today. 
Open models predict a tensor spectrum enhanced at small wavenumber $k\ltorder 1$, where $k=1$ corresponds to the curvature scale, but
our constraint on $\Omega _K$ and cosmic variance imply that
this aspect is likely unobservable.

\subsection{Relaxing the assumption of the late-time cosmological concordance model}

The joint constraints on $n_\mathrm{s}$ and $r$ shown in Fig.~\ref{fig:nsvsr} are one of the central
results of this paper.  However, they are derived assuming the standard $\Lambda$CDM cosmology
at late times (i.e., $z \lesssim 10^4$).  It is therefore natural to ask how 
robust our conclusions are to
changes of the late time cosmological model. We discuss two classes of models:
firstly, changes to the $\Lambda$CDM energy content; and secondly, a more general
reionization model. These extensions can lead to degeneracies of the additional
parameters with $n_\mathrm{s}$ or $r$.\footnote{ 
We considered a further generalization, which also causes the joint 
constraints on $n_s$ and $r$ to change slightly. We allowed
the amplitude of the lensing contribution to the temperature power 
spectrum $A_\mathrm{L}$ 
to vary as a free parameter. In this case we find the following 
{\it Planck}+WP+BAO constraints: 
$n_s=0.972 \pm 0.006$, $A_\mathrm{L}=1.24^{+0.10}_{-0.11},$ $r<0.15$ at 95\%~CL.}

\subsubsection{Extensions to the energy content}

We consider the $\Lambda$CDM+$r$+$N_\mathrm{eff}$, $\Lambda$CDM+$r$+$Y_\mathrm{P}$,
$\Lambda$CDM+$r$+$\sum m_\nu,$ and $\Lambda$CDM+$r$+$w$ extensions of the standard model. 
This selection is motivated by the impact on the CMB damping tail of the first two and the effect on the
Sachs-Wolfe plateau at low multipoles for the latter two.  The resulting contours are shown in
Fig.~\ref{fig:nsvsr_ext}. 
While the lower limit on $n_\mathrm{s}$ is stable under all extensions considered here, the models that
alter the high-$\ell$ part of the spectrum permit significantly bluer spectral tilts, and accordingly also
lead to a weaker bound on the tensor-to-scalar ratio.    
By allowing $N_\mathrm{eff}$ to float, we obtain $n_{\mathrm{s}} = 0.9764 \pm 0.0106$ and
$r_{0.002} < 0.15$ at 95\% CL.\footnote{Selected non-standard values for $N_\mathrm{eff}$ deserve further 
investigation. For the additional fractional contribution motivated by 
a Goldstone boson $\Delta N_\mathrm{eff}= 0.39$ \citep{Weinberg:2013kea},
we obtain $n_{\mathrm{s}} = 0.9726 \pm 0.0057$ and
$r_{0.002} < 0.14$ at 95\% CL for {\it Planck}+WP+BAO+high-$\ell$. For an additional species of neutrinos, we obtain 
$n_{\mathrm{s}} = 0.9947 \pm 0.0056$ and $r_{0.002} < 0.17$ at 95\% CL for {\it Planck}+WP+BAO+high-$\ell$.} 
For $\Lambda$CDM+$r$+$Y_\mathrm{P}$ we obtain
$n_{\mathrm{s}} = 0.9810 \pm 0.0111$ and $r_{0.002} < 0.18$ at 95\% CL.
The models modifying the large-scale part of the power spectrum, on the other hand, do not lead to a
notable degradation of constraints on either $n_\mathrm{s}$ or $r$ ($n_{\mathrm{s}} = 0.9648 \pm
0.0061$ and $r_{0.002} < 0.13$ at 95\% CL for $\Lambda$CDM+$r$+$\sum m_\nu$, and $n_{\mathrm{s}}
= 0.9601 \pm 0.0070$ and $r_{0.002} < 0.11$ at 95\% CL for $\Lambda$CDM+$r$+$w$).

\begin{figure}[!ht]
\centering
\includegraphics[width=8.8cm]{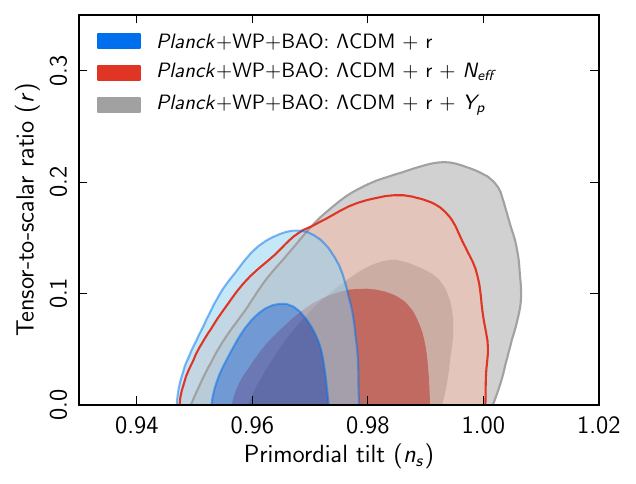}
\includegraphics[width=8.8cm]{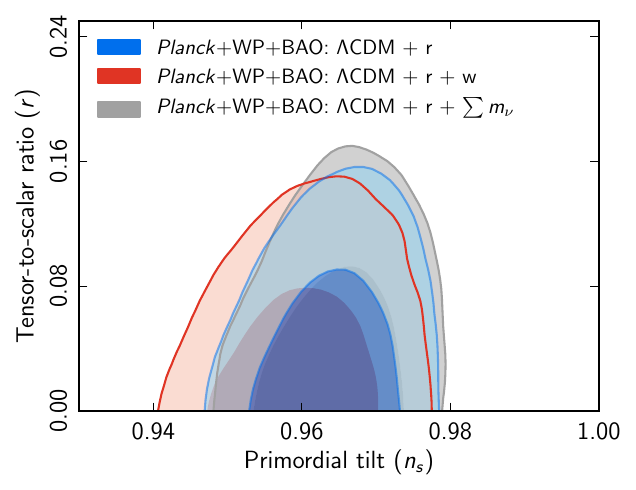}
\caption{Marginalized joint 68\% and 95\%~CL regions for \Planck+WP+BAO data for $\Lambda$CDM+$r$+$N_\mathrm{eff}$
and $\Lambda$CDM+$r$+$Y_\mathrm{P}$ ({\it top}); and $\Lambda$CDM+$r$+$\sum m_\nu$ and $\Lambda$CDM+$r$+$w$ ({\it bottom}).
Shown for comparison are the corresponding contours for the $\Lambda$CDM+$r$ model.}
\label{fig:nsvsr_ext}
\end{figure}

\subsubsection{General reionization scenario}
In the standard rapid reionization scenario typically used
in CMB analysis, the Universe is assumed to be completely transparent after recombination,
but the ionization fraction increases from zero to one over a duration $\Delta z\approx 1$ at a certain redshift $z_\mathrm{reion},$ which is
the only unknown parameter of the reionization model. This model is obviously
simplistic, but for CMB analysis it works quite well because the CMB has little
sensitivity to the details of how the ionization fraction changes from 0 to 1. 
In this section we study to what extent allowing more general reionization scenarios
may alter some of the conclusions concerning the constraints on $n_\mathrm{s}$ and $r$
as well as on $\tau$.
As discussed in \cite{Mortonson:2007hq} and \cite{Mortonson:2008rx}, CMB anisotropies 
constrain reionization almost entirely by using the
shape of the large-scale {\it EE} power spectrum, and the power is redistributed from larger to smaller
scales for reionization processes which take place
during a non-negligible redshift interval, since they start at an earlier epoch.

We use the method developed by \cite{Mortonson:2007hq}  
to describe and constrain the reionization history.
A complete principal component basis serves to describe the effect 
of reionization on the large-scale $E$-mode polarization power spectrum.
Following \cite{Mortonson:2007hq} we bin the ionization history $x_{\mathrm e}(z_{i})$ using $95$ equal width bins with $\Delta z=0.25$ 
ranging from $z_\mathrm{min}=6$ to $z_\mathrm{max}=30$. 
For the redshifts $z< z_\mathrm{min}$ we assume values for $x_{\mathrm e}$ 
which take into account first (and possibly second) helium ionization and 
complete hydrogen ionization ($x_{\mathrm e}=1.16$ for $z<3$ and $x_{\mathrm e}=1.08$ for $3<z<6$).
For  $z>30$ 
we fix $x_{\mathrm e}=2\times 10^{-4}$ as the value of $x_{\mathrm e}$ 
expected before reionization (and after primordial recombination).
Any reionization history can be parameterized as a free function of redshift by decomposing 
the ionization fraction as $x_{\mathrm e}(z)=x_{\mathrm e}^{\mathrm f}(z)
+\sum_{\mu}m_{\mu}S_{\mu}(z)$, 
where the principal 
components, $S_{\mu}(z)$,  are the eigenfunctions of the
Fisher matrix computed by taking the derivatives of the {\it EE} polarization power spectrum with 
respect to $x_{\mathrm e}(z)$ and $x_{\mathrm e}^{\mathrm f}(z)$ is a fiducial ionization fraction. 
Following \cite{Mortonson:2007hq} we consider here the first five eigenfunctions, 
$S_{\mu}(z)$  with $\mu=1,...,5$, 
which will be varied with the other cosmological parameters. 

In Fig.~\ref{fig3} we plot the 68\% and 95\%~CL regions for $n_\mathrm{s}$ and $r.$ 
The constraint on the tensor-to-scalar ratio is not significantly affected by this additional marginalization, increasing 
to $r_{0.002} < 0.13$  at 95\%~CL.
The scalar spectral index is increased to $n_\mathrm{s} = 0.9650\pm 0.0080$, 
compared with $n_\mathrm{s} = 0.9603\pm 0.0073$ obtained with the rapid reionization scheme.
This is the same trend as noted in \cite{Pandolfi:2010mv} using {\it WMAP} data, but the effect is less significant due to the improved measurement of the temperature spectrum by \Planck.
The larger freedom in the reionization history
increases the width of the posterior on the derived optical depth, 
which is still partially degenerate with the scalar spectral index of primordial perturbations. 
The $n_\mathrm{s}=1$ model is still excluded at high significance; however, 
we find $\Delta \chi_\mathrm{eff}^2 = 12.5$ compared to the $\Lambda$CDM model.

\begin{figure}
\centering
\includegraphics[width=8.8cm]{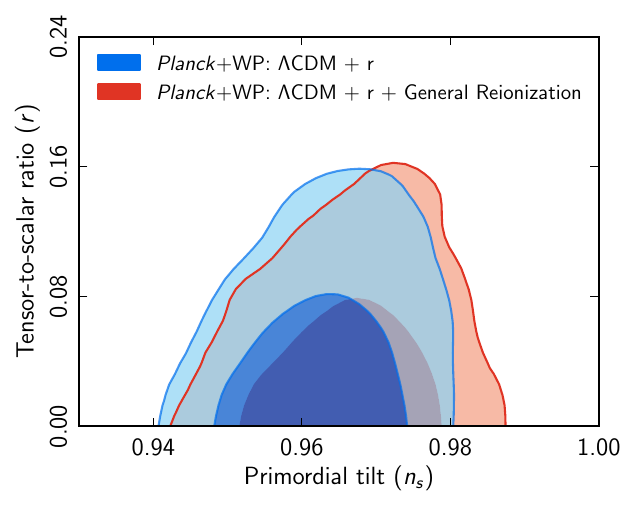}
\caption{Marginalized joint 68\% and 95\%~CL regions for {\it Planck}+WP data for 
$\Lambda$CDM+$r$ for instantaneous and general reionization.}
\label{fig3}
\end{figure}


}

\section{Inflationary model comparison}
\label{sec:modelcomp}

{
{

\def\baSect.{\begin{eqnarray}}
\def\ea{\end{eqnarray}}

In Sect.~\ref{akaSsectionFourOne}
several representative families of parameterized models
for the inflationary potential were analysed within the
slow-roll approximation in the neighbourhood of the pivot scale
$k_*$. Approximate constraints were 
applied to reject models for which there is no plausible scenario
for entropy generation. In this
section we revisit some of the parametric models defined
in Sect.~\ref{akaSsectionFourOne}. 
Here, however, the modes for the first order
perturbations, as described in Sect.~\ref{FluctGeneration}, are integrated
numerically. Thus there is no slow-roll approximation, and the issue of the
existence of a plausible scenario for entropy generation
is examined more carefully. We perform a statistical
model comparison between the competing parameterized
potentials, both within the framework of
Bayesian model comparison and in terms of the relative
likelihoods of the best fit models from each parameterized
family.

As noted in Sect.~\ref{EntropyGeneration}, considerable uncertainty
surrounds what occurred during the epoch of entropy generation,
particularly with respect to the energy scale at which
entropy generation ends and the average
equation of state between that epoch and the end of inflation.
For this reason, we explore a number of scenarios for this intermediate era 
during which entropy generation takes place.
The models compared in this section include
inflation with power law potentials, defined in Eq.~\ref{PowerLawPot:Eq}, 
with several representative values for $n$;
natural inflation, in Eq.~\ref{NatInf}; and hilltop inflation, with
$p=4$ and $\lambda = 4 \Lambda^4/\mu^4$ in Eq.~\ref{newinf}.

The free parameters in these inflationary potentials may vary over several orders of 
magnitude corresponding to unknown scales in high energy particle physics. Consequently a 
logarithmic prior is a sensible choice for these parameters. However, there is no 
theoretical guidance on how to truncate these model priors. We therefore adopt broad priors initially
and then truncate them as follows in order to compare all the models 
on an equal footing. 

The strongest constraint on the inflationary parameter space comes from the amplitude of the 
primordial power spectrum. This is a free parameter in most models, and successful structure 
formation in a universe dominated by cold dark matter has long been known to require primordial 
fluctuations with $\mathcal{R} \approx 10^{-5}$, or $A_{\mathrm{s}}\approx 10^{-10}$ \citep[see 
e.g.,][]{Zeldovich:1972zz,Linde1990Bk}. We can therefore immediately reject models for which 
$A_{\mathrm{s}}$ is far from this value, so regions of parameter space which do not yield 
$10^{-11} \le A_{\mathrm{s}} \le 10^{-7}$ are {\it a priori} excluded. This range is generous 
relative to estimates of $A_{\mathrm{s}}$ prior to \Planck\ \citep[e.g.,][]{komatsu2010}, but 
the results do not depend strongly on the range chosen. This effectively truncates the 
logarithmic priors on the model parameters, leaving a parameter subspace compatible with basic 
structure formation requirements.

For the single parameter models this requirement defines the range of $\lambda$ in 
Eq.~\ref{PowerLawPot:Eq}. However, for generic multi-parameter models, an $A_{\mathrm s}$-based 
cut may select a nontrival region of parameter space, as happens for the two cases considered 
here (see e.g., Fig.~1 in \cite{2012PhRvD..85j3533E}).  Without the $A_{\mathrm s}$-based cut in 
the prior, the parameter volume for both natural and hilltop inflation would be rectangular, and 
the corresponding Bayesian evidence values computed for these models would be lowered.

\begin{table}
\begin{center}
\begin{tabular}{lc}
\hline
\hline
Model & Priors   \\
\hline 
$n=2/3$ & $-13<\log _{10} (\lambda)<-7$  \\ 
$n=1$ & $-13<\log _{10} (\lambda)<-7$  \\ 
$n=2$ & $-13.5<\log _{10} (m^2)<-8$  \\ 
$n=4$ & $-16<\log _{10} (\lambda)<-10$  \\ 
&\\
Natural & $-5<\log _{10} (\Lambda)<0$ \\
 & $0<\log _{10} (f)<2.5$ \\ 
Hilltop & $-8<\log _{10} (\Lambda)<-1$    \\
 $p=4$ & $-17<\log _{10} (\lambda)<-10$   \\
&\\
$\Lambda$CDM & $0.9 < n_{\mathrm s} < 1.02$  \\
& $ 3.0 < \ln \left[10^{10} A_s \right] <  3.2$\\
\hline
\end{tabular}
\end{center}

\begin{center}
\begin{tabular}{lc}
\hline
\hline
Matching   & Prior \\
\hline 
$N_*$ &   $20\le N^* \le 90$  \\
\hline
\end{tabular}
\end{center}
\caption{Model priors. Dimensionful quantities are expressed in units with 
$M_\mathrm{pl}$ set to unity. 
The matching parameter $N_*$, computed at  $k_*= 0.05$~Mpc$^{-1}$, 
allows us to marginalize over the uncertainty in connecting the 
inflationary era to astrophysical scales. Additional cuts are 
made in these parameter ranges to select out physically relevant subspaces 
compatible with basic structure formation requirements and different 
entropy generation scenarios, as described in the text, resulting in non-rectangular model 
priors in some cases. We also marginalize over the concordance cosmological parameters and 
foreground parameters of the \Planck\ likelihood, as described in the text.
}
\label{tab:priors}
\end{table}

As discussed in Sect.~\ref{EntropyGeneration}, specifying an inflationary 
potential does not enable us to predict the late time CMB angular power spectra. The subsequent 
expansion history and
details of the epoch of entropy generation are required to relate the value of 
the inflaton field at Hubble radius crossing to comoving wavenumbers in today's 
Universe, through Eq.~\ref{eq:nefolds}. Physically, the fundamental parameter that 
sets the observable perturbation spectrum is the value of the field $\phi _*$ 
at which the pivot mode leaves the Hubble radius. 
It can be rescaled by a shift $\phi _* \rightarrow \phi _* + \phi _0$, and the 
range over which $\phi $ changes during inflation varies greatly between models. 
Consequently we treat the remaining number of $e$-folds, $N_*$, after the pivot 
scale leaves the Hubble radius as a free parameter with a wide uniform prior, since 
this quantity has a consistent interpretation across models. The pivot scale 
used to compute $N_*$ is $k_*=0.05$~Mpc${}^{-1}$. 
However, given our ignorance  
concerning the epoch of entropy generation, a multitude of entropy generation 
scenarios for each inflationary potential can occur. Some possibilities are as follows, with 
parameters referring to Eq.~\ref{eq:nefolds}.

\begin{enumerate}
\item
{\sl Instantaneous entropy generation scenario}. 
\hfill \break 
At the end of inflation, all the energy in the inflaton field is instantaneously 
converted into radiation.
\item
{\sl Restrictive entropy generation scenario (narrow range for $\wint$)}.
\hfill \break
$\rho_\mathrm{th}^{1/4} =10^9$ GeV, and $\wint\in [-1/3,1/3].$
\item
{\sl Permissive entropy generation scenario (wide range for $\wint$)}.
\hfill \break 
$\rho_\mathrm{th}^{1/4}=10^3$ GeV, and $\wint\in [-1/3,1].$
\end{enumerate}
The equations of state with $\wint$  in the range $[1/3,1]$ appear less 
plausible, but models with these values have been put forward \citep{Pallis:2005bb}, so this 
possibility cannot be completely excluded. 
Moreover the $\wint$ parameterization captures a variety of scenarios in which the 
post-inflationary Universe is thermalized, but not radiation dominated, 
including phases of coherent oscillations \citep{Martin:2010kz,Easther:2010mr}, 
resonance \citep{Traschen:1990sw,Kofman:1994rk,Kofman:1997yn,Allahverdi:2010xz}, kination 
\citep{Spokoiny:1993kt,Chung:2007vz}, secondary or thermal inflation 
\citep{Lyth:1995ka}, moduli domination \citep{Banks:1993en,deCarlos:1993jw}, 
primordial black hole domination \citep{Anantua:2008am}, or a frustrated cosmic 
string network \citep{Burgess:2005sb}, all of which lead to an expansion rate 
differing from that of a radiation dominated universe.

\begin{table*}
\begin{center}
\begin{tabular}{lrrrrrrrrrr}
\hline
\hline
Model&
\multicolumn{2}{c}{Instantaneous }&
\multicolumn{2}{c}{Restrictive}&
\multicolumn{2}{c}{Permissive}\\
&
\multicolumn{2}{c}{entropy generation}&
\multicolumn{2}{c}{entropy generation}&
\multicolumn{2}{c}{entropy generation}\\
\\
&   
$\ln [{\cal E}/{\cal E_0}]$ & $\chisq$  & 
$\ln [{\cal E}/{\cal E_0}]$ & $\chisq$  & 
$\ln [{\cal E}/{\cal E_0}]$ & $\chisq$  \\ 
\hline 
$n=4$  &  $-14.9$ & $25.9$ & $-18.8$ & $27.2$ & $-13.2$ & $17.4$ \\
$n=2$  &  $-4.7$ & $5.4$ &  $-7.3$ & $6.3$ & $-6.2$ & $5.0$  \\
$n=1$  &  	$-4.1$ & $3.3$ & $-5.4$  & $2.8$ & $-4.9$ &  $2.1$ \\
$n=2/3$  &  $-4.7$ & $5.1$ & $-5.2$ & $3.1$ & $-5.2$ & $2.3$ \\
Natural &  $-6.6$ & $5.2$ & $-8.9$ & $5.5$ &  $-8.2$ & $5.0$ \\
Hilltop  &	$-7.1$  & $6.1$  & $-9.1$ & $7.1$ & $-6.6$ & $2.4$  \\
\hline
$\Lambda $CDM&  $-4940.7$ &  $9808.4$ & $ ...  $ & $ ...  $ & $ ...  $ & $ ...  $ \\
\hline
\end{tabular}
\end{center}
\caption{Inflationary model comparison results.
For each model and set of assumptions concerning entropy generation [(1), (2), (3)], the 
natural logarithm of the Bayesian evidence ratio as well as $\chisq$ for the 
best fit model in each category are indicated, relative to the $\Lambda$CDM concordance 
model (denoted by subscript ``0''). $\ln {\cal E}_0$ and $-2\ln{\cal L_0}$ are 
given for the reference model.
}
\label{tab:modelselectionresults}
\end{table*}

At the other extreme, the decision to exclude $\wint<-1/3,$ as done here, is 
not completely justifiable. We cannot, for example, rule out a first order phase 
transition at a lower energy scale that would drive $\wint$ below $-1/3$, but 
here we neglect this possibility. Our analysis does not preclude a 
secondary period of inflation, but does require that the 
{\it average} expansion 
during the post-inflationary regime parameterized by $\wint$ should not be 
inflationary. This caveat should be kept in mind.

For some of the parameterized models, tighter constraints can, in principle, be 
placed on $\wint$. It has been argued \citep[see e.g.,][]{Liddle:2003as} that 
for the $\lambda \phi ^4$ potential, the uncertainties concerning 
entropy generation 
contribute almost no uncertainty in the determination of $\phi _*$. This is because 
according to the virial theorem, a field sloshing about $\phi =0$ in a quartic 
potential has the same average $w$, namely $w=1/3$, as the radiation equation of 
state. More generally, for a potential of the form $\phi^n$ around the minimum, 
$w_\mathrm{vir}= \langle w\rangle =(n-2)/(n+2)$ \citep{1983PhRvD..28.1243T}; 
therefore, one may argue that $\wint$ should be restricted to the interval 
whose endpoints are $1/3$ and $w_\mathrm{vir}$. This approach was taken by 
\cite{Martin:2010kz} in obtaining Bayesian constraints on the reheating 
temperature for monomial potentials from the CMB. However this scenario 
requires a carefully tuned potential that has approximately a $\phi^n$ shape, both at 
large field values and near the origin far below the inflationary scale. 
Typically, potentials for which $V(\phi) \sim \phi^n$ at large field values can 
have very different shapes near the origin. Thus, following 
\cite{2012PhRvD..85j3533E}, in this paper we explore a broader range 
of $\wint$ for these models (including the cases above as subsets) in order to 
obtain data-driven constraints on $\wint$.

In this paper we focus on the three representative scenarios itemized above, 
referred to hereafter as scenarios (1), (2), and (3). Our algorithm draws a 
value of $N_*$ and then given the value of $\rho_\mathrm{th}$, computes 
$\wint$, which is a derived parameter. Models for which $\wint$ lies outside 
the specified range of each scenario under consideration are excluded.

The full set of priors for the inflationary physics is given in 
Table~\ref{tab:priors}. Dimensionful quantities are expressed in units with 
reduced Planck mass $M_\mathrm{pl}$ set to unity. 

Due to the nontrivial likelihood surfaces and the large dimensionality of the parameter spaces 
explored in this section, we use \ModeCode\ coupled to \MultiNest\ {\tt v3.0}\footnote{Made 
available ahead of public release to the \Planck\ Collaboration by Farhan Feroz and Mike 
Hobson \citep{2013arXiv1306.2144F}.} to map out the parameter space. In addition to the standard nested sampling (NS) 
algorithm, \MultiNest\ {\tt v3.0} enables nested importance sampling (NIS), resulting in 
substantial speed gains\footnote{We have carried out 
extensive tests of NIS versus NS, and chosen the 
following settings for the computations presented here: NIS on, constant efficiency mode on, 
$300$ live points, tolerance and efficiency parameters set to $0.5$ and $0.02$, respectively.} 
and significant enhancements in the accuracy of the Bayesian evidence computation compared 
to NS alone for 
the same computational setup.

\subsection{Results}

Table~\ref{tab:modelselectionresults} presents model comparison results for the ensemble of 
parameterized potential families described above. We report the Bayesian evidence ({\it 
model averaged likelihood}) ratio, which provides a self-consistent framework for calculating 
the {\it betting odds} between models (see Sect.~\ref{sec:modelsel}). The uncertainty in these 
logarithmic evidence values is approximately $0.2$. 
We also report the $\chisq$ values computed from the $2 
\ln {\cal L}_\mathrm{max}$ values found by the sampler.

The monomial models have a single parameter potential, and the natural and hilltop inflation 
models have two parameters each. All entropy generation scenarios except case (1) contribute one 
additional parameter to the inflationary sector. The evidence ratios and $\chisq$ values are 
presented with respect to the $\Lambda$CDM cosmological model.

None of the inflationary models tested here fit the data as well as the $\Lambda$CDM model.  
This mostly reflects that there is no evidence in the data for $r$ different from zero. 
Furthermore, the priors listed in Table~\ref{tab:priors} for the $\Lambda$CDM primordial sector 
are purely phenomenological, roughly corresponding to ranges somewhat broader than {\it WMAP} 
constraints. Narrowing them around the best fit model arbitrarily increases the evidence in its 
favour. Instead it is instructive to compare the relative evidence for the inflationary models 
presented.

Table~\ref{tab:modelselectionresults} shows that the $\lambda \phi^4$ model is decisively ruled 
out by \Planck, confirming previous analyses by the {\it WMAP} team 
\citep{2003ApJS..148..213P,spergel2007,dunkley2009,komatsu2010} based on the model track plotted 
on the $n_{\mathrm s}\textrm{-}r$ plane. Recent model selection analyses \citep{Martin:2010hh, 
2012PhRvD..85j3533E} with {\it WMAP} 7-year data found that the model was already 
disfavoured by odds of about $400$:$1$ against. With \Planck, the odds against this model are 
{\em at least} $500\, 000$:$1$ compared to $\Lambda$CDM  for a broad range of entropy generation 
scenarios. The same conclusion is confirmed by the extremely poor $\chisq$ values for the model. 
Given the strength of our results, in the flexible setting of the permissive entropy generation 
scenario, it is possible not just to rule out models where the potential is of the quartic form 
in the full range from the origin to the inflationary scales, but also a general class where the 
potential is of the $n=4$ form in the $\phi$-range where the cosmological perturbations are 
generated, but exhibits a different shape near the origin.

Two other large field models, the quadratic potential and natural inflation, are somewhat
disfavoured by the \Planck\ data, especially when broader entropy generation scenarios are 
considered. Compared with the $\Lambda$CDM model, these models are disfavoured by $\chisq \sim 
5$--$6$ depending on the entropy generation scenario. This reflects the analysis of 
Sect.~\ref{sec:slow-roll}, where the overlap of the model predictions and the data constraints 
on the $n_{\mathrm s}\textrm{-}r$ plane is seen to be mostly outside the joint $68$\% CL contour. 
However, from the Bayesian evidence point of view, it is too early to declare these models 
incompatible with the data. To make this judgement, it is more prudent to compare these 
models to the $n=1$ case, which has the best evidence with respect to $\Lambda$CDM, rather 
than to $\Lambda$CDM itself, which provides our reference point for the evidence calculation, 
but has arbitrary prior ranges. In their simplest forms---instantaneous entropy generation---the 
$n=2$ and natural inflation models are only disfavoured by odds of about $1$--$12$:$1$ against, 
which does not rise to a high level of significance.\footnote{In comparison, odds of $~150$:$1$ 
are considered highly significant in this context.}

\begin{figure*}
\includegraphics[height=0.25\textwidth]{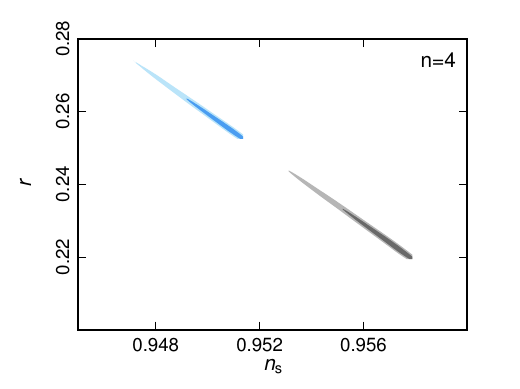}
\includegraphics[height=0.25\textwidth]{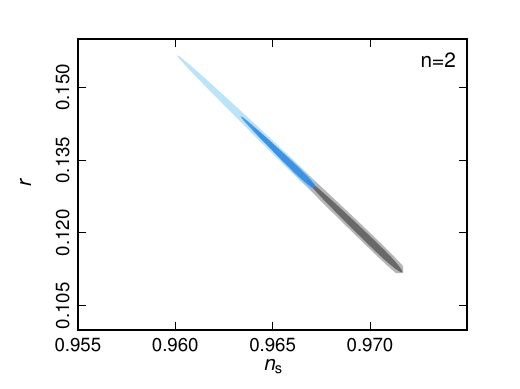}
\includegraphics[height=0.25\textwidth]{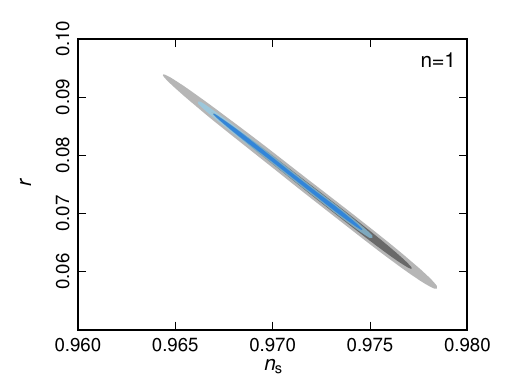} \\
\includegraphics[height=0.25\textwidth]{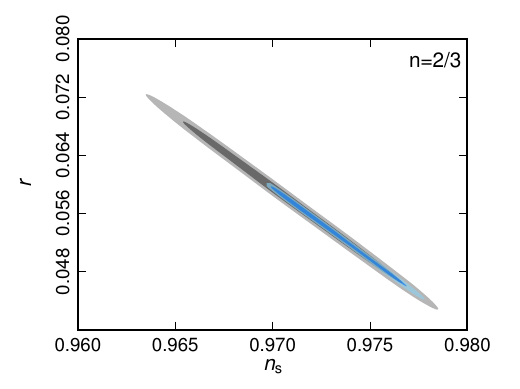}
\includegraphics[height=0.25\textwidth]{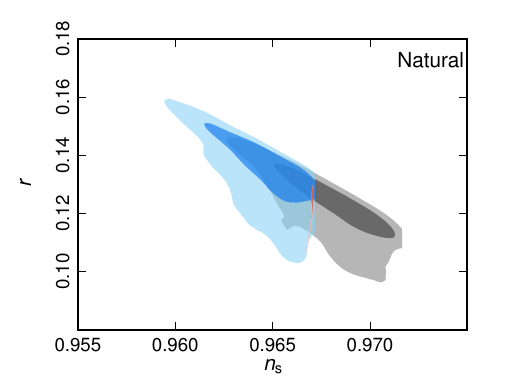}
\includegraphics[height=0.25\textwidth]{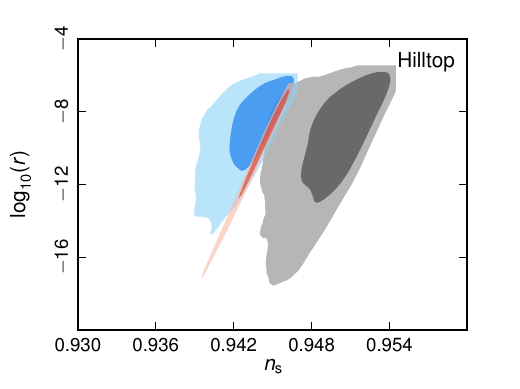}
\caption{Constraints on $n_{\mathrm s}$ vs.\ $r$ at $k_*=0.002$ Mpc$^{-1}$ for the inflationary 
models considered (i.e., power law potentials with \mbox{$n=2/3,1,2$}, and $4$, natural 
inflation, and hilltop inflation), showing joint 68\% and 95\%~CL. Blue and grey 
distributions correspond to the restrictive and permissive entropy generation scenarios, 
respectively. The instantaneous entropy generation case corresponds to the thin (red) 
contours in the natural and hilltop panels; for the single parameter models, 
this case corresponds to the lowest-$r$ extremity of the restrictive case. The difference between the 
{\it natural inflation} region in Fig.~\ref{fig:nsvsr} and the natural inflation constraints 
shown here is due to the strong projection effect described in the text.}
\label{fig:modelComp_nsr}
\end{figure*}

\begin{figure*}
\includegraphics[height=0.25\textwidth]{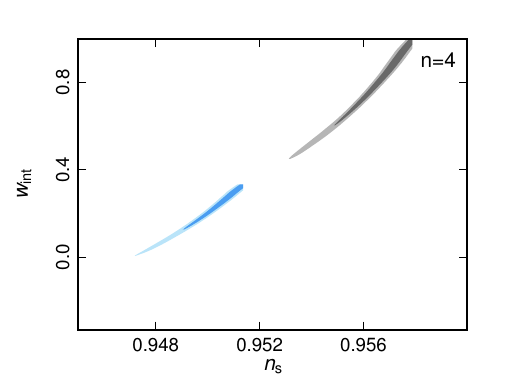}
\includegraphics[height=0.25\textwidth]{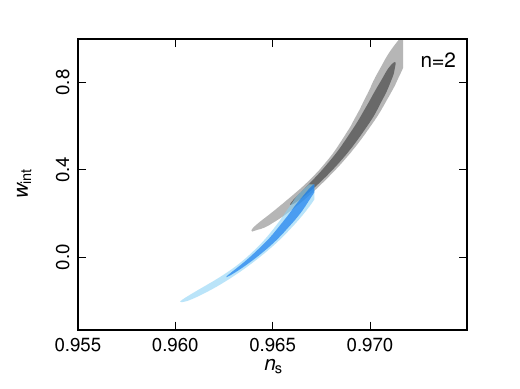} 
\includegraphics[height=0.25\textwidth]{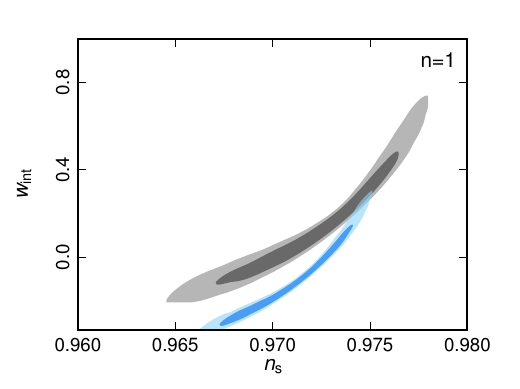} \\
\includegraphics[height=0.25\textwidth]{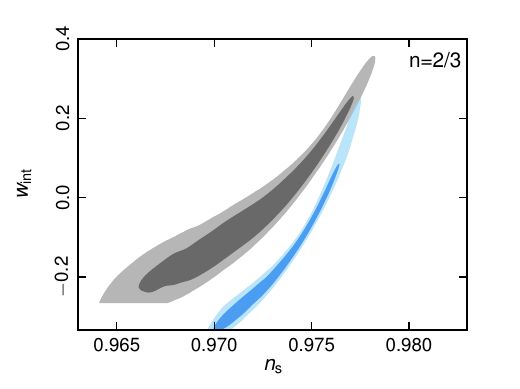} 
\includegraphics[height=0.25\textwidth]{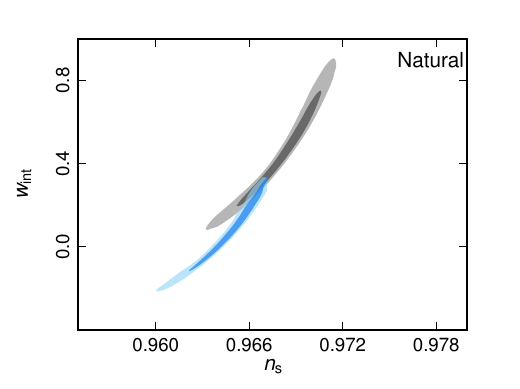}
\includegraphics[height=0.25\textwidth]{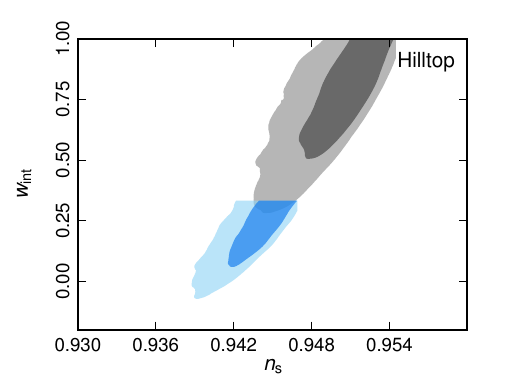}
\caption{Constraints on $n_{\mathrm s}$ vs.\ $\wint$ at $k_*=0.002$ Mpc$^{-1}$ for the 
inflationary models considered, as in Fig. \ref{fig:modelComp_nsr}. The instantaneous entropy generation
case 
(1) corresponds to $\wint=1/3$.}
\label{fig:modelComp_nsw}
\end{figure*}
    
\begin{figure*}
\includegraphics[height=0.25\textwidth]{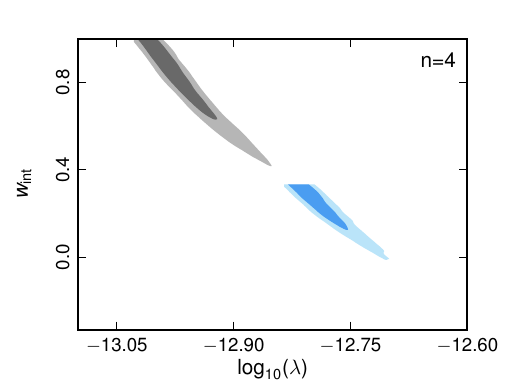}
\includegraphics[height=0.25\textwidth]{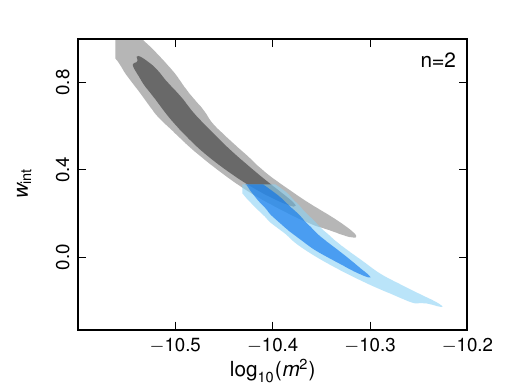} 
\includegraphics[height=0.25\textwidth]{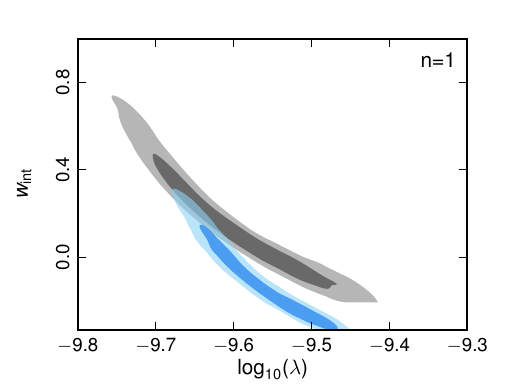} \\
\includegraphics[height=0.25\textwidth]{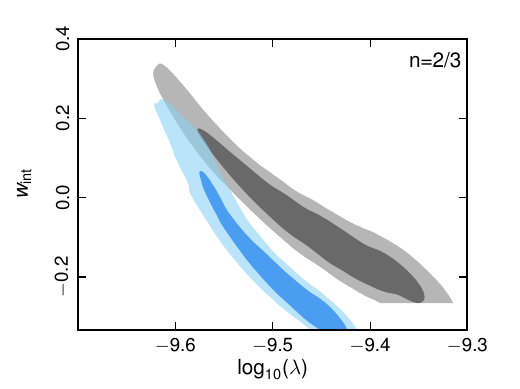} 
\includegraphics[height=0.25\textwidth]{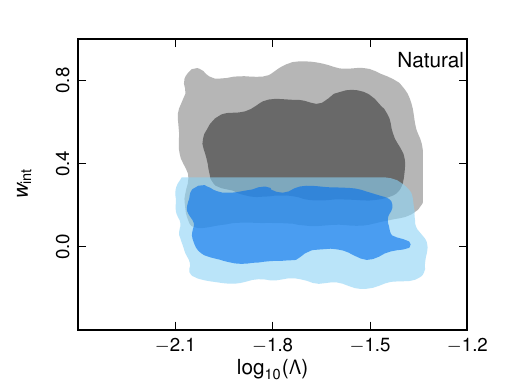}
\includegraphics[height=0.25\textwidth]{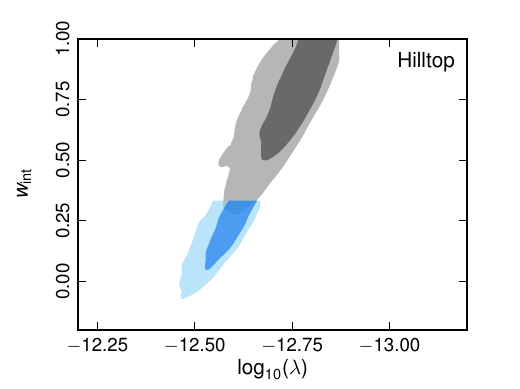}
\caption{Constraints on $\log _{10} (\lambda )$ vs.\ $\wint$ for the inflationary models 
considered, as in Figs. \ref{fig:modelComp_nsr} and \ref{fig:modelComp_nsw}.}
\label{fig:modelComp_lambdaw}
\end{figure*}

\begin{figure}
\includegraphics[height=0.38\textwidth]{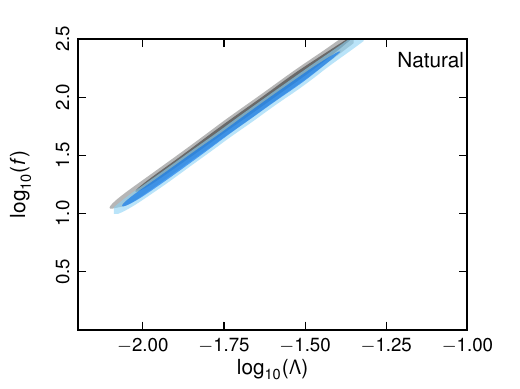} \\
\includegraphics[height=0.38\textwidth]{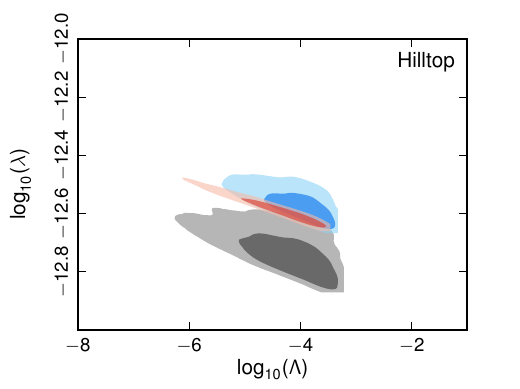} 
\caption{Potential parameters for natural inflation and hilltop 
inflation, as in Figs. \ref{fig:modelComp_nsr}, \ref{fig:modelComp_nsw}, and 
\ref{fig:modelComp_lambdaw}. On the
natural inflation panel, instantaneous entropy generation corresponds to a thin diagonal along 
the top edge of entropy generation case (3).
}
\label{fig:modelComp_potparams}
\end{figure}

The models most compatible with the \Planck\ data in the set considered here are the two 
interesting {\it axion monodromy} potentials, $n=1$ \citep{McAllister:2008hb} and $n=2/3$ 
\citep{Silverstein:2008sg}, which are motivated by inflationary model building in the context of 
string theory. The $p=4$ hilltop model presents an interesting case. This model was previously 
found to be compatible with {\it WMAP} 7-year data, performing almost as well as the monodromy 
potentials \citep{2012PhRvD..85j3533E}. However it exhibits significant tension with the 
\Planck\ data, both in terms of evidence ratios and the maximum likelihood. The only exception 
is the entropy generation scenario (3) which has odds of greater than $1000$:$1$ against compared to 
$\Lambda$CDM, and yet the maximum likelihood is not significantly different from the $n=1$ case. 
This indicates that while the extra freedom allowed by the least restrictive
entropy generation scenario 
improves the best fit, this prior is not very predictive of the data. However, the result seems
counterintuitive and merits further comment; we will consider this question further at the end of this 
section when parameter estimation results are discussed.

We now turn to parameter constraints. Figure~\ref{fig:modelComp_nsr} presents marginalized joint 
constraints from \Planck +WP alone on the derived parameters $n_{{\mathrm s},0.002}$ and 
$r_{0.002}.$ Figure~\ref{fig:modelComp_nsw} shows the corresponding joint constraints on 
$n_{{\mathrm s},0.002}$ and $\wint$, again a derived parameter.
Figure~\ref{fig:modelComp_lambdaw} shows joint constraints on $\wint$ and the potential parameter 
$\log _{10} (\lambda)$ ($\log _{10} (\Lambda)$ in the case of natural inflation). It is instructive to 
consider the three sets of figures together. The restrictive and permissive entropy generation 
scenarios are shown on all panels; the instantaneous entropy generation 
case is shown for the two parameter 
models, natural inflation and hilltop, in Fig.~\ref{fig:modelComp_nsr}---for the monomial 
potentials this case corresponds to the lowest-$r$ extremity of the restrictive case 
constraints, and to $\wint=1/3$ in the other two figures.

The quartic potential conflicts with the data because it predicts a high tensor-to-scalar ratio. 
Hence the model maximizes its likelihood by pushing towards the lower-$r$, bluer-$n_{\mathrm s}$ 
limits of its parameter space, which corresponds to increasing $\wint$ as much as allowed by the 
entropy generation prior. The contours terminate at the lowest-$r$ limit when each entropy 
generation case hits 
its $\wint$ prior upper limit (i.e., $\wint = 1/3$ and $\wint = 1$ for restrictive and 
permissive entropy generation, respectively). For each case, the lower limit on $\wint$, 
corresponding to the reddest-$n_{\mathrm s}$ extremity of the confidence contours, is 
data-driven. The quadratic potential encounters the same difficulty but at a less extreme level.

The two axion monodromy potentials are compatible with a wide range of entropy generation 
scenarios. The instantaneous entropy generation scenario is 
compatible with the data for both models. For 
restrictive entropy generation in the $n=1$ model, we obtain a data-driven upper limit on 
$\wint$, which just touches the $\wint = 1/3$ case at the 95\% CL. At the lower limit, the 
$\wint$ posterior is truncated by the prior, as for the $n=2/3$ case. For the latter, there is a 
data-driven upper limit on $\wint$ which is controlled by the upper limit on $n_{\mathrm s}$. 
For permissive entropy generation, the upper and lower limits on $\wint$ for both models are 
data-driven, corresponding to the upper and lower limits on $n_{\mathrm s},$ respectively.

The constraints on natural inflation require some interpretation. The relationship between the 
empirical $n_{\mathrm s}$ and $r$ parameters and the potential parameters for natural inflation 
is discussed in detail by \cite{Savage:2006tr} and \cite{Mortonson:2010er} along with parameter 
constraints derived from {\it WMAP} 3-year \citep{spergel2007} and {\it WMAP} 7-year \citep{larson2010}, 
respectively. In this model, there is a degeneracy between $f$ and $\Lambda$ in the limit where 
these parameters are large so that natural inflation resembles the quadratic model. The priors 
are chosen to exclude most of this region. The priors on $\log _{10} (f)$ and $\log _{10} (\Lambda)$ 
still allow a region of nearly degenerate models that contribute to a {\it ridge} seen in the natural 
inflation panel of Fig.~\ref{fig:modelComp_nsr}. These models closely match the values of 
$n_{\mathrm s}$ and $r$ seen in the quadratic potential constraints.  The marginalized 
constraints on $n_{\mathrm s}$ and $r$ depend strongly on the prior on $\log _{10} (f)$ due to the 
projection of a large number of degenerate models onto this ridge. Therefore the apparent 
preference for this region of parameter space over models with lower values of $r$ is largely 
due to this effect and is not driven by the data. This highly nonlinear mapping between the 
logarithmic priors on the potential parameters and the power law parameters (which are derived 
parameters in this analysis) leads to a strong projection effect, which accounts for the 
difference in visual appearance between these contours and the region labelled {\it natural 
inflation} in Fig.~\ref{fig:nsvsr}.

Generally, for fixed $N_*$, decreasing $\Lambda$ and $f$ reduces both $n_{\mathrm s}$ and $r$. 
Thus natural inflation models can have lower values of $r$ than the quadratic potential without 
increasing $n_{\mathrm s}$ and $N_*$. This feature means that the potential parameters for this 
model are relatively uncorrelated with $\wint$, in contrast with the other models considered 
here, as illustrated in the Fig.~\ref{fig:modelComp_lambdaw}. Nevertheless we obtain data-driven 
bounds on $\wint$ in the permissive entropy generation case as well as a lower bound in the restrictive 
entropy generation case. Both bounds overlap with the instantaneous generation limit.

The upper panel of Fig.~\ref{fig:modelComp_potparams} shows our lower limit of $f\gtrsim 10.0$ 
$M_\mathrm{pl}$ (95\% CL), compared with the {\it WMAP}7 limit $f\gtrsim 5.0$ $M_\mathrm{pl}$ 
(95\% CL) reported by \cite{Mortonson:2010er}. Indeed, the \Planck\ limit is in agreement with 
the \Planck\ prediction presented in that work. There is a hint of an upper limit on $f$ as 
well, driven by the fact that this corresponds to the quadratic inflation limit, which is in 
tension with the data. However this is only a $1\sigma$ effect.

The $p=4$ hilltop model has two distinct branches: a {\it small-field} scenario, where 
$r\lesssim 0.001$ and $n_{\mathrm s}< 0.95$; and a {\it large-field} limit in which $V(\phi) 
\sim \phi$ \citep{Adshead:2010mc}. Physically, the small-field limit is consistent with the Lyth 
bound, and we can select it by fixing $\log _{10} (\Lambda)< -2.5$ in the prior. We observe that the 
data select out this small field branch, which requires explanation given that we know that the 
$n=1$ model (the limit of the large field branch) is perfectly compatible with the data. In 
fact, this can be tested by restricting the $\log _{10} (\Lambda)$ prior by hand to the large-field 
branch. The hilltop model constitutes a difficult sampling problem, as is apparent from Fig. 4 
in \cite{2012PhRvD..85j3533E}. An examination of the progress
of the nested sampler for the case of 
the full hilltop prior reveals the reason for the small-field branch being selected out. In 
this case, the posterior for the large-field branch is extremely thin compared to the model 
prior in this regime---much thinner than the posterior for the small-field branch in comparison 
to its respective prior. Therefore this region occupies very little probability mass, and is 
dropped in preference to the more predictive small-field branch. This high likelihood but 
extremely thin ridge is also responsible for the counterintuitive result reported in the 
model selection analysis, where the model was found to have a good $\chisq$ with respect to 
$\Lambda$CDM, and yet be highly disfavoured by the Bayesian evidence. In summary, confirming the 
results of Sect.~\ref{sec:slow-roll}, this model is in agreement with the data in the limit 
where it overlaps the linear $n=1$ model. But since this region of high likelihood occupies a 
very small fraction of the prior, this model is heavily penalized by the Bayesian evidence for 
failing to predict the data over most of its prior space.

}

\section{Observable window of inflation}
\label{sec:numerical}

{

Section~\ref{akaSsectionFourOne}
presented an analysis of several representative inflationary 
potentials within the framework of the slow-roll approximation and their 
compatibility with the \Planck\ data.
The results are summarized in Fig.~\ref{fig:nsvsr}. In that case
the full potential is considered in order
to identify a plausible range for the location
of $\phi_{*}$ on the potential $V(\phi ).$ This requires
a complete story. In other words, the potential must be specified
starting above the point where the largest observable scales
first exited the Hubble radius, and extending 
to the minimum of the potential.  

In this section we explore another approach. We adopt the
point of view that we are interested in reconstructing
the inflationary potential only over the observable range---that is, 
the interval of $\phi $ corresponding to the scales 
observable today in the CMB. We constrain the potential over the range where 
these scales 
exited the Hubble radius during inflation as well as a few $e$-folds before and after.
The cosmological perturbations are not imprinted
instantaneously at the moment of Hubble radius crossing, but rather gradually 
over a few $e$-folds. We expand around $\phi _{*}$, taking the view that
a plausible extension of the potential outside this observable range is
always possible, so that one has precisely the number of  $e$-folds
of inflation needed for $\phi _{*}$ to correspond to $k_{*}$ today.

The argument is that one can always end inflation abruptly by
imposing a sharp waterfall feature where needed, or prolong inflation 
by inserting a sufficiently long
plateau into the potential by hand, for example, for models with a large tilt.
A foreseeable objection 
to this approach is that the extensions of the potential required 
outside the observable window may render the potential {\it unnatural}. 
This possibility should be kept in mind, although 
{\it naturalness} is an elusive and uncomfortably subjective concept.
The analysis in this section does not rely on the slow-roll
approximation. Instead each $k$ mode is integrated exactly by
numerical integration, as described in Sect.~\ref{FluctGeneration}, under the 
assumption of a canonical kinetic term.  

\begin{figure}[t]
\includegraphics[width=0.99\columnwidth]{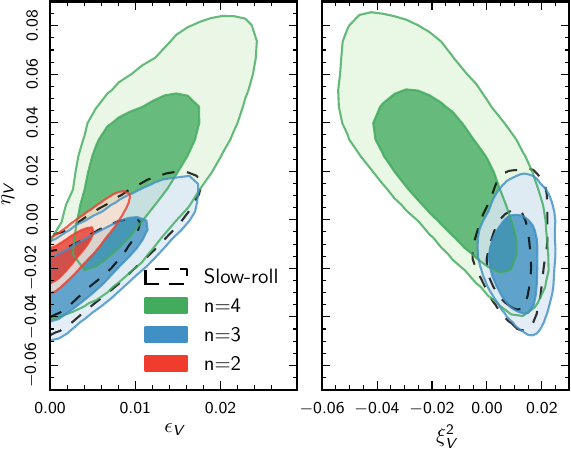}
\caption{Posterior distribution for the first three slow-roll potential parameters using \Planck+WP data.
In the $n=2 - 4$ cases, the inflaton potential is expanded to $n$th order, 
and the spectrum is obtained by fitting the numerically computed power spectrum to the data, with no slow-roll approximation 
and no assumption about the extrapolation of the potential outside 
the observable window. \label{fig:R}}
\end{figure}

\begin{figure}[t]
\includegraphics[width=0.99\columnwidth]{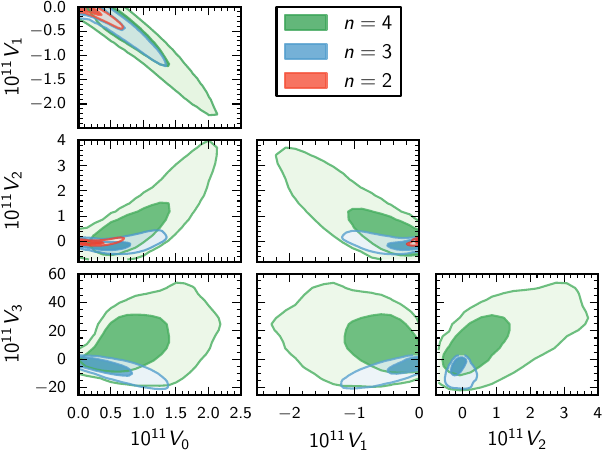}
\caption{Posterior distribution for 
the Taylor expansion coefficients, $V_i$, of the inflaton potential. The potential is expanded 
to $n$th order, assuming a flat prior on $\epsilon_V$, $\eta_V$, $\xi^2_V$, and $\varpi_V^3$.
The coefficients $V_i$ are expressed in natural units 
(where $\sqrt{8 \pi} M_\mathrm{pl}=1).$
The contours show only half of the allowed regions for potential parameters: the other half is symmetric, with
opposite signs for $V_1$ and $V_3$. \label{fig:V}
}
\end{figure}

\begin{figure}[t]
\includegraphics[width=0.99\columnwidth]{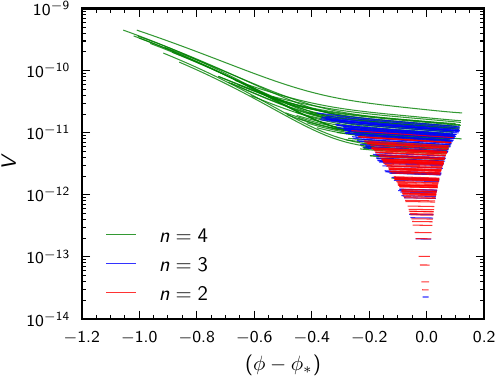}
\caption{Observable range of the best fitting inflaton potentials, when $V(\phi)$ is Taylor expanded to 
$n$th order around the pivot value $\phi_*$, in natural units (where $\sqrt{8 \pi} M_\mathrm{pl}=1),$ assuming 
a flat prior on $\epsilon_V$, $\eta_V$, $\xi^2_V$, and $\varpi_V^3$, and using \Planck+WP data. Potentials obtained 
under the transformation $(\phi-\phi_*)\rightarrow(\phi_*-\phi)$ leave the same observable signature and are 
also allowed. The sparsity of potentials with a small $V_0=V(\phi_*)$ is explained by the flat prior on $\epsilon_V$ rather 
than on $\ln(V_0).$ In fact, $V_0$ is unbounded from below. \label{fig:Vphi}
}
\end{figure}
  
Two complementary approaches 
to reconstruct the potential
have been explored in the literature.
The first approach, followed in this paper, expands the potential $V(\phi )$ directly in powers
of $(\phi -\phi _{*}).$ In this case the numerical integration
of the slow-roll solution must start sufficiently early so that any
initial transient has had a chance to decay, and one is in the 
attractor solution when the dynamics of the largest observable modes
in the Universe today start to have an interesting evolution. This is
the approach followed, e.g., by
\cite{Lesgourgues:2007gp} and \cite{Mortonson:2010er} using
publicly available 
codes.\footnote{\tt \tiny http://wwwlapp.in2p3.fr/valkenbu/inflationH/,
http://zuserver2.star.ucl.ac.uk/\~{ }hiranya/ModeCode/.}

A second approach expands $H(\phi )$ as a Taylor series
in $(\phi -\phi _{*}).$ As discussed in Sect.~\ref{FluctGeneration}, this has the advantage that $H(\phi )$ 
determines both the potential
$V(\phi )$ and the solution $\phi (t),$ so the issue of having
to start sufficiently early in order to allow the initial transient to decay
is avoided. This method was used, for example, in 
\cite{Kinney:2002qn}, \cite{Kinney:2006qm}, \cite{Peiris:2006ug}, \cite{Easther:2006tv}, \cite{Peiris:2006sj}, and \cite{Peiris:2008be}
using analytic and semi-analytic approximations,
and in 
\cite{Lesgourgues:2007aa}, \cite{Powell:2007gu}, \cite{Hamann:2008pb}, and \cite{Norena:2012rs}
using a fully numerical approach.}

These approaches could lead to results that differ from those in 
Sect.~\ref{akaSsectionFourOne}.
Firstly, if the running of the index is large, the slow-roll approximation 
taken to second order is not necessarily accurate for all models 
allowed by the data. The relation between the spectral 
parameters ($\ln A_{\mathrm s}$, $n_{\mathrm s}$, 
$\mathrm{d} n_\mathrm{s} / \mathrm{d}\ln k$, $r$) and the underlying
inflationary potential $V(\phi)$ is therefore 
uncertain. Secondly, for spectra with a large running,
there is no guarantee that an inflationary model giving 
such a spectrum exists. All allowed
models have $r \ll 1$, so these models are consistent with 
$\epsilon_V(k_*) \ll 1$ at the pivot scale. However, towards the 
edge of the observable range, the potential 
may become incompatible with $\epsilon_V(k)< 1$
(i.e., with the requirement of inflationary expansion). These possible 
pitfalls are avoided using the methods in this section, since the data have been fit directly by the candidate
$V(\phi)$ or $H(\phi)$, computed numerically without any slow
roll assumptions over the entire observable range.

\begin{table}
\centering
\begin{tabular}{c | c c c |c}
\hline
\hline
& \multicolumn{3}{c|}{from $V(\phi)$} & from \\
$n$ & 2 & 3 & 4 & slow-roll  \\
\hline
&&&&\\[-0.9ex]
$\epsilon_V$  & $<0.0078$ & $<0.015$ & $<0.021$ & $<0.015$ \\[1.1ex]
$\eta_V$      & $-0.011_{-0.015}^{+0.018}$ & $-0.016_{-0.025}^{+0.028}$   & $0.022_{-0.047}^{+0.052}$   & $-0.014_{-0.022}^{+0.030}$ \\[1.1ex]
$\xi^2_V$     & -- & $0.011_{-0.011}^{+0.012}$ & $-0.015_{-0.032}^{+0.031}$    & $0.009_{-0.011}^{+0.011}$ \\[1.1ex]
$\varpi^3_V$  & -- & -- & $0.016_{-0.019}^{+0.018}$ & -- \\[1.1ex]
$\Delta \chi_\mathrm{eff}^2$ & 0  & $-0.7$ & $-3.7$ & $-0.9$ \\
\hline
\end{tabular}
\caption{Numerical reconstruction of potential parameters, compared to results with the slow-roll
approximation, when tensors and running are
included (\Planck+WP 95\%~CL, with $k_{*}=0.05~$Mpc$^{-1}$). The effective $\chi^2$
value is given relative to the model with a quadratic potential.
\label{tab:R}}
\end{table}

\begin{table}
\centering
\begin{tabular}{c | c c c }
\hline
\hline
& \multicolumn{3}{c}{from $V(\phi)$} \\
$n$ & 2 & 3 & 4 \\
\hline
&&&\\[-0.9ex]
$\ln[10^{10} A_{\mathrm s}]$   & $3.087_{-0.050}^{+0.050}$ & $3.115_{-0.063}^{+0.066}$  & $3.130_{-0.066}^{+0.071}$  \\[1.1ex]
$n_{\mathrm s}$               & $0.961_{-0.015}^{+0.015}$ & $0.958_{-0.016}^{+0.017}$  & $0.954_{-0.018}^{+0.018}$ \\[1.1ex]
$100\,  \mathrm{d} n_\mathrm{s} / \mathrm{d}\ln k$       & $-0.05_{-0.14}^{+0.13}$   & $-2.2_{-2.3}^{+2.2}$ & $-0.61_{-3.1}^{+3.1}$ \\[1.1ex]
$100\, \mathrm{d}^2 n_\mathrm{s} / \mathrm{d}\ln k^2$ & $-0.01_{-0.75}^{+0.73}$ & $-0.3_{-1.2}^{+1.0}$ & $6.3_{-7.8}^{+8.6}$ \\[1.1ex]
$r$                 & $<0.12$                   & $<0.22$                    & $<0.35$                    \\
\hline
\end{tabular}
\caption{The scalar amplitude, tilt, running, running of the running, and tensor-to-scalar 
ratio inferred from a numerical reconstruction of the inflaton potential (\Planck+WP 95\%~CL, with $k_{*}=0.05~$Mpc$^{-1}$).
\label{tab:VvsHvsPS}}
\end{table}

We define a class of models over the observable range based on
the expectation that the potential should be smooth.
$V(\phi)$ is approximated by a Taylor expansion up to order $n,$ 
and we explore the cases $n=2,$ $3,$ and $4.$
For each $V(\phi)$, we integrate over inflationary fluctuations 
using the inflation module implemented in {\tt CLASS}\footnote{%
\tt \tiny http://class-code.net}~\citep{Lesgourgues:2011re,Blas:2011rf} 
as described in~\cite{Lesgourgues:2007gp}. 
Potentials are rejected for which the attractor solution cannot be reached 
when the largest observable scales cross the Hubble radius. 
The parameters sampled are the potential
and its derivatives at the pivot scale when 
$k_*$ crosses the Hubble radius during inflation.

To avoid parameter degeneracies, we impose uniform priors on $\epsilon_V, \eta_V, \xi^2_V,$ and 
$\varpi^3_V$ at the pivot field value $\phi _{*}.$ 
The advantage of uniform priors on these parameters is that---to the extent that the slow-roll conditions are
satisfied---these coefficients relate linearly to observable quantities such as 
$n_\mathrm{s}$, $r$, $\mathrm{d} n_\mathrm{s} / \mathrm{d}\ln k,$ and $\mathrm{d}^2 n_\mathrm{s} / \mathrm{d}\ln k^2$.
Figure~\ref{fig:R} and Table~\ref{tab:R} show the posterior
probability for these coefficients, and Fig.~\ref{fig:V} shows the
posterior probability for the Taylor series coefficients $V_i$. In
Fig.~\ref{fig:Vphi}, we show the observable range of the
best fitting inflaton potentials (for a sample extracted randomly from the converged Markov chains). 
The edges of the observable range correspond to Hubble crossing for the minimum and maximum
values of $k$ used in the Boltzmann code. 
We stress here that the \Planck\ data suggest 
a flat potential when the lowest order slow-roll primordial
spectra are considered, as analysed in Sect.~\ref{akaSsectionFourOne}. 
However, when the restrictions 
to lowest order in the slow-roll approximation
are relaxed, the inflaton potential can differ markedly from a plateau-like potential, 
as the green curves in Fig.~\ref{fig:Vphi} show.

When fitting $V(\phi)$ for each model in parameter space, we compute
($n_{\mathrm s}$, $d n_\mathrm{s} / d\ln k$, $d^2 n_\mathrm{s} / d\ln k^2$, and $r$) at the pivot scale {\it a
posteriori} directly from the numerical primordial spectra. The results are shown in
Table~\ref{tab:VvsHvsPS} and can be compared to those of Table~\ref{tab:alpha} in Sect.~\ref{sec:slow-roll}.
We can also use the results of
Sect.~\ref{sec:slow-roll} for the $\Lambda $CDM+$r$+$\mathrm{d} n_\mathrm{s}
/ \mathrm{d}\ln k$ model to infer the potential parameters ($\epsilon_V$,
$\eta_V$, and $\xi_V^2$) using the second-order slow-roll expressions, and
compare different approaches in the space of potential parameters
(see Fig.~\ref{fig:R} and the last column in Table~\ref{tab:R}).

The model with a quadratic potential in the observable window ($n=2$) leads
to bounds on  $\epsilon_V$, $\eta_V$, $n_{\mathrm s}$, and $r$ very close to the $\Lambda $CDM+$r$ case. This is 
not a surprise since such potentials cannot give values of $n_\mathrm{s}$ and $r$ 
compatible with the data and at the same time 
a large running. A significant $\mathrm{d} n_\mathrm{s} / \mathrm{d}\ln k$ can be generated
only in the presence of a large $\xi_V^3$ (i.e., with a significant $V^{\prime \prime \prime }$). 
Since quadratic potentials produce little running, they are faithfully described by 
the slow-roll approximation.

The model with a cubic term ($n=3$) has the freedom to generate a large running, $\mathrm{d} n_\mathrm{s} / \mathrm{d}\ln k$. 
Indeed one can check that the results for the $n=3$ model are close to those of the 
$\Lambda $CDM+$r$+$\mathrm{d} n_\mathrm{s} / \mathrm{d}\ln k$ model presented in Sect.~\ref{sec:slow-roll}. The agreement between these two models 
remains very good, despite the fact that in the presence of a large running, the slow-roll approximation 
can become inaccurate. The running in a potential with $n=3$ is not exactly scale invariant; 
this is not captured by the $\Lambda $CDM+$r$+$\mathrm{d} n_\mathrm{s} / \mathrm{d}\ln k$ parameterization. 

The $n=4$ model has even more freedom, allowing a considerable running
of the running $\mathrm{d}^2 n_\mathrm{s} / \mathrm{d}\ln k^2$ (to the extent that inflation holds during
the observable $e$-folds). In that case, the spectrum is better fitted
when the two parameters $r$ and $\mathrm{d}^2 n_\mathrm{s} / \mathrm{d}\ln k^2$ are non-zero. In
Fig.~\ref{fig:Vphi}, we see that most $n=4$ potentials have a long
and steep tail for $\phi<\phi_*$, with a kink around $\phi_*-0.4$ (in
natural units). This shape generates a significant running on the
largest observable scales, while preserving a smaller running on
smaller scales. With such a feature in the scalar
primordial spectrum at large scales combined with a non-zero contribution from tensor
fluctuations, the best fit model for $n=4$ has a temperature spectrum
very close to that of the minimal $\Lambda$CDM model for $\ell>40$, but not for
smaller multipoles. The amplitude of the Sachs-Wolfe plateau is
smaller. This allows the large-scale data points from \Planck\ to be fitted
slightly better. However, the case $r=\mathrm{d} n_\mathrm{s} / \mathrm{d}\ln k=0$ still lies at
the edge of the 95\% CL, and the minimum effective $\chi
^2$ of this model is smaller than in the $n=2$ case by only 3.7.

A comparison of the $n=3$ and $n=4$ results clearly shows that the process of 
expanding the inflaton potential to various orders and fitting it to the data does 
not converge (at least not by $n=4$). Given the $1$--$2\sigma$ preference of \Planck\ data 
for a nonzero running and running of the running, we find that a model-independent reconstruction 
of the inflaton potential is not possible under the assumptions of this section. 
In other words, as long as we assume that $V(\phi)$ can be described during and after observable 
inflation by a polynomial of order 2 or 3, we can put strong bounds on $\epsilon_V$ and $\eta_V$. 
But if we introduce more derivatives to describe the observable part of the potential 
and allow complete freedom to extrapolate $V(\phi)$ outside this region, 
the constraints can be easily evaded.

}

\section{Primordial power spectrum reconstruction}
\label{sec:pk}

{

\def\ba{\begin{eqnarray}}
\def\ea{\end{eqnarray}}
\def\L{\mathcal{L}}
\def\f{\mathbf{f}}
\def\ns{0.9603}
\def\As{2.20 \times 10^{-9}}
\def\kpiv{0.05~\textrm{Mpc}^{-1}}
\def\alphas{-0.013 \pm 0.009}

\newcommand{\camspec}{{\tt CamSpec }}
\newcommand{\plik}{{\tt Plik}} 
\newcommand{\WMAP}{\textit{WMAP}}

\begin{figure*}
\begin{center}
\includegraphics[scale=.9]{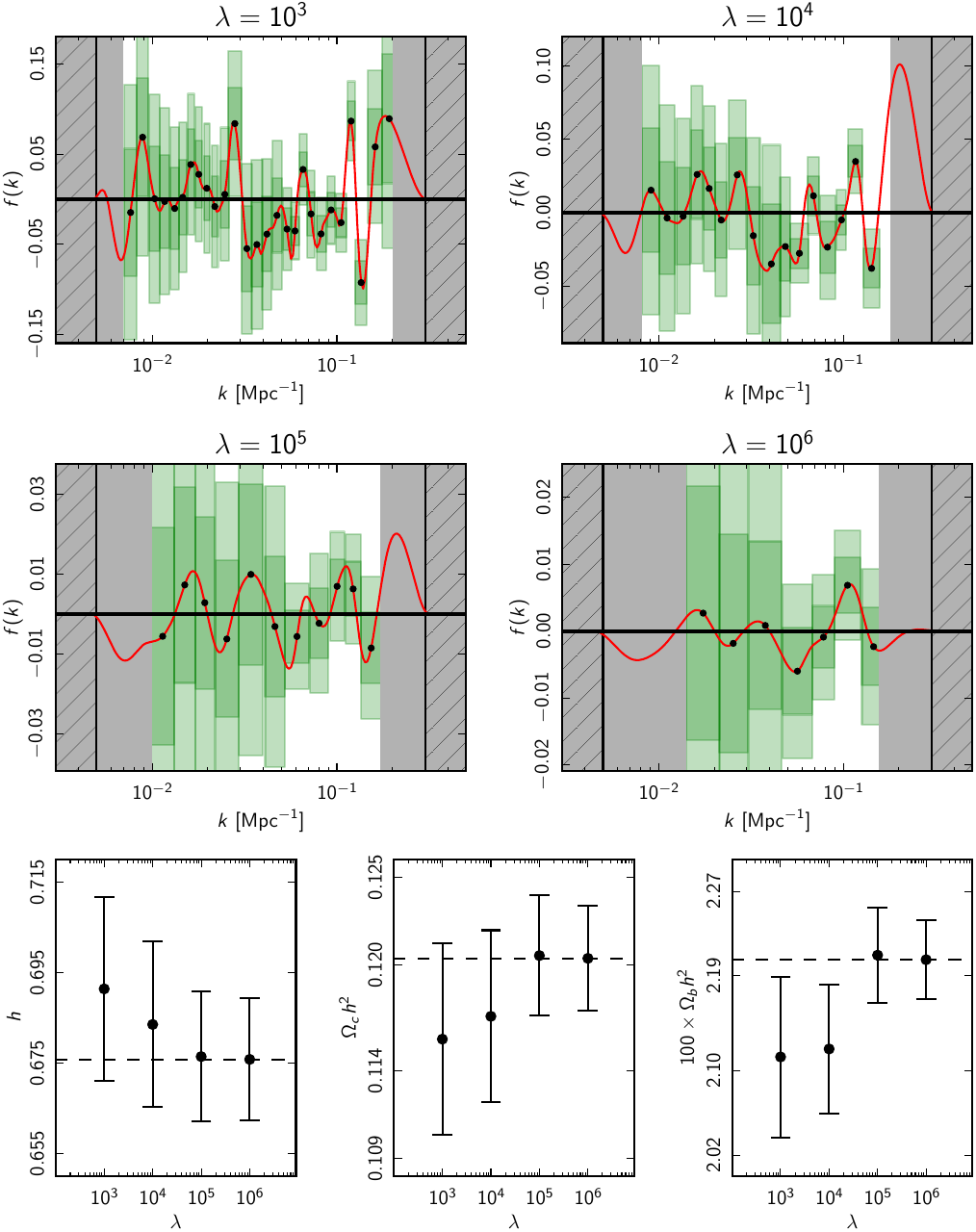}\\
\end{center}
\caption{
\Planck\ primordial power spectrum feature search results.
{\it Top four panels}: The reconstructed power spectrum at four values for 
the smoothing parameter $\lambda$. The red curves indicate the maximum 
likelihood configuration for the fractional deviation 
$f(k)$ of the power 
spectrum relative to a power law fiducial model (with $A_\mathrm{s}=\As$ and $n_\mathrm{s}=\ns$) for the penalized 
likelihood. The error bars have a width corresponding to the {\it minimum reconstructible width} 
(the minimum width for a Gaussian feature so that the mean square deviation
of the expectation value of the reconstruction differs by less than 10\%) and a vertical extent showing the $1\sigma$ and 
$2\sigma$ limits for the fractional deviation averaged over the box. The grey hashed regions
at the far left and right show where 
the fixing prior (i.e., $\alpha$) sets $f(k)=0.$ 
The inner grey regions show where the reconstruction bias is so great that the 
minimum reconstructible width is undefined. Mock features in this region produce reverberations
over the entire interval.
With $\lambda = 10^{3}$ and $\lambda =10^{4}$, we find 
statistically significant fluctuations around $k \sim 0.1~$Mpc$^{-1}$. {\it Lower panels}: The 
$1\sigma$ error bars for three combinations of cosmological parameters at the four values of $\lambda$.
The maximum likelihood value for the fiducial model is indicated by
the dashed line for comparison.
}
\label{recons}
\end{figure*}

In this section we report on a search for features in the primordial
power spectrum. In the basic six parameter model studied in the companion \Planck\
paper \cite{planck2013-p11}, the primordial power spectrum $\mathcal{P_R}(k)$, which 
includes only the adiabatic
mode, is modelled using the power law \mbox{$\mathcal{P_R}(k)=A_{\mathrm s}~(k/k_*)^{n_{\mathrm s}-1}$}, for which
the best fit values are \mbox{$A_{\mathrm s}=\As$} and $n_{\mathrm s}=\ns$ for a
pivot scale \mbox{$k_*=\kpiv$}. An extension of this parameterization
is also considered allowing for a running of 
the spectral index ($\mathrm{d} n_{\mathrm s}/\mathrm{d} \ln k = \alphas$). But in
all cases considered it was assumed that the power spectrum is smooth and without
bumps, sharp features, or wiggles. In this section we investigate
whether any statistically significant evidence for features is present in the
data when these assumptions are relaxed. Allowing an arbitrary function for the input
power spectrum is not an option because in this case the recovered primordial
power spectrum is dominated by small-scale noise. Instead we consider here
a penalized likelihood approach where a preference for smooth power spectra
is imposed. Section~\ref{sec:oscillations} 
of this paper pursues a complementary
approach where several parametric models for wiggles and features are explored
to see whether a statistically significantly better fit can be obtained.

An extensive literature exists on how to search for features in the power spectrum
using a wide range of methods. The following papers and the references therein
provide a sampling of the literature on non-parametric reconstruction: 
Richardson-Lucy deconvolution 
\citep{1974AJ79745L,RICHARDSON:72,Hamann:2009bz,Shafieloo:2003gf,Shafieloo:2007tk}, 
deconvolution \citep{TocchiniValentini:2004ht,TocchiniValentini:2005ja,
Ichiki:2009zz,Nagata:2008tk,Nagata:2008zj}, 
smoothing splines \citep{Verde:2008zza,Peiris:2009wp,Sealfon:2005em,Gauthier:2012aq}, linear 
interpolation \citep{Hannestad:2003zs,Bridle:2003sa}, and Bayesian model selection 
\citep{2009MNRAS.400.1075B, Vazquez:2012ux}. The approach pursued here follows \cite{TocchiniValentini:2005ja}
and \cite{Gauthier:2012aq} most closely.
More technical details and extensive tests validating the method 
can be found in the latter reference.

Let $\mathcal{P}_0(k)=A_{\mathrm s}(k/k_*)^{n_{\mathrm s}-1}$ be the best fit power spectrum 
of the six parameter model. We define a general Ansatz for the power spectrum in terms
of a fractional variation, $f(k)$, relative to this fiducial model, so that
\begin{equation} 
\mathcal{P_R}(k) = \mathcal{P}_{0}(k) \Bigl[1 + f(k)\Bigr] .
\end{equation}
Any features are then described in terms of $f(k)$. 

In this analysis we use the \Planck+WP likelihood supplemented by the following roughness penalty or 
prior, which is
added to $-2\ln \mathcal{L}$:
\begin{equation}
\begin{aligned}
&
\mathbf{f}^{T}
\mathbf{R}(\lambda,\alpha)
\mathbf{f} =
 \lambda  \int
\mathrm{d}\kappa~
\left(
\frac{
\partial^2f(\kappa)
}{
\partial \kappa^2
}
\right) ^2
 \cr
&\qquad 
+
\alpha \int _{-\infty }^{\kappa _{\mathrm{min}}}
\mathrm{d}\kappa~f^2(\kappa)
+
\alpha \int ^{+\infty }_{\kappa _{\mathrm{max}}} \mathrm{d}\kappa~f^2(\kappa) \,,
\label{Priors}
\end{aligned}
\end{equation}
where $\kappa = \ln k$
and $\kappa _{\mathrm{min}}$
and 
$\kappa _{\mathrm{max}}$
delimit the scales probed by the data. 
The first regularization term penalizes any deviation from a straight 
line of the function $f(\kappa)$. The second and third terms drive the 
$f(\kappa )$ to zero where there are effectively no constraints from the data. The 
value of $\lambda $ controls the smoothness of the reconstruction, but the precise 
value of $\alpha $ is less important. It must be large enough to force $f(k)$ towards
zero when $\kappa <\kappa _{\mathrm{min}}$ and  $\kappa >\kappa _{\mathrm{max}}$ but not so large as to
render the matrices ill-conditioned. We use $\alpha =10^4.$

We represent $f(k)$ using a cubic 
B-spline on a grid of points in $k$-space uniformly spaced in $\kappa $ with step size 
$\Delta \kappa = 0.025$ and extended from $\kappa = -12.5$ to $\kappa = -0.3$ 
giving us a total of 485 knots so that $\f = \{f_{i}\}_{i=1}^{485}$. The density of grid points 
is sufficiently large so that artefacts near the scale of the knot spacing are suppressed 
for the values of $\lambda$ used here. 
Given the large number of dimensions, it is not practical to explore the likelihood using MCMC methods. 
However, for the power spectrum parameters, the predicted 
$C_{\ell}$s are related by a linear transformation given fixed 
cosmological parameters, allowing us (for fixed cosmological and nuisance 
parameters) to find the maximum likelihood solution using the 
Newton-Raphson method.\footnote{Since the \Planck\ likelihood
uses a quadratic approximation, the maximum likelihood solution can be found by 
solving a linear system. The Newton-Raphson method, however, is needed for other CMB 
likelihoods which include terms beyond quadratic order.}
We define 
\begin{equation}
\mathcal{M}(\boldsymbol \Theta) =\underset{f_{i} \in [-1,1]}{\textrm{min}} \left\{-2 \ln 
\mathcal{L}(\boldsymbol \Theta,\mathbf{f}) + \mathbf{f}^{T} \mathbf{R}(\lambda,\alpha) 
\mathbf{f}\right\} ,
\end{equation}
where the vector $\boldsymbol \Theta $ represents the cosmological
parameters unrelated to the power spectrum and the foreground nuisance parameters.
We first minimize over $f_{i}$ using Newton-Raphson iteration and then 
in an outer loop minimize over $\boldsymbol \Theta $ using the downhill simplex algorithm.
To carry out this procedure, the \Planck\ likelihood code was modified to compute the gradient 
and Hessian of the likelihood with respect to the $C_{\ell}$s. 

The cosmological Boltzmann solver {\tt CAMB} 
was modified to accept the vector of primordial power spectrum  knots ${\textbf{f}}.$ 
By default {\tt CAMB} calculates 
the $C_{\ell}$s for a subset of $\ell$ and interpolates
to obtain the full multipole power 
spectrum. Instead we calculate the $C_{\ell}$s at each $\ell $ explicitly.

The boundaries $\kappa_{\mathrm{min}}$ and $\kappa_{\mathrm{max}}$ defining where $f(\kappa )$
is allowed to differ from zero are chosen to match the range of $\ell$ constrained by 
the high-$\ell $ likelihood. The likelihood 
includes $C_{\ell}$s between $\ell =50$ and $\ell=2500$, which roughly 
corresponds to $k \in [0.003,0.2]~$Mpc$^{-1}$. The low-$\ell$ likelihood covers 
$\ell=2$ to $\ell=49$, which roughly corresponds to 
$k \in [10^{-4}, 0.003] ~\textrm{Mpc}^{-1}$. In this range of $\ell$, cosmic variance is
large, making feature detection difficult. Calculating the gradient and 
Hessian of a pixel-based low-$\ell$ likelihood is computationally time consuming. 
We therefore use the low-$\ell$ likelihood  only to constrain the cosmological 
parameters. We choose $k_{\mathrm{min}} = 0.005 ~\textrm{Mpc}^{-1}$ and $k_{\mathrm{max}} = 0.3 ~\textrm{Mpc}^{-1}.$ 
Within this $k$ range, variations in 
the $C_{\ell}$s due to the $f_{i}$ are too small to affect 
the overall likelihood through the low-$\ell $ likelihood. 
We observed that the difference in the 
low-$\ell$ likelihood between the reconstructed $f(k)$ and $f(k)=0$ is small 
($<1\%$) compared to the difference in the high-$\ell$ likelihood.

The cosmological parameters $\tau$ and $A_{\mathrm s}$ are almost completely
degenerate for the temperature anisotropy except at very low $\ell$, so we 
fix $\tau$ to its best fit value for the fiducial model.
The likelihood contains additional nuisance parameters that 
model foreground components and beam shapes, as discussed in \cite{planck2013-p08}. 
Many of the nuisance parameters, it can be argued, are unlikely 
to introduce spurious small-scale structure because they represent foreground models with a power 
law and thus smooth angular power spectrum.
However some nuisance parameters, in particular those describing 
beam uncertainties, could conceivably introduce artefacts into the 
reconstruction. Unfortunately, converging to the correct 
maximum likelihood reconstruction with all 
the beam shape parameters included is prohibitively time consuming. Therefore we fix the
nuisance parameters to their fiducial best fit values, leaving a more detailed
examination of this issue to future work. 

\begin{figure*}
\begin{center}
\includegraphics[width=16cm]{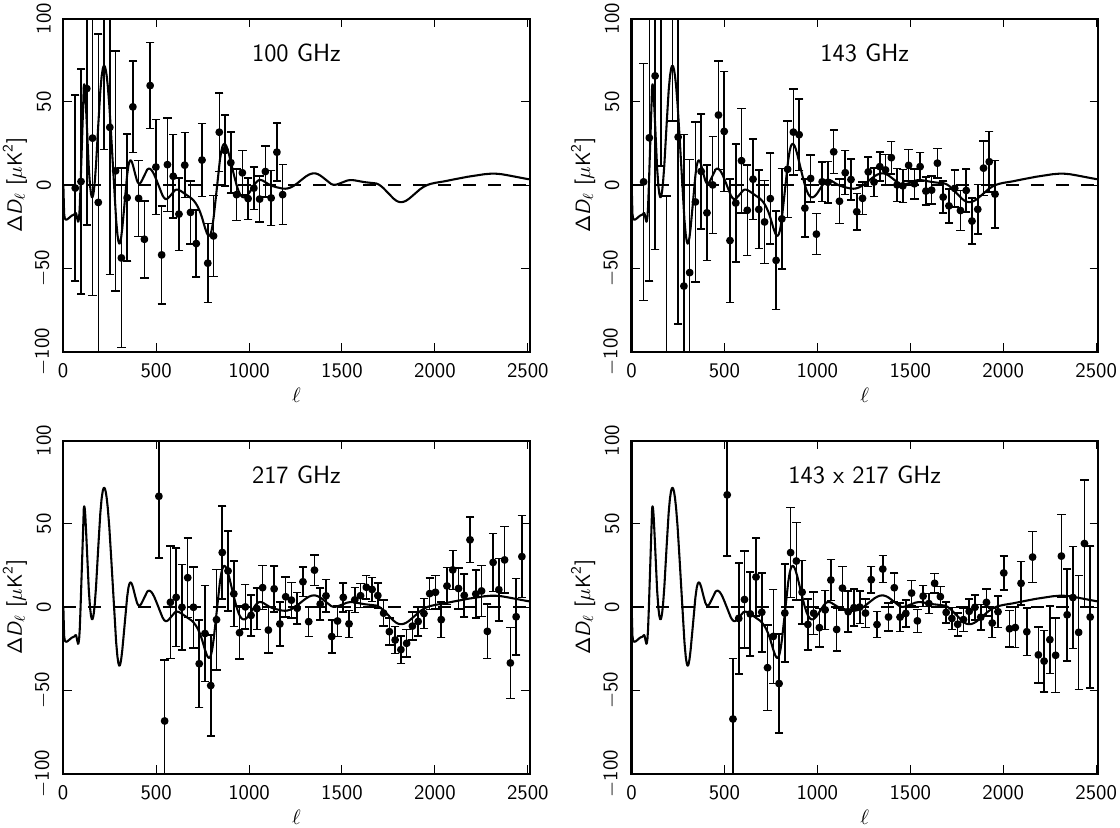}\\
\end{center}
\caption{
CMB multipole spectrum residuals for best fit primordial power spectrum reconstruction with 
smoothing parameter
$\lambda =10^3$. The panels show the $C_\ell $ spectrum residuals (compared to the best fit 
power law fiducial model represented by the horizontal straight dashed line) for the four auto- and 
cross-spectra included in the high-$\ell $ likelihood. Here 
$\cal \mathcal{D}_\ell =\ell (\ell +1)C_\ell /(2\pi ).$ 
The data points have been binned with $\Delta \ell =31$ and foregrounds subtracted 
according to the best fit foreground parameters. The solid black line shows the CMB spectrum 
residual for the maximum likelihood primordial power spectrum reconstruction with 
$\lambda =10^3.$
}
\label{reconsClplot}
\end{figure*}

\begin{figure}
\begin{center}
\includegraphics[width=9cm]{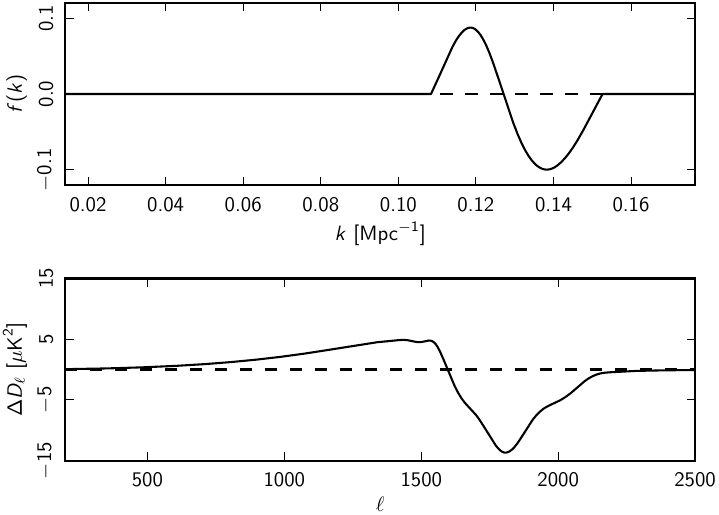}\\
\end{center}
\caption{
CMB multipole spectrum residual for the primordial power spectrum test feature. The test feature ({\it top}) is 
set to the anomalously large deviation of the primordial power spectrum reconstruction 
for $\lambda = 10^{3}$ in the interval $0.1 \textrm{ Mpc}^{-1} < k < 0.15 \textrm{ Mpc}^{-1}$, and is zero 
elsewhere. {\it Bottom}: The angular spectrum corresponding to this feature. We observe a large dip at $\ell \approx 1800$.
}
\label{ClsResidualTest}
\end{figure}

We found that simultaneously allowing extra degrees of freedom for small-scale structure and variations
in the cosmological parameters changes the best fit fiducial model---that is, $A_{\mathrm s}$ and
$n_{\mathrm s}$---so that the variations with respect to the fiducial model no longer visibly gave the best straight line fit.
Therefore we allow $A_{\mathrm s}$ and $n_{\mathrm s}$ to vary, so that the fiducial model is indeed the best straight
line fit through the plotted data points. Detailed investigation showed that neither the priors, nor
low-$\ell ,$ nor high-$\ell $ data play a significant role in determining the best fit fiducial model.
This effect is small and within the error bars for $A_{\mathrm s}$ and $n_{\mathrm s}$ 
established assuming the fiducial model. To summarize, we maximize the likelihood with 
respect to the control points $f_{i}$ and the three cosmological parameters 
$h$, $\Omega_\mathrm{c} h^{2}$, and $\Omega_\mathrm{b} h^{2}$. We 
then update the fiducial model ($A_\mathrm{s}$ and $n_\mathrm{s}$) 
at each iteration by finding the best fit power law through the current best fit reconstruction.

Once the maximum likelihood solution has been found, the second derivatives about this solution are 
readily calculated by extracting the relevant matrices for most of the components, and 
estimating the remaining components using finite differences. The second derivative matrix is used to 
estimate the error on the reconstructed $f_{i}$ and the three cosmological parameters $h$, $\Omega_\mathrm{c} 
h^{2}$, and $\Omega_\mathrm{b} h^{2}$. Monte Carlo simulations of a fiducial data set with a simplified
CMB likelihood including some of the non-Gaussianities\footnote{The simplified likelihood 
$-2 \ln \mathcal{L}_{\mathrm {toy}} = \sum_{\ell=2}^{\ell_{\mathrm{max}}}  (2 \ell + 1) 
\biggl(\frac{C_{\ell}^{\mathrm{obs}}}{C_{\ell} + N_{\ell}} - \ln \frac{C_{\ell}^{\mathrm{obs}}}{C_{\ell} + N_{\ell}} -1\biggr) $
assumes full sky coverage and isotropic instrument noise.} suggest that this approximation of the error
is accurate for $\lambda \gtorder 10^{3}$.

Figure~\ref{recons} summarizes our results, showing the estimated $f(k)$ in bins 
and the corresponding  1$\sigma$ and 2$\sigma$ errors. Errors in $k$ are also shown to represent the 
minimum reconstructible width evaluated at the middle of each bin. This is the minimum 
width that a Gaussian feature must have to be reconstructed 
with a small enough bias such that the mean reconstruction differs by less than
10\% rms.  The minimum 
reconstructible width is closely related to the correlation length, so that the errors
between adjacent bins are weakly correlated and the total  
number of bins represents roughly the effective number of independent degrees 
of freedom. 

While the plots with a significant roughness penalty---that is, with $\lambda =10^6$ and $\lambda =10^5$---do
not show any statistically significant evidence for features standing out above
the noise of the reconstruction, 
for a smaller roughness penalty---that is, for $\lambda =10^4$ and $\lambda =10^3$---a nominally statistically 
significant feature is clearly visible around  $k\approx 0.13~\textrm{Mpc}^{-1}.$
We do not understand the origin of this feature, which may be primordial or may arise as a foreground or other
systematic error in the high-$\ell $ portion of 
the likelihood. It should be noted that most of the robustness tests described in the likelihood
paper assume smooth power spectra.  
The maximum excursions are locally at 3.2$\sigma $ and  3.9$\sigma $ for 
$\lambda =10^4$ and $\lambda =10^3,$ respectively. In each of these two cases
we correct for the {\it look elsewhere} effect by calculating the probability that one 
of the plotted error bars deviates by the same number of or more standard deviations. This calculation 
is carried out using the covariance matrix of the plotted error bars. We obtain  $p=1.74\% $
and $p=0.21\%$, which corresponds to 2.4$\sigma$ and  
3.1$\sigma$, respectively. Additional simulations were
carried out to validate the method by generating mock data according to the fiducial
model and measuring the errors of the reconstruction obtained. These investigations
confirm the error model. These tests were carried out both with and without test features.
It can be argued that foregrounds are unlikely to explain 
the observed feature because all the foreground models involve smooth power law templates,
whereas this feature is localized in multipole number. 
It is important to assess by means of a more extensive set
of simulations whether the statistical significance assigned
to this result is accurate.

We investigate which CMB angular multipoles correspond to this 
apparent feature. Figure~\ref{reconsClplot} shows the $C_\ell $ residual from the 
reconstructed power spectrum with the best fit power law power spectrum subtracted
together with the data for each of the frequency map correlation combinations used in the 
{\tt CamSpec} likelihood. We observe a smooth dip around $\ell \approx 1800,$ which is  
significant compared to the error bars, in particular for the 217~GHz map. To determine whether
this dip is in fact responsible for large deviation in the reconstruction, 
we take the $\lambda=10^{3}$ best fit 
reconstruction and set $f(k)=0$ everywhere except for $0.1 \textrm{ Mpc}^{-1} < k < 0.15 
\textrm{ Mpc}^{-1}$---the region where the large deviation is located---and 
calculate the corresponding $C_{\ell}$ spectrum. Figure~\ref{ClsResidualTest} plots
the $C_{\ell}$ residuals of this test feature, which
show a large dip at around $\ell \approx 1800$, thus demonstrating 
that the dip in the $C_{\ell}$ residual of the data centred at $\ell \approx 1800$ is responsible
for the large excursions in the primordial power spectrum reconstructions. 

\underline{Note added}: The broad dip around $\ell= 1800$ in the
temperature power spectrum, seen in the 217 GHz channel,
has been shown to result from residuals that were
strongest in the first survey. In work done
after submission of this paper, this feature was shown to be associated
with imperfectly subtracted electromagnetic interference
generated by the drive electronics of the 4\,K cooler and picked
up by the detector read-out electronics. In a recent study, more aggressive
measures were applied to remove 4\,K contaminated
data. When this censured data was propagated all
the way to the maps and the power spectrum, the amplitude of
the feature was lowered to below the noise for the first survey.
A more complete account and analysis will
appear in the next round of {\it Planck} cosmology papers.

}

\section{Parametric searches for primordial power spectrum features}
\label{sec:oscillations}

{
In this section we continue to investigate deviations 
of the primordial power spectrum from a smooth, featureless function, in this case 
by testing a set of theoretically motivated models.

\subsection{Models and priors}

We consider three models describing features in the primordial power spectrum:
adding a global oscillation, a localized oscillation, or a cutoff to the large-scale power spectrum.

\subsubsection{Wiggles model}
Due to the exponential growth of the scale factor during inflation, a
periodically recurring event in proper time which affects the amplitude of curvature
perturbations would produce features that are periodic in $\ln k$.  
This occurs, for instance, for non Bunch-Davies initial
conditions \citep{Easther:2001fi,Danielsson:2002kx,Martin:2003kp,Bozza:2003pr}, or,
e.g., in the axion monodromy model \citep{Silverstein:2008sg}, as a
consequence of instanton induced corrections to the
potential \citep{Flauger:2009ab}.  In these scenarios the
primordial spectrum has an oscillation superimposed on an underlying smooth spectrum.

Here we consider the following parameterization of the primordial
spectrum (referred to as the {\em wiggles} model):
\begin{equation}
\mathcal{P_R}(k) = \mathcal{P}_{0}(k) \; 
\left\{ 1 + \alpha_\mathrm{w} \, \sin \left[ \omega \ln \left( \frac{k}{k_*} \right) + \varphi \right] \right\},
\end{equation}
with amplitude $\alpha_\mathrm{w}$,  frequency $\omega$,
and phase $\varphi$ to quantify the superimposed oscillations.  The
underlying smooth spectrum has the standard power law form
\begin{equation}
  \mathcal{P}_{0}(k) = A_\mathrm{s} \, \left( \frac{k}{k_*} \right)^{n_\mathrm{s} - 1}.
\end{equation}

The prior ranges for the wiggles model parameters are given in 
Table~\ref{tab:wiggles_priors}.  The obvious prior for the phase $\varphi$ is uniform over 
the interval $(0,2\pi).$  We also choose a uniform prior
on $\alpha_\mathrm{w}$ (a logarithmic prior on $\alpha_\mathrm{w}$ introduces considerable
dependence of the resulting marginalized posteriors on the lower limit
and does not contain the smooth spectrum as a limiting case).  
The sensitivity to primordial wiggles is limited at high frequencies 
by the width of the transfer function~\citep{Hamann:2008yx} 
and at low frequencies by the requirement of at least one full 
oscillation in the observable part of the power spectrum.
Since \Planck\  data are sensitive to wavenumbers over a range of
roughly four orders of magnitude, this condition implies $\omega \gtrsim 0.5$.
Here we restrict the analysis to $\omega < 100$ and assume a uniform prior. 
Larger values of the frequency are theoretically possible, e.g., in axion
monodromy models~\citep{Flauger:2009ab}, but the amplitude of the oscillations
in the $C_\ell$s will be suppressed with respect to the primordial one. 
A comprehensive search for higher frequency oscillations is currently underway.

\subsubsection{Step inflation model}

If the slow roll of inflation is briefly interrupted, for instance by a phase transition 
\citep{Starobinsky:1992ts,Hunt:2004vt},
a burst of resonant particle production~\citep{Chung:1999ve}, 
or a step in the inflaton potential~\citep{Adams:2001vc}, 
or if the speed of sound changes suddenly \citep{Achucarro:2010da},
a localized oscillatory 
feature is superimposed on the scalar primordial power spectrum.
We adopt the approximate parameterization for such a feature from a 
step in the potential, introduced by~\cite{Adshead:2011jq}, with
\begin{equation}
 \mathcal{P_R}(k) = \exp \left[ \ln \mathcal{P}_{0}(k) 
+ \frac{\mathcal{A}_\mathrm{f}}{3} \, \sqrt{\eta_\mathrm{f}/\mathrm{Gpc}} 
\frac{k \eta_\mathrm{f}/x_d}
{\sinh (k \eta_\mathrm{f}/x_\mathrm{d})}  W'(k \eta_\mathrm{f})\right],
\end{equation}
where
\begin{equation}
 W'(x) = \left( -3 + \frac{9}{x^2} \right) \cos 2x + \left( 15 - \frac{9}{x^2} \right) \frac{\sin 2x}{2x}.
\end{equation}
As in the wiggles model, we choose a uniform prior on the amplitude parameter 
$\mathcal{A}_\mathrm{f}$ (see Table~\ref{tab:wiggles_priors}).  
The parameter $\eta_\mathrm{f}$ determines both the frequency of the feature and its location, 
which is required to lie in the observable range.  The damping envelope of the feature is set
by the ratio $\eta_\mathrm{f}/x_\mathrm{d}$.  We impose uniform priors on the logarithms of
$\eta_\mathrm{f}$ and $x_\mathrm{d}$.

\subsubsection{Cutoff model \label{sec:cutoff}}

A number of models have been suggested to explain the apparent lack of power in the quadrupole
and octupole of the {\it WMAP} temperature power spectrum. Typically in these models, the onset
of a slow-roll phase coincides with the time when the largest observable scales exited the Hubble radius 
during inflation.  This naturally suppresses the primordial power spectrum at large scales \citep[see, e.g.,][]{Sinha:2005mn}.
We consider a phenomenological parameterization of a cutoff proposed in \cite{Contaldi:2003zv}, given by 
\begin{equation}
  \mathcal{P_R}(k) = \mathcal{P}_{0}(k) 
\left\{ 1 - \exp \left[- \left( \frac{k}{k_\mathrm{c}} \right)^{\lambda_\mathrm{c}} \right] \right\}.
\end{equation}
We apply uniform priors on $\lambda_\mathrm{c}$, which determines the steepness of the cutoff, 
and on the logarithm of the cutoff scale $k_\mathrm{c}$.

\begin{table}
\centering
\begin{tabular}{cll}
\hline
\hline
Model & Parameter & Prior range \\ 
\hline
\multirow{3}{*}{Wiggles} &$\alpha_\mathrm{w}$  & [0, 0.2] \\ 
& $\omega$  & [0.5, 100]  \\
&$\varphi$ & [0,$2\pi$] \\
\\
\multirow{3}{*}{Step inflation} &$\mathcal{A}_{\mathrm f}$  & [0, 0.2] \\ 
& $\ln \left(\eta_{\mathrm f}/\mathrm{Mpc} \right)$  & [0, 12]  \\
&$\ln x_{\mathrm d}$ & [-1, 5] \\
\\
\multirow{2}{*}{Cutoff} &$\ln \left( k_\mathrm{c}/\mathrm{Mpc}^{-1} \right)$  & [-12, -4] \\ 
& $\lambda_\mathrm{c}$  & [0, 15]  \\
\hline

\end{tabular}
\vspace{.4cm}
\caption{\label{tab:wiggles_priors} Prior ranges imposed for the 
wiggles, step inflation, and cutoff model parameters.}
\end{table}

\subsection{Method}

To achieve the necessary numerical precision for models with features in the primordial spectra,
we modify the standard settings of the {\tt CAMB} numerical code in order to calculate
$C_\ell$ at each $\ell $ rather than interpolating and refine
the grid in wavenumber for the numerical integration.
These changes significantly slow down the computation. 
In the models considered here, the likelihood function has
characteristics that make sampling difficult, such as extended plateaus and multiple isolated maxima, 
which render the Metropolis-Hastings algorithm inefficient.  
We therefore use the nested sampling algorithm implemented in the
{\tt MultiNest} add-on~\citep{Feroz:2007kg,Feroz:2008xx} to {\tt CosmoMC}~\citep{Lewis:2002ah},
which can also calculate the Bayesian evidence and the likelihood profiles.

The signatures of the feature models considered here are unique and cannot be mimicked
by varying other parameters, which lead to smooth variations of the power spectrum (with the exception of
highly tuned very low frequency oscillations that can change the acoustic peak structure).  We thus
restrict ourselves to varying only the parameters describing the features and keep all remaining
cosmological and nuisance parameters fixed to their $\Lambda$CDM best fit values.\footnote{An {\it a posteriori}
maximization of the likelihood in a narrow parameter range around the best fit feature model parameters, including
a variation of all remaining cosmological and nuisance parameters, shows that the change in the best fit 
$\chi^2_\mathrm{eff}$ is merely $\mathcal{O}(1)$ and hence does not affect our conclusions.}

\subsection{Results}

For all three models we find that including these additional features improves the quality of the fit 
with respect to a pure power law spectrum.
For the \Planck +WP data, we show the best fit primordial curvature power spectra and 
temperature angular power spectrum residuals
in Fig.~\ref{fig:bestfitspectra}, and report the best fit parameter values in Table~\ref{tab:wiggles_results}.
Since in all three cases the likelihood functions do not tend to zero in all directions of the 
respective parameter spaces, the Bayesian quantities (i.e., posterior distributions and Bayes factors)
depend strongly on the choice of prior.  For this reason, we also quote two prior-independent 
quantities, the effective $\chi ^2$ (i.e., $-2 \Delta \ln \mathcal{L}_\mathrm{max} =
2 \ln \mathcal{L}_\mathrm{max} 
- 2 \ln \mathcal{L}_\mathrm{max}^{\Lambda\mathrm{CDM}})$ and the profile $-2 \Delta \ln \mathcal{L}_\mathrm{max}$ 
as a function of selected model parameters plotted alongside the marginalized posteriors in 
Fig.~\ref{fig:wiggles_1d}, which illustrates the unconventional shape of the likelihood functions.

\begin{figure}[t]
\centering
\includegraphics[height=8.8cm,angle=270]{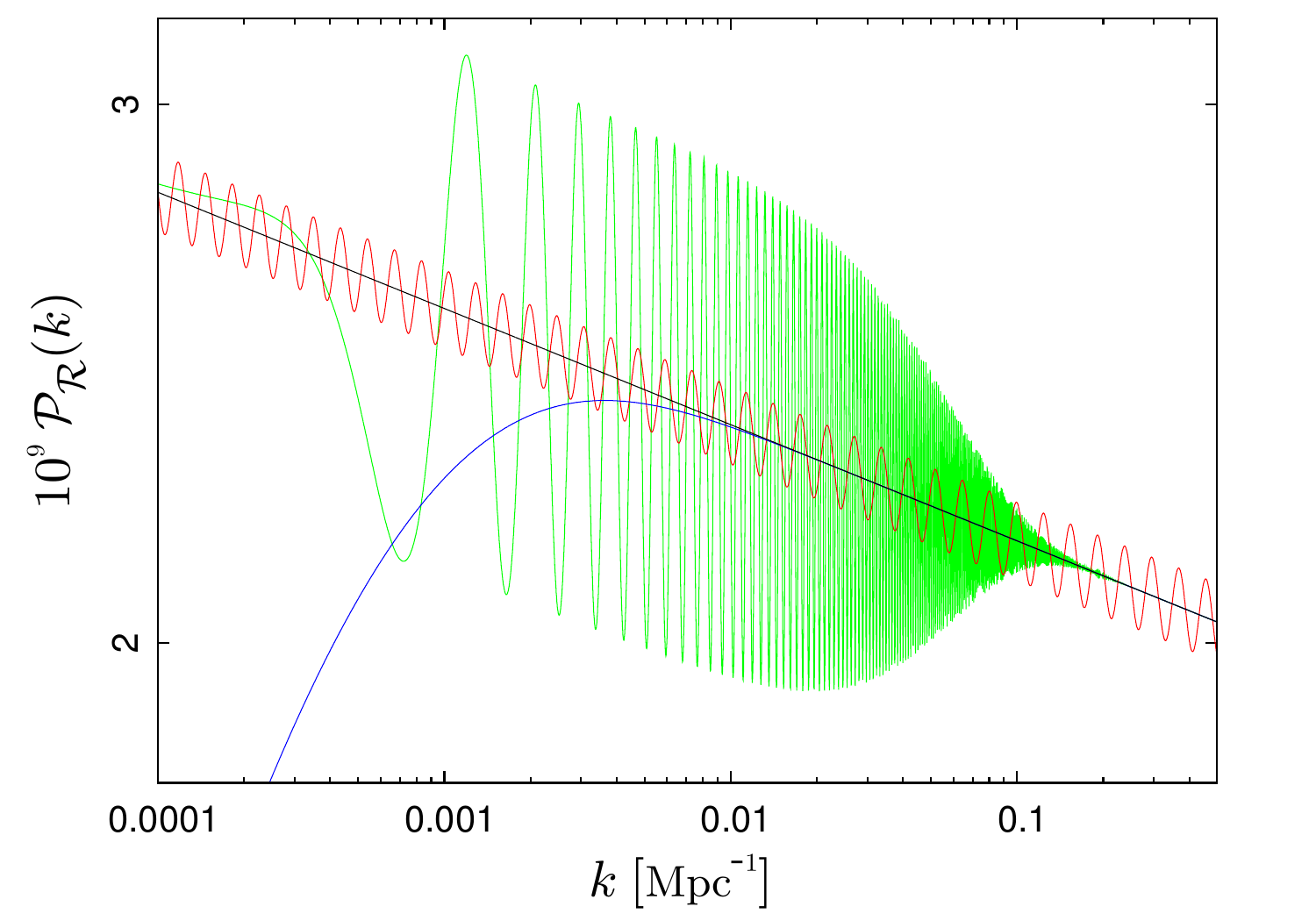}
\includegraphics[height=8.8cm,angle=270]{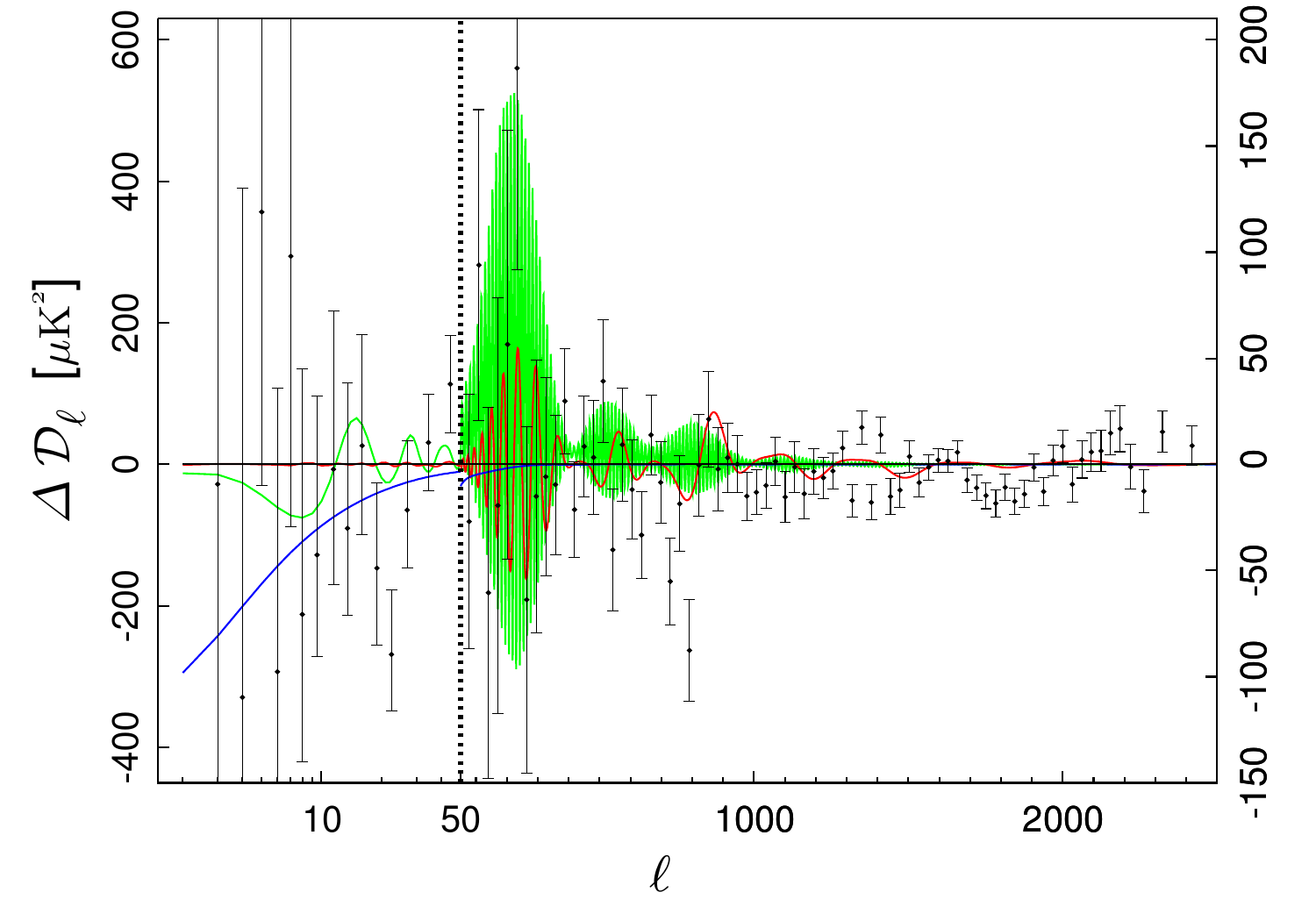}
\includegraphics[height=8.8cm,angle=270]{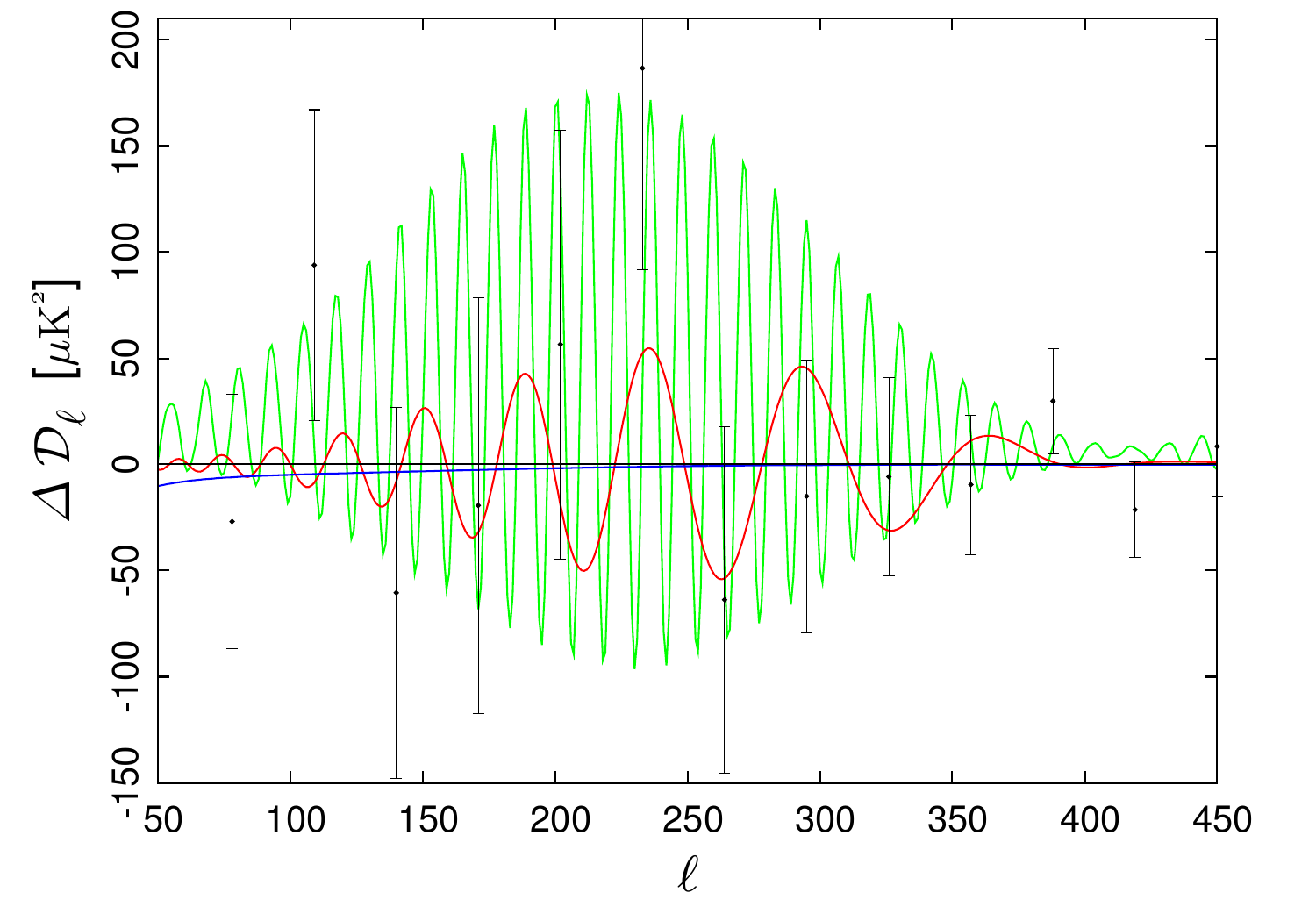}
\caption{\label{fig:bestfitspectra} 
{\it Top}: Best fit primordial spectrum of curvature perturbations for the power law ({\it black}), 
wiggles ({\it red}), step inflation ({\it green}), and cutoff ({\it blue})
models. {\it Centre}: Residuals of the temperature angular power spectrum. Note that the scale of the 
vertical axis changes at $\ell = 50$.
{\it Bottom}: Zoom of region around the first acoustic peak.}
\end{figure}

For the wiggles model, oscillations around the first acoustic peak and in the $700 < \ell < 900$ range improve the fit to the data,
whereas for the best fit step inflation model the spectrum between the Sachs-Wolfe plateau and the first acoustic peak is fit better.  
Quantitatively, the cutoff model improves the fit only modestly, with $\Delta \chi_\mathrm{eff}^2 \approx 3$, but both the wiggles
and step inflation models lead to a larger improvement, with $\Delta \chi_\mathrm{eff}^2 \approx 10$, at the cost of three new parameters. 
Already for pre-{\Planck \ data, improvements of $\Delta \chi_\mathrm{eff}^2 \approx 10$ have
been reported in related analyses (e.g., \citealt{2003ApJS..148..213P,Martin:2003sg,Elgaroy:2003hp,Covi:2006ci,Meerburg:2011gd,Benetti:2012wu,Peiris:2013opa}).   
Note that in the step inflation model, the best fit does not coincide with the maximum of the marginalized posterior probability,
indicating that some degree of fine tuning is necessary to reach the maximum of the likelihood.  The maximum of the marginalized
posterior at $\ln \left( \eta_\mathrm{f}/\mathrm{Mpc} \right) \approx 7.2$ actually reproduces the feature at $\ell \approx 20-40$ found
previously in {\it WMAP} data~\citep{2003ApJS..148..213P}.  The secondary peak at
$\ln \left( \eta_\mathrm{f}/\mathrm{Mpc} \right) \approx 4$  corresponds to a feature at multipoles $\ell \approx 1800,$ where the
analysis of Sect.~\ref{sec:pk} found a feature.  However the model does not 
account for this feature well, yielding an
improvement of only $\Delta \chi^2_\mathrm{eff} \approx 3$. 

Whether or not these findings can be considered statistically significant or arise simply from overfitting noisy data
is not a trivial question (see, for instance, the discussion in \cite{Bennett:2010jb}).  From a frequentist statistics point of view,
an answer would require the rather involved procedure of repeating the analysis on a large set of simulations.  In
designing the test statistic, special care would need to be taken in making sure to take into account the
{\it look elsewhere} effect (i.e., the fact that a particular observed anomaly may be very unlikely, whereas the probability of
observing {\it some} anomaly may be much larger).  From a Bayesian statistics point of view, it is the Bayesian evidence
that can tell us how probable the extended models are, compared to the baseline power law primordial power spectrum.

For the models and the choice of prior probabilities considered here, the Bayesian evidence in fact favours, albeit
weakly, the simple power law spectrum over the more complex models.  The reason is that the Bayesian evidence punishes
a lack of predictivity in these models. Most of the parameter space volume is not compatible
with the data. A good match to observations is obtained within only a small subregion.  Nonetheless, the observed features
remain interesting since if they are real, they will also leave traces in other observabless, most notably, in the
$E$-mode polarization spectrum, where the signatures of features in the primordial spectrum are actually less washed
out than in the temperature spectrum~\citep{2009PhRvD..79j3519M}. The forthcoming \Planck\ 
polarization data will
prove very useful in this regard.  Additionally, since strong deviations from power law 
behaviour typically indicate nonlinear
physics, these models generically also predict a non-Gaussian signal potentially observable in the bispectrum~\citep{planck2013-p09a}.  
However, the best fit wiggles and step inflation models
have oscillations with a frequency too high to be accessible to bispectrum analysis at present.

\begin{table}
\centering
\begin{tabular}{cccll}
\hline
\hline
Model & $-2 \Delta \ln \mathcal{L}_\mathrm{max}$ & $\ln B_{0X}$ & Parameter & Best fit value \\ 
\hline
\multirow{3}{*}{Wiggles} &\multirow{3}{*}{$-9.0$} &\multirow{3}{*}{1.5} &$\alpha_\mathrm{w}$  & $0.0294$ \\ 
& & & $\omega$  & $28.90$  \\
& & &$\varphi$ & $0.075\; \cdot 2 \pi$ \\
\\
\multirow{3}{*}{Step inflation} &\multirow{3}{*}{$-11.7$} &\multirow{3}{*}{0.3} 
&$\mathcal{A}_{\mathrm f}$  & $0.102$ \\ 
& & & $\ln \left(\eta_{\mathrm f}/\mathrm{Mpc} \right)$  & $8.214$  \\
& & &$\ln x_{\mathrm d}$ & $4.47$ \\
\\
\multirow{2}{*}{Cutoff} &\multirow{2}{*}{$-2.9$} &\multirow{2}{*}{0.3} &$\ln \left( k_\mathrm{c}/\mathrm{Mpc}^{-1} \right)$  & $-8.493$ \\ 
& & & $\lambda_\mathrm{c}$  & $0.474$  \\
\hline
\end{tabular}
\vspace{.4cm}
\caption{\label{tab:wiggles_results} Improvement in fit and logarithm of the Bayes factor 
$B_{0X}$ with respect to power law 
$\Lambda$CDM and best fit parameter values for the wiggles, step inflation, and cutoff models.  
The larger $\ln B_{0X}$, the greater the preference for a featureless power law spectrum.}
\end{table}

\begin{figure}[t]
  \centering
\includegraphics[height=8.8cm,angle=270]{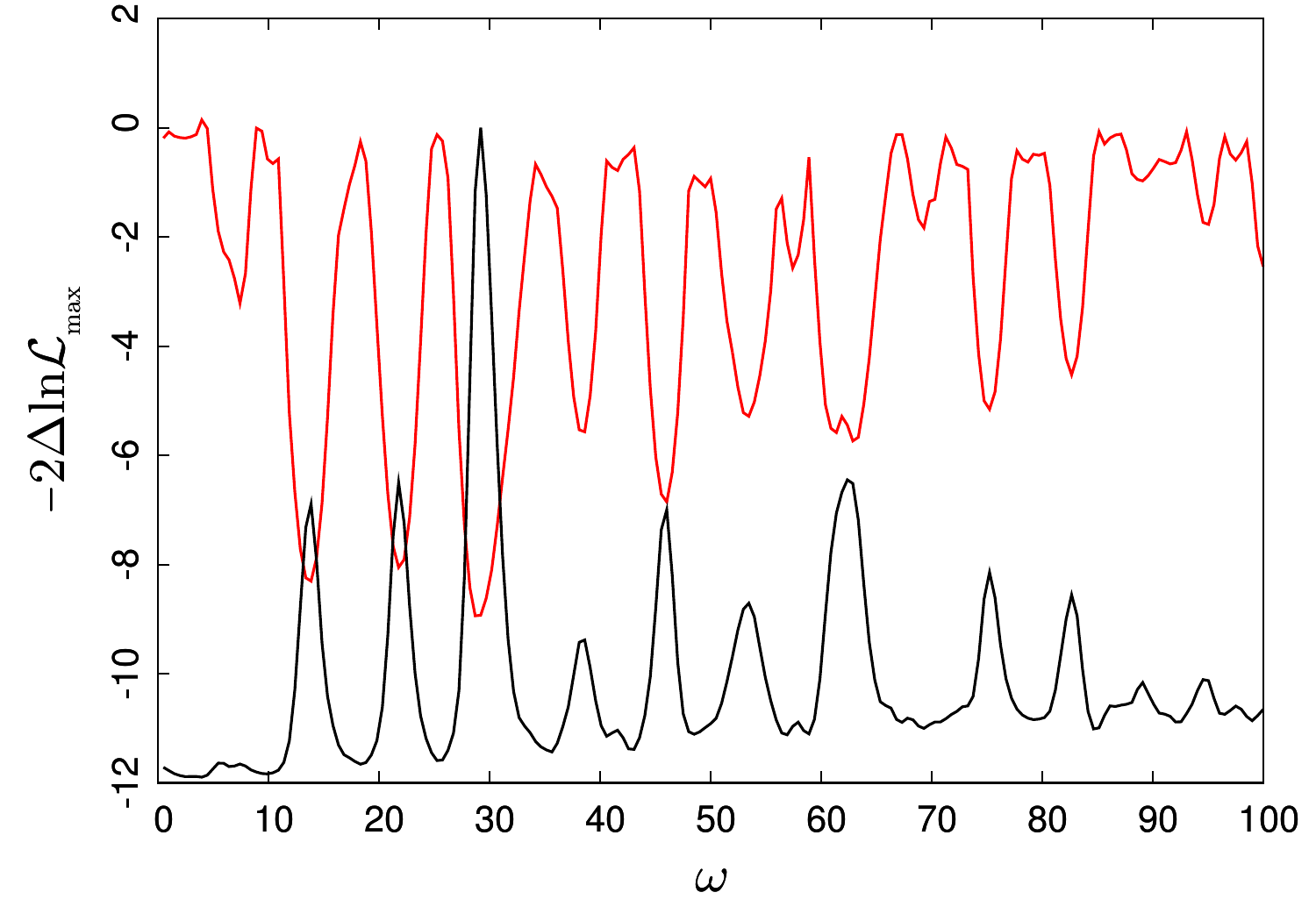}
\includegraphics[height=8.8cm,angle=270]{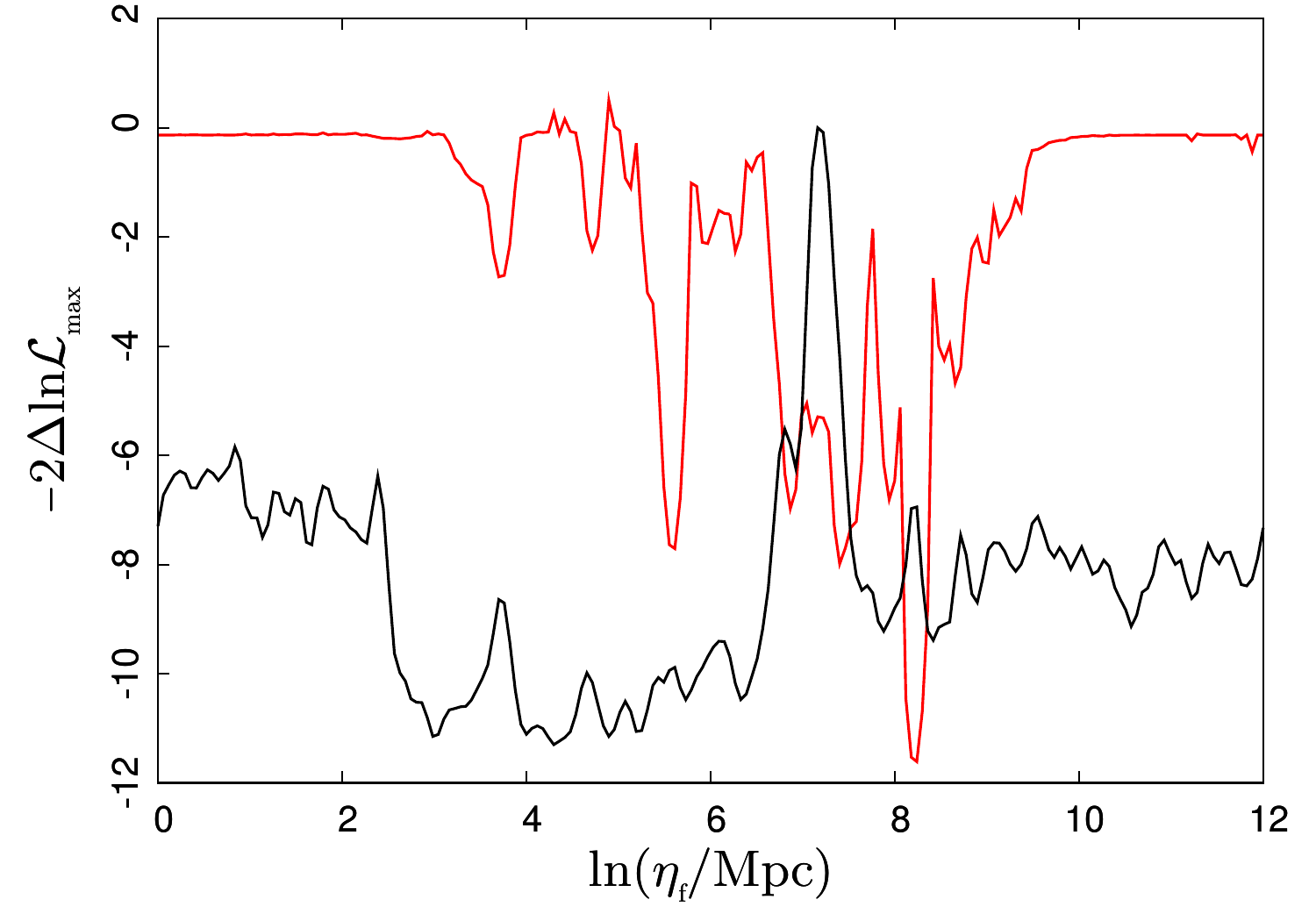}
\includegraphics[height=8.8cm,angle=270]{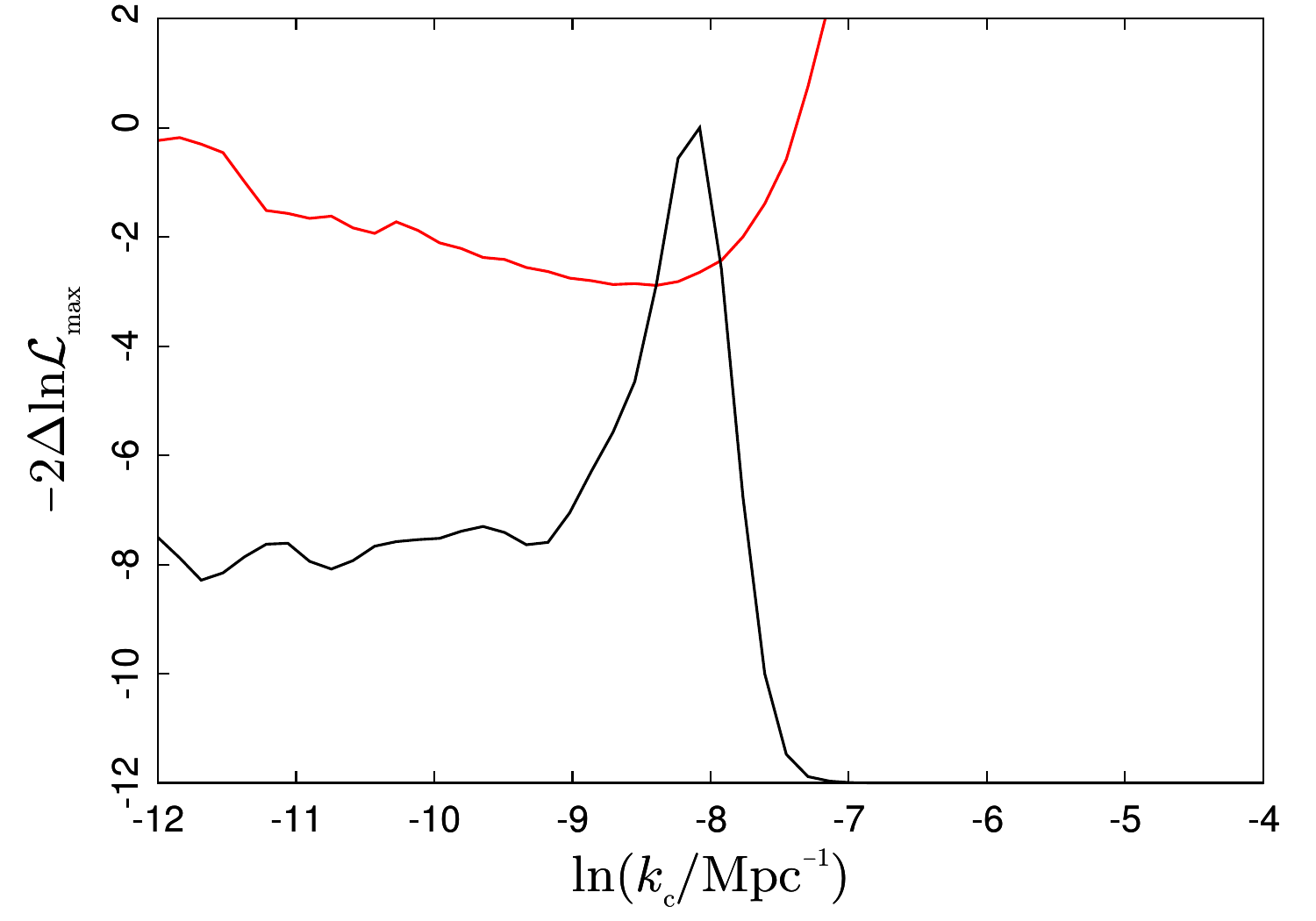}
\caption{\label{fig:wiggles_1d} 
Marginalized posterior 
probability ({\it red}) and profile $-2 \Delta \ln \mathcal{L}_\mathrm{max}$ 
({\it black}) for selected parameters of the wiggles ({\it top}), step inflation ({\it middle}),
and cutoff model ({\it bottom}).}
\end{figure}

}

\section{Combined analysis with {\it Planck} $f_\mathrm{NL}$ constraints for single field inflation}

\label{sec:putfnl}

{
In the previous sections we have analysed inflationary models with a canonical kinetic term. This 
led to the tensor-to-scalar consistency condition requiring $n_\mathrm{t}= -r/8$.  It is 
interesting to consider more general classes of inflationary models characterized by a non-standard 
kinetic term \citep{1999PhLB..458..219G} 
or more general higher-derivative operators \citep{Kobayashi:2010cm}. 
An interesting subclass of these models 
are those in which the Lagrangian is a general function of the scalar inflaton field and its first 
derivative: $\mathfrak{L}=P(\phi,X)$, where $X=- g^{\mu \nu} \partial_\mu \phi \partial_\nu \phi/2$. 
A more general extension is provided by the so-called effective field theory of inflation 
\citep{Cheung:2007st}, which has a richer phenomenology.

We restrict our analysis to the first class of 
models ~\citep{1999PhLB..458..219G,2007JCAP...01..002C}, which includes $k$-inflation 
models~\citep{1999PhLB..458..209A,1999PhLB..458..219G}, and Dirac-Born-Infield (DBI) models 
introduced in the context of brane inflation \citep{2004PhRvD..70j3505S,2004PhRvD..70l3505A}. 
In this class of models inflation can take place with a steep potential or can be driven by the 
kinetic term. One of the main features of inflationary models with a non-standard kinetic term is 
that the inflaton fluctuations can propagate at a sound speed $c_\mathrm{s} <1$. As shown in 
previous analyses \citep[e.g.,][]{2007PhRvD..76j3517P,2009JCAP...04..019P,2008PhRvD..78h3513L,2009PhRvD..79b3503A} 
there are strong degeneracies between the parameters determining the observable power spectra. 
Constraints on primordial non-Gaussianity can help break this degeneracy, and we show how \Planck's  
combined measurement of the power spectrum and the nonlinearity parameter 
$f_\mathrm{NL}$ \citep{planck2013-p09a} improves constraints on this class of models.

In models with a non-standard kinetic term the sound speed of the inflaton is given by 
$c_\mathrm{s}^2=P_{,X}/(P_{,X}+2 X P_{,XX})$ \citep{1999PhLB..458..219G}, so that in the canonical 
models, where $P(\phi,X)=V(\phi)-X$, one finds $c_\mathrm{s}=1$, while in general a non-trivial 
$c_\mathrm{s}<1$ corresponds to deviations from this standard case. Therefore, in these models, new 
parameters, such as the sound speed and its running, appear in the expressions for the inflationary 
observables. For the running of the sound speed it is useful to define an additional slow-roll 
parameter
\begin{equation}
\label{slowparameters}
s \equiv \frac{\dot{c_\mathrm{s}}}{c_\mathrm{s} H}\, .
\end{equation}
For values of the slow-roll parameters much less than unity, the leading order scalar power spectrum 
is modified
\citep{1999PhLB..458..219G} to
\begin{equation}
\label{sps}
A_\mathrm{s} \approx \frac{1}{8 \pi^2 M^2_\mathrm{pl}} \frac{H^2}{c_\mathrm{s} \epsilon_1}\, ,
\end{equation}
which is evaluated at $k c_\mathrm{s}=aH$. The scalar spectral index gets an 
additional contribution from the running of the sound speed,
\begin{equation}
n_\mathrm{s}-1=-2 \epsilon_1-\epsilon_2-s\, .
\end{equation} 

The gravitational sector remains unaltered by the non-trivial inflaton sound speed, retaining the 
same form as for the standard slow-roll models. Therefore the usual consistency relation is 
modified to $r \approx -8 n_\mathrm{t} c_\mathrm{s}$ with $n_{\mathrm t}=-2 \epsilon_1$ as usual 
\citep{1999PhLB..458..219G}. The more accurate relation employed in this analysis is 
\begin{equation}
r = 16 \epsilon_1 c_\mathrm{s}^{(1+\epsilon_1)/(1-\epsilon_1)}.
\label{c1}
\end{equation}
This accounts for the difference in freeze-out between the scalar and tensor perturbations 
\citep{2007PhRvD..76j3517P,2009JCAP...04..019P,2008PhRvD..78h3513L,2009PhRvD..79b3503A} taking place 
at $k c_\mathrm{s}=a H$ for the scalar fluctuations, and at $k=a H$ for the tensor modes.

Limiting ourselves to the predictions at lowest order in the slow-roll parameters, there are 
clearly degeneracies between the parameters ($A_\mathrm{s},c_\mathrm{s},\epsilon_1, \epsilon_2,s$), 
which make the constraints on the inflationary power spectra observables less stringent in terms of 
these microscopic parameters. However, for models where the inflaton field has a 
non-standard kinetic term with $c_\mathrm{s} \ll 1$, a high level of primordial non-Gaussianity of 
the scalar perturbations is generated \citep[see, e.g.,][]{2007JCAP...01..002C}. In these models 
primordial non-Gaussianity is produced by the higher-derivative interaction terms that arise when 
expanding the kinetic part of the Lagrangian, $P(\phi,X)$. The amplitude of the non-Gaussianity, 
defined by the nonlinearity parameter $f_\mathrm{NL}$, receives two dominant contributions, arising 
from the inflaton interaction terms $(\dot{\delta \phi})^3$ and $\dot{\delta \phi} (\nabla \delta 
\phi)^2$. Each of them produces non-Gaussianity shapes similar to the so-called 
{\it equilateral} type~(\citealt{2004JCAP...08..009B}), i.e., a signal that peaks for equilateral 
triangles $k_1=k_2=k_3$. However, they are sufficiently distinct that the total signal can be very 
different from the equilateral one~\citep{2010JCAP...01..028S}. The nonlinearity parameter of the 
second interaction term is $f_{\mathrm{NL}}=(85/324)(1-c_\mathrm{s}^{-2})$, while the other is 
determined by a second independent amplitude \citep{2007JCAP...01..002C,2010JCAP...01..028S}. 
Constraints on the primordial non-Gaussianity, presented in the companion paper 
\citet{planck2013-p09a}, thus allow us to construct a {\em lower} limit for the sound speed 
$c_\mathrm{s}$. This helps reduce degeneracies in the parameter space of inflationary models with 
non-standard kinetic terms. In particular, without the limits on the sound speed coming from the 
constraints on primordial non-Gaussianity, it is not possible to derive an upper limit on the 
parameter $\epsilon_1$, because the relation between the tensor-to-scalar ratio and $\epsilon_1$ 
also involves the sound speed (see, e.g., Eq.~\ref{c1}).

In this paper, we consider three cases. One is a general analysis as described above, where we 
focus on the simplest case of a constant speed of sound with $s=0$. From the \Planck\ limits on 
primordial non-Gaussianity in general single field models of inflation \citep{planck2013-p09a}, the 
most conservative constraint on the sound speed is
\begin{equation}
c_\mathrm{s} \geq 0.02 \quad (95\%~\mathrm{CL}) \, .
\label{cs_prior}
\end{equation}
In this large parameter space, we assume a uniform prior \mbox{$0.02 \leq c_\mathrm{s} \leq 1$} in 
Eq. \ref{c1} within the HFF formalism described in the Appendix. We show the joint constraints on 
$\epsilon_1$ and $\epsilon_2$ in Fig.~\ref{canonvscs}. By including the 95$\%$ CL constraint on 
$c_\mathrm{s}$ from Eq. \ref{cs_prior}, \Planck+WP constrain $\epsilon_1 < 0.053$. Such 
constraints can be compared with the restricted case of $c_\mathrm{s}=1$, also shown in 
Fig.~\ref{canonvscs}, with $\epsilon_1 < 0.008$ at 95\% CL.

\begin{figure}[!h]
\includegraphics[width=8.8cm]{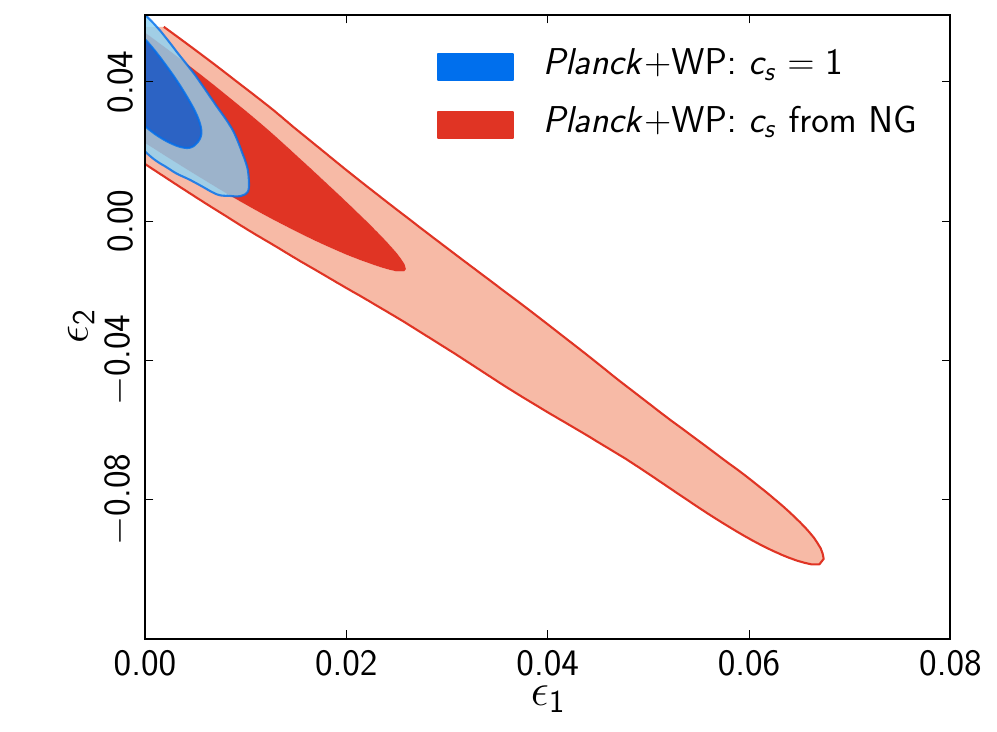}
\caption{Marginalized joint 68\%\ and 95\% CL for $(\epsilon_1\,, \epsilon_2)$
for {\it Planck}+WP data comparing the canonical Lagrangian case with $c_\mathrm{s}=1$ to the 
case of varying $c_\mathrm{s}$ with a uniform prior $0.02 < c_\mathrm{s} <1$ derived from the {\it 
Planck} non-Gaussianity measurements.}
\label{canonvscs}
\end{figure}

The other two cases analysed correspond to specific models where the inflaton has a 
non-standard kinetic term. The degeneracy between the different slow-roll parameters is broken 
because these models specifically predict that $s=0$, or $s \propto \epsilon_2$. 
As an example, we first consider the case where the action takes the Dirac-Born-Infield (DBI) form
\begin{equation}
P(\phi,X)=- f(\phi)^{-1} \sqrt{1-2f(\phi) X}+f(\phi)^{-1}-V(\phi)\, .
\end{equation}
Here $V(\phi)$ is the potential and $f(\phi)$ is the warp factor determined by the geometry of the 
extra dimensions.  For DBI models a stronger bound on $c_\mathrm{s}$ is derived 
\citep{planck2013-p09a}: $c_\mathrm{s} > 0.07$ at 95\% CL. With the uniform prior $0.07 < 
c_\mathrm{s} < 1$ and $s=0$, {\it Planck} + WP constrain $\epsilon_1 < 0.042$ at 95\% CL.

An important case is $f(\phi) \approx 
\lambda/\phi^4$ (for details, see~\cite{2004PhRvD..70j3505S},
\cite{2004PhRvD..70l3505A}, \cite{2007JCAP...01..002C}, and references therein). There are two 
possibilities. First, in {\sl ultraviolet} (UV) DBI models, the inflaton field moves under a 
quadratic potential $V(\phi)\approx m^2 \phi^2/2$ from the UV side of the warped space to the 
infrared side, with $m \gg M_\mathrm{pl} / \sqrt{\lambda}$. It is known that this case is already 
at odds with observations if theoretical internal consistency of the model and constraints on 
power spectra and primordial non-Gaussianity are taken into account 
\citep{2007PhRvD..75l3508B,2007JCAP...07..002L,2007JCAP...05..004B,2007PhRvD..76j3517P,2008PhRvD..77b3527B}. 
It is therefore interesting to look at the other case, namely infrared DBI 
models~(\citealt{2005PhRvD..71f3506C,2005JHEP...08..045C}) where the inflaton field moves from the 
IR to the UV side, and the inflaton potential is
\begin{equation}
V(\phi)=V_0-\frac{1}{2} \beta H^2 \phi^2\, ,
\end{equation}
with a wide range of values allowed for $\beta$ in principle, \mbox{$0.1 < \beta < 
10^9$}~\citep{2008PhRvD..77b3527B}. Here we focus on a minimal version of the IR DBI models where 
string effects are neglected, so that the usual field theory computation of the primordial 
curvature perturbation holds.  For IR DBI models accounting for such effects and a more involved 
treatment of the dynamics, see \cite{2005JHEP...08..045C}, \cite{2005PhRvD..72l3518C}, and \cite{2008PhRvD..77b3527B}. 
In this minimal IR DBI model, one finds ~\citep{2005PhRvD..72l3518C,2007JCAP...01..002C} 
$c_\mathrm{s} \approx (\beta N_*/3)^{-1}$, $n_\mathrm{s}-1=-4/N_*$, and $dn_\mathrm{s}/d\ln k=-4/N_*^2$ 
(in this model one can verify that $s\approx 1/N_*\approx \epsilon_2/3$). Here primordial 
non-Gaussianity of the equilateral type is generated with an amplitude 
$f^\mathrm{DBI}_\mathrm{NL}=-(35/108)\, [(\beta^2\, N_*^2/9)-1]$.

If we consider $60 \leq N_* \leq 90$, then the predicted spectral index lies within the range $0.93 
\leq n_\mathrm{s} \leq 0.96$, which is consistent with the \textit{Planck} measurement of the 
spectral index at the 3$\sigma$ level, for $N_* \geq 60$. The 
constraints on non-Gaussianity give $f^\mathrm{DBI}_{\mathrm{NL}}=11 \pm 69$ 
at $68 \%$ CL \citep{planck2013-p09a}. Combining these constraints with 
the power spectrum constraints, marginalizing over $60 \leq N_* \leq 90$, we obtain
\begin{equation}
\beta \leq 0.7 \quad (95\%~\mathrm{CL}).
\end{equation}
This strongly limits the allowed parameter space of these models.

As a final example, we consider a class of power-law $k$-inflation models characterized by the 
Lagrangian~\citep{1999PhLB..458..209A} \begin{equation} P(\phi,X)=\frac{4}{9} \frac{4-3 
\gamma}{\gamma^2} \frac{1}{\phi^2}(-X+X^2).\, \end{equation} In this case, for small values of 
$\gamma$ one finds: $c_\mathrm{s}^2\approx \gamma/8$, ${\mathcal P_R}=2 H^2/(3 \gamma 
c_\mathrm{s} 8 \pi^2 M^2_{\mathrm{pl}}) (k/k_0)^{-3 \gamma}$, $n_\mathrm{s}-1=- 3 \gamma$. The 
sound speed is a constant ($s=0$), with constant $\gamma$. The primordial non-Gaussianity in this 
model has an amplitude $f^\mathrm{equil}_\mathrm{NL}=-170/(81 \gamma)$. Therefore, all the 
inflationary observables depend essentially on a single parameter $\gamma$. Imposing a prior of $0 
< \gamma < 2/3$ from the non-Gaussianity constraint $f^\mathrm{equil}_\mathrm{NL}=-42 \pm 75$ at $68 \%$ CL
\citep{planck2013-p09a}, we obtain $\gamma \geq 0.05$ at $95 \%$ CL. At the same time, our 
measurement of the spectral index constrains $0.01 \leq \gamma \leq 0.02$ at $95 \%$ CL. This 
class of $k$-inflation models is therefore excluded by the combined constraints on primordial 
non-Gaussianity and the power spectrum.

}

\section{Isocurvature modes}
\label{sec:isometho}

{

\newcommand{\msc}[1]{\textcolor{red}{MS: #1 endMS.}}

\def\ba{\begin{eqnarray}}
\def\ea{\end{eqnarray}}

\subsection{Theoretical background}

In this section we explore the constraints 
imposed by \Planck\ on scenarios where the primordial cosmological
perturbations were not entirely {\it adiabatic}. These scenarios 
also include {\it isocurvature} modes, possibly correlated
among themselves as well as with the adiabatic mode. The adiabatic
mode is characterized by the property that at very early
times the universe obeyed a common, spatially uniform equation
of state and all components initially shared a common velocity field. 
For the adiabatic mode the density perturbations in the various
components (i.e., baryons, CDM, photons, and neutrinos) are locked
together. Here {\it baryons} include their accompanying leptons, 
assumed tightly coupled to maintain charge neutrality.

Isocurvature modes arise from spatial variations in the equation of state 
or from relative velocities between the components. 
To analyse how the CMB perturbations were imprinted, it is most convenient to
define isocurvature modes at a sufficiently late time, such that the 
relevant components, according to our present best understanding,
consisted of baryons, photons, CDM, and neutrinos. Under this hypothesis,
in addition to the adiabatic mode there are four possible non-decaying
isocurvature modes: the baryon, CDM, and neutrino density isocurvature modes, 
and the neutrino velocity isocurvature mode (see, e.g., \cite{Bucher:1999re} for a discussion and further references). 

The impact of isocurvature modes on the CMB was first studied in detail
by \cite{1970ApJ...162..815P} and \cite{Efstathiou:1986,Efstathiou:1987}, who contemplated the
possibility that isocurvature perturbations rather than adiabatic
perturbations were the sole source of cosmological perturbations.
\cite{Linde:1985yf}, \cite{Polarski:1994rz}, \cite{Linde:1996gt},
and \cite{GarciaBellido:1995qq} pointed out various scenarios in which isocurvature 
perturbations could be generated within the context of inflation. 
\cite{Bucher:1999re} carried out a systematic study of isocurvature modes
from a phenomenological perspective, pointing out the relevance of two
additional modes: the neutrino density and velocity modes.
\cite{Lyth:2001nq}, \cite{Moroi:2001ct}, and \cite{Bartolo:2002vf} studied 
an interesting so-called curvaton scenario, in which adiabatic fluctuations
from inflation contribute negligibly, but quantum fluctuations in a transverse
direction modulate the density of decaying particles, leading to isocurvature
perturbations correlated with the adiabatic mode. 

Several authors have studied the constraints on isocurvature modes 
imposed by previous microwave background experiments, including \cite{Stompor:1995py}, 
\cite{Langlois:2000ar}, \cite{Amendola:2001ni}, \cite{2003ApJS..148..213P}, \cite{Valiviita:2003ty}, 
\cite{Bucher:2004an}, \cite{Moodley:2004nz}, \cite{Beltran:2004uv}, \cite{KurkiSuonio:2004mn}, 
\cite{Dunkley:2005va}, \cite{Bean:2006qz}, \cite{Trotta:2006ww}, \cite{Keskitalo:2006qv}, 
and \cite{komatsu2009}. A more complete set of references may be found in \cite{Valiviita:2012ub}.

Before proceeding we must define precisely how to characterize
these isocurvature modes on super-Hubble scales during
the epoch after entropy generation, during which we assume that the stress-energy 
content of the universe can be modelled as a multi-component fluid 
composed of baryons, CDM particles, photons, and neutrinos. 
If we assume that the evolution of the universe during this epoch
was adiabatic (used here in the sense of thermodynamically reversible), then
the entropy per unit comoving volume is conserved and serves as
a useful reference with respect to which the abundances of the other
components can be expressed. 

The baryon isocurvature mode may be expressed in terms of fractional fluctuations in the baryon-to-entropy
ratio, which is conserved on super-Hubble scales during this epoch. 
The CDM and neutrino density isocurvature (NDI) modes may be defined analogously. 
The neutrino velocity isocurvature (NVI) mode refers to fluctuations in the neutrino velocity relative 
to the average bulk velocity of the cosmic fluid.
For the CMB, the baryon and CDM isocurvature modes yield
almost identical angular spectra because the deficit of one is balanced by an excess of the other, so we do not consider them separately here.
In this way the {\it primordial} isocurvature modes may be defined
as dimensionless stochastic variables ${\cal I}_{\mathrm {CDI}}$, ${\cal I}_{\mathrm{NDI}}$,
${\cal I}_{\mathrm{NVI}}$, like the variable ${\cal R}$ describing the adiabatic
mode.\footnote{The symbol ${\cal S}$ is sometimes used in the literature
to denote the isocurvature modes, also known as {\it entropy} perturbations. 
To prevent confusion we avoid this terminology 
because isocurvature modes are unrelated
to any notion of thermodynamic entropy.}  
In this basis, the CDI mode can be seen as an {\it effective} isocurvature mode, encoding both CDM and baryon isocurvature 
fluctuations through
${\cal I}_{{\mathrm{CDI}}}^\mathrm{effective}={\cal I}_{{\mathrm{CDI}}}+(\Omega_{\mathrm b} / \Omega_{{\mathrm{c}}}){\cal I}_{{\mathrm{BI}}}$ \citep{Gordon:2002gv}.

Within this framework, Gaussian fluctuations for the most general cosmological
perturbation are described by a $4\times 4$ positive definite matrix-valued power
spectrum of the form
\ba 
\pmb{\cal P}(k)=
\begin{pmatrix}
{\cal P}_{ {\cal R} ~~~      {\cal R}      }(k)&
{\cal P}_{ {\cal R} ~~~      {\cal I}_{{\mathrm{CDI}}}}(k)&
{\cal P}_{ {\cal R} ~~~      {\cal I}_{{\mathrm{NDI}}}}(k)&
{\cal P}_{ {\cal R} ~~~      {\cal I}_{{\mathrm{NVI}}}}(k)\\
{\cal P}_{ {\cal I}_{{\mathrm{CDI}}} {\cal R}      }(k)&
{\cal P}_{ {\cal I}_{{\mathrm{CDI}}} {\cal I}_{{\mathrm{CDI}}}}(k)&
{\cal P}_{ {\cal I}_{{\mathrm{CDI}}} {\cal I}_{{\mathrm{NDI}}}}(k)&
{\cal P}_{ {\cal I}_{{\mathrm{CDI}}} {\cal I}_{{\mathrm{NVI}}}}(k)\\
{\cal P}_{ {\cal I}_{{\mathrm{NDI}}} {\cal R}      }(k)&
{\cal P}_{ {\cal I}_{{\mathrm{NDI}}} {\cal I}_{{\mathrm{CDI}}}}(k)&
{\cal P}_{ {\cal I}_{{\mathrm{NDI}}} {\cal I}_{{\mathrm{NDI}}}}(k)&
{\cal P}_{ {\cal I}_{{\mathrm{NDI}}} {\cal I}_{{\mathrm{NVI}}}}(k)\\
{\cal P}_{ {\cal I}_{{\mathrm{NVI}}} {\cal R}      }(k)&
{\cal P}_{ {\cal I}_{{\mathrm{NVI}}} {\cal I}_{{\mathrm{CDI}}}}(k)&
{\cal P}_{ {\cal I}_{{\mathrm{NVI}}} {\cal I}_{{\mathrm{NDI}}}}(k)&
{\cal P}_{ {\cal I}_{{\mathrm{NVI}}} {\cal I}_{{\mathrm{NVI}}}}(k)\\
\end{pmatrix}
.
\label{pmatrix}
\ea
Following the conventions used in {\tt CAMB}~\citep{Lewis:2002ah,camb_notes}
and {\tt CLASS}~\citep{Lesgourgues:2011re,Blas:2011rf},
the {\it primordial} isocurvature modes are normalized as follows in the synchronous gauge: for the CDI mode, 
${\cal P}_{{\cal I}{\cal I}}(k)$ is the primordial power spectrum of the density contrast 
$\delta \rho _{{\mathrm{CDM}}}/\rho _{\mathrm{CDM}};$ for the NDI mode it is that of 
$\delta \rho _{\nu }/\rho _{\nu };$ and for the NVI mode, that of the neutrino velocity $v_\nu$ 
times 4/3.
\footnote{In other words, of the neutrino perturbation dipole, \mbox{$F_{\nu 1}=4\theta_v/(3k)$} 
in the notation of~\cite{Ma:1995ey}.}

If isocurvature modes are present, the most plausible mechanism
for exciting them involves inflation with a multi-component
inflaton field. To have an interesting spectrum on the large scales
probed by the CMB, isocurvature modes require long-range
correlations. Inflation with a multi-component
inflaton provides a well motivated scenario for establishing
such correlations. Inflation with a single-component scalar field
can excite only the adiabatic mode. In models of inflation 
with light (compared to the Hubble expansion rate)
transverse directions, the scalar field along these transverse
directions becomes disordered in a way described by an approximately
scale-invariant spectrum. If the inflaton has $M$ light components,
there are $(M-1)$ potential isocurvature modes during inflation.
Whether or not the fluctuations along these transverse directions are
subsequently transformed into the late-time isocurvature modes
described above depends on the details of what happens after inflation, 
as described more formally below.

\begin{figure}
\centering
\includegraphics[width=\columnwidth]{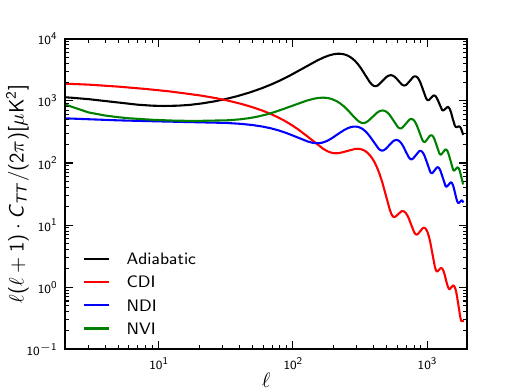}
\includegraphics[width=\columnwidth]{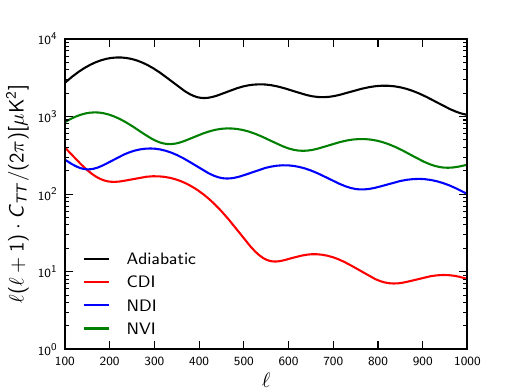}
\caption{$C_{TT}$ anisotropy shapes for the three isocurvature modes.
{\it Top}: The shapes of the CDM isocurvature mode, neutrino density isocurvature mode, and neutrino velocity isocurvature mode 
are shown together with the adiabatic mode. 
The modes have the same amplitude parameters ($\mathcal{P}_{\cal R\cal R}$ for the adiabatic mode and $\mathcal{P}_{\cal I\cal I}$ for each isocurvature mode). 
{\it Bottom}: The narrower multipole range illustrates the relative phases of the acoustic oscillations for these modes.
}
\label{fig:adia_vs_iso}
\end{figure}

As explained for example in \cite{Langlois:1999dw}, \citet{Gordon:2000hv}, \cite{GrootNibbelink:2000vx,
GrootNibbelink:2001qt}, and \cite{Byrnes:2006fr}, for inflationary models
where the inflaton follows a curved trajectory, correlations
are generically established between the isocurvature and
curvature degrees of freedom. To lowest order in the
slow-roll approximation, this leads to a situation where
the adiabatic perturbation is the sum of several components
each of differing spectral index. 

The post-inflationary evolution determines
how the isocurvature fluctuations generated during inflation 
transmute into the three specific isocurvature modes studied here. 
Little is known about the details of what
happens during the epoch of entropy generation, but to linear
order we may express how the fields ${\cal R}_\mathrm{inf}$ (i.e.,
the curvature perturbation at the end of inflation) and
the transverse components of the inflaton field
$\sigma _1,\ldots ,\sigma _{M-1}$ (i.e., the components
orthogonal to the slow-roll direction) transform
into curvature perturbations and late-time isocurvature modes
at the end of the epoch of entropy generation as the 
linear transformation
\begin{equation}
\begin{pmatrix}
{\cal R}_\mathrm{out}\\
{\cal I}_{a}\\
\end{pmatrix}
=
\begin{pmatrix}
1 & \Sigma _A \\
\textbf{0} & M_{aA}\\
\end{pmatrix}
\begin{pmatrix}
{\cal R}_\mathrm{inf}\\
\sigma _{A}\\
\end{pmatrix},
\end{equation}
where $a=$ BI, CDI, NDI, NVI, while $A=1,\ldots ,(M-1)$
labels the transverse components of the $N$ component inflaton field.
Physically, the fluctuations along the transverse directions modulate particle production
during the epoch of entropy generation.

The neutrino density isocurvature can be excited in much the same way
as the CDM and baryon isocurvature mode because at least within
the standard electroweak model, in which there are no leptonic
flavour changing processes, $L_e,$ $L_\mu ,$ and $L_\tau $ are
separately conserved. Known non-perturbative processes such as
the sphaleron can trade lepton and baryon asymmetries with
each other and alter flavour asymmetries, but they cannot erase
such asymmetries altogether. Plausible models
generating the neutrino density mode are therefore possible~\citep{Bucher:1999re,Gordon:2003hw}, but
as for the neutrino velocity isocurvature mode, to date
no plausible generation mechanism has been put forth.

This extension of the adiabatic $\Lambda$CDM 
model to non-adiabatic initial conditions represents an important test of inflation.
Single-field inflation can produce only adiabatic perturbations,
since exciting isocurvature perturbations 
requires additional degrees of freedom during inflation.  
Therefore a detection of primordial isocurvature perturbations
would point to more complicated models of inflation.

\subsection{Adiabatic with one isocurvature mode and free spectral
indices}

In this paper we investigate three of the four possible isocurvature
modes of the $\Lambda$CDM scenario, since the baryon and CDM
isocurvature perturbations are indistinguishable in the CMB angular
power spectra.  The CDM, neutrino density, and neutrino velocity isocurvature
perturbations lead to different power spectra for CMB anisotropies, as
shown in Fig.~\ref{fig:adia_vs_iso}. We limit ourselves
to studying one isocurvature mode at a time, in the presence of a
curvature perturbation. More general combinations with
two or three isocurvature modes may be contemplated, but without
the \Planck\ high frequency polarization likelihood, it is difficult
to constrain such scenarios, so we postpone a discussion of this case
to the next release.

Theoretically, one expects the power spectra of
the isocurvature modes and their correlations to exhibit
near but not necessarily exact scale invariance. 
As a general test of adiabaticity, it is nevertheless interesting to compare a more general
model to the \Planck\ data, assuming that the adiabatic, isocurvature, 
and cross-correlation spectra obey power laws with
free spectral indices. Blue values of the spectral indices are 
particularly interesting from the point of view of testing
adiabaticity, because the acoustic peaks arising
from two of the isocurvature modes are out
of phase with the adiabatic peaks by roughly $\pi /2$ 
near the first acoustic peak. This is not true for the neutrino velocity isocurvature mode however.

In the literature,
models with one isocurvature as well as the adiabatic mode (possibly correlated)
are often parameterized by specifying the $2 \times 2$ correlation matrix at a certain
pivot scale $k_0$ wth components 
${\cal P}_{\cal R\cal R},$
${\cal P}_{\cal R\cal I},$
${\cal P}_{\cal I\cal I}$ 
along with their respective spectral indices
$n_{\cal R\cal R},$
$n_{\cal R\cal I},$
$n_{\cal I\cal I}$ 
\citep[e.g.,][]{Amendola:2001ni,Beltran:2004uv}.
We do not follow this approach because in the absence
of a statistically significant detection, the posterior
distributions for the spectral indices are difficult
to interpret and sensitive to how the prior is chosen. 
We instead adopt a parameterization where
${\cal P}_{\mathrm{ab}}$ is specified at two scales $k=k_1$ and $k=k_2$
and interpolated geometrically according to\footnote{
Although the 
models spanned by  the one-scale and two-scale parameterizations
are the same, the priors for these parameterizations are related
by a non-constant Jacobian and therefore do not coincide.}
\begin{align}
\begin{split}
{\cal P}_{\mathrm{ab}}(k)=&\exp
\Biggl[
\left(
\frac%
{\ln (k  )-\ln (k_2)}
{\ln (k_1)-\ln (k_2)}
\right)
\ln( {\cal P}^{(1)}_{\mathrm{ab}}) \\
&\qquad
+
\left(
\frac%
{\ln (k  )-\ln (k_1)}
{\ln (k_2)-\ln (k_1)}
\right)
\ln( {\cal P}^{(2)}_{\mathrm{ab}})
\Biggr] ,
\label{matInter}
\end{split}
\end{align}
where ${\mathrm{a,b}}={\cal I}, {\cal R}$ and 
${\cal I}={\cal I}_{{\mathrm{CDI}}}$, ${\cal I}_{{\mathrm{NDI}}},$ or ${\cal I}_{{\mathrm{NVI}}}$.
We set \mbox{$k_1 = 2\times 10^{-3}\textrm{ Mpc}^{-1}$} 
and $k_2 = 0.1\textrm{ Mpc}^{-1},$ so that 
$[k_1, k_2]$ spans most of the range in $k$ constrained by \Planck\ data.
A uniform prior for the components 
$\mathcal{P}_{\cal R\cal R}^{(1)},$ 
$\mathcal{P}_{\cal I\cal I}^{(1)},$ 
$\mathcal{P}_{\cal R\cal I}^{(1)},$
$\mathcal{P}_{\cal R\cal R}^{(2)},$ 
$\mathcal{P}_{\cal I\cal I}^{(2)},$ 
$\mathcal{P}_{\cal R\cal I}^{(2)}$ is assumed, where 
auto-correlation amplitudes 
$\mathcal{P}_{\cal R\cal R}^{(1)},$ 
$\mathcal{P}_{\cal I\cal I}^{(1)},$ 
$\mathcal{P}_{\cal R\cal R}^{(2)},$ 
$\mathcal{P}_{\cal I\cal I}^{(2)}$ 
are positive, although the cross-correlation amplitudes
$\mathcal{P}_{\cal R\cal I}^{(1)},$
$\mathcal{P}_{\cal R\cal I}^{(2)}$
may take both signs subject to the constraints
\begin{equation}
(\mathcal{P}_{\cal R\cal I}^{(1)})^2< \mathcal{P}_{\cal R\cal R}^{(1)} \mathcal{P}_{\cal I\cal I}^{(1)},\quad
(\mathcal{P}_{\cal R\cal I}^{(2)})^2< \mathcal{P}_{\cal R\cal R}^{(2)} \mathcal{P}_{\cal I\cal I}^{(2)}
\label{PD:Constraint}
\end{equation} 
to ensure positive definiteness.  For the logarithm of the
off-diagonal elements in Eq.~\ref{matInter} to be real, we must have
\mbox{${\cal P}_{\cal R\cal I}^{(1)}, {\cal P}_{\cal R\cal I}^{(2)}>0$.} This Ansatz can be trivially modified to admit the case
${\cal P}_{\cal R\cal I}^{(1)}, {\cal P}_{\cal R\cal I}^{(2)}<0$ by
inserting appropriate minus signs, but this parameterization does not
allow the case where the sign of the correlation changes. In practice
we deal with this by assuming a uniform prior not on ${\cal P}_{\cal
  RI}^{(2)}$, but on its absolute value, and then we impose ${\cal
  P}_{\cal RI}^{(2)} = \mathrm{sign}({\cal P}_{\cal RI}^{(1)}) \times
|{\cal P}_{\cal RI}^{(2)}|$.  The constraints in
Eq.~\ref{PD:Constraint} ensure that 
$\det\left({\cal P}_{\mathrm{ab}}(k)\right)$ 
is positive definite within the interval
$[k_1,k_2],$ but generically positive definiteness is violated
sufficiently far outside this interval, either for very small or very
large $k.$ Where this happens we reduce the magnitude of 
${\cal P}_{\cal R\cal I}$ so that there is either total correlation or
anti-correlation. The kinks thus introduced lie
outside the range $[k_1,k_2].$ Within 
this range, the
spectral indices 
$n_{\cal R\cal R},$
$n_{\cal R\cal I},$
$n_{\cal I\cal I}$ 
are scale-independent.
Finally, our
sign conventions are such that positive values for ${\cal P}_{\cal
  RI}^{(1,2)}$ correspond to a positive contribution of the
cross-correlation term to the Sachs-Wolfe component of the total temperature spectrum.

When the constraining power of the data is weak, a crucial question
is to what extent the posterior distribution results from the data
rather than from the prior distribution. 
The parameterization above is not the only
one that could have been adopted, and other
possible priors are typically 
related by a non-constant Jacobian. 
For each model, we indicate the log-likelihood for the best fit model, in order
to allow model comparison.

The \Planck+WP results for the three isocurvature modes 
using this two-scale parameterization 
are shown in Fig.~\ref{TwoDimConsolidated}
and included in the summary Table~\ref{tab:how_much_ic}. 
The power spectra ${\cal P}_\mtc{RR}(k),$ ${\cal P}_\mtc{RI}(k),$
and ${\cal P}_\mtc{II}(k)$ are normalized according
to the {\it primordial} values of the fields 
${\cal R}({\textbf{x}})$ and ${\cal I}({\textbf{x}})$
defined above. It is
interesting to consider how much isocurvature power is allowed
expressed as a fraction of the power in three bands
spanning the CMB temperature spectrum observed by \Planck. To this end,
we define the following derived quantities 
\begin{align}
\alpha _\mathcal{RR}(\ell _\mathrm{min},\ell _\mathrm{max})
=&\frac%
{(\Delta T)^2_\mathcal{RR}(\ell _\mathrm{min},\ell _\mathrm{max})}
{(\Delta T)^2_\mathrm{tot}(\ell _\mathrm{min},\ell _\mathrm{max})},\\
\alpha _\mathcal{II}(\ell _\mathrm{min},\ell _\mathrm{max})
=&\frac%
{(\Delta T)^2_\mathcal{II}(\ell _\mathrm{min},\ell _\mathrm{max})}
{(\Delta T)^2_\mathrm{tot}(\ell _\mathrm{min},\ell _\mathrm{max})},\\
\alpha _\mathcal{RI}(\ell _\mathrm{min},\ell _\mathrm{max})
=&\frac%
{(\Delta T)^2_\mathcal{RI}(\ell _\mathrm{min},\ell _\mathrm{max})}
{(\Delta T)^2_\mathrm{tot}(\ell _\mathrm{min},\ell _\mathrm{max})},
\label{eq:FracDef}
\end{align}
where
\begin{equation}
{(\Delta T)^2_X(\ell _\mathrm{min},\ell _\mathrm{max})}=
\sum _{\ell =\ell _\mathrm{min}}^{\ell _\mathrm{max}}
(2\ell +1)C_{X, \ell }^{TT}.
\end{equation}
The 95\% confidence limits from the one-dimensional posterior
distributions for these fractional contributions in the full range
$(\ell _\mathrm{min},\ell _\mathrm{max})=(2,2500)$ are shown in
Table~\ref{tab:how_much_ic}.  The
range of allowed values for $\alpha_\mathcal{RR}(2,2500)$ is a measure
of the adiabaticity of fluctuations in the CMB.
The posterior distributions of the fractions $\alpha _\mathcal{II},$ $\alpha _\mathcal{RI}$ in three
multipole ranges are shown in Fig.~\ref{OneDimConsolidated}.
We also report the primordial isocurvature fraction, defined as 
\begin{equation}
\beta _\mathrm{iso}(k)=\frac{\mathcal{P}_\mathcal{II}(k)}{\mathcal{P}_\mathcal{RR}(k)+\mathcal{P}_\mathcal{II}(k)}
\label{PrimFrac}
\end{equation}
at three values of $k$. 
Table~\ref{tab:how_much_ic}
also shows the effective $\chi^2=-2\ln {\cal L}_\mathrm{max}$ for all models, compared to
the minimal six-parameter $\Lambda$CDM model. 
In Fig.~\ref{IsoBestFits} we show the ratio of temperature spectra for the best fit mixed model to the adiabatic model.

\begin{figure}[ht]
\begin{center}
\includegraphics[width=0.96\columnwidth]{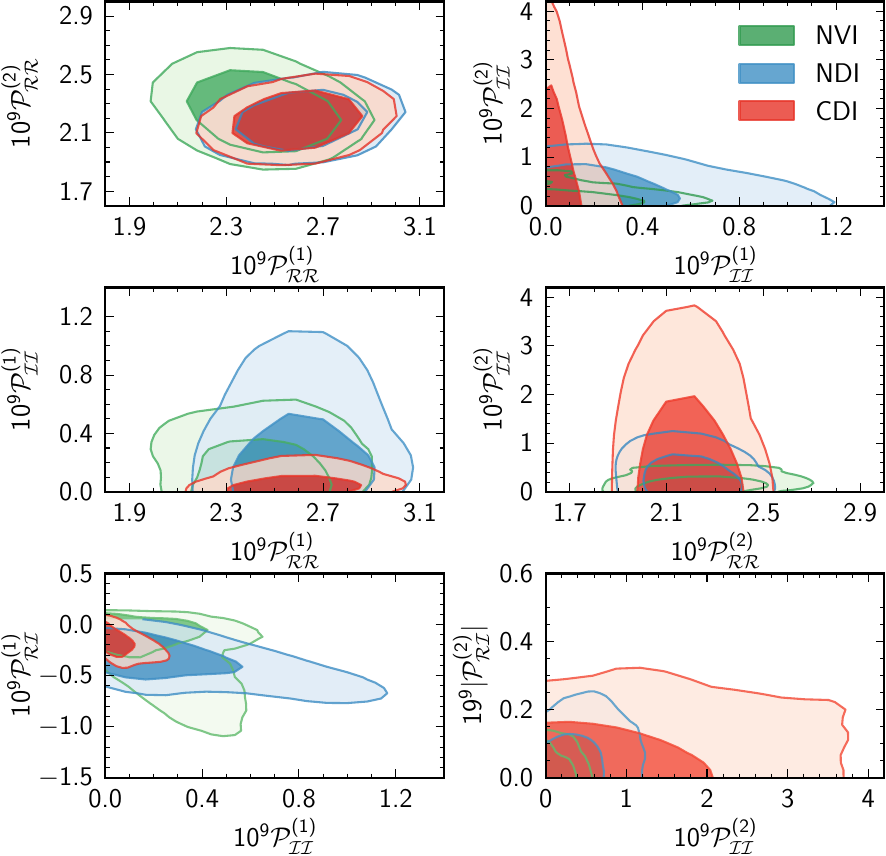}
\end{center}
\caption{
Two dimensional distributions for power in isocurvature modes using \Planck+WP data. }
\label{TwoDimConsolidated}
\end{figure}

\begin{figure}[ht]
\begin{center}
\includegraphics[width=0.96\columnwidth]{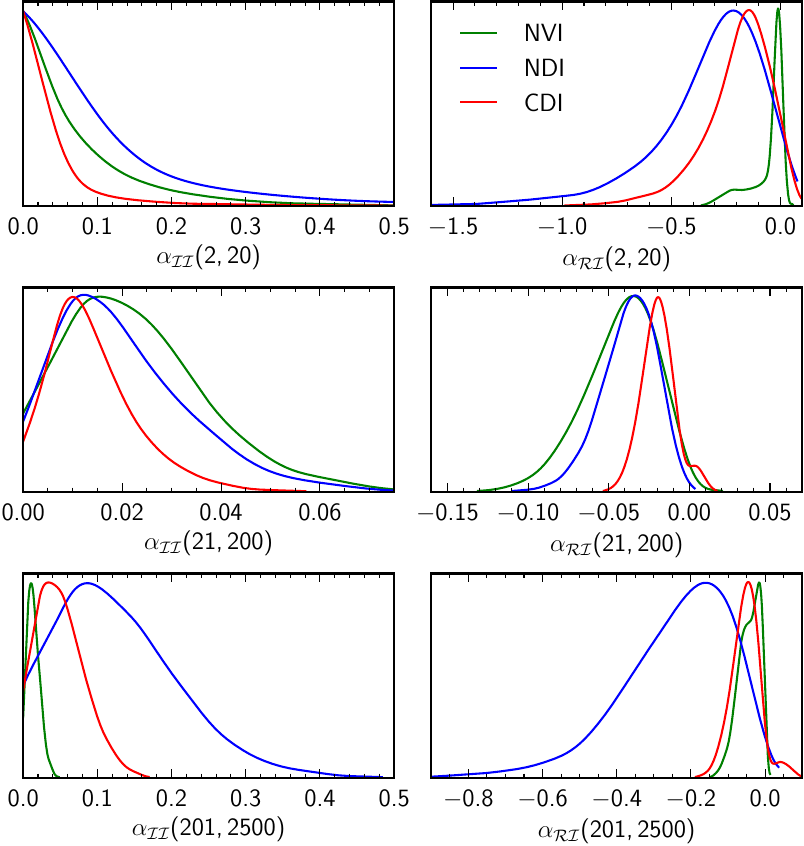}
\end{center}
\caption{
Fractional contribution of isocurvature modes to the power spectrum. We show the distributions
$\alpha _\mathcal{II}(2,20)$, $\alpha _\mathcal{RI}(2,20)$,
$\alpha _\mathcal{II}(21,200)$, $\alpha _\mathcal{RI}(21,200)$,
$\alpha _\mathcal{II}(201,2500)$, and $\alpha _\mathcal{RI}(201,2500),$ 
defined in Eq.~\ref{eq:FracDef}, for the CDI, NDI, and NVI modes as constrained by the
\Planck+WP data.
}
\label{OneDimConsolidated}
\end{figure}

\begin{figure}[ht]
\begin{center}
\includegraphics[width=0.96\columnwidth]{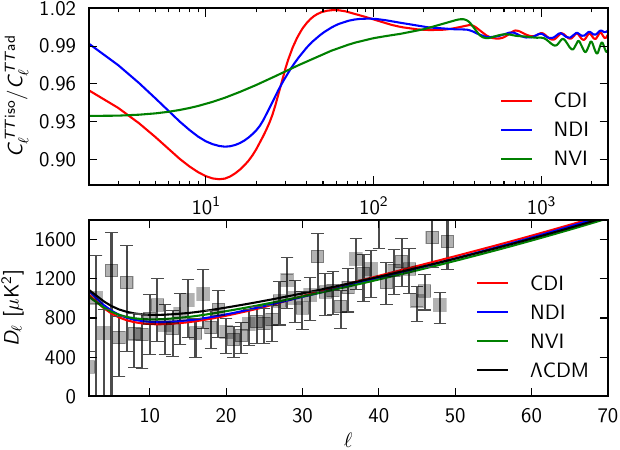}
\end{center}
\caption{
Temperature spectrum of best fit models with a mixture of adiabatic and isocurvature modes. 
{\it Top:} Spectrum of the best fit mixed models relative to that of the pure adiabatic model. {\it Bottom:} 
Zoom of the Sachs-Wolfe plateau of the best fit temperature spectrum $D_\ell=[\ell(\ell+1)/2\pi] C_\ell^{TT}$ for 
each of the three cases plus the pure adiabatic model, shown together with \Planck\ low-$\ell$ data points.
}
\label{IsoBestFits}
\end{figure}

\begin{table*}
\centering
\begin{tabular}{l|cccccccc}
\hline
\hline
Model \phantom{$\Biggr( $} &
$\beta_\mathrm{iso}(k_{\mathrm{low}})$ &
$\beta_\mathrm{iso}(k_{\mathrm{mid}})$ &
$\beta_\mathrm{iso}(k_{\mathrm{high}})$ &
$\alpha _{\cal R\cal R}^{(2,2500)}$ &
$\alpha _{\cal I\cal I}^{(2,2500)}$ &
$\alpha _{\cal R\cal I}^{(2,2500)}$ &
$\Delta n$ & 
$-2 \Delta \ln \mathcal{L}_\mathrm{max}$ \\
\hline
General model:
& & & & & \cr
$\quad $ CDM isocurvature
& 0.075 & 0.39 & 0.60
& [0.98:1.07] & 0.039 & [-0.093:0.014] & 4 & -4.6 \cr 
$\quad $ ND isocurvature
& 0.27 & 0.27 & 0.32
& [0.99:1.09] & 0.093 & [-0.18:0] & 4 & -4.2 \cr 
$\quad $ NV isocurvature
& 0.18 & 0.14 & 0.17
& [0.96:1.05] & 0.068 & [-0.090:0.026] & 4 & -2.5 \cr 
\hline
Special CDM isocurvature cases:
& & & & & \cr
$\quad $ Uncorrelated, $n_\mathcal{II}=1$ (``axion'')
& 0.036 & 0.039 & 0.040
& [0.98:1] & 0.016 & -- & 1 & 0 \cr
$\quad $ Fully correlated, $n_\mathcal{II}=n_\mathcal{RR}$ (``curvaton'')
& 0.0025 & 0.0025 & 0.0025
& [0.97:1] & 0.0011 & [0:0.028] & 1 & 0 \cr
$\quad $ Fully anti-correlated, $n_\mathcal{II}=n_\mathcal{RR}$
& 0.0087 & 0.0087 & 0.0087
& [1:1.06] & 0.0046 & [-0.067:0] & 1 & -1.3 \cr
\hline
\end{tabular}
\vspace{.4cm}
\caption{\label{tab:how_much_ic} Isocurvature mode constraints
from {\it Planck}+WP data. For each model,
we report the 95\% CL upper bound on the fractional primordial contribution of
isocurvature modes at three comoving wavenumbers ($k_{\mathrm{low}}= 0.002~$Mpc$^{-1}$,
$k_{\mathrm{mid}}= 0.05~$Mpc$^{-1} ,$ and $k_{\mathrm{high}}= 0.10~$Mpc$^{-1}$),
and the 95\% CL bounds on the fractional contribution 
$\alpha_\mathcal{RR},$ 
$\alpha_\mathcal{II},$ 
and $\alpha_\mathcal{RI}$ to the total CMB temperature anisotropy in the range
$2\leq\ell\leq2500$.  We also report $-2 \Delta \ln \cal{L}_\mathrm{max}$ for the
best fitting model in each case, relative to the best fit 6-parameter $\Lambda$CDM model,
with the number of additional parameters $\Delta n$. In the Gaussian approximation,
$-2 \Delta \ln \mathcal{L}_\mathrm{max}$ corresponds to $\Delta \chi^2$.
The general models have six parameters that specify the primordial correlation matrix at
two scales $k_1$ and $k_2$, thus allowing all spectral indices to vary (so, four parameters more than the pure adiabatic model).}
\end{table*}

The results for $\alpha _\mathcal{RR}(2,2500)$ show that the
nonadiabatic contribution to the temperature variance can
be as large as 7\% (9\%, 5\%) in the CDI (NDI, NVI) model (at 95\%~CL).
These results are driven by the fact that on large scales, for $\ell \leq
40$, the \Planck\ data points on average have a slightly smaller
amplitude than the best fitting $\Lambda$CDM model. Hence the data prefer a
significant amount of anticorrelated isocurvature modes, leading to a
reduction of amplitude of the Sachs-Wolfe plateau and to a decrease of the effective
$\chi^2$ by up to 4.6.\footnote{For the three general models, the posterior 
distribution is actually multimodal. Here we are referring to models contributing to 
the main peak in the posterior, with the highest maximum likelihood. There is another peak 
with a smaller maximum likelihood, appearing in Fig.~\ref{OneDimConsolidated} 
as a small bump for positive values of the cross-correlation amplitude. In this paper,
we do not carry out a separate investigation for models contributing to
this secondary peak.} This situation explains the rather loose bounds
on the derived parameter $\alpha_\mathcal{II}(2,20)$, as shown in
Fig.~\ref{OneDimConsolidated}.

A comparison of $\mathcal{P}_\mathcal{II}^{(1)}$ and $\mathcal{P}_\mathcal{II}^{(2)}$
shows that best fitting models have an isocurvature spectral index $n_\mathcal{II}$ 
close to 1.7 for CDI, 1.1 for NDI, and 1.0 for NVI modes.

For CDI and NDI, the amplitude of acoustic peaks quickly decreases
with increasing $\ell$, so that the constraints are entirely driven by
small $\ell$s. Since the same value of the primordial amplitude
$\mathcal{P}_\mathcal{II}^{(1)}$ leads to different plateau amplitudes for the
two isocurvature models (see Fig.~\ref{fig:adia_vs_iso}), the bounds
on $\mathcal{P}_\mathcal{II}^{(1)}$ and $\mathcal{P}_\mathcal{RI}^{(1)}$ are consistently
stronger for CDI than for NDI. For NVI, the acoustic peak amplitude is
larger than the plateau amplitude. In NVI models, the data cannot
allow for a too large amplitude of correlated isocurvature modes
at small $\ell$, because the total spectrum would be distorted at
larger $\ell$.  This possibility is strongly disfavoured by the data,
which are consistent with the peak location predicted by a pure
adiabatic model. Hence in the NVI case we obtain slightly stronger bounds and a smaller reduction of the effective $\chi^2$.

The fact that the data prefer models with a significant contribution
from CDI or NDI modes should be interpreted with care.  The detection
of a shift in the phase of acoustic oscillations would bring
unambigous evidence in favour of isocurvature modes. With
\Planck\ data, we are not in this situation. The evidence is driven by
a small deficit of amplitude in the Sachs-Wolfe plateau, which could
have several different possible explanations (such as a deficit in the
large-scale primordial power spectrum, as already seen in the previous
sections).
However, multi-field inflationary scenarios can produce the mixture of
curvature and isocurvature fluctuations which
we have found to provide a good fit to the \Planck\ data.

\subsection{Special cases}

The six-parameter models of the previous subsection including
one isocurvature mode and the adiabatic mode make no
assumptions about the spectral indices of each mode or  
the degree of correlation between the isocurvature mode and the 
adiabatic mode. This leads to a large number of additional degrees 
of freedom. There are both theoretical and phenomenological motivations
for choosing special values for some of the parameters,
leading to special cases with just one more degree of freedom with
respect to the adiabatic case.
The results are reported in Table~\ref{tab:how_much_ic}, 
for uncorrelated perturbations with $n_\mathcal{II}=1$, and
fully correlated or anti-correlated perturbations with $n_\mathcal{II}=n_\mathcal{RR}$. 
In the general case, anti-correlated
isocurvature perturbations slightly improve the fit to the \Planck\ data.
We consider below the implications of our results for two    
important cases: the axion and curvaton scenarios.

\subsubsection{Constraints on axion isocurvature}

The axion field was proposed to solve the strong CP problem and 
constitutes a well-motivated dark matter candidate.
(See for example \cite{Preskill:1982cy}, \cite{Turner:1989vc},  
\cite{Peccei:2006as}, \cite{Sikivie:2006ni}, \cite{Raffelt:2006cw}, and \cite{Kim:2008hd}). 
The axion is the psuedo-Goldstone boson of the broken Peccei-Quinn (PQ) symmetry. Under certain
assumptions, the axion field may induce 
significant isocurvature perturbations \citep{Turner:1983sj,Axenides:1983hj,Steinhardt:1983ia,Linde:1984ti,Linde:1985yf,Seckel:1985tj,Kofman:1985zx,Lyth:1989pb,Linde:1990yj,Turner:1990uz,Linde:1991km,Lyth:1991ub}.
If inflation takes place after PQ symmetry breaking, the quantum fluctuations of the inflaton are responsible for primordial 
curvature perturbations, while those of the axion field generate primordial entropy 
perturbations.  After the QCD transition, when one of the vacua
becomes preferred giving the axion field a mass, the axions
behave as cold dark matter. This way of producing axionic dark matter is called the misalignment angle mechanism.
In such a scenario, the CMB anisotropy may include significant power from 
CDM isocurvature fluctuations. In that case, the fraction 
$\beta _\mathrm{iso}\equiv \mathcal{P_{II}}/(\mathcal{P_{RR}}+\mathcal{P_{II}})$
of CDM isocurvature modes is related to the energy scale of inflation, $H_\mathrm{inf}$,
through \citep{Lyth:1989pb,Beltran:2006sq,Bae:2008ue,Hamann:2009yf}
\begin{equation}
H_\mathrm{inf} = \frac{0.96\times 10^{7}~\mathrm{GeV}}{R_\mathrm{a}} 
\left(\frac{\beta _\mathrm{iso}}{0.04}\right)^{1/2} \left(\frac{\Omega_\mathrm{a}}{0.120}\right)^{1/2} 
\left(\frac{f_\mathrm{a}}{10^{11}~\mathrm{GeV}}\right)^{0.408},
\end{equation}
where $\Omega _\mathrm{a}$ is the relic axion density, $R_\mathrm{a}$ the fraction of CDM consisting 
of axions, and $f_\mathrm{a}$ the PQ symmetry breaking scale.
In this model, CDM isocurvature perturbations should be totally uncorrelated 
with adiabatic perturbations and have a spectral index $n_\mathcal{II}$ very close to one 
since in the first-order slow-roll approximation the index reads $(1-2\epsilon_V)$. 
Since the sensitivity of the data to $n_\mathcal{II}$ is very limited \citep{Beltran:2006sq}, we assume for simplicity that $n_\mathcal{II}=1$.

Within the general parametrization presented in Eq.~\ref{matInter}, 
we can select the axion case by imposing $\mathcal{P}_\mathcal{RI}^{(1,2)}=0$, as well as the condition
\beq
\mathcal{P}_\mathcal{II}^{(2)}=\mathcal{P}_\mathcal{II}^{(1)},
\eeq
corresponding to $n_\mathcal{II}=1$. We therefore have 
three independent parameters, 
$\mathcal{P}_\mathcal{RR}^{(1)},$ 
$\mathcal{P}_\mathcal{RR}^{(2)},$ and 
$\mathcal{P}_\mathcal{II}^{(1)}$, and we sample these parameters with uniform prior distributions. The fraction $\beta_\mathrm{iso}(k_*)$ with \mbox{$k_*=0.05$~Mpc$^{-1}$} 
is then a derived parameter. Since the data constrain $\beta_\mathrm{iso}\ll 1$,
the relation between $\beta_\mathrm{iso}$ and $\mathcal{P}_\mathcal{II}^{(1)}$ is nearly linear, 
so the primordial isocurvature fraction is sampled with a close to uniform prior. 

Constraints on this model are shown in Table~\ref{tab:how_much_ic}. We find 
\beq
\beta_\mathrm{iso} < 0.039 \quad (95\%~\mathrm{CL}, \mathit{Planck}\mathrm{+WP} ), 
\eeq
at the scale $k_\mathrm{mid}=0.05\,\mathrm{Mpc}^{-1}$, 
with a best fit value of zero. Hence there is no evidence for axion generated 
isocurvature perturbations. This limit significantly improves the previous CMB bounds.
At the scale $k=0.002$~Mpc$^{-1}$, our result reads $\beta_\mathrm{iso} < 0.036$, to be compared to 
$\beta_\mathrm{iso}<0.15$ for {\it WMAP} 9-year alone, or $\beta_\mathrm{iso}<0.061$ for {\it WMAP}+ACT+SPT at 
95\% CL \citep{Hinshaw:2012fq}. This bound can be used to exclude  
regions of parameter space composed of $f_\mathrm{a}$, $R_\mathrm{a}$, and the energy scale 
of inflation, but cannot be used to obtain a 
model-independent bound on $f_\mathrm{a}$. However, if we assume (i) that the PQ symmetry is 
broken during inflation, 
(ii) that it is not restored by the quantum fluctuations of the inflaton 
(which imposes $H_\mathrm{inf}/(2\pi)<f_\mathrm{a}$), nor by thermal fluctuations 
in case of a very efficient reheating stage, and (iii) that all the CDM 
consists of axions produced by the misalignment angle, then we can derive an 
upper bound on the energy scale of inflation as
\begin{equation}
H_\mathrm{inf} \leq 0.87
\times10^{7}~\mathrm{GeV} 
\left(\frac{f_\mathrm{a}}{10^{11}~\mathrm{GeV}}\right)^{0.408}\, (95 \% ~\mathrm{CL})~.
\end{equation}

\subsubsection{Constraints on the curvaton scenario}

In the simplest one-field inflationary models curvature
perturbations arise from quantum fluctuations in the inflaton field,
but this is not the only way to generate curvature perturbations. 
Isocurvature perturbations may seed curvature perturbations outside the Hubble 
radius~\citep{Polarski:1994rz,Langlois:1999dw,Gordon:2000hv}, so it is possible that
a significant component of the observed adiabatic mode could be
strongly correlated with an isocurvature mode. This happens for
instance in the curvaton
scenario~\citep{Mollerach:1989hu,Enqvist:2001zp,Moroi:2001ct,Lyth:2001nq,Lyth:2002my,Gordon:2002gv}.
The curvaton is an extra light scalar field acquiring a spectrum of
fluctuations on cosmological scales during inflation. Depending on its
density evolution and decay history, this field could be responsible
for part of the observed adiabatic perturbations, or all of them, or
for a mixture of correlated adiabatic and isocurvature perturbations.

We focus here on the simplest viable version of this scenario 
in which the curvaton decays into CDM particles while contributing a non-negligible fraction 
\begin{equation}
r_D=\frac{3 \rho_\mathrm{curvaton}}{3 \rho_\mathrm{curvaton}+4\rho_\mathrm{radiation}}
\end{equation}
to the total energy density of the universe. If the curvaton dominates at decay time 
($r_D=1$), its primordial fluctuations seed curvature perturbations equivalent to a pure 
adiabatic mode. If $r_D<1$, curvaton fluctuations are only partially converted into adiabatic 
perturbations, while CDM particles carry CDI perturbations, which are fully correlated with the 
adiabatic perturbations since they share a common origin. We recall that with our conventions, 
``fully correlated'' means that the cross-correlation term contributes constructively
to the Sachs-Wolfe component of the total temperature spectrum. Some authors define the 
correlation with the opposite sign and call this case ``fully anti-correlated'' 
\citep[e.g.,][]{komatsu2010,Hinshaw:2012fq}. In this model, the CDI fraction is related to $r_D$ by~\citep{Gordon:2002gv}
\begin{equation}
\frac{{\cal I}_\mathrm{CDI}}{\cal R} = \frac{3(1-r_D)}{r_D}.
\end{equation}
In our notation this is equivalent to
\begin{equation}
\beta_\mathrm{iso} = \frac{9(1-r_D)^2}{r_D^2+9(1-r_D)^2}.
\end{equation}
Within the general parametrization presented in 
Eq.~\ref{matInter}, we can satisfy this case by imposing 
\beq
\frac{\mathcal{P}_\mathcal{RI}^{(1)}}{\sqrt{\mathcal{P}_\mathcal{RR}^{(1)}\mathcal{P}_\mathcal{II}^{(1)}}}
=\frac{\mathcal{P}_\mathcal{RI}^{(2)}}{\sqrt{\mathcal{P}_\mathcal{RR}^{(2)}\mathcal{P}_\mathcal{II}^{(2)}}}=1,
\label{eq:curvaton}
\eeq 
together with the condition
\beq
\mathcal{P}_\mathcal{II}^{(2)}=\frac{\mathcal{P}_\mathcal{II}^{(1)}\mathcal{P}_\mathcal{RR}^{(2)}}{\mathcal{P}_\mathcal{RR}^{(1)}},
\eeq 
corresponding to $n_\mathcal{II}=n_\mathcal{RR}$.
As in the axion case, this results in three independent parameters 
$\mathcal{P}_\mathcal{RR}^{(1,2)}$ and $\mathcal{P}_\mathcal{II}^{(1)},$ which we sample with uniform priors.
The constraints for this model are shown in Table~\ref{tab:how_much_ic}. 
The best fit model is still the pure adiabatic case, and the upper bound
\beq
\beta_\mathrm{iso}<0.0025 \quad (95\%~\mathrm{CL}, \mathit{Planck}\mathrm{+WP})
\eeq
is scale independent, since the adiabatic and 
isocurvature tilts are assumed to be equal. 
This is a significant improvement over the {\it WMAP} 
9-year bounds, $\beta_\mathrm{iso}<0.012$ for {\it WMAP} alone, 
or $\beta_\mathrm{iso}<0.0076$ for {\it WMAP}+ACT+SPT at 95\%~CL \citep{Hinshaw:2012fq}. 
We conclude that in this scenario, the curvaton should decay when it dominates the energy density of the universe, with $r_D>0.983$.

The nonlinearity parameter in the curvaton model studied here is
\citep{Bartolo:2003jx, Bartolo:2004ty}
\begin{equation}
f_{\mathrm{NL}}^\mathrm{local}=\frac{5}{4 r_D} - \frac{5}{3} - \frac{5 r_D}{6},
\end{equation}
assuming a quadratic potential for the curvaton field \citep{Sasaki:2006kq}. In
the pure adiabatic case ($r_D=1$) this leads to $f_{\mathrm{NL}}^\mathrm{local} = -5/4$.
The constraint $0.98 < r_D < 1$ then corresponds to $-1.25 < f_{\mathrm{NL}}^\mathrm{local} <
-1.21$. Taking into account the \Planck\  result $f_{\mathrm{NL}}^\mathrm{local} = 2.7 \pm 5.8$ \citep{planck2013-p09a}, we conclude
that the \Planck\  data are consistent with the scenario where the
curvaton decays into CDM when it dominates the energy density of the
universe, and its fluctuations are almost entirely converted into adiabatic
ones.

}

\section{Conclusions}
\label{sec:conclusions}

This paper establishes the status of cosmic inflation in the context of the 
first release of the \Planck\ cosmological results,
which includes the temperature data from the first $2.6$ sky surveys. 
CMB polarization as measured by \Planck\ will be the subject of a future release.
We find that standard slow-roll single-field inflation is compatible with the \Planck\ data.
This result is confirmed by other papers of this series.
\Planck\ in combination with {\it WMAP} 9-year 
large angular scale polarization (WP) yields $\Omega_K = -0.006 \pm 0.018$ at 95\% CL by combining temperature and 
lensing information \citep{planck2013-p11,planck2013-p12}.
The bispectral non-Gaussianity parameter $f_\mathrm{NL}$ measured by \Planck\ is consistent with zero \citep{planck2013-p09a}.
These results are compatible with zero spatial curvature and a small value of $f_\mathrm{NL},$
as predicted in the simplest slow-roll inflationary models.

A key \Planck\ result is the measurement of the scalar perturbation spectral index.
\Planck+WP data give $n_\mathrm{s}=0.9603 \pm 0.0073$ (and $n_\mathrm{s}= 0.9629 \pm 0.0057$ 
when combined with BAO).
This result disfavours the Harrison-Zeldovich (HZ) $n_\mathrm{s}=1$ model at more than 5$\sigma$.
Even in extended cosmological models, the HZ spectrum cannot be reconciled with the data.
Allowing a general reionization scenario yields $\Delta \chi_\mathrm{eff}^2 = 12.5$ with respect to
$\Lambda$CDM for \Planck+WP data.  
When the primordial helium abundance or the effective number of
neutrino species are allowed to vary, the best fit of the HZ model to a combination of  \Planck+WP and BAO data 
is still worse by $\Delta \chi_\mathrm{eff}^2 = 4.6$ and $8.0$, respectively.

We find no evidence for \Planck\ data preferring a generalization of a simple power law spectrum to include 
a running of the spectral index ($\mathrm{d} n_{\mathrm s}/\mathrm{d}\ln k = -0.0134 \pm 0.0090$)
or a running of the running
($\mathrm{d}^2 n_{\mathrm s}/\mathrm{d}\ln k^2 = 0.020^{+0.016}_{+0.015}$ with \Planck+WP).
In a model admitting tensor fluctuations,
the $95\%$ CL bound on the tensor-to-scalar ratio is 
\mbox{$r_{0.002} < 0.12$ ($< 0.11$)} using \Planck+WP (plus high-$\ell$
CMB data). This bound on $r$ implies an upper limit for the inflation 
energy scale of $1.9 \times 10^{16}~$GeV}, or equivalently, for the Hubble parameter $H_* < 3.7 \times 10^{-5}~ M_{\mathrm{pl}}$, at $95\%$ CL.

The degeneracy between $n_\mathrm{s}$ and $r,$ which plagued previous CMB measurements,
is now removed by the \Planck\ precision in the determination of the highest acoustic peaks.  Inflaton potentials with a concave 
shape are favoured and occupy most of the 95\% confidence region allowed by
\Planck+WP in the $n_\mathrm{s}\textrm{-}r$ plane. Models with an exponential potential, a monomial potential with a power larger than two, or
hybrid models driven by a quadratic term are disfavoured at more than 95\% confidence. 
The quadratic large-field model, in the past often cited as the simplest inflationary model, 
now lies at the edge of the 95\%~CL contours allowed by \Planck+WP+high-$\ell$ CMB data.

A Bayesian parameter estimation and model comparison analysis of a representative
sample of single-field slow-roll models shows that \Planck\ is able to discriminate
between these models with results that are robust even when a broad set of entropy generation
scenarios are allowed. In addition to confirming the exclusion of the $\phi ^4$ potential,
the Bayesian evidence computed from the \Planck\ data provides significant odds (logarithms of the Bayes factor of about
$-5$ or lower relative to $\Lambda$CDM)
against large-field models compatible with previous cosmological data, such as the $\phi^2$ potential, and two-parameter potentials
such as natural inflation and the hilltop potential. As presented in Sect.~\ref{sec:modelcomp}, 
\Planck\ establishes strong constraints on the parameter values
of specific inflationary scenarios. For example, the scale parameter of the natural inflation potential 
is constrained to be $\log(f/M_\mathrm{pl}) \gtrsim 1.1$ (95\% CL), improving upon the {\it WMAP} 7-year 
limit on $f$ by a factor of two. The \Planck\ data
limit the possibilities for the unexplored physics between the end of inflation and
the beginning of the radiation dominated era. Data-driven constraints are obtained on $w_\mathrm{int}$, the effective
equation of state in the post-inflationary era. Particularly for the disfavoured models listed above,
their parameters are pushed to unnatural values ($w_\mathrm{int} \gtrsim 1/3$) in order to become more compatible with the data.

Using an essentially exact numerical calculation of the predicted 
primordial spectrum, we
reconstruct the observable window of the inflaton potential, 
expanding the potential as a Taylor series up to a fixed order. For 
an observable potential
described by a polynomial of order three, the
reconstruction agrees well with the slow-roll predictions. If a quartic term
is allowed, the result deviates from the slow-roll prediction because
the \Planck\ data favour a slightly smaller amplitude for the
Sachs-Wolfe plateau relative to the $\ell>40$ part of the power
spectrum than the best fitting minimal $\Lambda$CDM model with a power law
primordial spectrum. A potential with a fourth-order polynomial can
fit this feature, thus reducing the effective $\chi_\mathrm{eff}^2$ by approximately four.

A penalized likelihood reconstruction of the primordial power spectrum 
shows hints of structure at modest 
statistical significance. However, recent work after submission suggests that this 
feature can be explained by electromagnetic interference.
Parameterized models producing superimposed oscillations (possibly motivated by deviations
from the Bunch-Davies vacuum state, axion monodromy, or a sharp step in the inflaton potential) improve 
the $\chi_\mathrm{eff}^2$ by roughly $10$, where three extra parameters have been added.  
However, a Bayesian model comparison analysis does not strongly favour the model with oscillations over 
the standard featureless power spectrum.  With \Planck\ polarization data, a more 
conclusive result on superimposed oscillations is expected.

We combine power spectrum constraints
with those on the nonlinearity parameter
$f_\mathrm{NL}$ \citep{planck2013-p09a} to constrain single-field
inflation with generalized Lagrangians, in which
non-Gaussianities are larger than those predicted by the simplest
slow-roll inflationary models. We show how the limits on the
inflation sound speed derived in \cite{planck2013-p09a} are crucial
to constrain slow-roll parameters for generalized Lagrangians.
We also show how particular examples of DBI Inflation and $k$-inflation can be constrained by this combination of \Planck\ data.

We test the hypothesis that the primordial cosmological perturbations were exclusively
adiabatic.  We analyse all nonsingular (i.e., nondecaying) 
isocurvature modes arbitrarily correlated to the adiabatic mode, 
using a parameterization where the isocurvature contributions are specified at two scales.
The oscillatory pattern in the \Planck\ temperature spectrum is 
compatible with purely adiabatic perturbations, and therefore constrains any isocurvature contribution to be small at those multipoles.
As a consequence, axion and curvaton scenarios, in which the CDM isocurvature mode is 
uncorrelated or fully correlated with the adiabatic
mode, are not favoured by \Planck. The upper bounds on the isocurvature fraction at
$k=0.05$ Mpc$^{-1}$ are $0.039$ for the axion, and $0.0025$ for the curvaton, at 95\% CL. 
However general models with an arbitrarily correlated mixture of
adiabatic and (CDM or neutrino) isocurvature modes have the freedom to
lower the Sachs-Wolfe plateau relative to the high-$\ell$ spectrum,
and reduce the effective $\chi_\mathrm{eff}^2$ by more than four.

The simplest inflationary models have passed an exacting test with the \Planck\ data. 
The full mission data including \Planck's polarization measurements will help answer further fundamental questions,
helping to probe nonsmooth power spectra and the energy scale of inflation as well as extensions to more complex models.

\begin{acknowledgements}
The development of \Planck\ has been supported by: ESA; CNES and CNRS/INSU-IN2P3-INP (France); ASI, CNR, and INAF (Italy); 
NASA and DoE (USA); STFC and UKSA (UK); CSIC, MICINN, JA and RES (Spain); Tekes, AoF and CSC (Finland); DLR and MPG (Germany); 
CSA (Canada); DTU Space (Denmark); SER/SSO (Switzerland); RCN (Norway); SFI (Ireland); FCT/MCTES (Portugal); and PRACE (EU). 
A description of the Planck Collaboration and a list of its members, including the technical or scientific activities in which 
they have been involved, can be found at \url{http://www.sciops.esa.int/index.php?project=planck&page=Planck_Collaboration}.
We gratefully acknowledge CINECA (\texttt{http://www.cineca.it/}) 
under the agreement LFI/CINECA and IN2P3 Computer Center 
(\texttt{http://cc.in2p3.fr}) for providing a significant 
amount of the computing resources and services needed for this work.
\end{acknowledgements}
\begin{raggedright}
\bibliographystyle{aa_arxiv}

\bibliography{Planck_bib,references}

\def\eprinttmppp@#1arXiv:@{#1}
\providecommand{\arxivlink[1]}{\href{http://arxiv.org/abs/#1}{arXiv:#1}}
\def\eprinttmp@#1arXiv:#2 [#3]#4@{\ifthenelse{\equal{#3}{x}}{\ifthenelse{
\equal{#1}{}}{\arxivlink{\eprinttmppp@#2@}}{\arxivlink{#1}}}{\arxivlink{#2}
  [#3]}}
\providecommand{\eprintlink}[1]{\eprinttmp@#1arXiv: [x]@}
\providecommand{\eprint}[1]{\eprintlink{#1}}
\providecommand{\adsurl}[1]{\href{#1}{ADS}}
\begin{thebibliography}{336}
\expandafter\ifx\csname natexlab\endcsname\relax\def\natexlab#1{#1}\fi

\bibitem[{Abbott \& Wise(1984)}]{Abbott:1984fp}
Abbott, L. \& Wise, M.~B. 1984, Nucl. Phys., B244, 541

\bibitem[{Ach\'ucarro {et~al.}(2011)Ach\'ucarro, Gong, Hardeman, Palma, \&
  Patil}]{Achucarro:2010da}
Ach\'ucarro, A., Gong, J.-O., Hardeman, S., Palma, G.~A., \& Patil, S.~P. 2011,
  JCAP, 1101, 030

\bibitem[{{Acquaviva} {et~al.}(2003){Acquaviva}, {Bartolo}, {Matarrese}, \&
  {Riotto}}]{2003NuPhB.667..119A}
{Acquaviva}, V., {Bartolo}, N., {Matarrese}, S., \& {Riotto}, A. 2003, Nucl.
  Phys. B, 667, 119

\bibitem[{Adams {et~al.}(1993)Adams, Bond, Freese, Frieman, \&
  Olinto}]{Adams:1992bn}
Adams, F.~C., Bond, J.~R., Freese, K., Frieman, J.~A., \& Olinto, A.~V. 1993,
  Phys. Rev., D47, 426

\bibitem[{{Adams} {et~al.}(2001){Adams}, {Cresswell}, \&
  {Easther}}]{2001PhRvD..64l3514A}
{Adams}, J., {Cresswell}, B., \& {Easther}, R. 2001, \prd, 64, 123514

\bibitem[{Adams {et~al.}(2001)Adams, Cresswell, \& Easther}]{Adams:2001vc}
Adams, J.~A., Cresswell, B., \& Easther, R. 2001, Phys. Rev., D64, 123514

\bibitem[{Adshead {et~al.}(2012)Adshead, Dvorkin, Hu, \& Lim}]{Adshead:2011jq}
Adshead, P., Dvorkin, C., Hu, W., \& Lim, E.~A. 2012, Phys. Rev., D85, 023531

\bibitem[{Adshead {et~al.}(2011)Adshead, Easther, Pritchard, \&
  Loeb}]{Adshead:2010mc}
Adshead, P., Easther, R., Pritchard, J., \& Loeb, A. 2011, JCAP, 1102, 021

\bibitem[{{Agarwal} \& {Bean}(2009)}]{2009PhRvD..79b3503A}
{Agarwal}, N. \& {Bean}, R. 2009, \prd, 79, 023503

\bibitem[{{Akaike}(1974)}]{1974ITAC...19..716A}
{Akaike}, H. 1974, IEEE Trans. on Automatic Control, 19, 716

\bibitem[{Albrecht \& Steinhardt(1982)}]{Albrecht:1982wi}
Albrecht, A. \& Steinhardt, P.~J. 1982, Phys. Rev. Lett., 48, 1220

\bibitem[{{Alishahiha} {et~al.}(2004){Alishahiha}, {Silverstein}, \&
  {Tong}}]{2004PhRvD..70l3505A}
{Alishahiha}, M., {Silverstein}, E., \& {Tong}, D. 2004, \prd, 70, 123505

\bibitem[{Allahverdi {et~al.}(2010)Allahverdi, Brandenberger, Cyr-Racine, \&
  Mazumdar}]{Allahverdi:2010xz}
Allahverdi, R., Brandenberger, R., Cyr-Racine, F.-Y., \& Mazumdar, A. 2010,
  Ann. Rev. Nucl. Part. Sci., 60, 27

\bibitem[{Amendola {et~al.}(2002)Amendola, Gordon, Wands, \&
  Sasaki}]{Amendola:2001ni}
Amendola, L., Gordon, C., Wands, D., \& Sasaki, M. 2002, Phys.Rev.Lett., 88,
  211302

\bibitem[{Anantua {et~al.}(2009)Anantua, Easther, \& Giblin}]{Anantua:2008am}
Anantua, R., Easther, R., \& Giblin, J.~T. 2009, Phys. Rev. Lett., 103, 111303

\bibitem[{{Anderson} {et~al.}(2012){Anderson}, {Aubourg}, {Bailey}, {Bizyaev},
  {Blanton}, {Bolton}, {Brinkmann}, {Brownstein}, {Burden}, {Cuesta}, {da
  Costa}, {Dawson}, {de Putter}, {Eisenstein}, {Gunn}, {Guo}, {Hamilton},
  {Harding}, {Ho}, {Honscheid}, {Kazin}, {Kirkby}, {Kneib}, {Labatie},
  {Loomis}, {Lupton}, {Malanushenko}, {Malanushenko}, {Mandelbaum}, {Manera},
  {Maraston}, {McBride}, {Mehta}, {Mena}, {Montesano}, {Muna}, {Nichol},
  {Nuza}, {Olmstead}, {Oravetz}, {Padmanabhan}, {Palanque-Delabrouille}, {Pan},
  {Parejko}, {P{\^a}ris}, {Percival}, {Petitjean}, {Prada}, {Reid}, {Roe},
  {Ross}, {Ross}, {Samushia}, {S{\'a}nchez}, {Schlegel}, {Schneider},
  {Sc{\'o}ccola}, {Seo}, {Sheldon}, {Simmons}, {Skibba}, {Strauss}, {Swanson},
  {Thomas}, {Tinker}, {Tojeiro}, {Maga{\~n}a}, {Verde}, {Wagner}, {Wake},
  {Weaver}, {Weinberg}, {White}, {Xu}, {Y{\`e}che}, {Zehavi}, \&
  {Zhao}}]{Anderson:2012sa}
{Anderson}, L., {Aubourg}, E., {Bailey}, S., {et~al.} 2012, \mnras, 427, 3435

\bibitem[{{Armend{\'a}riz-Pic{\'o}n} {et~al.}(1999){Armend{\'a}riz-Pic{\'o}n},
  {Damour}, \& {Mukhanov}}]{1999PhLB..458..209A}
{Armend{\'a}riz-Pic{\'o}n}, C., {Damour}, T., \& {Mukhanov}, V. 1999, Phys.
  Lett. B, 458, 209

\bibitem[{Audren {et~al.}(2012)Audren, Lesgourgues, Benabed, \&
  Prunet}]{Audren:2012wb}
Audren, B., Lesgourgues, J., Benabed, K., \& Prunet, S. 2012, JCAP, 1302, 001

\bibitem[{Aver {et~al.}(2012)Aver, Olive, \& Skillman}]{Aver:2011bw}
Aver, E., Olive, K.~A., \& Skillman, E.~D. 2012, JCAP, 1204, 004

\bibitem[{Axenides {et~al.}(1983)Axenides, Brandenberger, \&
  Turner}]{Axenides:1983hj}
Axenides, M., Brandenberger, R.~H., \& Turner, M.~S. 1983, Phys. Lett., B126,
  178

\bibitem[{{Babich} {et~al.}(2004){Babich}, {Creminelli}, \&
  {Zaldarriaga}}]{2004JCAP...08..009B}
{Babich}, D., {Creminelli}, P., \& {Zaldarriaga}, M. 2004, JCAP, 8, 9

\bibitem[{Bae {et~al.}(2008)Bae, Huh, \& Kim}]{Bae:2008ue}
Bae, K.~J., Huh, J.-H., \& Kim, J.~E. 2008, JCAP, 0809, 005

\bibitem[{Banks {et~al.}(1994)Banks, Kaplan, \& Nelson}]{Banks:1993en}
Banks, T., Kaplan, D.~B., \& Nelson, A.~E. 1994, Phys. Rev., D49, 779

\bibitem[{Bardeen {et~al.}(1983)Bardeen, Steinhardt, \&
  Turner}]{Bardeen:1983qw}
Bardeen, J.~M., Steinhardt, P.~J., \& Turner, M.~S. 1983, Phys. Rev., D28, 679

\bibitem[{Barrow(1990)}]{Barrow:1990vx}
Barrow, J.~D. 1990, Phys. Lett., B235, 40

\bibitem[{Barrow \& Liddle(1993)}]{Barrow:1993zq}
Barrow, J.~D. \& Liddle, A.~R. 1993, Phys. Rev., D47, 5219

\bibitem[{Bartolo {et~al.}(2004{\natexlab{a}})Bartolo, Komatsu, Matarrese, \&
  Riotto}]{Bartolo:2004if}
Bartolo, N., Komatsu, E., Matarrese, S., \& Riotto, A. 2004{\natexlab{a}},
  Phys. Rept., 402, 103

\bibitem[{Bartolo \& Liddle(2002)}]{Bartolo:2002vf}
Bartolo, N. \& Liddle, A.~R. 2002, Phys.Rev., D65, 121301

\bibitem[{Bartolo {et~al.}(2004{\natexlab{b}})Bartolo, Matarrese, \&
  Riotto}]{Bartolo:2004ty}
Bartolo, N., Matarrese, S., \& Riotto, A. 2004{\natexlab{b}}, Phys. Rev. Lett.,
  93, 231301

\bibitem[{Bartolo {et~al.}(2004{\natexlab{c}})Bartolo, Matarrese, \&
  Riotto}]{Bartolo:2003jx}
Bartolo, N., Matarrese, S., \& Riotto, A. 2004{\natexlab{c}}, Phys. Rev., D69,
  043503

\bibitem[{Barvinsky {et~al.}(2008)Barvinsky, Kamenshchik, \&
  Starobinsky}]{Barvinsky:2008ia}
Barvinsky, A.~O., Kamenshchik, A.~Y., \& Starobinsky, A.~A. 2008, JCAP, 0811,
  021

\bibitem[{{Baumann} \& {McAllister}(2007)}]{2007PhRvD..75l3508B}
{Baumann}, D. \& {McAllister}, L. 2007, \prd, 75, 123508

\bibitem[{{Bean} {et~al.}(2008){Bean}, {Chen}, {Peiris}, \&
  {Xu}}]{2008PhRvD..77b3527B}
{Bean}, R., {Chen}, X., {Peiris}, H., \& {Xu}, J. 2008, \prd, 77, 023527

\bibitem[{Bean {et~al.}(2006)Bean, Dunkley, \& Pierpaoli}]{Bean:2006qz}
Bean, R., Dunkley, J., \& Pierpaoli, E. 2006, Phys. Rev., D74, 063503

\bibitem[{{Bean} {et~al.}(2007){Bean}, {Shandera}, {Tye}, \&
  {Xu}}]{2007JCAP...05..004B}
{Bean}, R., {Shandera}, S.~E., {Tye}, S.-H.~H., \& {Xu}, J. 2007, JCAP, 5, 4

\bibitem[{Beltran {et~al.}(2007)Beltran, Garc\'ia-Bellido, \&
  Lesgourgues}]{Beltran:2006sq}
Beltran, M., Garc\'ia-Bellido, J., \& Lesgourgues, J. 2007, Phys. Rev., D75,
  103507

\bibitem[{Beltran {et~al.}(2004)Beltran, Garc\'ia-Bellido, Lesgourgues, \&
  Riazuelo}]{Beltran:2004uv}
Beltran, M., Garc\'ia-Bellido, J., Lesgourgues, J., \& Riazuelo, A. 2004, Phys.
  Rev., D70, 103530

\bibitem[{Benetti {et~al.}(2013)Benetti, Pandolfi, Lattanzi, Martinelli, \&
  Melchiorri}]{Benetti:2012wu}
Benetti, M., Pandolfi, S., Lattanzi, M., Martinelli, M., \& Melchiorri, A.
  2013, Phys. Rev., D87, 023519

\bibitem[{Bennett {et~al.}(2011)Bennett, Hill, Hinshaw, Larson, Smith,
  {et~al.}}]{Bennett:2010jb}
Bennett, C., Hill, R., Hinshaw, G., {et~al.} 2011, \apjs, 192, 17

\bibitem[{Bennett {et~al.}(2013)Bennett, Larson, Weiland, Jarosik, Hinshaw,
  {et~al.}}]{Bennett:2012fp}
Bennett, C., Larson, D., Weiland, J., {et~al.} 2013, \apjs, 208, 20

\bibitem[{Beutler {et~al.}(2011)Beutler, Blake, Colless, Jones, Staveley-Smith,
  {et~al.}}]{Beutler:2011hx}
Beutler, F., Blake, C., Colless, M., {et~al.} 2011, MNRAS, 416, 3017

\bibitem[{Bezrukov \& Gorbunov(2012)}]{Bezrukov:2011gp}
Bezrukov, F. \& Gorbunov, D. 2012, Phys. Lett., B713, 365

\bibitem[{Bezrukov \& Shaposhnikov(2008)}]{Bezrukov:2007ep}
Bezrukov, F. \& Shaposhnikov, M. 2008, Phys. Lett., B659, 703

\bibitem[{Bezrukov \& Shaposhnikov(2009)}]{Bezrukov:2009db}
Bezrukov, F. \& Shaposhnikov, M. 2009, JHEP, 07, 089

\bibitem[{Bin\'etruy \& Gaillard(1986)}]{Binetruy:1986ss}
Bin\'etruy, P. \& Gaillard, M.~K. 1986, Phys. Rev., D34, 3069

\bibitem[{Blas {et~al.}(2011)Blas, Lesgourgues, \& Tram}]{Blas:2011rf}
Blas, D., Lesgourgues, J., \& Tram, T. 2011, JCAP, 1107, 034

\bibitem[{Boubekeur \& Lyth(2005)}]{Boubekeur:2005zm}
Boubekeur, L. \& Lyth, D. 2005, JCAP, 0507, 010

\bibitem[{Bozza {et~al.}(2003)Bozza, Giovannini, \& Veneziano}]{Bozza:2003pr}
Bozza, V., Giovannini, M., \& Veneziano, G. 2003, JCAP, 0305, 001

\bibitem[{{Bridges} {et~al.}(2009){Bridges}, {Feroz}, {Hobson}, \&
  {Lasenby}}]{2009MNRAS.400.1075B}
{Bridges}, M., {Feroz}, F., {Hobson}, M.~P., \& {Lasenby}, A.~N. 2009, \mnras,
  400, 1075

\bibitem[{Bridle {et~al.}(2003)Bridle, Lewis, Weller, \&
  Efstathiou}]{Bridle:2003sa}
Bridle, S., Lewis, A., Weller, J., \& Efstathiou, G. 2003, MNRAS, 342, L72

\bibitem[{Brout {et~al.}(1978)Brout, Englert, \& Gunzig}]{Brout:1977ix}
Brout, R., Englert, F., \& Gunzig, E. 1978, Annals Phys., 115, 78

\bibitem[{Bucher \& Cohn(1997)}]{Bucher:1997xs}
Bucher, M. \& Cohn, J. 1997, Phys. Rev., D55, 7461

\bibitem[{Bucher {et~al.}(2004)Bucher, Dunkley, Ferreira, Moodley, \&
  Skordis}]{Bucher:2004an}
Bucher, M., Dunkley, J., Ferreira, P., Moodley, K., \& Skordis, C. 2004, Phys.
  Rev. Lett., 93, 081301

\bibitem[{Bucher {et~al.}(1995)Bucher, Goldhaber, \& Turok}]{Bucher:1994gb}
Bucher, M., Goldhaber, A.~S., \& Turok, N. 1995, Phys. Rev., D52, 3314

\bibitem[{Bucher {et~al.}(2000)Bucher, Moodley, \& Turok}]{Bucher:1999re}
Bucher, M., Moodley, K., \& Turok, N. 2000, Phys. Rev., D62, 083508

\bibitem[{Bucher \& Turok(1995)}]{Bucher:1995ga}
Bucher, M. \& Turok, N. 1995, Phys. Rev., D52, 5538

\bibitem[{Bunch \& Davies(1978)}]{Bunch:1978yq}
Bunch, T. \& Davies, P. 1978, Proc. Roy. Soc. Lond., A360, 117

\bibitem[{Burgess {et~al.}(2005)Burgess, Easther, Mazumdar, Mota, \&
  Multamaki}]{Burgess:2005sb}
Burgess, C., Easther, R., Mazumdar, A., Mota, D.~F., \& Multamaki, T. 2005,
  JHEP, 0505, 067

\bibitem[{Byrnes \& Wands(2006)}]{Byrnes:2006fr}
Byrnes, C.~T. \& Wands, D. 2006, Phys. Rev., D74, 043529

\bibitem[{Casadio {et~al.}(2006)Casadio, Finelli, Kamenshchik, Luzzi, \&
  Venturi}]{Casadio:2006wb}
Casadio, R., Finelli, F., Kamenshchik, A., Luzzi, M., \& Venturi, G. 2006,
  JCAP, 0604, 011

\bibitem[{{Chen}(2005{\natexlab{a}})}]{2005JHEP...08..045C}
{Chen}, X. 2005{\natexlab{a}}, JHEP, 8, 45

\bibitem[{{Chen}(2005{\natexlab{b}})}]{2005PhRvD..71f3506C}
{Chen}, X. 2005{\natexlab{b}}, \prd, 71, 063506

\bibitem[{{Chen}(2005{\natexlab{c}})}]{2005PhRvD..72l3518C}
{Chen}, X. 2005{\natexlab{c}}, \prd, 72, 123518

\bibitem[{{Chen}(2010)}]{2010AdAst2010E..72C}
{Chen}, X. 2010, Adv. in Astr., 2010

\bibitem[{{Chen} {et~al.}(2007){Chen}, {Huang}, {Kachru}, \&
  {Shiu}}]{2007JCAP...01..002C}
{Chen}, X., {Huang}, M.-x., {Kachru}, S., \& {Shiu}, G. 2007, JCAP, 1, 2

\bibitem[{Cheung {et~al.}(2008)Cheung, Creminelli, Fitzpatrick, Kaplan, \&
  Senatore}]{Cheung:2007st}
Cheung, C., Creminelli, P., Fitzpatrick, A.~L., Kaplan, J., \& Senatore, L.
  2008, JHEP, 0803, 014

\bibitem[{Chung {et~al.}(2007)Chung, Everett, \& Matchev}]{Chung:2007vz}
Chung, D.~J., Everett, L.~L., \& Matchev, K.~T. 2007, Phys. Rev., D76, 103530

\bibitem[{Chung {et~al.}(2000)Chung, Kolb, Riotto, \& Tkachev}]{Chung:1999ve}
Chung, D.~J., Kolb, E.~W., Riotto, A., \& Tkachev, I.~I. 2000, Phys. Rev., D62,
  043508

\bibitem[{Coleman \& De~Luccia(1980)}]{Coleman:1980aw}
Coleman, S.~R. \& De~Luccia, F. 1980, Phys. Rev., D21, 3305

\bibitem[{Conley {et~al.}(2011)Conley, Guy, Sullivan, Regnault, Astier,
  {et~al.}}]{Conley:2011ku}
Conley, A., Guy, J., Sullivan, M., {et~al.} 2011, \apjs, 192, 1

\bibitem[{Contaldi {et~al.}(2003)Contaldi, Peloso, Kofman, \&
  Linde}]{Contaldi:2003zv}
Contaldi, C.~R., Peloso, M., Kofman, L., \& Linde, A.~D. 2003, JCAP, 0307, 002

\bibitem[{Cort\^{e}s \& Liddle(2009)}]{Cortes:2009ej}
Cort\^{e}s, M. \& Liddle, A.~R. 2009, Phys. Rev., D80, 083524

\bibitem[{Cort\^{e}s {et~al.}(2007)Cort\^{e}s, Liddle, \&
  Mukherjee}]{Cortes:2007ak}
Cort\^{e}s, M., Liddle, A.~R., \& Mukherjee, P. 2007, Phys. Rev., D75, 083520

\bibitem[{Covi {et~al.}(2006)Covi, Hamann, Melchiorri, Slosar, \&
  Sorbera}]{Covi:2006ci}
Covi, L., Hamann, J., Melchiorri, A., Slosar, A., \& Sorbera, I. 2006, Phys.
  Rev., D74, 083509

\bibitem[{Cox(1946)}]{Cox:1946}
Cox, R.~T. 1946, Am. J. Phys, 14, 1

\bibitem[{Danielsson(2002)}]{Danielsson:2002kx}
Danielsson, U.~H. 2002, Phys. Rev., D66, 023511

\bibitem[{Das {et~al.}(2013)Das, Louis, Nolta, Addison, Battistelli,
  {et~al.}}]{Das:2013zf}
Das, S., Louis, T., Nolta, M.~R., {et~al.} 2013, JCAP, submitted,
  \eprint{1301.1037}

\bibitem[{de~Carlos {et~al.}(1993)de~Carlos, Casas, Quevedo, \&
  Roulet}]{deCarlos:1993jw}
de~Carlos, B., Casas, J.~A., Quevedo, F., \& Roulet, E. 1993, Phys. Lett.,
  B318, 447

\bibitem[{Dodelson {et~al.}(1997)Dodelson, Kinney, \& Kolb}]{Dodelson:1997hr}
Dodelson, S., Kinney, W.~H., \& Kolb, E.~W. 1997, Phys. Rev., D56, 3207

\bibitem[{Dunkley {et~al.}(2005)Dunkley, Bucher, Ferreira, Moodley, \&
  Skordis}]{Dunkley:2005va}
Dunkley, J., Bucher, M., Ferreira, P., Moodley, K., \& Skordis, C. 2005, Phys.
  Rev. Lett., 95, 261303

\bibitem[{Dunkley {et~al.}(2013)Dunkley, Calabrese, Sievers, Addison,
  Battaglia, {et~al.}}]{Dunkley:2013vu}
Dunkley, J., Calabrese, E., Sievers, J., {et~al.} 2013, JCAP, 7, 25

\bibitem[{Dunkley {et~al.}(2011)Dunkley, Hlozek, Sievers, Acquaviva, Ade,
  {et~al.}}]{Dunkley:2010ge}
Dunkley, J., Hlozek, R., Sievers, J., {et~al.} 2011, \apj, 739, 52

\bibitem[{{Dunkley} {et~al.}(2009){Dunkley}, {Komatsu}, {Nolta}, {Spergel},
  {Larson}, {Hinshaw}, {Page}, {Bennett}, {Gold}, {Jarosik}, {Weiland},
  {Halpern}, {Hill}, {Kogut}, {Limon}, {Meyer}, {Tucker}, {Wollack}, \&
  {Wright}}]{dunkley2009}
{Dunkley}, J., {Komatsu}, E., {Nolta}, M.~R., {et~al.} 2009, \apjs, 180, 306

\bibitem[{Dvali {et~al.}(1994)Dvali, Shafi, \& Schaefer}]{Dvali:1994ms}
Dvali, G.~R., Shafi, Q., \& Schaefer, R.~K. 1994, Phys. Rev. Lett., 73, 1886

\bibitem[{Easther {et~al.}(2011)Easther, Flauger, \& Gilmore}]{Easther:2010mr}
Easther, R., Flauger, R., \& Gilmore, J.~B. 2011, JCAP, 1104, 027

\bibitem[{Easther {et~al.}(2001)Easther, Greene, Kinney, \&
  Shiu}]{Easther:2001fi}
Easther, R., Greene, B.~R., Kinney, W.~H., \& Shiu, G. 2001, Phys. Rev., D64,
  103502

\bibitem[{Easther \& Peiris(2006)}]{Easther:2006tv}
Easther, R. \& Peiris, H. 2006, JCAP, 0609, 010

\bibitem[{{Easther} \& {Peiris}(2012)}]{2012PhRvD..85j3533E}
{Easther}, R. \& {Peiris}, H.~V. 2012, \prd, 85, 103533

\bibitem[{Efstathiou \& Bond(1986)}]{Efstathiou:1986}
Efstathiou, G. \& Bond, J.~R. 1986, MNRAS, 218, 103

\bibitem[{Efstathiou \& Bond(1987)}]{Efstathiou:1987}
Efstathiou, G. \& Bond, J.~R. 1987, MNRAS, 227, 33P

\bibitem[{Elgar{\o}y {et~al.}(2003)Elgar{\o}y, Hannestad, \&
  Haugb{\o}lle}]{Elgaroy:2003hp}
Elgar{\o}y, {\O}., Hannestad, S., \& Haugb{\o}lle, T. 2003, JCAP, 0309, 008

\bibitem[{Enqvist \& Sloth(2002)}]{Enqvist:2001zp}
Enqvist, K. \& Sloth, M.~S. 2002, Nucl. Phys., B626, 395

\bibitem[{Eriksen {et~al.}(2007)Eriksen, Huey, Saha, {et~al.}}]{Eriksen:2006xr}
Eriksen, H.~K., Huey, G., Saha, R., {et~al.} 2007, \apj, 656, 641

\bibitem[{Fabbri \& Pollock(1983)}]{Fabbri:1983us}
Fabbri, R. \& Pollock, M. 1983, Phys. Lett., B125, 445

\bibitem[{Fakir \& Unruh(1990)}]{Fakir:1990eg}
Fakir, R. \& Unruh, W. 1990, Phys. Rev., D41, 1783

\bibitem[{Feroz \& Hobson(2008)}]{Feroz:2007kg}
Feroz, F. \& Hobson, M. 2008, MNRAS, 384, 449

\bibitem[{Feroz {et~al.}(2009)Feroz, Hobson, \& Bridges}]{Feroz:2008xx}
Feroz, F., Hobson, M., \& Bridges, M. 2009, MNRAS, 398, 1601

\bibitem[{{Feroz} {et~al.}(2013){Feroz}, {Hobson}, {Cameron}, \&
  {Pettitt}}]{2013arXiv1306.2144F}
{Feroz}, F., {Hobson}, M.~P., {Cameron}, E., \& {Pettitt}, A.~N. 2013,
  \eprint{1306.2144}

\bibitem[{Finelli {et~al.}(2010)Finelli, Hamann, Leach, \&
  Lesgourgues}]{Finelli:2009bs}
Finelli, F., Hamann, J., Leach, S.~M., \& Lesgourgues, J. 2010, JCAP, 1004, 011

\bibitem[{Fixsen(2009)}]{Fixsen:2009ug}
Fixsen, D. 2009, \apj, 707, 916

\bibitem[{Flauger {et~al.}(2010)Flauger, McAllister, Pajer, Westphal, \&
  Xu}]{Flauger:2009ab}
Flauger, R., McAllister, L., Pajer, E., Westphal, A., \& Xu, G. 2010, JCAP,
  1006, 009

\bibitem[{{Freese} {et~al.}(1990){Freese}, {Frieman}, \&
  {Olinto}}]{1990PhRvL..65.3233F}
{Freese}, K., {Frieman}, J.~A., \& {Olinto}, A.~V. 1990, Phys. Rev. Lett., 65,
  3233

\bibitem[{Freivogel {et~al.}(2006)Freivogel, Kleban, Rodr\'iguez~Mart\'inez, \&
  Susskind}]{Freivogel:2005vv}
Freivogel, B., Kleban, M., Rodr\'iguez~Mart\'inez, M., \& Susskind, L. 2006,
  JHEP, 0603, 039

\bibitem[{Garc\'ia-Bellido \& Wands(1996)}]{GarciaBellido:1995qq}
Garc\'ia-Bellido, J. \& Wands, D. 1996, Phys. Rev., D53, 5437

\bibitem[{Garriga {et~al.}(1998)Garriga, Montes, Sasaki, \&
  Tanaka}]{Garriga:1997wz}
Garriga, J., Montes, X., Sasaki, M., \& Tanaka, T. 1998, Nucl. Phys., B513, 343

\bibitem[{Garriga {et~al.}(1999)Garriga, Montes, Sasaki, \&
  Tanaka}]{Garriga:1998he}
Garriga, J., Montes, X., Sasaki, M., \& Tanaka, T. 1999, Nucl. Phys., B551, 317

\bibitem[{{Garriga} \& {Mukhanov}(1999)}]{1999PhLB..458..219G}
{Garriga}, J. \& {Mukhanov}, V.~F. 1999, Phys. Lett. B, 458, 219

\bibitem[{Gauthier \& Bucher(2012)}]{Gauthier:2012aq}
Gauthier, C. \& Bucher, M. 2012, JCAP, 1210, 050

\bibitem[{Gong \& Stewart(2001)}]{Gong:2001he}
Gong, J.-O. \& Stewart, E.~D. 2001, Phys. Lett., B510, 1

\bibitem[{Gordon \& Lewis(2003)}]{Gordon:2002gv}
Gordon, C. \& Lewis, A. 2003, Phys. Rev., D67, 123513

\bibitem[{Gordon \& Malik(2004)}]{Gordon:2003hw}
Gordon, C. \& Malik, K.~A. 2004, Phys. Rev., D69, 063508

\bibitem[{Gordon {et~al.}(2001)Gordon, Wands, Bassett, \&
  Maartens}]{Gordon:2000hv}
Gordon, C., Wands, D., Bassett, B.~A., \& Maartens, R. 2001, Phys. Rev., D63,
  023506

\bibitem[{Gott(1982)}]{Gott:1982zf}
Gott, J. 1982, Nature, 295, 304

\bibitem[{Gott \& Statler(1984)}]{Gott:1984ps}
Gott, J. \& Statler, T. 1984, Phys. Lett., B136, 157

\bibitem[{Gratton {et~al.}(2000)Gratton, Hertog, \& Turok}]{Gratton:1999hv}
Gratton, S., Hertog, T., \& Turok, N. 2000, Phys. Rev., D62, 063501

\bibitem[{Gratton {et~al.}(2002)Gratton, Lewis, \& Turok}]{Gratton:2001gw}
Gratton, S., Lewis, A., \& Turok, N. 2002, Phys. Rev., D65, 043513

\bibitem[{Gratton \& Turok(1999)}]{Gratton:1999ya}
Gratton, S. \& Turok, N. 1999, Phys. Rev., D60, 123507

\bibitem[{Grishchuk(1975)}]{Grishchuk:1974ny}
Grishchuk, L. 1975, Sov. Phys. JETP, 40, 409

\bibitem[{Groot~Nibbelink \& van Tent(2000)}]{GrootNibbelink:2000vx}
Groot~Nibbelink, S. \& van Tent, B. 2000, \eprint{hep-ph/0011325}

\bibitem[{Groot~Nibbelink \& van Tent(2002)}]{GrootNibbelink:2001qt}
Groot~Nibbelink, S. \& van Tent, B. 2002, Class. Quant. Grav., 19, 613

\bibitem[{Guth(1981)}]{Guth:1980zm}
Guth, A.~H. 1981, Phys. Rev., D23, 347

\bibitem[{Guth \& Nomura(2012)}]{Guth:2012ww}
Guth, A.~H. \& Nomura, Y. 2012, Phys. Rev., D86, 023534

\bibitem[{Guth \& Pi(1982)}]{Guth:1982ec}
Guth, A.~H. \& Pi, S. 1982, Phys. Rev. Lett., 49, 1110

\bibitem[{Habib {et~al.}(2002)Habib, Heitmann, Jungman, \&
  Molina-Paris}]{Habib:2002yi}
Habib, S., Heitmann, K., Jungman, G., \& Molina-Paris, C. 2002, Phys. Rev.
  Lett., 89, 281301

\bibitem[{Hamann {et~al.}(2009)Hamann, Hannestad, Raffelt, \&
  Wong}]{Hamann:2009yf}
Hamann, J., Hannestad, S., Raffelt, G.~G., \& Wong, Y. Y.~Y. 2009, JCAP, 0906,
  022

\bibitem[{Hamann {et~al.}(2008{\natexlab{a}})Hamann, Hannestad, Sloth, \&
  Wong}]{Hamann:2008yx}
Hamann, J., Hannestad, S., Sloth, M.~S., \& Wong, Y. Y.~Y. 2008{\natexlab{a}},
  JCAP, 0809, 015

\bibitem[{Hamann {et~al.}(2008{\natexlab{b}})Hamann, Lesgourgues, \&
  Mangano}]{Hamann:2007sb}
Hamann, J., Lesgourgues, J., \& Mangano, G. 2008{\natexlab{b}}, JCAP, 0803, 004

\bibitem[{Hamann {et~al.}(2008{\natexlab{c}})Hamann, Lesgourgues, \&
  Valkenburg}]{Hamann:2008pb}
Hamann, J., Lesgourgues, J., \& Valkenburg, W. 2008{\natexlab{c}}, JCAP, 0804,
  016

\bibitem[{Hamann {et~al.}(2010)Hamann, Shafieloo, \& Souradeep}]{Hamann:2009bz}
Hamann, J., Shafieloo, A., \& Souradeep, T. 2010, JCAP, 1004, 010

\bibitem[{Hamimeche \& Lewis(2008)}]{Hamimeche:2008ai}
Hamimeche, S. \& Lewis, A. 2008, Phys. Rev., D77, 103013

\bibitem[{Hannestad(2004)}]{Hannestad:2003zs}
Hannestad, S. 2004, JCAP, 0404, 002

\bibitem[{Harrison(1970)}]{Harrison:1969fb}
Harrison, E.~R. 1970, Phys. Rev., D1, 2726

\bibitem[{Hawking(1982)}]{Hawking:1982cz}
Hawking, S. 1982, Phys. Lett., B115, 295

\bibitem[{Hawking \& Turok(1998)}]{Hawking:1998bn}
Hawking, S. \& Turok, N. 1998, Phys. Lett., B425, 25

\bibitem[{Hertog \& Turok(2000)}]{Hertog:1999kg}
Hertog, T. \& Turok, N. 2000, Phys. Rev., D62, 083514

\bibitem[{{Hinshaw} {et~al.}(2013){Hinshaw}, {Larson}, {Komatsu}, {Spergel},
  {Bennett}, {Dunkley}, {Nolta}, {Halpern}, {Hill}, {Odegard}, {Page}, {Smith},
  {Weiland}, {Gold}, {Jarosik}, {Kogut}, {Limon}, {Meyer}, {Tucker}, {Wollack},
  \& {Wright}}]{Hinshaw:2012fq}
{Hinshaw}, G., {Larson}, D., {Komatsu}, E., {et~al.} 2013, ApJS, 208, 19

\bibitem[{Hou {et~al.}(2013)Hou, Keisler, Knox, Millea, \&
  Reichardt}]{Hou:2011ec}
Hou, Z., Keisler, R., Knox, L., Millea, M., \& Reichardt, C. 2013, Phys. Rev.,
  D87, 083008

\bibitem[{Hou {et~al.}(2012)Hou, Reichardt, Story, Follin, Keisler,
  {et~al.}}]{Hou:2012xq}
Hou, Z., Reichardt, C., Story, K., {et~al.} 2012, \eprint{1212.6267}

\bibitem[{Hunt \& Sarkar(2004)}]{Hunt:2004vt}
Hunt, P. \& Sarkar, S. 2004, Phys. Rev., D70, 103518

\bibitem[{Ichikawa \& Takahashi(2006)}]{Ichikawa:2006dt}
Ichikawa, K. \& Takahashi, T. 2006, Phys. Rev., D73, 063528

\bibitem[{Ichiki \& Nagata(2009)}]{Ichiki:2009zz}
Ichiki, K. \& Nagata, R. 2009, Phys. Rev., D80, 083002

\bibitem[{{Jaynes} \& {Bretthorst}(2003)}]{Jaynes2003}
{Jaynes}, E.~T. \& {Bretthorst}, G.~L. 2003, {Probability Theory} (Cambridge,
  UK: Probability Theory, by E.~T.~Jaynes and Edited by G.~Larry Bretthorst,
  Cambridge University Press)

\bibitem[{Jeffreys(1998)}]{Jeffreys1998BK}
Jeffreys, H. 1998, Theory of Probability, 3rd edn. (Oxford University Press)

\bibitem[{Kazanas(1980)}]{Kazanas:1980tx}
Kazanas, D. 1980, \apj, 241, L59

\bibitem[{Keskitalo {et~al.}(2007)Keskitalo, Kurki-Suonio, Muhonen, \&
  Valiviita}]{Keskitalo:2006qv}
Keskitalo, R., Kurki-Suonio, H., Muhonen, V., \& Valiviita, J. 2007, JCAP,
  0709, 008

\bibitem[{Kim \& Carosi(2010)}]{Kim:2008hd}
Kim, J.~E. \& Carosi, G. 2010, Rev. Mod. Phys., 82, 557

\bibitem[{Kinney(2002)}]{Kinney:2002qn}
Kinney, W.~H. 2002, Phys. Rev., D66, 083508

\bibitem[{Kinney {et~al.}(2006)Kinney, Kolb, Melchiorri, \&
  Riotto}]{Kinney:2006qm}
Kinney, W.~H., Kolb, E.~W., Melchiorri, A., \& Riotto, A. 2006, Phys. Rev.,
  D74, 023502

\bibitem[{{Kinney} \& {Riotto}(2006)}]{2006JCAP...03..011K}
{Kinney}, W.~H. \& {Riotto}, A. 2006, JCAP, 3, 11

\bibitem[{Kleban \& Schillo(2012)}]{Kleban:2012ph}
Kleban, M. \& Schillo, M. 2012, JCAP, 1206, 029

\bibitem[{Knox \& Turner(1994)}]{Knox:1994qj}
Knox, L. \& Turner, M.~S. 1994, Phys. Rev. Lett., 73, 3347

\bibitem[{Kobayashi \& Takahashi(2011)}]{Kobayashi:2010pz}
Kobayashi, T. \& Takahashi, F. 2011, JCAP, 1101, 026

\bibitem[{Kobayashi {et~al.}(2010)Kobayashi, Yamaguchi, \&
  Yokoyama}]{Kobayashi:2010cm}
Kobayashi, T., Yamaguchi, M., \& Yokoyama, J. 2010, Phys. Rev. Lett., 105,
  231302

\bibitem[{Kofman(1986)}]{Kofman:1985zx}
Kofman, L. 1986, Phys. Lett., B173, 400

\bibitem[{Kofman {et~al.}(1994)Kofman, Linde, \& Starobinsky}]{Kofman:1994rk}
Kofman, L., Linde, A.~D., \& Starobinsky, A.~A. 1994, Phys. Rev. Lett., 73,
  3195

\bibitem[{Kofman {et~al.}(1997)Kofman, Linde, \& Starobinsky}]{Kofman:1997yn}
Kofman, L., Linde, A.~D., \& Starobinsky, A.~A. 1997, Phys. Rev., D56, 3258

\bibitem[{{Komatsu} {et~al.}(2009){Komatsu}, {Dunkley}, {Nolta}, {Bennett},
  {Gold}, {Hinshaw}, {Jarosik}, {Larson}, {Limon}, {Page}, {Spergel},
  {Halpern}, {Hill}, {Kogut}, {Meyer}, {Tucker}, {Weiland}, {Wollack}, \&
  {Wright}}]{komatsu2009}
{Komatsu}, E., {Dunkley}, J., {Nolta}, M.~R., {et~al.} 2009, \apjs, 180, 330

\bibitem[{{Komatsu} {et~al.}(2011){Komatsu}, {Smith}, {Dunkley}, {Bennett},
  {Gold}, {Hinshaw}, {Jarosik}, {Larson}, {Nolta}, {Page}, {Spergel},
  {Halpern}, {Hill}, {Kogut}, {Limon}, {Meyer}, {Odegard}, {Tucker}, {Weiland},
  {Wollack}, \& {Wright}}]{komatsu2010}
{Komatsu}, E., {Smith}, K.~M., {Dunkley}, J., {et~al.} 2011, \apjs, 192, 18

\bibitem[{Kosowsky \& Turner(1995)}]{Kosowsky:1995aa}
Kosowsky, A. \& Turner, M.~S. 1995, Phys. Rev., D52, 1739

\bibitem[{Kurki-Suonio {et~al.}(2005)Kurki-Suonio, Muhonen, \&
  Valiviita}]{KurkiSuonio:2004mn}
Kurki-Suonio, H., Muhonen, V., \& Valiviita, J. 2005, Phys. Rev., D71, 063005

\bibitem[{Langlois(1999)}]{Langlois:1999dw}
Langlois, D. 1999, Phys. Rev., D59, 123512

\bibitem[{Langlois \& Riazuelo(2000)}]{Langlois:2000ar}
Langlois, D. \& Riazuelo, A. 2000, Phys. Rev., D62, 043504

\bibitem[{{Larson} {et~al.}(2011){Larson}, {Dunkley}, {Hinshaw}, {Komatsu},
  {Nolta}, {Bennett}, {Gold}, {Halpern}, {Hill}, {Jarosik}, {Kogut}, {Limon},
  {Meyer}, {Odegard}, {Page}, {Smith}, {Spergel}, {Tucker}, {Weiland},
  {Wollack}, \& {Wright}}]{larson2010}
{Larson}, D., {Dunkley}, J., {Hinshaw}, G., {et~al.} 2011, \apjs, 192, 16

\bibitem[{Leach {et~al.}(2002)Leach, Liddle, Martin, \& Schwarz}]{Leach:2002ar}
Leach, S.~M., Liddle, A.~R., Martin, J., \& Schwarz, D.~J. 2002, Phys. Rev.,
  D66, 023515

\bibitem[{Lesgourgues(2011)}]{Lesgourgues:2011re}
Lesgourgues, J. 2011, astro-ph/1104.2932

\bibitem[{Lesgourgues {et~al.}(2008)Lesgourgues, Starobinsky, \&
  Valkenburg}]{Lesgourgues:2007aa}
Lesgourgues, J., Starobinsky, A.~A., \& Valkenburg, W. 2008, JCAP, 0801, 010

\bibitem[{Lesgourgues \& Valkenburg(2007)}]{Lesgourgues:2007gp}
Lesgourgues, J. \& Valkenburg, W. 2007, Phys. Rev., D75, 123519

\bibitem[{Lewis(2011)}]{camb_notes}
Lewis, A. 2011, \url{http://cosmologist.info/notes/CAMB.pdf}

\bibitem[{Lewis \& Bridle(2002)}]{Lewis:2002ah}
Lewis, A. \& Bridle, S. 2002, Phys. Rev., D66, 103511

\bibitem[{Liddle(2007)}]{Liddle:2007fy}
Liddle, A.~R. 2007, MNRAS, 377, L74

\bibitem[{Liddle \& Leach(2003)}]{Liddle:2003as}
Liddle, A.~R. \& Leach, S.~M. 2003, Phys. Rev., D68, 103503

\bibitem[{Liddle \& Lyth(1993)}]{Liddle:1993fq}
Liddle, A.~R. \& Lyth, D.~H. 1993, Phys. Rept., 231, 1

\bibitem[{{Lidsey} \& {Huston}(2007)}]{2007JCAP...07..002L}
{Lidsey}, J.~E. \& {Huston}, I. 2007, JCAP, 7, 2

\bibitem[{Lifshitz(1946)}]{Lifshitz:1945du}
Lifshitz, E. 1946, J. Phys. (USSR), 10, 116

\bibitem[{Lifshitz \& Khalatnikov(1963)}]{Lifshitz:1963ps}
Lifshitz, E. \& Khalatnikov, I. 1963, Adv. Phys., 12, 185

\bibitem[{Linde {et~al.}(2011)Linde, Noorbala, \& Westphal}]{Linde:2011nh}
Linde, A., Noorbala, M., \& Westphal, A. 2011, JCAP, 1103, 013

\bibitem[{Linde(1982)}]{Linde:1981mu}
Linde, A.~D. 1982, Phys. Lett., B108, 389

\bibitem[{Linde(1983)}]{Linde:1983gd}
Linde, A.~D. 1983, Phys. Lett., B129, 177

\bibitem[{Linde(1984)}]{Linde:1984ti}
Linde, A.~D. 1984, JETP Lett., 40, 1333

\bibitem[{Linde(1985)}]{Linde:1985yf}
Linde, A.~D. 1985, Phys. Lett., B158, 375

\bibitem[{Linde(1990)}]{Linde1990Bk}
Linde, A.~D. 1990, {Particle physics and inflationary cosmology} (Harwood)

\bibitem[{Linde(1991)}]{Linde:1991km}
Linde, A.~D. 1991, Phys. Lett., B259, 38

\bibitem[{Linde(2003)}]{Linde:2003hc}
Linde, A.~D. 2003, JCAP, 0305, 002

\bibitem[{Linde \& Lyth(1990)}]{Linde:1990yj}
Linde, A.~D. \& Lyth, D.~H. 1990, Phys. Lett., B246, 353

\bibitem[{Linde \& Mukhanov(1997)}]{Linde:1996gt}
Linde, A.~D. \& Mukhanov, V.~F. 1997, Phys. Rev., D56, 535

\bibitem[{{Lorenz} {et~al.}(2008){Lorenz}, {Martin}, \&
  {Ringeval}}]{2008PhRvD..78h3513L}
{Lorenz}, L., {Martin}, J., \& {Ringeval}, C. 2008, \prd, 78, 083513

\bibitem[{Lucchin \& Matarrese(1985)}]{Lucchin:1984yf}
Lucchin, F. \& Matarrese, S. 1985, Phys. Rev., D32, 1316

\bibitem[{Lucchin {et~al.}(1986)Lucchin, Matarrese, \&
  Pollock}]{Lucchin:1985ip}
Lucchin, F., Matarrese, S., \& Pollock, M. 1986, Phys. Lett., B167, 163

\bibitem[{{Lucy}(1974)}]{1974AJ79745L}
{Lucy}, L.~B. 1974, Ast. J., 79, 745

\bibitem[{Lyth(1992)}]{Lyth:1991ub}
Lyth, D. 1992, Phys. Rev., D45, 3394

\bibitem[{Lyth(1990)}]{Lyth:1989pb}
Lyth, D.~H. 1990, Phys. Lett., B236, 408

\bibitem[{Lyth(1997)}]{Lyth:1996im}
Lyth, D.~H. 1997, Phys. Rev. Lett., 78, 1861

\bibitem[{Lyth \& Riotto(1999)}]{Lyth:1998xn}
Lyth, D.~H. \& Riotto, A. 1999, Phys. Rept., 314, 1

\bibitem[{Lyth \& Stewart(1990)}]{Lyth:1990dh}
Lyth, D.~H. \& Stewart, E.~D. 1990, Phys. Lett., B252, 336

\bibitem[{Lyth \& Stewart(1996)}]{Lyth:1995ka}
Lyth, D.~H. \& Stewart, E.~D. 1996, Phys. Rev., D53, 1784

\bibitem[{Lyth {et~al.}(2003)Lyth, Ungarelli, \& Wands}]{Lyth:2002my}
Lyth, D.~H., Ungarelli, C., \& Wands, D. 2003, Phys. Rev., D67, 023503

\bibitem[{Lyth \& Wands(2002)}]{Lyth:2001nq}
Lyth, D.~H. \& Wands, D. 2002, Phys. Lett., B524, 5

\bibitem[{Ma \& Bertschinger(1995)}]{Ma:1995ey}
Ma, C.-P. \& Bertschinger, E. 1995, \apj, 455, 7

\bibitem[{{Maldacena}(2003)}]{2003JHEP...05..013M}
{Maldacena}, J. 2003, JHEP, 5, 13

\bibitem[{Martin \& Brandenberger(2003)}]{Martin:2003kp}
Martin, J. \& Brandenberger, R. 2003, Phys. Rev., D68, 063513

\bibitem[{Martin \& Ringeval(2004)}]{Martin:2003sg}
Martin, J. \& Ringeval, C. 2004, Phys. Rev., D69, 083515

\bibitem[{Martin \& Ringeval(2010)}]{Martin:2010kz}
Martin, J. \& Ringeval, C. 2010, Phys. Rev., D82, 023511

\bibitem[{Martin {et~al.}(2011)Martin, Ringeval, \& Trotta}]{Martin:2010hh}
Martin, J., Ringeval, C., \& Trotta, R. 2011, Phys. Rev., D83, 063524

\bibitem[{Martin \& Schwarz(2003)}]{Martin:2002vn}
Martin, J. \& Schwarz, D.~J. 2003, Phys. Rev., D67, 083512

\bibitem[{McAllister {et~al.}(2010)McAllister, Silverstein, \&
  Westphal}]{McAllister:2008hb}
McAllister, L., Silverstein, E., \& Westphal, A. 2010, Phys. Rev., D82, 046003

\bibitem[{{Meerburg} {et~al.}(2012){Meerburg}, {Wijers}, \& {van der
  Schaar}}]{Meerburg:2011gd}
{Meerburg}, P.~D., {Wijers}, R.~A.~M.~J., \& {van der Schaar}, J.~P. 2012,
  \mnras, 421, 369

\bibitem[{{Mehta} {et~al.}(2012){Mehta}, {Cuesta}, {Xu}, {Eisenstein}, \&
  {Padmanabhan}}]{Mehta:2012hh}
{Mehta}, K.~T., {Cuesta}, A.~J., {Xu}, X., {Eisenstein}, D.~J., \&
  {Padmanabhan}, N. 2012, \mnras, 427, 2168

\bibitem[{Mollerach(1990)}]{Mollerach:1989hu}
Mollerach, S. 1990, Phys. Rev., D42, 313

\bibitem[{Moodley {et~al.}(2004)Moodley, Bucher, Dunkley, Ferreira, \&
  Skordis}]{Moodley:2004nz}
Moodley, K., Bucher, M., Dunkley, J., Ferreira, P., \& Skordis, C. 2004, Phys.
  Rev., D70, 103520

\bibitem[{Moroi \& Takahashi(2001)}]{Moroi:2001ct}
Moroi, T. \& Takahashi, T. 2001, Phys. Lett., B522, 215

\bibitem[{{Mortonson} {et~al.}(2009){Mortonson}, {Dvorkin}, {Peiris}, \&
  {Hu}}]{2009PhRvD..79j3519M}
{Mortonson}, M.~J., {Dvorkin}, C., {Peiris}, H.~V., \& {Hu}, W. 2009, \prd, 79,
  103519

\bibitem[{Mortonson \& Hu(2008{\natexlab{a}})}]{Mortonson:2007hq}
Mortonson, M.~J. \& Hu, W. 2008{\natexlab{a}}, \apj, 672, 737

\bibitem[{Mortonson \& Hu(2008{\natexlab{b}})}]{Mortonson:2008rx}
Mortonson, M.~J. \& Hu, W. 2008{\natexlab{b}}, \apj, 686, L53

\bibitem[{Mortonson {et~al.}(2011)Mortonson, Peiris, \&
  Easther}]{Mortonson:2010er}
Mortonson, M.~J., Peiris, H.~V., \& Easther, R. 2011, Phys. Rev., D83, 043505

\bibitem[{Mukhanov(1985)}]{Mukhanov:1985rz}
Mukhanov, V.~F. 1985, JETP Lett., 41, 493

\bibitem[{Mukhanov(1988)}]{Mukhanov:1988jd}
Mukhanov, V.~F. 1988, Sov. Phys. JETP, 67, 1297

\bibitem[{Mukhanov \& Chibisov(1981)}]{Mukhanov:1981xt}
Mukhanov, V.~F. \& Chibisov, G. 1981, JETP Lett., 33, 532

\bibitem[{Mukhanov \& Chibisov(1982)}]{Mukhanov:1982nu}
Mukhanov, V.~F. \& Chibisov, G. 1982, Sov. Phys. JETP, 56, 258

\bibitem[{Mukhanov {et~al.}(1992)Mukhanov, Feldman, \&
  Brandenberger}]{Mukhanov:1990me}
Mukhanov, V.~F., Feldman, H., \& Brandenberger, R.~H. 1992, Phys. Rept., 215,
  203

\bibitem[{Muslimov(1990)}]{Muslimov:1990be}
Muslimov, A. 1990, Class. Quant. Grav., 7, 231

\bibitem[{Nagata \& Yokoyama(2008)}]{Nagata:2008tk}
Nagata, R. \& Yokoyama, J. 2008, Phys. Rev., D78, 123002

\bibitem[{Nagata \& Yokoyama(2009)}]{Nagata:2008zj}
Nagata, R. \& Yokoyama, J. 2009, Phys. Rev., D79, 043010

\bibitem[{Norena {et~al.}(2012)Norena, Wagner, Verde, Peiris, \&
  Easther}]{Norena:2012rs}
Norena, J., Wagner, C., Verde, L., Peiris, H.~V., \& Easther, R. 2012, Phys.
  Rev., D86, 023505

\bibitem[{Okada {et~al.}(2010)Okada, Rehman, \& Shafi}]{Okada:2010jf}
Okada, N., Rehman, M.~U., \& Shafi, Q. 2010, Phys. Rev., D82, 043502

\bibitem[{Okamoto \& Hu(2003)}]{Okamoto:2003zw}
Okamoto, T. \& Hu, W. 2003, Phys. Rev., D67, 083002

\bibitem[{Olive(1990)}]{Olive:1989nu}
Olive, K.~A. 1990, Phys. Rept., 190, 307

\bibitem[{{Padmanabhan} {et~al.}(2012){Padmanabhan}, {Xu}, {Eisenstein},
  {Scalzo}, {Cuesta}, {Mehta}, \& {Kazin}}]{Padmanabhan:2012hf}
{Padmanabhan}, N., {Xu}, X., {Eisenstein}, D.~J., {et~al.} 2012, \mnras, 427,
  2132

\bibitem[{Page {et~al.}(2007)Page, Hinshaw, Komatsu, {et~al.}}]{Page:2006hz}
Page, L., Hinshaw, G., Komatsu, E., {et~al.} 2007, \apjs, 170, 335

\bibitem[{Pahud {et~al.}(2007)Pahud, Liddle, Mukherjee, \&
  Parkinson}]{Pahud:2007gi}
Pahud, C., Liddle, A.~R., Mukherjee, P., \& Parkinson, D. 2007, MNRAS, 381, 489

\bibitem[{Pallis(2006)}]{Pallis:2005bb}
Pallis, C. 2006, Nucl. Phys., B751, 129

\bibitem[{Pandolfi {et~al.}(2010)Pandolfi, Giusarma, Kolb, Lattanzi,
  Melchiorri, {et~al.}}]{Pandolfi:2010mv}
Pandolfi, S., Giusarma, E., Kolb, E.~W., {et~al.} 2010, Phys. Rev., D82, 123527

\bibitem[{Peccei(2008)}]{Peccei:2006as}
Peccei, R. 2008, Lect. Notes Phys., 741, 3

\bibitem[{Peebles \& Yu(1970)}]{1970ApJ...162..815P}
Peebles, P.~J.~E. \& Yu, J.~T. 1970, \apj, 162, 815

\bibitem[{{Peiris} {et~al.}(2007){Peiris}, {Baumann}, {Friedman}, \&
  {Cooray}}]{2007PhRvD..76j3517P}
{Peiris}, H., {Baumann}, D., {Friedman}, B., \& {Cooray}, A. 2007, \prd, 76,
  103517

\bibitem[{Peiris \& Easther(2006{\natexlab{a}})}]{Peiris:2006ug}
Peiris, H. \& Easther, R. 2006{\natexlab{a}}, JCAP, 0607, 002

\bibitem[{Peiris \& Easther(2006{\natexlab{b}})}]{Peiris:2006sj}
Peiris, H. \& Easther, R. 2006{\natexlab{b}}, JCAP, 0610, 017

\bibitem[{Peiris {et~al.}(2013)Peiris, Easther, \& Flauger}]{Peiris:2013opa}
Peiris, H., Easther, R., \& Flauger, R. 2013, JCAP, 1309, 018

\bibitem[{Peiris \& Easther(2008)}]{Peiris:2008be}
Peiris, H.~V. \& Easther, R. 2008, JCAP, 0807, 024

\bibitem[{{Peiris} {et~al.}(2003){Peiris}, {Komatsu}, {Verde}, {Spergel},
  {Bennett}, {Halpern}, {Hinshaw}, {Jarosik}, {Kogut}, {Limon}, {Meyer},
  {Page}, {Tucker}, {Wollack}, \& {Wright}}]{2003ApJS..148..213P}
{Peiris}, H.~V., {Komatsu}, E., {Verde}, L., {et~al.} 2003, \apjs, 148, 213

\bibitem[{Peiris \& Verde(2010)}]{Peiris:2009wp}
Peiris, H.~V. \& Verde, L. 2010, Phys.Rev., D81, 021302

\bibitem[{{\sorthelp{Planck Collaboration 2013A}}{Planck Collaboration
  I}(2014)}]{planck2013-p01}
{\sorthelp{Planck Collaboration 2013A}}{Planck Collaboration I}. 2014, \aap, in
  press, \eprint{1303.5062}

\bibitem[{{\sorthelp{Planck Collaboration 2013B}}{Planck Collaboration
  II}(2014)}]{planck2013-p02}
{\sorthelp{Planck Collaboration 2013B}}{Planck Collaboration II}. 2014, \aap,
  in press, \eprint{1303.5063}

\bibitem[{{\sorthelp{Planck Collaboration 2013C}}{Planck Collaboration
  III}(2014)}]{planck2013-p02a}
{\sorthelp{Planck Collaboration 2013C}}{Planck Collaboration III}. 2014, \aap,
  in press, \eprint{1303.5064}

\bibitem[{{\sorthelp{Planck Collaboration 2013D}}{Planck Collaboration
  IV}(2014)}]{planck2013-p02d}
{\sorthelp{Planck Collaboration 2013D}}{Planck Collaboration IV}. 2014, \aap,
  in press, \eprint{1303.5065}

\bibitem[{{\sorthelp{Planck Collaboration 2013E}}{Planck Collaboration
  V}(2014)}]{planck2013-p02b}
{\sorthelp{Planck Collaboration 2013E}}{Planck Collaboration V}. 2014, \aap, in
  press, \eprint{1303.5066}

\bibitem[{{\sorthelp{Planck Collaboration 2013F}}{Planck Collaboration
  VI}(2014)}]{planck2013-p03}
{\sorthelp{Planck Collaboration 2013F}}{Planck Collaboration VI}. 2014, \aap,
  in press, \eprint{1303.5067}

\bibitem[{{\sorthelp{Planck Collaboration 2013G}}{Planck Collaboration
  VII}(2014)}]{planck2013-p03c}
{\sorthelp{Planck Collaboration 2013G}}{Planck Collaboration VII}. 2014, \aap,
  in press, \eprint{1303.5068}

\bibitem[{{\sorthelp{Planck Collaboration 2013H}}{Planck Collaboration
  VIII}(2014)}]{planck2013-p03f}
{\sorthelp{Planck Collaboration 2013H}}{Planck Collaboration VIII}. 2014, \aap,
  in press, \eprint{1303.5069}

\bibitem[{{\sorthelp{Planck Collaboration 2013I}}{Planck Collaboration
  IX}(2014)}]{planck2013-p03d}
{\sorthelp{Planck Collaboration 2013I}}{Planck Collaboration IX}. 2014, \aap,
  in press, \eprint{1303.5070}

\bibitem[{{\sorthelp{Planck Collaboration 2013J}}{Planck Collaboration
  X}(2014)}]{planck2013-p03e}
{\sorthelp{Planck Collaboration 2013J}}{Planck Collaboration X}. 2014, \aap, in
  press, \eprint{1303.5071}

\bibitem[{{\sorthelp{Planck Collaboration 2013K}}{Planck Collaboration
  XI}(2014)}]{planck2013-p06b}
{\sorthelp{Planck Collaboration 2013K}}{Planck Collaboration XI}. 2014, \aap,
  in press, \eprint{1312.1300}

\bibitem[{{\sorthelp{Planck Collaboration 2013L}}{Planck Collaboration
  XII}(2014)}]{planck2013-p06}
{\sorthelp{Planck Collaboration 2013L}}{Planck Collaboration XII}. 2014, \aap,
  in press, \eprint{1303.5072}

\bibitem[{{\sorthelp{Planck Collaboration 2013M}}{Planck Collaboration
  XIII}(2014)}]{planck2013-p03a}
{\sorthelp{Planck Collaboration 2013M}}{Planck Collaboration XIII}. 2014, \aap,
  in press, \eprint{1303.5073}

\bibitem[{{\sorthelp{Planck Collaboration 2013N}}{Planck Collaboration
  XIV}(2014)}]{planck2013-pip88}
{\sorthelp{Planck Collaboration 2013N}}{Planck Collaboration XIV}. 2014, \aap,
  in press, \eprint{1303.5074}

\bibitem[{{\sorthelp{Planck Collaboration 2013O}}{Planck Collaboration
  XV}(2014)}]{planck2013-p08}
{\sorthelp{Planck Collaboration 2013O}}{Planck Collaboration XV}. 2014, \aap,
  in press, \eprint{1303.5075}

\bibitem[{{\sorthelp{Planck Collaboration 2013P}}{Planck Collaboration
  XVI}(2014)}]{planck2013-p11}
{\sorthelp{Planck Collaboration 2013P}}{Planck Collaboration XVI}. 2014, \aap,
  in press, \eprint{1303.5076}

\bibitem[{{\sorthelp{Planck Collaboration 2013Q}}{Planck Collaboration
  XVII}(2014)}]{planck2013-p12}
{\sorthelp{Planck Collaboration 2013Q}}{Planck Collaboration XVII}. 2014, \aap,
  in press, \eprint{1303.5077}

\bibitem[{{\sorthelp{Planck Collaboration 2013R}}{Planck Collaboration
  XVIII}(2014)}]{planck2013-p13}
{\sorthelp{Planck Collaboration 2013R}}{Planck Collaboration XVIII}. 2014,
  \aap, in press, \eprint{1303.5078}

\bibitem[{{\sorthelp{Planck Collaboration 2013S}}{Planck Collaboration
  XIX}(2014)}]{planck2013-p14}
{\sorthelp{Planck Collaboration 2013S}}{Planck Collaboration XIX}. 2014, \aap,
  in press, \eprint{1303.5079}

\bibitem[{{\sorthelp{Planck Collaboration 2013T}}{Planck Collaboration
  XX}(2014)}]{planck2013-p15}
{\sorthelp{Planck Collaboration 2013T}}{Planck Collaboration XX}. 2014, \aap,
  in press, \eprint{1303.5080}

\bibitem[{{\sorthelp{Planck Collaboration 2013U}}{Planck Collaboration
  XXI}(2014)}]{planck2013-p05b}
{\sorthelp{Planck Collaboration 2013U}}{Planck Collaboration XXI}. 2014, \aap,
  in press, \eprint{1303.5081}

\bibitem[{{\sorthelp{Planck Collaboration 2013V}}{Planck Collaboration
  XXII}(2014)}]{planck2013-p17}
{\sorthelp{Planck Collaboration 2013V}}{Planck Collaboration XXII}. 2014, \aap,
  in press, \eprint{1303.5082}

\bibitem[{{\sorthelp{Planck Collaboration 2013W}}{Planck Collaboration
  XXIII}(2014)}]{planck2013-p09}
{\sorthelp{Planck Collaboration 2013W}}{Planck Collaboration XXIII}. 2014,
  \aap, in press, \eprint{1303.5083}

\bibitem[{{\sorthelp{Planck Collaboration 2013X}}{Planck Collaboration
  XXIV}(2014)}]{planck2013-p09a}
{\sorthelp{Planck Collaboration 2013X}}{Planck Collaboration XXIV}. 2014, \aap,
  in press, \eprint{1303.5084}

\bibitem[{{\sorthelp{Planck Collaboration 2013Y}}{Planck Collaboration
  XXV}(2014)}]{planck2013-p20}
{\sorthelp{Planck Collaboration 2013Y}}{Planck Collaboration XXV}. 2014, \aap,
  in press, \eprint{1303.5085}

\bibitem[{{\sorthelp{Planck Collaboration 2013ZA}}{Planck Collaboration
  XXVI}(2014)}]{planck2013-p19}
{\sorthelp{Planck Collaboration 2013ZA}}{Planck Collaboration XXVI}. 2014,
  \aap, in press, \eprint{1303.5086}

\bibitem[{{\sorthelp{Planck Collaboration 2013ZB}}{Planck Collaboration
  XXVII}(2014)}]{planck2013-pipaberration}
{\sorthelp{Planck Collaboration 2013ZB}}{Planck Collaboration XXVII}. 2014,
  \aap, in press, \eprint{1303.5087}

\bibitem[{{\sorthelp{Planck Collaboration 2013ZC}}{Planck Collaboration
  XXVIII}(2014)}]{planck2013-p05}
{\sorthelp{Planck Collaboration 2013ZC}}{Planck Collaboration XXVIII}. 2014,
  \aap, in press, \eprint{1303.5088}

\bibitem[{{\sorthelp{Planck Collaboration 2013ZD}}{Planck Collaboration
  XXIX}(2014)}]{planck2013-p05a}
{\sorthelp{Planck Collaboration 2013ZD}}{Planck Collaboration XXIX}. 2014,
  \aap, in press, \eprint{1303.5089}

\bibitem[{{\sorthelp{Planck Collaboration 2013ZE}}{Planck Collaboration
  XXX}(2014)}]{planck2013-pip56}
{\sorthelp{Planck Collaboration 2013ZE}}{Planck Collaboration XXX}. 2014, \aap,
  in press, \eprint{1309.0382}

\bibitem[{{\sorthelp{Planck Collaboration 2013ZF}}{Planck Collaboration
  XXXI}(2014)}]{planck2013-p01a}
{\sorthelp{Planck Collaboration 2013ZF}}{Planck Collaboration XXXI}. 2014, In
  preparation

\bibitem[{Polarski \& Starobinsky(1994)}]{Polarski:1994rz}
Polarski, D. \& Starobinsky, A.~A. 1994, Phys. Rev., D50, 6123

\bibitem[{Powell \& Kinney(2007)}]{Powell:2007gu}
Powell, B.~A. \& Kinney, W.~H. 2007, JCAP, 0708, 006

\bibitem[{{Powell} {et~al.}(2009){Powell}, {Tzirakis}, \&
  {Kinney}}]{2009JCAP...04..019P}
{Powell}, B.~A., {Tzirakis}, K., \& {Kinney}, W.~H. 2009, JCAP, 4, 19

\bibitem[{Preskill {et~al.}(1983)Preskill, Wise, \& Wilczek}]{Preskill:1982cy}
Preskill, J., Wise, M.~B., \& Wilczek, F. 1983, Phys. Lett., B120, 127

\bibitem[{Raffelt(2008)}]{Raffelt:2006cw}
Raffelt, G.~G. 2008, Lect. Notes Phys., 741, 51

\bibitem[{Ratra \& Peebles(1994)}]{Ratra:1994dm}
Ratra, B. \& Peebles, P. 1994, \apj, 432, L5

\bibitem[{Ratra \& Peebles(1995)}]{Ratra:1994vw}
Ratra, B. \& Peebles, P. 1995, Phys. Rev., D52, 1837

\bibitem[{Reichardt {et~al.}(2012)Reichardt, Shaw, Zahn, Aird, Benson,
  {et~al.}}]{Reichardt:2011yv}
Reichardt, C., Shaw, L., Zahn, O., {et~al.} 2012, \apj, 755, 70

\bibitem[{Richardson(1972)}]{RICHARDSON:72}
Richardson, W.~H. 1972, J. Opt. Soc. Am., 62, 55

\bibitem[{Riess {et~al.}(2011)Riess, Macri, Casertano, Lampeitl, Ferguson,
  {et~al.}}]{Riess:2011yx}
Riess, A.~G., Macri, L., Casertano, S., {et~al.} 2011, \apj, 730, 119

\bibitem[{Rubakov {et~al.}(1982)Rubakov, Sazhin, \& Veryaskin}]{Rubakov:1982df}
Rubakov, V., Sazhin, M., \& Veryaskin, A. 1982, Phys. Lett., B115, 189

\bibitem[{Salopek {et~al.}(1989)Salopek, Bond, \& Bardeen}]{Salopek:1988qh}
Salopek, D., Bond, J., \& Bardeen, J.~M. 1989, Phys. Rev., D40, 1753

\bibitem[{Sasaki(1986)}]{Sasaki:1986hm}
Sasaki, M. 1986, Prog. Theor. Phys., 76, 1036

\bibitem[{Sasaki \& Stewart(1996)}]{Sasaki:1995aw}
Sasaki, M. \& Stewart, E.~D. 1996, Prog. Theor. Phys., 95, 71

\bibitem[{Sasaki {et~al.}(1997)Sasaki, Tanaka, \& Yakushige}]{Sasaki:1997ex}
Sasaki, M., Tanaka, T., \& Yakushige, Y. 1997, Phys. Rev., D56, 616

\bibitem[{Sasaki {et~al.}(2006)Sasaki, Valiviita, \& Wands}]{Sasaki:2006kq}
Sasaki, M., Valiviita, J., \& Wands, D. 2006, Phys. Rev., D74, 103003

\bibitem[{Sato(1981)}]{Sato1981}
Sato, K. 1981, MNRAS, 195, 467

\bibitem[{Savage {et~al.}(2006)Savage, Freese, \& Kinney}]{Savage:2006tr}
Savage, C., Freese, K., \& Kinney, W.~H. 2006, Phys. Rev., D74, 123511

\bibitem[{Schwarz(1978)}]{Schwarz:1978xx}
Schwarz, G. 1978, Annals of Statistics, 6, 461

\bibitem[{Sealfon {et~al.}(2005)Sealfon, Verde, \& Jimenez}]{Sealfon:2005em}
Sealfon, C., Verde, L., \& Jimenez, R. 2005, Phys. Rev., D72, 103520

\bibitem[{Seckel \& Turner(1985)}]{Seckel:1985tj}
Seckel, D. \& Turner, M.~S. 1985, Phys. Rev., D32, 3178

\bibitem[{{Senatore} {et~al.}(2010){Senatore}, {Smith}, \&
  {Zaldarriaga}}]{2010JCAP...01..028S}
{Senatore}, L., {Smith}, K.~M., \& {Zaldarriaga}, M. 2010, JCAP, 1, 28

\bibitem[{Shafieloo \& Souradeep(2004)}]{Shafieloo:2003gf}
Shafieloo, A. \& Souradeep, T. 2004, Phys. Rev., D70, 043523

\bibitem[{Shafieloo \& Souradeep(2008)}]{Shafieloo:2007tk}
Shafieloo, A. \& Souradeep, T. 2008, Phys. Rev., D78, 023511

\bibitem[{Sievers {et~al.}(2013)Sievers, Hlozek, Nolta, Acquaviva, Addison,
  {et~al.}}]{Sievers:2013wk}
Sievers, J.~L., Hlozek, R.~A., Nolta, M.~R., {et~al.} 2013, JCAP, 10, 060

\bibitem[{Sikivie(2008)}]{Sikivie:2006ni}
Sikivie, P. 2008, Lect. Notes Phys., 741, 19

\bibitem[{{Silverstein} \& {Tong}(2004)}]{2004PhRvD..70j3505S}
{Silverstein}, E. \& {Tong}, D. 2004, \prd, 70, 103505

\bibitem[{Silverstein \& Westphal(2008)}]{Silverstein:2008sg}
Silverstein, E. \& Westphal, A. 2008, Phys. Rev., D78, 106003

\bibitem[{Sinha \& Souradeep(2006)}]{Sinha:2005mn}
Sinha, R. \& Souradeep, T. 2006, Phys. Rev., D74, 043518

\bibitem[{{Spergel} {et~al.}(2007){Spergel}, {Bean}, {Dor{\'e}}, {Nolta},
  {Bennett}, {Dunkley}, {Hinshaw}, {Jarosik}, {Komatsu}, {Page}, {Peiris},
  {Verde}, {Halpern}, {Hill}, {Kogut}, {Limon}, {Meyer}, {Odegard}, {Tucker},
  {Weiland}, {Wollack}, \& {Wright}}]{spergel2007}
{Spergel}, D.~N., {Bean}, R., {Dor{\'e}}, O., {et~al.} 2007, \apjs, 170, 377

\bibitem[{{Spergel} {et~al.}(2003){Spergel}, {Verde}, {Peiris}, {Komatsu},
  {Nolta}, {Bennett}, {Halpern}, {Hinshaw}, {Jarosik}, {Kogut}, {Limon},
  {Meyer}, {Page}, {Tucker}, {Weiland}, {Wollack}, \& {Wright}}]{spergel2003}
{Spergel}, D.~N., {Verde}, L., {Peiris}, H.~V., {et~al.} 2003, \apjs, 148, 175

\bibitem[{Spokoiny(1984)}]{Spokoiny:1984bd}
Spokoiny, B. 1984, Phys. Lett., B147, 39

\bibitem[{Spokoiny(1993)}]{Spokoiny:1993kt}
Spokoiny, B. 1993, Phys. Lett., B315, 40

\bibitem[{Starobinsky(1979)}]{Starobinsky:1979ty}
Starobinsky, A.~A. 1979, JETP Lett., 30, 682

\bibitem[{Starobinsky(1980)}]{Starobinsky:1980te}
Starobinsky, A.~A. 1980, Phys. Lett., B91, 99

\bibitem[{Starobinsky(1982)}]{Starobinsky:1982ee}
Starobinsky, A.~A. 1982, Phys. Lett., B117, 175

\bibitem[{Starobinsky(1983)}]{Starobinsky:1983zz}
Starobinsky, A.~A. 1983, Sov. Astron. Lett., 9, 302

\bibitem[{Starobinsky(1985{\natexlab{a}})}]{Starobinsky:1985ww}
Starobinsky, A.~A. 1985{\natexlab{a}}, Sov. Astron. Lett., 11, 133

\bibitem[{Starobinsky(1985{\natexlab{b}})}]{Starobinsky:1986fxa}
Starobinsky, A.~A. 1985{\natexlab{b}}, JETP Lett., 42, 152

\bibitem[{Starobinsky(1992)}]{Starobinsky:1992ts}
Starobinsky, A.~A. 1992, JETP Lett., 55, 489

\bibitem[{Starobinsky(2005)}]{Starobinsky:2005ab}
Starobinsky, A.~A. 2005, JETP Lett., 82, 169

\bibitem[{Steinhardt \& Turner(1983)}]{Steinhardt:1983ia}
Steinhardt, P.~J. \& Turner, M.~S. 1983, Phys. Lett., B129, 51

\bibitem[{{Stewart} \& {Lyth}(1993)}]{stewart:1993}
{Stewart}, E.~D. \& {Lyth}, D.~H. 1993, Phys. Lett. B, 302, 171

\bibitem[{Stompor {et~al.}(1996)Stompor, Banday, \& Gorski}]{Stompor:1995py}
Stompor, R., Banday, A.~J., \& Gorski, K.~M. 1996, \apj, 463, 8

\bibitem[{Story {et~al.}(2013)Story, Reichardt, Hou, Keisler, Aird,
  {et~al.}}]{Story:2012wx}
Story, K., Reichardt, C., Hou, Z., {et~al.} 2013, Ap. J., 779, 86

\bibitem[{Suzuki {et~al.}(2012)Suzuki, Rubin, Lidman, Aldering, Amanullah,
  {et~al.}}]{Suzuki:2011hu}
Suzuki, N., Rubin, D., Lidman, C., {et~al.} 2012, \apj, 746, 85

\bibitem[{Tanaka \& Sasaki(1994)}]{Tanaka:1994qa}
Tanaka, T. \& Sasaki, M. 1994, Phys. Rev., D50, 6444

\bibitem[{Tocchini-Valentini {et~al.}(2005)Tocchini-Valentini, Douspis, \&
  Silk}]{TocchiniValentini:2004ht}
Tocchini-Valentini, D., Douspis, M., \& Silk, J. 2005, MNRAS, 359, 31

\bibitem[{Tocchini-Valentini {et~al.}(2006)Tocchini-Valentini, Hoffman, \&
  Silk}]{TocchiniValentini:2005ja}
Tocchini-Valentini, D., Hoffman, Y., \& Silk, J. 2006, MNRAS, 367, 1095

\bibitem[{Traschen \& Brandenberger(1990)}]{Traschen:1990sw}
Traschen, J.~H. \& Brandenberger, R.~H. 1990, Phys. Rev., D42, 2491

\bibitem[{Trotta(2007)}]{Trotta:2006ww}
Trotta, R. 2007, MNRAS, 375, L26

\bibitem[{Trotta \& Hansen(2004)}]{Trotta:2003xg}
Trotta, R. \& Hansen, S.~H. 2004, Phys. Rev., D69, 023509

\bibitem[{Tsujikawa \& Gumjudpai(2004)}]{Tsujikawa:2004my}
Tsujikawa, S. \& Gumjudpai, B. 2004, Phys. Rev., D69, 123523

\bibitem[{{Turner}(1983)}]{1983PhRvD..28.1243T}
{Turner}, M.~S. 1983, Phys. Rev. D, 28, 1243

\bibitem[{Turner(1990)}]{Turner:1989vc}
Turner, M.~S. 1990, Phys. Rept., 197, 67

\bibitem[{Turner \& Wilczek(1991)}]{Turner:1990uz}
Turner, M.~S. \& Wilczek, F. 1991, Phys. Rev. Lett., 66, 5

\bibitem[{Turner {et~al.}(1983)Turner, Wilczek, \& Zee}]{Turner:1983sj}
Turner, M.~S., Wilczek, F., \& Zee, A. 1983, Phys. Lett., B125, 35

\bibitem[{Valiviita \& Muhonen(2003)}]{Valiviita:2003ty}
Valiviita, J. \& Muhonen, V. 2003, Phys. Rev. Lett., 91, 131302

\bibitem[{Valiviita {et~al.}(2012)Valiviita, Savelainen, Talvitie,
  Kurki-Suonio, \& Rusak}]{Valiviita:2012ub}
Valiviita, J., Savelainen, M., Talvitie, M., Kurki-Suonio, H., \& Rusak, S.
  2012, \apj, 753, 151

\bibitem[{V\'azquez {et~al.}(2012)V\'azquez, Bridges, Hobson, \&
  Lasenby}]{Vazquez:2012ux}
V\'azquez, J.~A., Bridges, M., Hobson, M., \& Lasenby, A. 2012, JCAP, 06, 060

\bibitem[{Verde \& Peiris(2008)}]{Verde:2008zza}
Verde, L. \& Peiris, H.~V. 2008, JCAP, 0807, 009

\bibitem[{Vilenkin(2007)}]{Vilenkin:2006xv}
Vilenkin, A. 2007, J. Phys., A40, 6777

\bibitem[{Weinberg(2013)}]{Weinberg:2013kea}
Weinberg, S. 2013, Phys. Rev. Lett., 110, 241301

\bibitem[{Yamamoto {et~al.}(1995)Yamamoto, Sasaki, \& Tanaka}]{Yamamoto:1995sw}
Yamamoto, K., Sasaki, M., \& Tanaka, T. 1995, \apj, 455, 412

\bibitem[{Zeldovich(1972)}]{Zeldovich:1972zz}
Zeldovich, Y. 1972, MNRAS, 160, 1P

\end{thebibliography}
\end{raggedright}
\begin{appendix} 

\section{Sampling the Hubble flow functions}

In this Appendix we briefly review how to constrain slow-roll inflation by sampling the Hubble flow functions (HFFs)
and discuss how well the results agree with those derived by sampling directly
the parameters $\ln A_\mathrm{s}$, $n_\mathrm{s}$, $r$, and $\mathrm{d} n_\mathrm{s}/\mathrm{d} \ln k$.
This method fully exploits an analytic perturbative expansion in terms of the HFFs
for the primordial spectra of cosmological fluctuations during slow-roll inflation 
\citep{stewart:1993,Gong:2001he,Leach:2002ar}, which self-consistently extends to highest order the first terms presented in 
Eqs.~\ref{eq:as_def}-\ref{eq:betas_def}. Since $z''/z$ in Eq.~\ref{Scalar:Evolution} and $a''/a$ in Eq.~\ref{Tensor:Evolution} 
can be rewritten exactly in terms of the Hubble flow functions, 
the HFF hierarchy, rather than the potential hierarchy, is best suited for this purpose.
The slow-roll analytic power spectra have been calculated up to second order 
using the Green's function method \citep{Gong:2001he,Leach:2002ar}. 
Other approximations are available in the literature, including WKB \citep{Martin:2002vn}, the uniform approximation 
\citep{Habib:2002yi}, or the 
method of comparison equations \citep{Casadio:2006wb}. 

The dependence of the amplitudes in Eqs.~\ref{eq:as_def} and \ref{eq:at_def} in terms of HFF is given by  
\begin{equation}
\label{plex}
A_X = A_{X \, 0} e^{b_{X \, 0}},
\end{equation}
where $X={\mathrm{s}, \mathrm{t}}$, and $b_{\mathrm{s}0} \,, b_{\mathrm{t}0}$ are
\begin{align}
b_{\mathrm{s}0} =&
- 2\left(C + 1\right)\epsilon_1 - C \epsilon_2
+ \left(- 2C + {\textstyle\frac{\pi^2}{2}} - 7\right)              
 \epsilon_1^2 \nonumber \\
  & + \left({\textstyle\frac{\pi^2}{8}} - 1\right)
\epsilon_2^2 +
\left(- X^2 - 3 X + {\textstyle\frac{7\pi^2}{12}} - 7 + \Delta_{\mathrm{s}0} \right)
\epsilon_1\epsilon_2\\
 & + \left(-{\textstyle\frac 12} X^2 +
{\textstyle\frac{\pi^2}{24}} + \Delta_{\mathrm{s}0} \right)
\epsilon_2\epsilon_3 , \nonumber \\
\begin{split}
b_{\mathrm{t}0} =&
 - 2\left(C + 1\right)\epsilon_1
 + \left(- 2C + {\textstyle\frac{\pi^2}{2}} - 7\right)
 \epsilon_1^2 \\
   & + \left(-C^2 - 2C + {\textstyle\frac{\pi^2}{12}} - 2 + \Delta_{\mathrm{t}0}\right)
 \epsilon_1\epsilon_2,
\end{split}
\end{align}
with $C \equiv \ln 2+\gamma_\mathrm{E}-2\approx-0.7296$ ($\gamma_\mathrm{E}$ 
is the Euler-Mascheroni constant). At second order the coefficients of the expansion depend 
on the particular approximation scheme:
$X=C$ and $\Delta_{\mathrm{s}0}=\Delta_{\mathrm{t}0}=0$ apply for 
the Green's function method (GFM, \cite{Gong:2001he,Leach:2002ar}),
and $X=D=1/3 -\ln 3$, $\Delta_{\mathrm{s}0} = (D-C)(D+\ln 2)-1/18$, $\Delta_{\mathrm{t}0} = 2D(D-C)-1/9$ apply for
the method of comparison equations (MCE, \cite{Casadio:2006wb}). As predicted by the consistency relation, 
$A_{h \, 0} = 16 \epsilon_1 A_{\cal{R} \, 0}$. 

The full perturbative expressions up to second order in HFF for the spectral indices and the running of the indices are
\begin{eqnarray}
n_\mathrm{s} - 1 &=& - 2 \epsilon_1 - \epsilon_2 - 2
\epsilon_1^2 -\left(2\,C+3\right)\,\epsilon_1\,\epsilon_2 - C \epsilon_2 \epsilon_3, \\ 
\mathrm{d} n_\mathrm{s}/\mathrm{d} \ln k &=& - 2 \epsilon_1 \epsilon_2 - \epsilon_2 \epsilon_3, \\
\label{eqn:bs2}
n_\mathrm{t} & =& - 2\epsilon_1 - 2\epsilon_1^2
-2\,\left(C+1\right)\,\epsilon_1\,\epsilon_2 , \\                  
\label{eqn:bt1}
\mathrm{d} n_\mathrm{t}/\mathrm{d} \ln k
&=& - 2\epsilon_1\epsilon_2.
\label{eqn:bt2}
\end{eqnarray}

\begin{figure}
\includegraphics[width=88mm,angle=0]{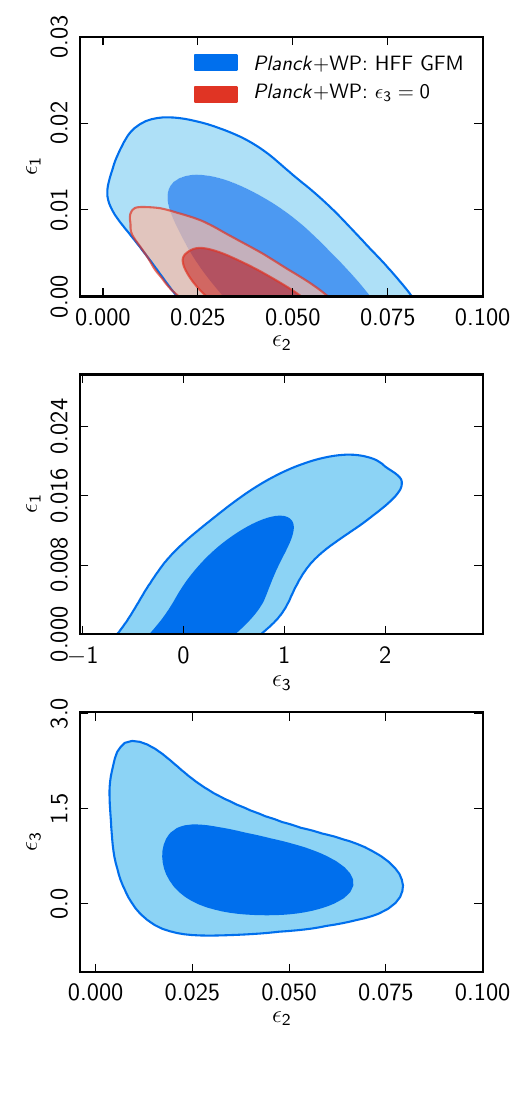}
\caption{\Planck\ constraints on the
  HFFs \mbox{$(\epsilon_1, \epsilon_2,\epsilon_3)$} 
  assuming either $\epsilon_3=0$ and
  the first-order slow-roll approximation for the computation of the
  primordial spectra, or $\epsilon_3 \neq 0$ and the
  second-order slow-roll approximation (HFF GFM).}
\label{fig:epsilon}
\end{figure}

\begin{figure}
\includegraphics[width=88mm,angle=0]{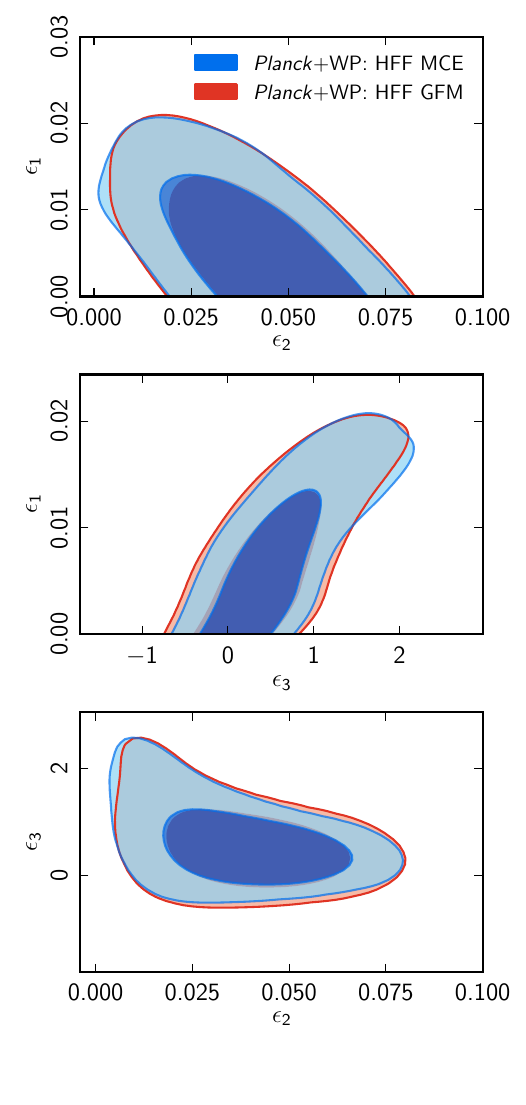}
\caption{Comparison of the \Planck\ constraints on the
HFFs \mbox{$(\epsilon_1, \epsilon_2,\epsilon_3)$} using 
the GFM and the MCE.}
\label{fig:GFMvsMCE}
\end{figure}

\begin{figure}
\includegraphics[width=88mm,angle=0]{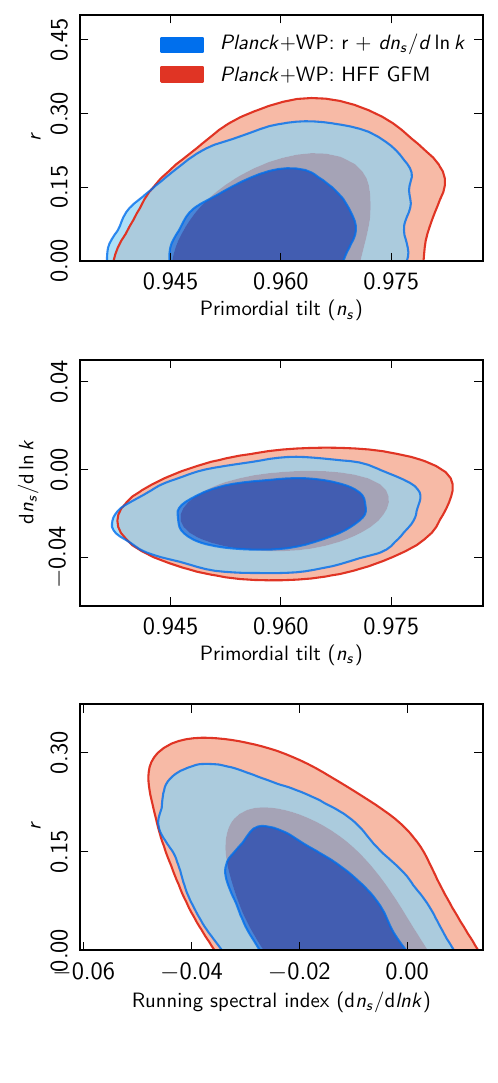}
\caption{\Planck\ constraints on the
spectral parameters $n_\mathrm{s}, \mathrm{d} n_\mathrm{s}/\mathrm{d} \ln k,$ and $r$. 
We compare constraints computed with the second-order slow-roll approximation, 
starting from flat priors on the HFF parameters at the pivot scale 
\mbox{$(\epsilon_1, \epsilon_2,\epsilon_3)$}, with those obtained directly from 
$n_\mathrm{s} \,, \mathrm{d} n_\mathrm{s}/\mathrm{d} \ln k,$ and $r$. 
In the latter case we  enforce the second-order consistency conditions for the
tensor-to-scalar ratio and for the running of the tensor spectral
index.}
\label{fig:physfromhff}
\end{figure}

We now constrain these parameters using the \Planck+WP data. If we restrict ourselves to first order, we obtain $\epsilon_1 < 0.0074$ at 95\% CL and $\epsilon_2 = 0.030^{+0.010}_{-0.009}$ at 68\%~CL. 
At second order with GFM, 
we obtain $\epsilon_1 < 0.013$ at 95\%~CL, $\epsilon_2 = 0.043^{+0.013}_{-0.014},$ and 
$\epsilon_3 = 0.36^{+0.19}_{-0.22}$ at 68\%~CL. The comparison of results at first  
and second order is shown in Fig.~\ref{fig:epsilon}.
Different approximation schemes lead to small differences, as Fig.~\ref{fig:GFMvsMCE} shows 
for the GFM versus the MCE.
Figure \ref{fig:physfromhff} shows the agreement between the physical parameters reconstructed from the 
HFF method and those directly sampled as in Sect.~\ref{running:section}.

\end{appendix}

\end{document}